\documentclass[10pt]{iopart}
\usepackage{graphicx,epsfig,iopams,latexsym,cite}
\newcommand{\be}{\begin{equation}}
\newcommand{\ee}{\end{equation}}
\newcommand{\ba}{\begin{eqnarray}}
\newcommand{\ea}{\end{eqnarray}}
\newcommand{\nn}{\nonumber}

\begin{document}

\title{Sgr~A$^\ast$ and General Relativity}

\author{Tim Johannsen}
\address{Perimeter Institute for Theoretical Physics, Waterloo, ON, N2L 2Y5, Canada}
\address{Department of Physics and Astronomy, University of Waterloo, Waterloo, ON, N2L 3G1, Canada}
\eads{\mailto{tjohannsen@pitp.ca}}

\begin{abstract}

General relativity has been widely tested in weak gravitational fields but still stands largely untested in the strong-field regime. According to the no-hair theorem, black holes in general relativity depend only on their masses and spins and are described by the Kerr metric. Mass and spin are the first two multipole moments of the Kerr spacetime and completely determine all higher-order moments. The no-hair theorem and, hence, general relativity can be tested by measuring potential deviations from the Kerr metric affecting such higher-order moments. Sagittarius~A$^\ast$ (Sgr~A$^\ast$), the supermassive black hole at the center of the Milky Way, is a prime target for precision tests of general relativity with several experiments across the electromagnetic spectrum. First, near-infrared (NIR) monitoring of stars orbiting around Sgr~A$^\ast$ with current and new instruments is expected to resolve their orbital precessions. Second, timing observations of radio pulsars near the Galactic center may detect characteristic residuals induced by the spin and quadrupole moment of Sgr~A$^\ast$. Third, the Event Horizon Telescope, a global network of mm and sub-mm telescopes, aims to study Sgr~A$^\ast$ on horizon scales and to image the silhouette of its shadow cast against the surrounding accretion flow using very-long baseline interferometric (VLBI) techniques. Both NIR and VLBI observations may also detect quasiperiodic variability of the emission from the accretion flow of Sgr~A$^\ast$. In this review, I discuss our current understanding of the spacetime of Sgr~A$^\ast$ and the prospects of NIR, timing, and VLBI observations to test its Kerr nature in the near future.

\end{abstract}


\section{Introduction}
\label{sec:Intro}

General relativity has been tested and confirmed by a variety of different experiments ranging from Eddington's solar eclipse expedition of 1919 to modern observations of double neutron stars~\cite{Will14}. These tests place tight limits on the properties of theories of gravity in the weak-field regime and leave little room for deviations from general relativity. Seldom, however, have these tests probed settings of strong spacetime curvature, where two theories of gravity can differ significantly in their predictions, even though both of them satisfy the current experimental constraints. In fact, general relativity still stands practically untested in the strong-field regime and it is incumbent to expand the scope of the current tests of general relativity~\cite{Psaltis08}. 

This can be illustrated by defining a parameter space spanned by the gravitational potential
\be
\varepsilon \equiv \frac{GM}{rc^2}
\ee
and the spacetime curvature
\be
\xi \equiv \frac{GM}{r^3 c^2}
\ee
of an object with mass $M$, where the coordinate $r$ measures the radial distance from the source and where $G$ and $c$ are the gravitational constant and the speed of light, respectively~\cite{Psaltis08}. While the full spacetime of the object is characterized by a metric $g_{\alpha\beta}$ with a corresponding Riemann tensor $R^\alpha_{~\beta\gamma\delta}$ and not simply by the potential $\varepsilon$ and the curvature $\xi$ defined above, the latter two quantities allow for an intuitive characterization of strong and weak gravitational fields using dimensional arguments.

\begin{figure}[h]
\centering{
\includegraphics[width=16cm]{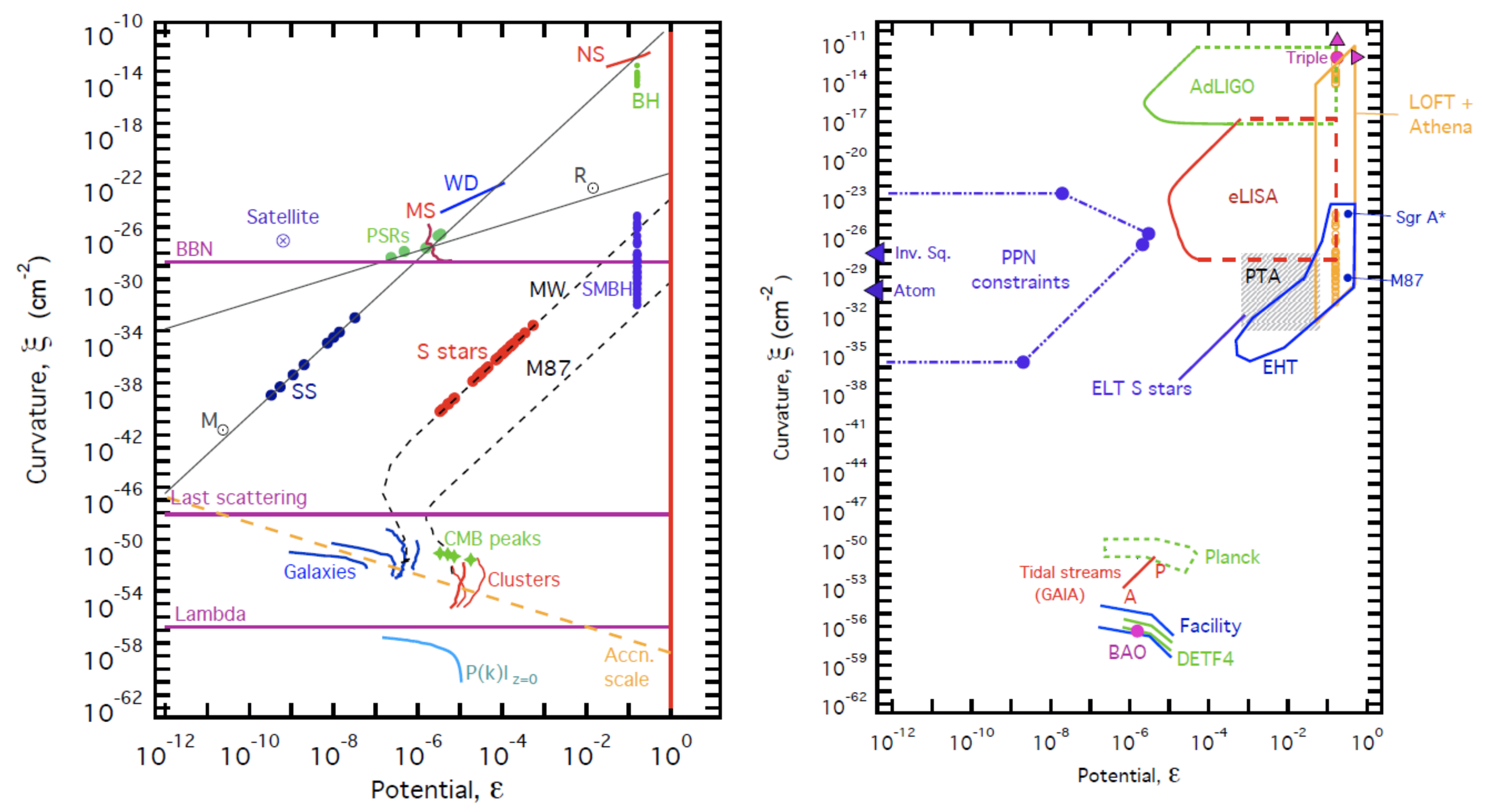}
}
\caption{Left: A parameter space for gravitational fields, spanned by the gravitational potential $\varepsilon$ and curvature $\xi$, showing the regimes probed by a wide range of astrophysical and cosmological systems. Right: The experimental version of the parameter space. Some of the label abbreviations are: SS = planets of the Solar System, MS = Main Sequence stars, WD = white dwarfs, PSRs = binary pulsars, NS = individual neutron stars, BH = stellar mass black holes, MW = Milky Way, SMBH = supermassive black holes, BBN = Big Bang nucleosynthesis, PPN = parameterized post-Newtonian region, Inv. Sq. = laboratory tests of the $1/r^2$ behavior of the gravitational force law, Atom = atom interferometry experiments to probe screening mechanisms, EHT = Event Horizon Telescope, ELT = Extremely Large Telescope, DETF4 = a hypothetical `stage 4' Dark Energy experiment, Facility = a futuristic large radio telescope such as the Square Kilometre Array. Taken from~\cite{Baker14}.}
\label{fig:parameterspace}
\end{figure}

Figure~\ref{fig:parameterspace} shows the regimes probed by a wide range of astrophysical and cosmological systems in this parameter space as well as a number of corresponding experiments. Only a small fraction of this parameter space has been accessed so far. In the solar system, where most tests of general relativity have been performed to date, typical values of the potential and the curvature can be vastly different from those found near compact objects or on cosmological scales~\cite{Psaltis08,Baker14}. To date, there have been only a few strong-field tests of general relativity in the context of either neutron stars (see Ref.~\cite{DeDeo03}; \cite{Psaltis08,Antoniadisetal13,Yagi14a,Yagi14b,Zhu15}) or black holes~\cite{Psaltis07,Braneworld1,Braneworld2,BambiBarausse11,BambiEfficiency,Braneworld3,Braneworld4,Bro14,Kong14}.

There are two basic approaches to test general relativity. Either, one assumes a particular theory of gravity and obtains constraints on potential deviations from general relativity within that theory. Almost all of the current strong-field tests fall into this category. Alternatively, one searches for generic deviations from general relativity and, thereby, hopes to gain insight into the underlying (but usually unknown) theory of gravity.

Both approaches are valid and have their place. However, tests of a specific theory of gravity are a priori limited by the fact that observations are interpreted within the narrow confines of that theory. Our lack of knowledge of a quantum theory of gravity as part of a grand unified theory of all fundamental interactions forces us to consider many different alternatives to general relativity. It is, therefore, advisable to search for deviations from general relativity using a broader setting that encompasses as many modified theories of gravity as possible. In this phenomenological approach, stars and compact objects have properties that differ from those in general relativity. These deviations can be expressed in terms of free parameters, which can, in principle, be determined by observations. The confirmation or exclusion of the predictions of different gravity theories can then lead to a greater understanding of the fundamental theory of gravity~\cite{Psaltis09}. The latter approach also has the practical advantage that different theories of gravity can be constrained at once without having to analyze (potentially large) data sets for each individual theory of gravity.

In this article, I focus on the regime probed by compact objects and by Sgr~A$^\ast$ in particular; see Ref.~\cite{Koyama15} for a recent summary of cosmological tests of gravity. Sgr~A$^\ast$ is a prime target for near-future tests of general relativity for several reasons. First, the highly-relativistic environment of Sgr~A$^\ast$ provides a laboratory to study some of the most extreme gravitational field strengths in the universe. The gravitational potential of Sgr~A$^\ast$ is $\sim5$ orders of magnitude larger than the gravitational potentials probed by current tests of general relativity, see Refs.~\cite{Psaltis09,Baker14} and Fig.~\ref{fig:parameterspace}.

Second, the mass and distance of Sgr~A$^\ast$ have been accurately measured by NIR observations of stars on orbits around Sgr~A$^\ast$~\cite{Ghez08,Gillessen09,Gillessen09b,Meyer12} and in the Galactic nuclear star cluster~\cite{Schoedel09,Do13,Chatzopoulos15}. The distance of Sgr~A$^\ast$ has also been inferred from parallax and proper motion measurements of masers throughout the Milky Way~\cite{Reid14}. These measurements robustly determine a mass of $\sim4\times10^6\,M_\odot$ and a distance of $\sim8~{\rm kpc}$ for Sgr~A$^\ast$.

\begin{figure}[ht]
\begin{center}
\psfig{figure=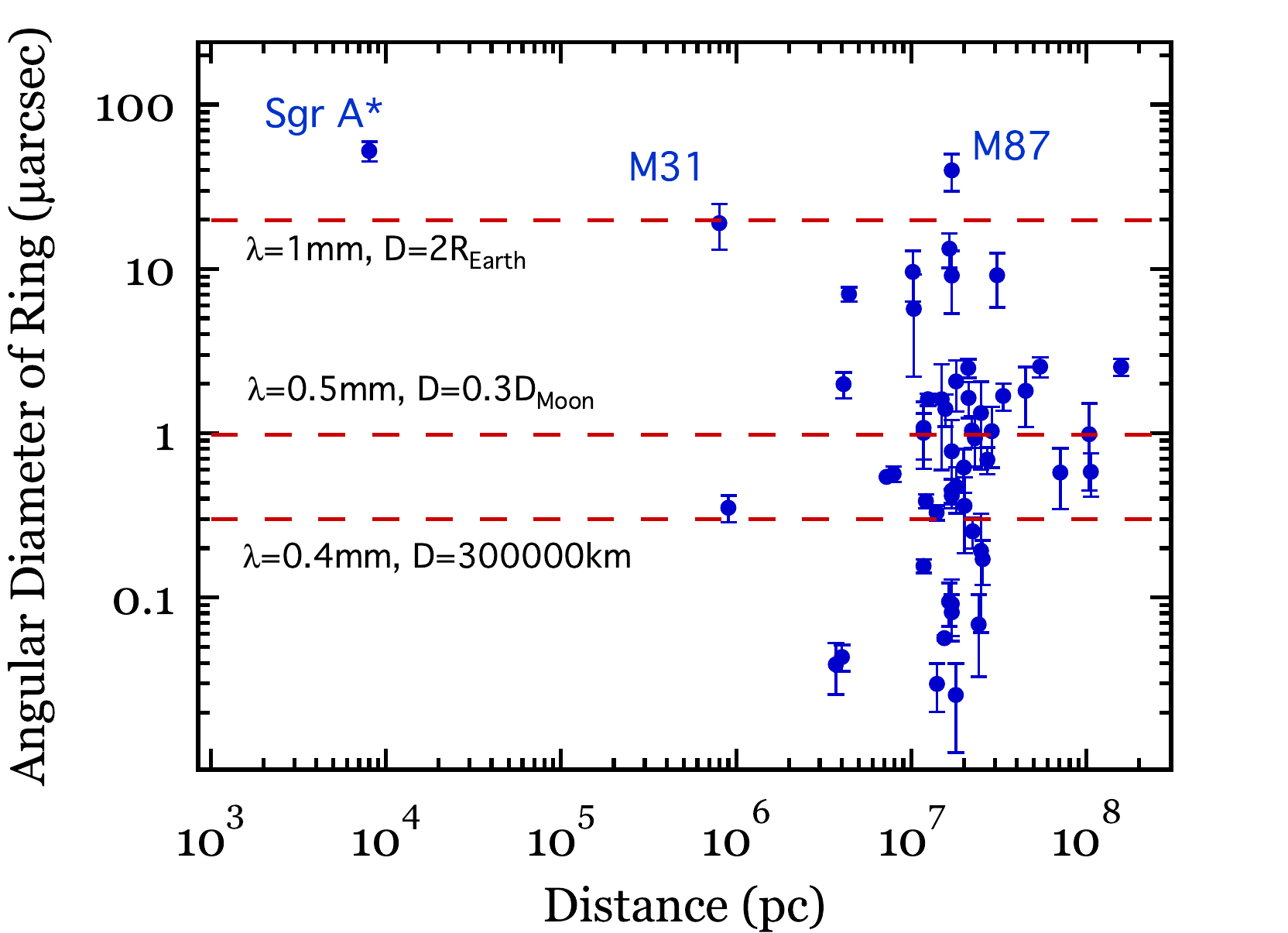,height=2.5in}
\psfig{figure=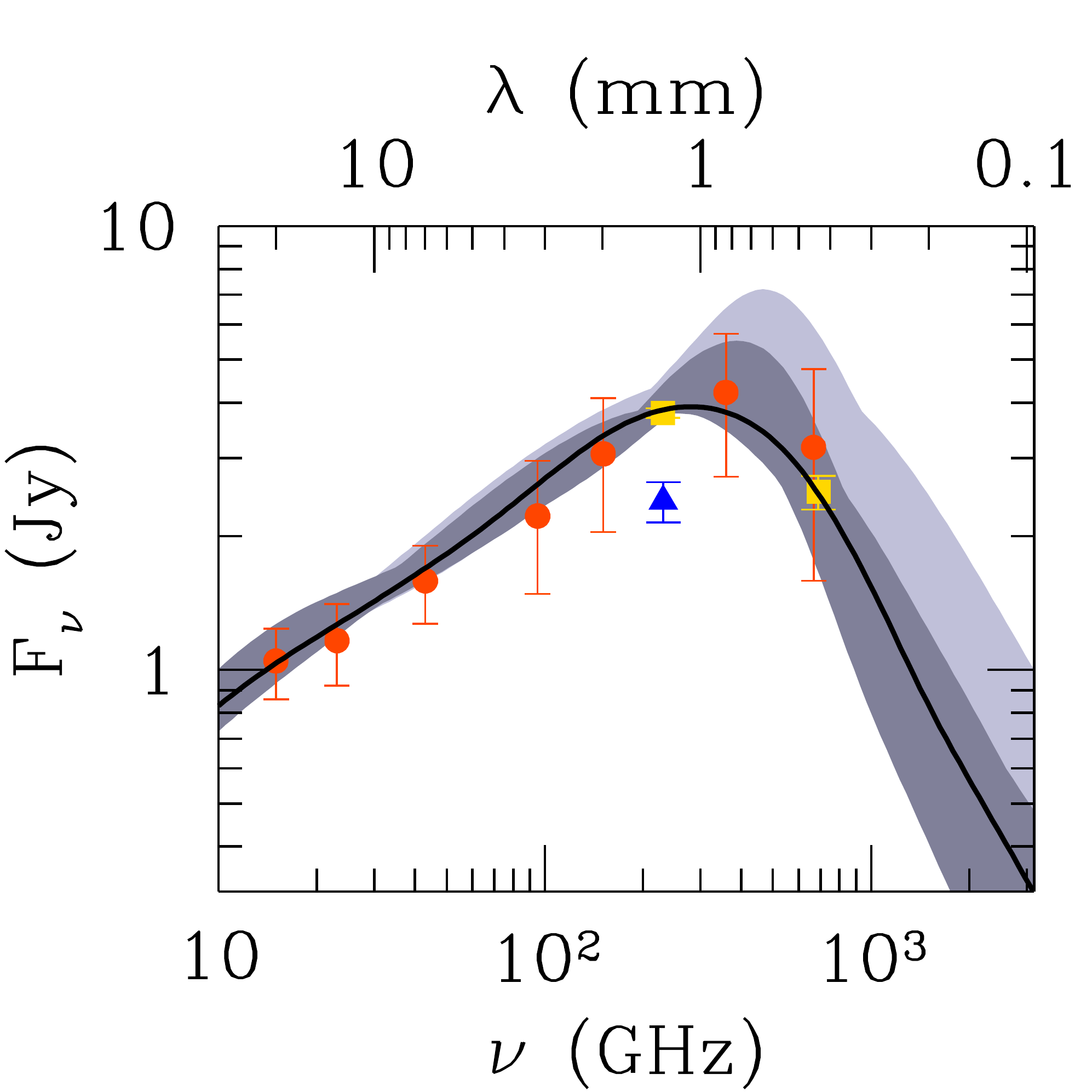,height=2.5in}
\end{center}
\caption{Left: Angular shadow diameters and distances of several supermassive black holes. The shadow of Sgr~A$^\ast$ has the largest angular diameter, closely followed by the one of the supermassive black hole at the center of M87 due to its high mass, making these sources ideal targets for the EHT. Taken from Ref.~\cite{SMBHmasses}. Right: The long-wavelength spectrum of Sgr~A$^\ast$. The gray filled areas show the envelopes of emission models fitting the data. Taken from Ref.~\cite{Bro09}.}
\label{fig:ehtsources}
\end{figure}

Third, the shadow of Sgr~A$^\ast$ has the largest opening angle of any black hole in the sky~\cite{SMBHmasses} making this source resolvable with mm/sub-mm VLBI observations. In addition, the emission of Sgr~A$^\ast$ becomes optically thin at wavelengths $\sim1~{\rm mm}$ (see Ref.~\cite{Bro09} and references therein), interstellar scattering becomes a subdominant effect, and the size of the long-wavelength emitting region is comparable to the horizon scale~\cite{Krichbaum98,Shen05,Bower}. Figure~\ref{fig:ehtsources} shows the angular sizes of the a number of supermassive black holes in the sky as well as the long-wavelength spectrum of Sgr~A$^\ast$. Reference~\cite{Doele08} resolved structures of Sgr~A$^\ast$ on scales of only $8r_g$, where $r_g \equiv GM/c^2$ is the gravitational radius of Sgr~A$^\ast$, and Refs.~\cite{Fish11} and \cite{JohnsonScience15} detected variability and polarized emission on event horizon scales, respectively. Similar observations of the supermassive black hole at the center of M87 have also detected structure on the order of $\approx5.5$ Schwarzschild radii~\cite{DoeleM87,Akiyama15} and measured a number of closure phases~\cite{Akiyama15}.

Three different types of experiments are aiming to test general relativity with observations of Sgr~A$^\ast$ in the near future. All of these operate at frequencies within atmospheric absorption windows (see, e.g., Ref.~\cite{abwindows}) and are ground-based. Orbital precessions of stars sufficiently close to Sgr~A$^\ast$~\cite{Will08} may be detected by continued NIR monitoring with current instruments and with the forthcoming instrument GRAVITY for the Very Large Telescope (VLT)~\cite{GRAVITY}. GRAVITY is also expected to detect quasi-periodic variability originating from density inhomogeneities orbiting in the accretion flow of Sgr~A$^\ast$~\cite{GRAVITY}. Future 30m-class optical telescopes such as the Thirty Meter Telescope (TMT~\cite{TMT}) and the European Extremely Large Telescope (E-ELT~\cite{EELT}) will possess further improved resolutions and sensitivities. Likewise, radio pulsars are thought to populate the stellar cluster at the Galactic center~\cite{Wharton12,Chennamangalam14}, spurred by the recent discovery of a magnetar at a distance of only $\sim0.1~{\rm pc}$ from Sgr~A$^\ast$~\cite{Kennea13,Mori13,Rea13,Eatough13,Zadeh15}. High-precision timing observations of such pulsars with exisiting 100m-class radio telescopes or future facilities such as the Square Kilometre Array (SKA~\cite{SKA}) may detect characteristic residuals in their spectra which depend on the general-relativistic properties of Sgr~A$^\ast$~\cite{WK99,Pfahl04,Liu12}. Finally, the Event Horizon Telescope (EHT), a global very-long baseline interferometer comprised by mm and sub-mm telescopes, is expected to perform detailed space- and time-resolved studies of the accretion flow of Sgr~A$^\ast$ and to take the first-ever image of a black hole~\cite{Doele09a,Doele09b,Fish09}.

Each of these tests of general relativity requires an appropriate theoretical framework. Observations of stars and (for the most part) pulsars probe Sgr~A$^\ast$ in the weak-field regime, i.e., at radii $r\gg r_g$. In this regime, it is sufficient to employ a parameterized post-Newtonian framework within which suitable corrections to Newtonian gravity in flat space can be calculated~\cite{Will93}. In the strong-field regime, however, i.e., at radii $r\sim r_g$, which are targeted by observations of the accretion flow of Sgr~A$^\ast$ with the EHT and NIR instruments such as GRAVITY, the parameterized post-Newtonian formalism can no longer be applied. Instead, a careful modeling of the underlying spacetime in terms of a Kerr-like metric (e.g., \cite{MN92,CH04,GB06,VH10,JPmetric,VYS11,Jmetric,CPR14,Lin16}) is required.

Strong-field tests of general relativity with black holes have also been proposed using gravitational-wave observations of extreme mass-ratio inspirals (EMRIs)~\cite{Ryan95,Ryan97a,Ryan97b,kludge,CH04,GB06,BC07,LL08,Brink08,Gair08,Apostolatos09,SY09,VH10,GY11,VYS11,RMG12} and of gravitational ringdown radiation of perturbed black holes after a merger with another object~\cite{Dreyer04,Berti06,Berti07}. See Refs.~\cite{GairLRR,YunesLRR} for recent reviews. Likewise, strong-field tests of general relativity have been suggested using other electromagnetic observations of accretion flows in terms of their continuum spectra~\cite{Torres02,Pun08,Harko09a,Harko09b,Harko10,BambiBarausse11,Krawcz12,BambiDiskCode,Xrayprobes,Moore15,Lin16,Krawcz16}, relativistically-broadened iron lines~\cite{PJ11,PaperIV,BambiIron,BambiIron2,pcGR2,Xrayprobes,Vincent14,Jiang15,Moore15,Jiang16}, variability~\cite{PaperIII,BambiQPO,Vincent14,Maselli15,BambiHotspot1,BambiHotspot2,Suvorov15}, X-ray polarization~\cite{Krawcz12,BambiPol,Krawcz16}, jets~\cite{BambiDiskJet}, and other accretion properties~\cite{BambiBarausse11,BambiEfficiency,Lukes3,Lukes4,Jedges,Ignacio15,Federico16}. See Ref.~\cite{TJReview} for a review.

All three approaches to test general relativity with observations of Sgr~A$^\ast$ in the electromagnetic spectrum are based on tests of the (general-relativistic) no-hair theorem, which I briefly review in Sec.~\ref{sec:NHT}. In Secs.~\ref{sec:stars} and \ref{sec:pulsars}, I discuss tests of general relativity with observations of stars and pulsars around Sgr~A$^\ast$, respectively. In Sec.~\ref{sec:metrics}, I review Kerr-like metrics and some of their properties. Table~\ref{tab:devparams} at the end of this section contains a list of the parameters and some of the essential properties of several Kerr-like metrics. In Sec.~\ref{sec:EHT}, I discuss tests of general relativity with EHT observations of Sgr~A$^\ast$. Quasiperiodic variability of the emission from the accretion flow of Sgr~A$^\ast$ may be probed by both NIR and VLBI observations, which I discuss in Sec.~\ref{sec:variability}. Sec.~\ref{sec:conclusions} contains my conclusions.

\section{The No-Hair Theorem}
\label{sec:NHT}

According to the general-relativistic no-hair theorem, isolated and stationary black holes are uniquely characterized by their masses $M$, spins $J$, and electric charges $Q$ and are described by the Kerr-Newman metric~\cite{Newman65}, which reduces to the Kerr metric~\cite{Kerr63} in the case of electrically neutral black holes. This metric is the unique stationary, axisymmetric, asymptotically flat, vacuum solution of the Einstein field equations which contains an event horizon but no closed timelike curves in the exterior domain~\cite{Israel67,Israel68,Carter71,Hawking72,Carter73,Robinson75,Mazur82}. The no-hair theorem relies on the cosmic censorship conjecture~\cite{Penrose69} as well as on the physically reasonable assumption that the exterior metric is free of closed timelike curves. See Refs.~\cite{Chrusciel12,Robinson12} for reviews. If these requirements are met, then all astrophysical black holes should be described by the Kerr metric\footnote{Note that astrophysical black holes are thought to be essentially electrically neutral, because any residual electric charge would quickly neutralize. Also note that the mathematical status of the no-hair theorem is not without controversy, principally in relation to the assumption (in the classical proof) of analyticity; see Sec.~3.4 in Ref.~\cite{Chrusciel12} for a discussion.}.

Black holes are commonly believed to be the final states of the evolution of sufficiently massive stars at the end of their lifecycle. The gravitational collapse of such stars leads to the formation of a black hole~\cite{OS39,Penrose65}, where any residual signature of the progenitor other than its mass and spin is radiated away by gravitational radiation~\cite{Price72a,Price72b}. This scenario provides an astrophysical mechanism with which black holes can be generated. Almost all nearby galaxies harbor dark objects of high mass and compactness at their centers~\cite{Kormendy95} including our own galaxy~\cite{Ghez08,Gillessen09} providing strong evidence that black holes are realized in nature. In addition, the measurement of the orbital parameters of many Galactic binaries supports the claim that they contain stellar-mass black holes (e.g., \cite{RMcC06}).

Despite the large amount of circumstantial evidence, there has been no direct proof, so far, for the existence of an actual event horizon. An event horizon, one of the most striking predictions of general relativity, is a virtual boundary that causally disconnects the interior of a black hole from the exterior universe. The presence of an event horizon in black-hole candidates has only been inferred indirectly from either the lack of observations of Type~I X-ray bursts~\cite{NGM97,NGM01,NH02,MNR04} (see the discussion in Ref.~\cite{Psaltis06}) or, in the case of the supermassives black holes at the centers of the Milky Way and the galaxy M87, the fact that these supermassive compact objects are greatly underluminous~\cite{Broderick09b,Broderick15}. These observations indicate the absence of a hard (stellar) surface which most likely identifies the compact objects as black holes.  In addition, the recent first direct observation of gravitational waves detected a waveform that is consistent with the inspiral of two black holes with masses of $\sim36M_\odot$ and $\sim29M_\odot$~\cite{GW1,GW2} (but see Ref.~\cite{Cardoso16}).

Astrophysical black holes, however, will not be perfectly stationary nor exist in perfect vacuum because of the presence of other objects or fields such as stars, accretion disks, or dark matter, which could alter the Kerr nature of the black hole (see, e.g., Ref.~\cite{Semerak15}). Nonetheless, under the assumption that such perturbations are so small to be practically unobservable, one can argue that astrophysical black holes are indeed described by the Kerr metric. This is the assumption I make throughout this article.

In Boyer-Lindquist coordinates, the Kerr metric $g_{\mu\nu}^{\rm K}$ is given by the expressions (setting $G=c=1$)
\ba
g_{tt}^{\rm K} &=&-\left(1-\frac{2Mr}{\Sigma}\right), \nn \\
g_{t\phi}^{\rm K} &=& -\frac{2Mar\sin^2\theta}{\Sigma}, \nn \\
g_{rr}^{\rm K} &=& \frac{\Sigma}{\Delta}, \nn \\
g_{\theta \theta}^{\rm K} &=& \Sigma, \nn \\
g_{\phi \phi}^{\rm K} &=& \left(r^2+a^2+\frac{2Ma^2r\sin^2\theta}{\Sigma}\right)\sin^2\theta,
\label{eq:kerr}
\ea
where
\ba
\Delta &\equiv& r^2-2Mr+a^2, \\
\Sigma &\equiv& r^2+a^2\cos^2 \theta
\label{eq:sigma}
\ea
and where $a\equiv J/M$ is the spin parameter. For values of the spin
\be
|a|\leq M,
\label{eq:kerrbound}
\ee
the event horizon of the black hole is located at the radius
\be
r_+^{\rm K} = M + \sqrt{M^2-a^2}.
\label{eq:kerrhor}
\ee

The alternative hypothesis that these dark compact objects are not described by the Kerr metric but perhaps by a solution of the Einstein equations with a naked singularity (e.g., \cite{MN92}) and, therefore, violate the no-hair theorem is still possible within general relativity. Alternatively, these dark objects might be stable stellar configurations consisting of exotic fields (e.g., boson stars~\cite{FLP87}, gravastars~\cite{MM01}, black stars~\cite{Barce08}) or black holes surrounded by a stable scalar field~\cite{Herdeiro14PRL} (c.f., Refs.~\cite{Arvanitaki11,Brito15}).

Finally, the fundamental theory of gravity may be different from general relativity in the strong-field regime, and the vacuum black-hole solution might not be described by the Kerr metric at all. In fact, black hole solutions in several theories of gravity other than general relativity have already been found including exact solutions of rotating black holes in Randall-Sundrum-type (RS2) braneworld gravity~\cite{RS2BH} and modified gravity (MOG)~\cite{MOG} as well as perturbative (i.e., black hole solutions valid for small deviations from the Kerr metric) and numerical solutions of rotating black holes in Einstein-dilaton-Gauss-Bonnet gravity (EdGB; \cite{Mignemi93,Kanti96,Kleihaus11,YS11,Pani11,AY14,M15}; see Ref.~\cite{Kleihaus15} for an asymptotic expansion of a nonperturbative solution), dynamical Chern-Simons gravity (dCS; \cite{YP09,Pani11,YYT12}), massive gravity~\cite{Babichev14a,Babichev14b}, Einstein-{\AE}ther gravity~\cite{BJS11,BS13,BSV16}, Ho\u{r}ava-Lifshitz gravity~\cite{BJS11,BS13}, as well as Horndeski gravity~\cite{MaselliBH15}. See Refs.~\cite{Babichevreview,Baraussereview,Bertireview,Herdeiro15} for reviews on this topic.

As a result, testing the no-hair theorem allows us not only to verify the identification of dark compact objects in the universe with Kerr black holes but to test the strong-field predictions of general relativity, as well. Unfortunately, such tests are slightly complicated by the fact that the Kerr metric is not unique to general relativity but also the most general black hole solution in a large class of scalar-tensor theories of gravity~\cite{PsaltisKerr,Sotiriou12,Graham14} (a similar property also holds for the Kerr-Newman metric~\cite{CruzDombriz09}).

Linear systems in flat space are often best described by a set of multipole moments. In theories of gravity like general relativity, however, space is curved due to the presence of stress-energy, and the resulting field equations are highly non-linear. It is, therefore, not immediately obvious that such a spacetime can actually be characterized by a set of multipole moments.

In Newtonian gravity, the potential $\Phi$ satisfies the Laplace equation
\begin{equation}
\nabla^2\Phi= \left\{ \begin{array}{ll}
              4\pi G\rho  & \mbox{(interior)}   \\
              0 & \mbox{(exterior)},\end{array} \right.
\end{equation}
where $\rho$ is the mass density. Therefore, the potential $\Phi$ can always be expanded in spherical harmonics $Y_{lm}$ as
\begin{equation}
\Phi=-G\sum_{l=0}^{\infty}\frac{4\pi}{2l+1}\sum_{m=-l}^{l}\frac{M_{lm}Y_{lm}}{r^{l+1}}
\label{newtonexp}
\end{equation}
with mass multipole moments
\begin{equation}
M_{lm}=\int_0^r r'^{l+2}dr'\oint d\Omega' Y_{lm}^*(\Omega')\rho(r',\Omega').
\label{newtonmult}
\end{equation}

In the curved space of general relativity, however, the vacuum Einstein equations
\begin{equation}
R_{\mu\nu}-\frac{1}{2}g_{\mu\nu}R=0
\label{einsteinvac}
\end{equation}
have to be solved for the spacetime metric $g_{\mu\nu}$ with the corresponding Ricci tensor $R_{\mu\nu}$ and Ricci scalar $R$. The Einstein equations are nonlinear and, therefore, cannot always be solved in terms of an expansion over orthonormal polynomials. Nonetheless, it can be shown that a multipole expansion of curved spacetime does indeed exist in certain cases (see Ref.~\cite{Thorne80} for a review).

For an asymptotically flat vacuum solution of the Einstein equations, (tensor) multipole moments can be defined based on a conformal compactification of 3-space, if the spacetime is also static~\cite{Geroch70} or, more generally, if it is stationary~\cite{Hansen74}. In both cases, such a set of multipole moments characterizes the spacetime uniquely~\cite{BeigSimon80,BeigSimon81} (see, also, Ref.~\cite{HauserErnst81} and references therein) and obeys an appropriate convergence condition~\cite{BackdahlHerberthson06}. If the spacetime is also axisymmetric, the multipole moments are given by a bi-infinite series of scalars $M_l$ and $S_l$, which are interpreted as mass and current multipole moments, respectively. The mass multipole moments are analogous to the multipole moments in Newtonian gravity given by expression (\ref{newtonmult}) and are nonzero only for even $l$. The current multipole moments are nonzero only for odd $l$ and arise from the fact that, in general relativity, all forms of stress-energy gravitate~\cite{Hansen74}. Stationary axisymmetric vacuum solutions of the Einstein equations can likewise be generated from a given set of multipole moments~\cite{Sibgatullin91,MankoSibgatullin93}.

In general relativity, black-hole spacetimes are asymptotically flat vacuum solutions of the Einstein equations. If these spacetimes are stationary, they must also be axisymmetric~\cite{Hawking72} and can, therefore, be described by a sequence of (Geroch-Hansen) multipole moments $\{M_l,S_l\}$. As a consequence of the no-hair theorem, the Kerr spacetime is the unique such black-hole solution within general relativity, and all multipole moments of order $l\geq2$ are determined only by the first two, i.e., by the mass $M=M_0$ and the spin $J=S_1$. This fact can be expressed mathematically with the relation~\cite{Geroch70,Hansen74}
\begin{equation}
M_{l}+{\rm i}S_{l}=M({\rm i}a)^{l}.
\label{eq:kerrmult}
\end{equation}

In the astrophysical context, the fact that the no-hair theorem requires the multipole moments of a stationary black hole to be locked by expression (\ref{eq:kerrmult}) allows for it to be tested quantitatively using observations of such black holes. Since the first two multipole moments (i.e., the mass and spin) already specify the entire spacetime, a promising strategy for testing the no-hair theorem, then, is to measure (at least) three multipole moments of the spacetime of a black hole~\cite{Ryan95}.

\section{NIR Monitoring of Stellar Orbits}
\label{sec:stars}

Observations of the Galactic center region over several decades have established the fact that its mass distribution can be well described by a compact supermassive object at the center, Sgr~A$^\ast$, surrounded by a dense nuclear star cluster with an extent of several parsecs. The most common explanation of the high mass and compactness of Sgr~A$^\ast$ is that this object is indeed a black hole. See Refs.~\cite{GenzelTownes87,Genzeletal94,Mezgeretal96,MeliaFalcke01,Reidrewiew09,Genzel10,Falcke13} for reviews.

\begin{figure}[ht]
\begin{center}
\psfig{figure=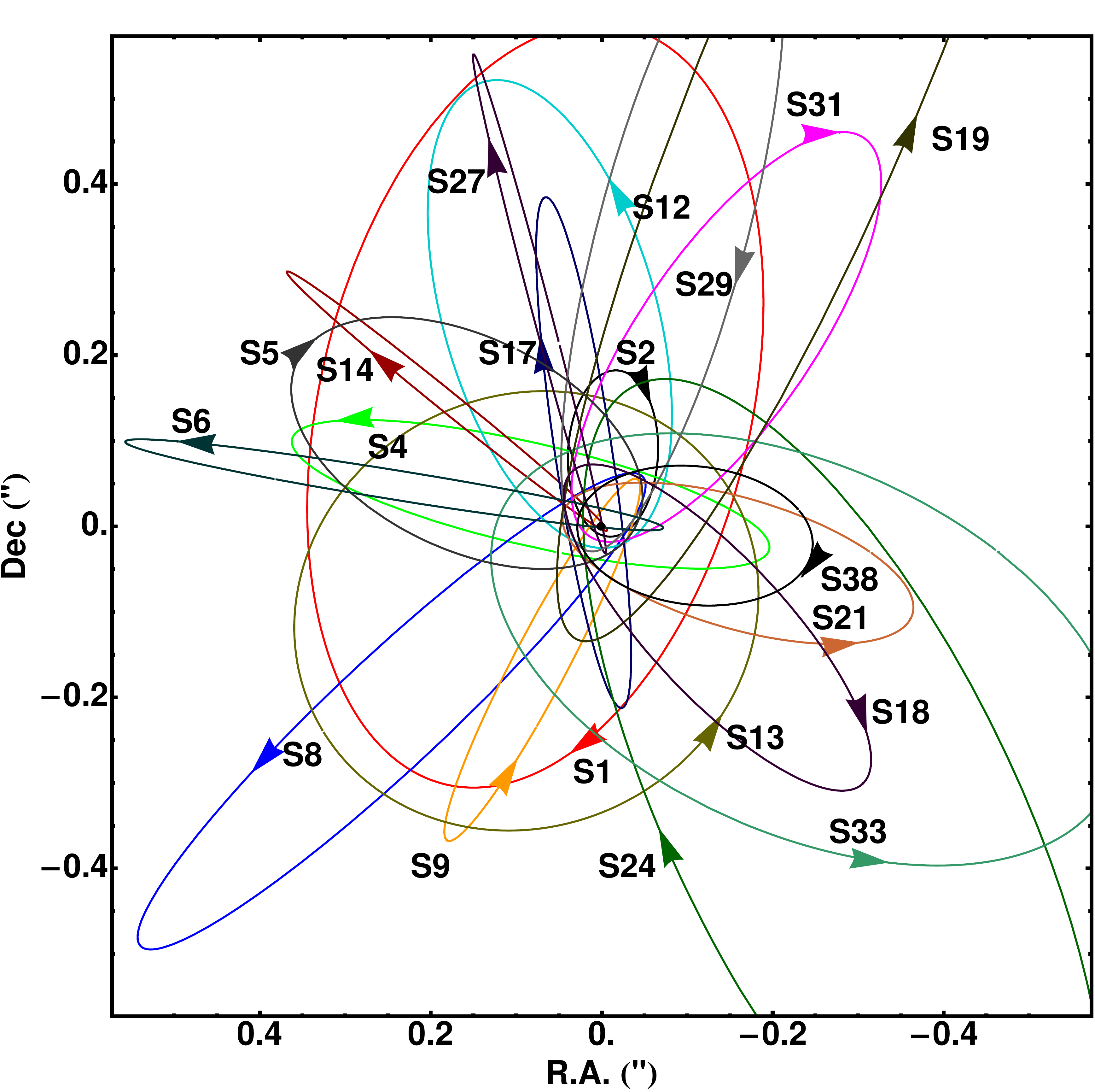,height=2.49in}
\psfig{figure=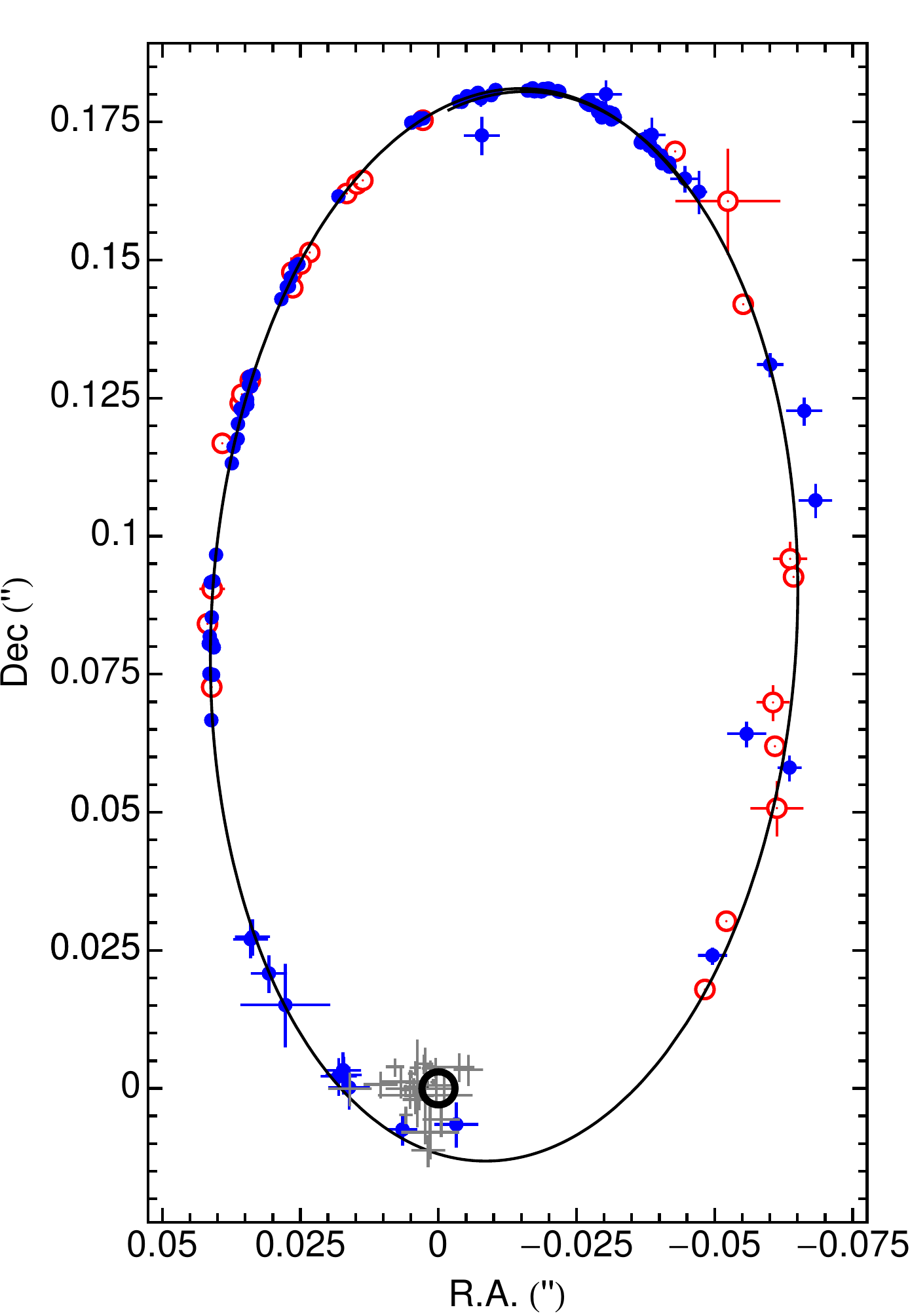,height=2.49in}
\psfig{figure=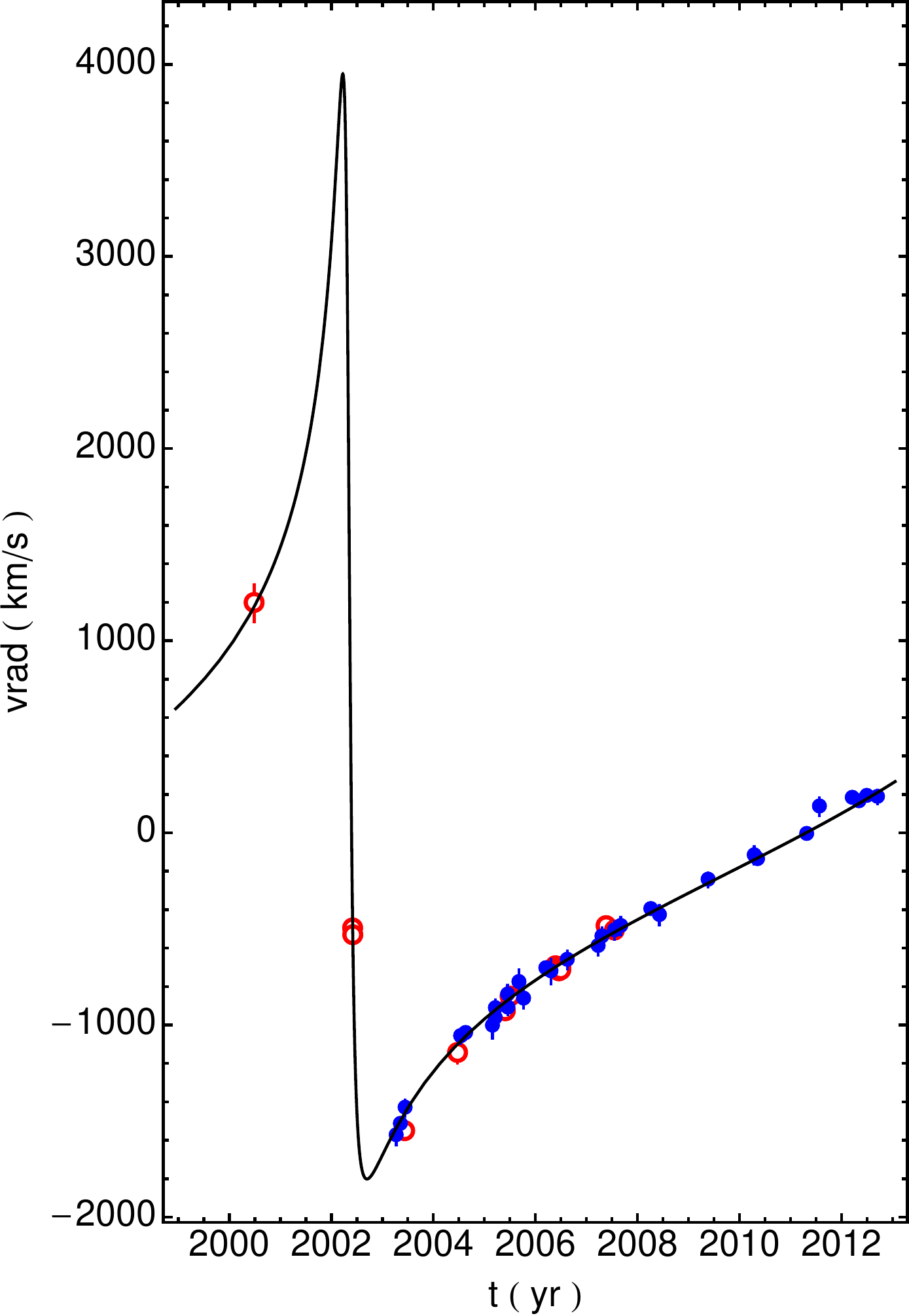,height=2.41in}
\end{center}
\caption{The left panel shows the orbits of 20 S-stars around Sgr~A$^\ast$ obtained from fits of NIR observations in 1992--2008 (taken from Ref.~\cite{Gillessen09}). The center and right panels show the orbit of the star S2 around Sgr~A$^\ast$ on the sky (black circle) and in radial velocity, respectively (taken from Refs.~\cite{Gillessen13}). Blue, filled circles denote the observations of Refs.~\cite{Gillessen09,Gillessen09b} (updated to 2012) with the New Technology Telescope (NTT) and the VLT, red circles denote the Keck observations of Ref.~\cite{Ghez08}, and grey crosses are the positions of infrared flares (see Sec.~\ref{sec:variability}). The orbit of S2 is not a closed ellipse due to the proper motion of Sgr~A$^\ast$, which, however, is consistent with the uncertainties of the NIR reference frame.}
\label{fig:stellarorbits}
\end{figure}

To date, numerous observations have led to precise measurements of the mass $M$ and distance $R_0$ of Sgr~A$^\ast$. References~\cite{Ghez08,Gillessen09,Gillessen09b,Meyer12} and Refs.~\cite{Schoedel09,Do13,Chatzopoulos15} inferred the mass and distance of Sgr~A$^\ast$ from NIR monitoring stars on orbits around Sgr~A$^\ast$, the so-called S-stars, and in the old Galactic nuclear star cluster, respectively. References~\cite{Ghez08,Meyer12} obtained the measurements $M=(4.1\pm0.4)\times10^6\,M_\odot$, $R_0=7.7\pm0.4~{\rm kpc}$, while combining the results of Refs.~\cite{Gillessen09,Chatzopoulos15} yields the measurements \mbox{$M=(4.23\pm0.14)\times10^6\,M_\odot$}, $R_0=8.33\pm0.11~{\rm kpc}$~\cite{Chatzopoulos15}. Figure~\ref{fig:stellarorbits} shows the orbits of 20 S-stars around Sgr~A$^\ast$ including the orbit of the star S2 which has been observed over the full span of its orbital period. The S-star orbits are consistent with Keplerian ellipses within the remaining uncertainties in the NIR coordinate system (c.f., Ref.~\cite{Menten97}). In addition, the distance of Sgr~A$^\ast$ has been obtained from parallax and proper motion measurements of masers throughout the Milky Way by Ref.~\cite{Reid14} finding $R_0=8.34\pm0.16~{\rm kpc}$. Although the stellar-orbit measurements presently disagree at an almost statistically-significant level, primarily due to the uncertainties in the NIR coordinate system, the above measurements robustly determine a mass of $\sim4\times10^6\,M_\odot$ and a distance of $\sim8~{\rm kpc}$ of Sgr~A$^\ast$.

The constraints on the mass and distance from the observations of stellar orbits will be improved in the near future by continued monitoring and by the use of the second-generation instrument GRAVITY for the VLT, which is expected to achieve astrometry with a precision of $\sim10~{\rm \mu as}$ and imaging with a $\sim4~{\rm mas}$ resolution~\cite{GRAVITY}. Once a 30m-class optical telescope such as the TMT~\cite{TMT} or the E-ELT~\cite{EELT} will become available, the mass and distance of Sgr~A$^\ast$ are likely to be determined with a precision of $\sim0.1\%$~\cite{Weinberg05}.

The monitoring of stellar orbits may also measure the spin and, perhaps, even the quadrupole moment of Sgr~A$^\ast$, thus testing the no-hair theorem via the relation in Eq.~(\ref{eq:kerrmult}). A star that is sufficiently close Sgr~A$^\ast$ experiences an acceleration ${\bf a}_{\rm star}$ which is given by the equation (see, e.g., Ref.~\cite{Will93})
\ba
{\bf a}_{\rm star} &=& -\frac{M{\bf x}}{r^3} + \frac{M{\bf x}}{r^3} \left (4 \frac{M}{r} - v^2 \right ) +4\frac{M{\dot r}}{r^2} {\bf v} - \frac{2J}{r^3} \left[ 2{\bf v}\times {\bf {\hat J}}
-3 {\dot r} {\bf n}\times {\bf {\hat J}} - \frac{ 3{\bf n}({\bf h} \cdot {\bf {\hat J}}) }{r} \right] \nn \\
&& + \frac{3}{2} \frac{Q}{r^4} \left[5{\bf n}({\bf n} \cdot {\bf {\hat J}})^2
- 2 ({\bf n} \cdot {\bf {\hat J}}){\bf {\hat J}} - {\bf n} \right],
\label{eq:EOM}
\ea
where $\bf x$ and $\bf v$ are the position and velocity of the star, ${\bf n} = {\bf x}/r$, ${\dot r} = {\bf n}\cdot {\bf v}$, ${\bf h}= {\bf x} \times {\bf v}$, ${\bf {\hat J}} = {\bf J}/|J|$, and ${\bf J}$ and $Q$ are the angular momentum vector and the quadrupole moment of Sgr~A$^\ast$, respectively. The first three terms of the acceleration in Eq.~(\ref{eq:EOM}) correspond to the Schwarzschild part of the metric at first post-Newtonian order, the next term is the frame-dragging effect induced by the spin of Sgr~A$^\ast$, and the final term is the effect of the quadrupole moment at Newtonian order. There are additional quadrupolar corrections to the acceleration in Eq.~(\ref{eq:EOM}) at first post-Newtonian order, but these will be much smaller, because they have a stronger dependence on the distance of the star from Sgr~A$^\ast$.

\begin{figure}[ht]
\begin{center}
\psfig{figure=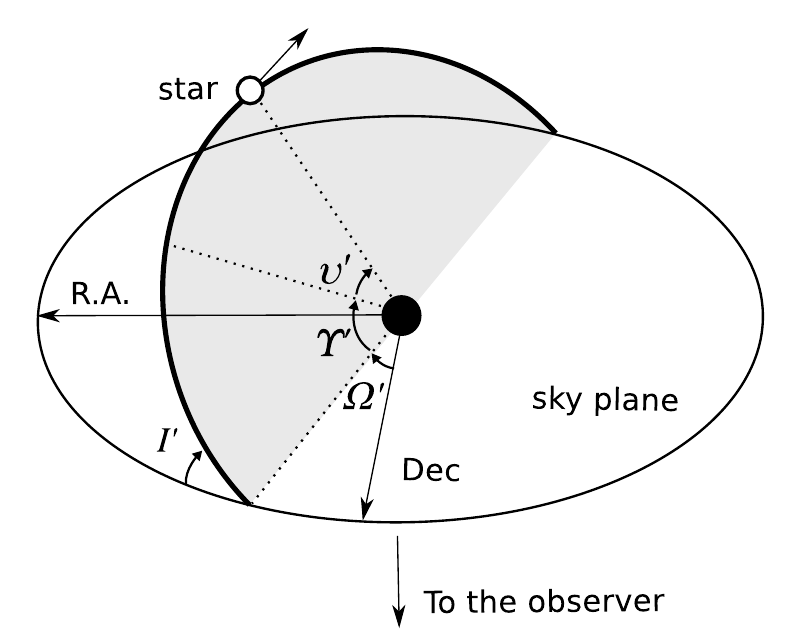,height=2.in}
\psfig{figure=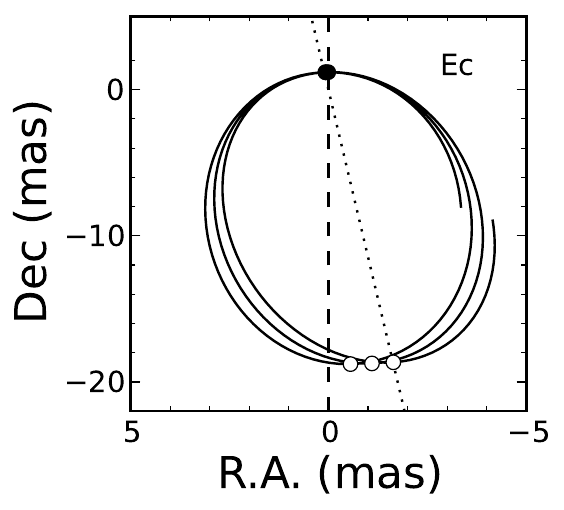,height=2.in}
\end{center}
\caption{The left panel shows the geometry of a star on an orbit around Sgr~A$^\ast$, which is described by six orbital elements, the semi-major axis $\tilde{a}$ (not shown), eccentricity $e$ (not shown), longitude of ascending node $\Omega'$, longitude of pericenter $\Upsilon'$, true anomaly $v'$, and inclination $I'$. The right panel shows the evolution of the apparent position in the sky (in terms of the right ascension ``R.A.'' and declination ``Dec'') of a hypothetical star over three full orbits around Sgr~A$^\ast$ (located at the origin) with a semi-major axis of $10~{\rm mas}$, an orbital eccentricity of $0.88$, and an inclination of $45^\circ$. The stellar trajectory includes the effects of the Schwarzschild corrections to the Newtonian potential of Sgr~A$^\ast$ as well as of frame dragging (assuming a spin value $\chi=0.99$), both of which cause the orbit to precess. Taken from Ref.~\cite{Zhang15}.}
\label{fig:Sgeometry}
\end{figure}

The corrections to the Newtonian gravitational potential of Sgr~A$^\ast$ cause the orbit of the star to precess. The Schwarzschild-type corrections in Eq.~(\ref{eq:EOM}) lead to a precession in the orbital plane of the star, while the corrections induced by the spin and quadrupole moment of Sgr~A$^\ast$ cause the orbit to precess both in and out of the orbital plane of the star\footnote{The spin-induced precession of the stellar orbit is commonly referred to as Lense-Thirring precession~\cite{LenseThirring18}, c.f., Eqs.~(\ref{eq:Phidot})--(\ref{eq:OmegaLT}).}. Using standard orbital perturbation theory, Ref.~\cite{Will08} (see, also, Refs.~\cite{Rubilar01,Zucker06,Nucita07,Iorio11b}) calculated the precessions per orbit of the pericenter angle $\Upsilon$, nodal angle $\Omega$, and inclination $i$ (see the left panel of Fig.~\ref{fig:Sgeometry}, identifying $\Upsilon=\Upsilon'$, $\Omega=\Omega'$, and $i=I'$) which are given by the expressions
\ba
\Delta {\Upsilon} &=& A_S - 2 A_J \cos \alpha - \frac{1}{2} A_Q (1-3\cos^2 \alpha),
\label{eq:dUpsilon} \\
\sin i \Delta \Omega &=& \sin \alpha \sin \beta (A_J - A_Q \cos \alpha) ,
\label{eq:dOmega} \\
\Delta i  &=& \sin \alpha \cos \beta (A_J - A_Q \cos \alpha).
\label{eq:dtheta}
\ea
Here~\cite{Will08,Merritt10}
\ba
A_S &=& \frac{6\pi}{c^2} \frac{GM}{(1-e^2)\tilde{a}} \approx 12.4' (1-e^2)^{-1} \left(\frac{M}{4\times10^6\,M_\odot}\right) \left(\frac{\tilde a}{{\rm mpc}}\right)^{-1}, \label{eq:AS}\\ 
A_J &=& \frac{4\pi \chi}{c^3} \left[\frac{GM}{(1-e^2)\tilde{a}}\right]^{3/2} \approx 0.115' (1-e^2)^{-3/2}\chi \left(\frac{M}{4\times10^6\,M_\odot}\right)^{3/2} \left(\frac{\tilde a}{{\rm mpc}}\right)^{-3/2}, \label{eq:AJ} \\
A_Q &=& \frac{3\pi \chi^2}{c^4} \left[\frac{GM}{(1-e^2)\tilde{a}}\right]^{2} \approx 1.19'\times 10^{-3} (1-e^2)^{-2}\chi^2 \left(\frac{M}{4\times10^6\,M_\odot}\right)^2 \left(\frac{\tilde a}{{\rm mpc}}\right)^{-2}, \label{eq:AQ}
\ea
where
\be
\chi\equiv cJ/GM^2
\label{eq:defchi}
\ee
is the dimensionless spin of Sgr~A$^\ast$, $\alpha$ and $\beta$ are the polar angles of the angular momentum vector ${\bf J}$ with respect to the orbital plane of the star, and $e$ and $\tilde{a}$ are the eccentricity and the semi-major axis of the orbit, respectively. In Eq.~(\ref{eq:AQ}), the quadrupole moment of Sgr~A$^\ast$ is assumed to be the quadrupole moment of a Kerr black hole, $-\chi^2G^2M^3/c^4$ [c.f., Eq.~(\ref{eq:kerrmult})]. Note that the expressions $A_S$, $A_J$, and $A_Q$ increase for orbits with higher eccentricities and that the expressions $A_J$ and $A_Q$ are proportional to $\chi$ and $\chi^2$, respectively. Therefore, orbital precessions are easier to detect for stars on highly eccentric orbits that approach Sgr~A$^\ast$ closely and for high values of the spin. The right panel of Fig.~\ref{fig:Sgeometry} illustrates the apparent position of a hypothetical star orbiting around Sgr~A$^\ast$ with a semi-major axis of $10~{\rm mas}$, an orbital eccentricity of $0.88$, and an inclination of $45^\circ$. The simulated stellar orbit is affected by the Schwarzschild and spin-induced corrections to the Newtonian potential of Sgr~A$^\ast$ (assuming a value of the spin $\chi=0.99$) which cause the orbit to precess.

\begin{figure}[ht]
\begin{center}
\psfig{figure=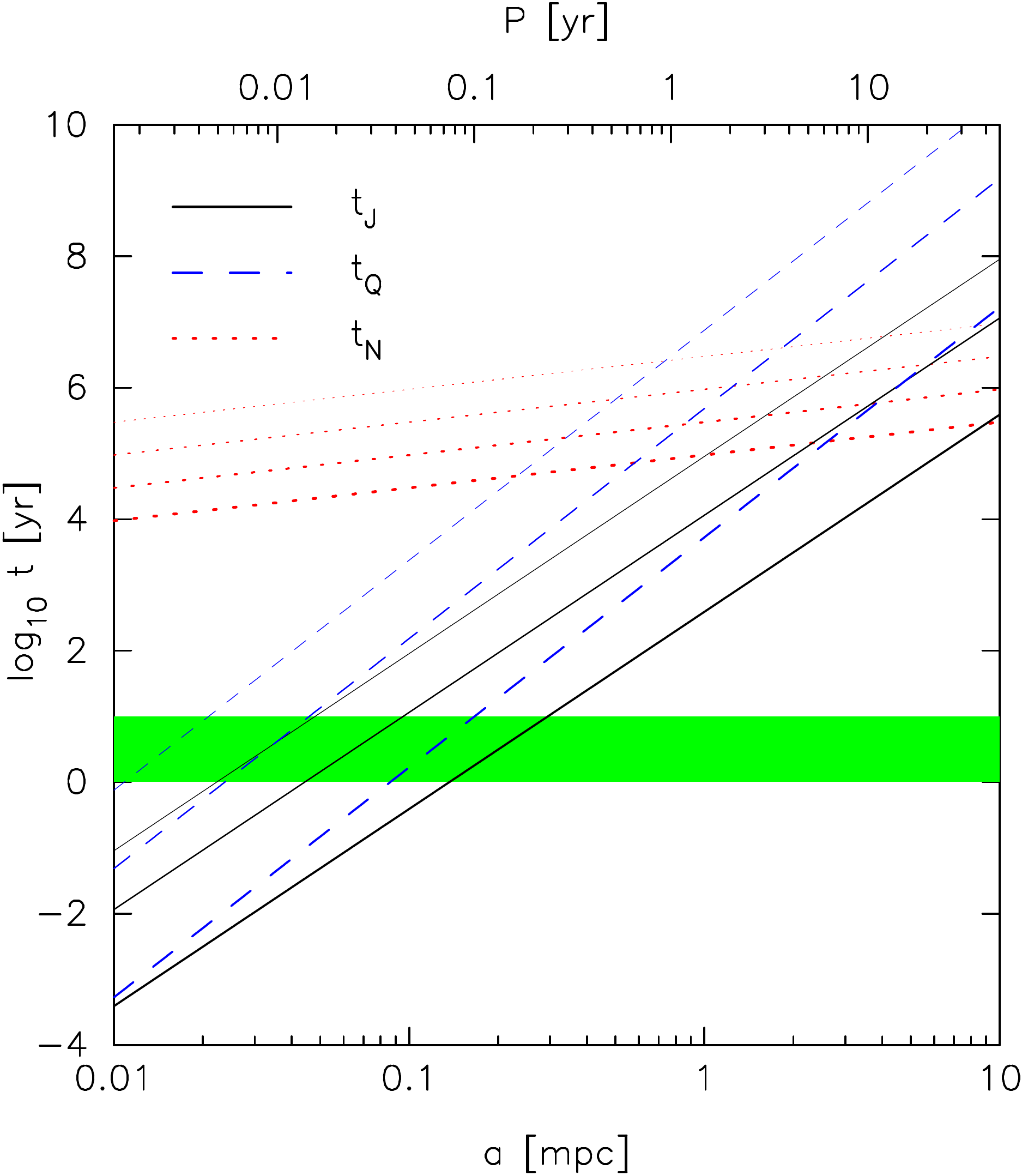,height=2.4in}
\psfig{figure=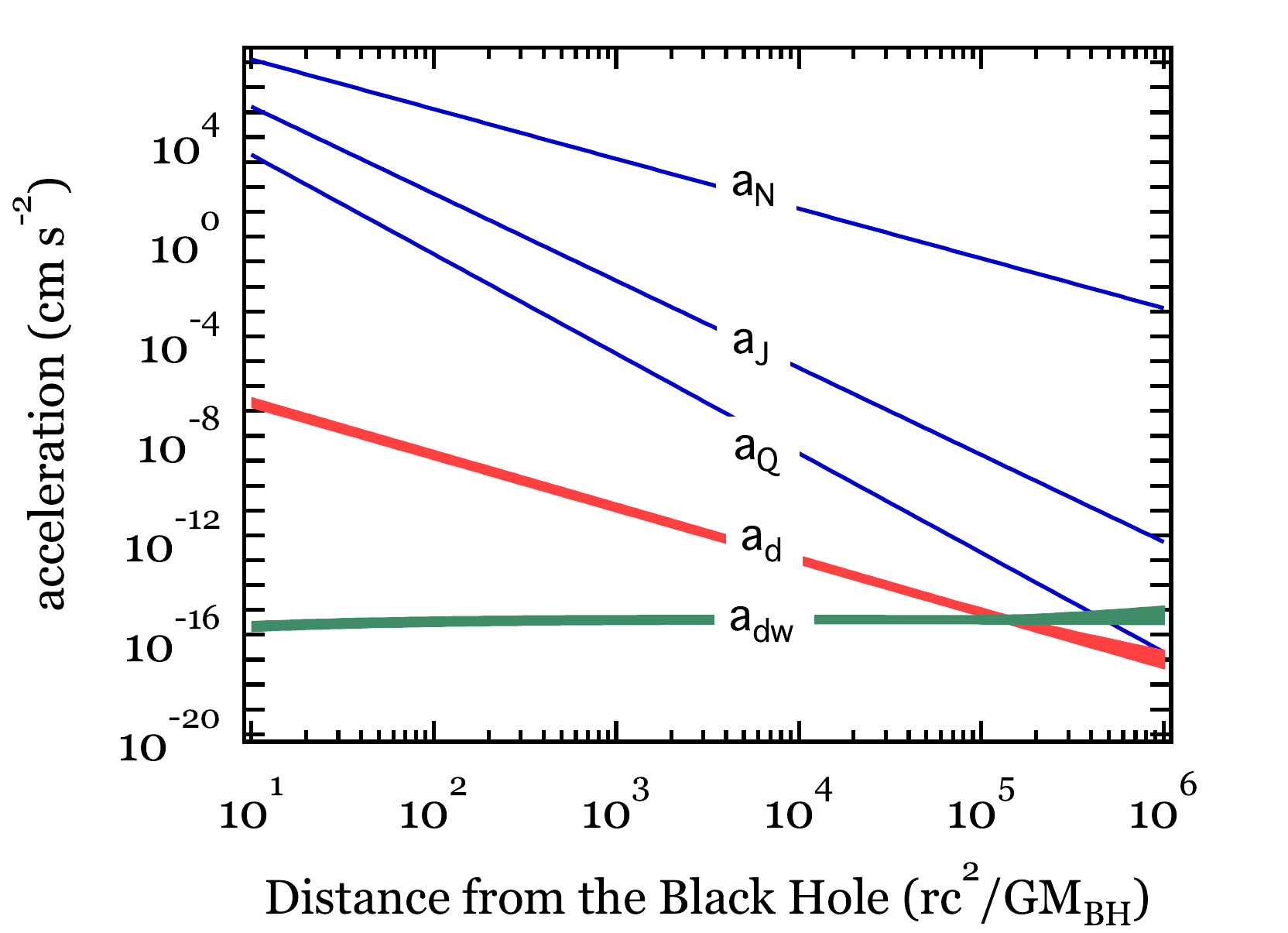,height=2.1in}
\end{center}
\caption{The left panel (taken from Ref.~\cite{Merritt10}) shows the characteristic timescales of the precession of orbital planes about Sgr~A$^\ast$ due to frame-dragging $(t_J)$, quadrupolar torque $(t_Q)$, and Newtonian perturbations from nearby $1\,M_\odot$ stars $(t_N)$ assuming that Sgr~A$^\ast$ is maximally spinning. For the timescales $t_J$ and $t_Q$, the line thickness corresponds to the orbital eccentricity ranging from $e=0.99$ (thickest) to $e=0.9$ and $e=0.5$ (thinnest). For the timescale $t_N$, the line thickness denotes the total distributed mass of perturbing stars within 1~mpc from Sgr~A$^\ast$ ranging from $10^3\,M_\odot$ (thickest) to $1\,M_\odot$ (thinnest), assuming that their density falls off as $r^{-1}$. The shaded (green) region corresponds to a time interval of observations over 1--10 years. The right panel (taken from Ref.~\cite{Psaltisdrag12}) shows the corresponding accelerations of a $10\,M_\odot$ star with a $10\,R_\odot$ radius including the accelerations due to hydrodynamic drag ($a_d$; red line) and the gravitational interaction of the star with its wake ($a_{\rm dw}$; green line) assuming a spin of $\chi=0.1$ for Sgr~A$^\ast$.}
\label{fig:timescales}
\end{figure}

Although the orbits of S-stars are predominently affected by the gravitational potential of the Galactic center, they may also be perturbed by other stars or astrophysical effects. In order to assess the magnitude of these effects, it is instructive to compare the timescales on which they contribute to the acceleration of the star. Ref.~\cite{Merritt10} defined characteristic timescales corresponding to the precession of the orbital plane of the star, which are given by the equations
\ba
t_{S} &\equiv& \left[\frac{A_S}{\pi P}\right]^{-1} = \frac{P\tilde{a}c^2}{6GM} (1-e^2) \nn \\
&& \approx 1.29\times 10^3 \left(1-e^2\right) \left(\frac{M}{4\times10^6\,M_\odot}\right)^{-3/2} \left(\frac{\tilde a}{{\rm mpc}}\right)^{5/2}\,{\rm yr}, \label{eq:tS} \\
t_J &=& \frac{P}{4\chi} \left[\frac{c^2 \tilde{a}(1-e^2)}{GM}\right]^{3/2} \nn \\
&& \approx 1.39\times 10^5 \left(1-e^2\right)^{3/2}\chi^{-1} \left(\frac{M}{4\times10^6\,M_\odot}\right)^{-2} \left(\frac{\tilde a}{{\rm mpc}}\right)^3\,{\rm yr}, \label{eq:tJ} \\
t_Q &=& \frac{P}{3\chi^2} \left[\frac{c^2 \tilde{a}(1-e^2)}{GM}\right]^2 \nn \\
&& \approx 1.34\times 10^7 \left(1-e^2\right)^2 \chi^{-2} \left(\frac{M}{4\times10^6\,M_\odot}\right)^{-5/2} \left(\frac{\tilde a}{{\rm mpc}}\right)^{7/2}\,{\rm yr}, \label{eq:tQ}
\ea
where
\be
P = \frac{2\pi \tilde{a}^{3/2}}{\sqrt{GM}} \approx 1.48 \left(\frac{M}{4\times10^6\,M_\odot}\right)^{-1/2} \left(\frac{\tilde a}{{\rm mpc}}\right)^{3/2}\,{\rm yr}
\ee
is the orbital period.

The left panel of Fig.~\ref{fig:timescales} shows these timescales together with the characteristic timescale of Newtonian perturbations of nearby stars for different eccentricities of the stellar orbit assuming that the spin of Sgr~A$^\ast$ is maximal. The right panel of Fig.~\ref{fig:timescales} shows the corresponding accelerations experienced by a star with a mass of $10\,M_\odot$ and a radius of $10\,R_\odot$ of an orbit around Sgr~A$^\ast$ assuming a black-hole spin of $\chi=0.1$, together with the accelerations of the star due to hydrodynamic drag and the gravitational interaction of the star with its wake. The latter two effects are much weaker than the effect of the quadrupole moment out to $\sim10^5$ Schwarzschild radii and, therefore, can be neglected for stars sufficiently close to Sgr~A$^\ast$~\cite{Psaltisdrag12}.

\begin{figure}[ht]
\begin{center}
\psfig{figure=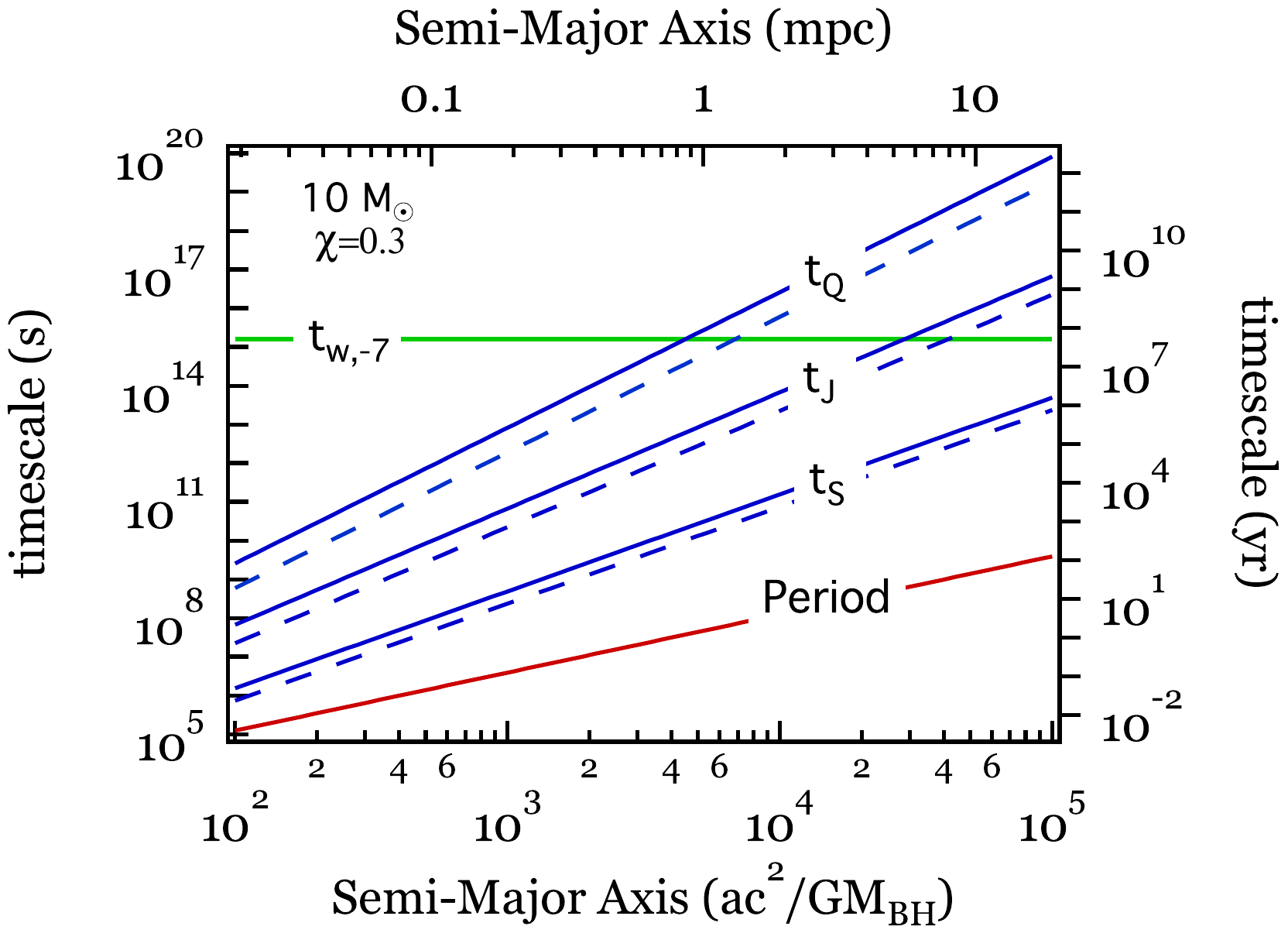,height=2.1in}
\psfig{figure=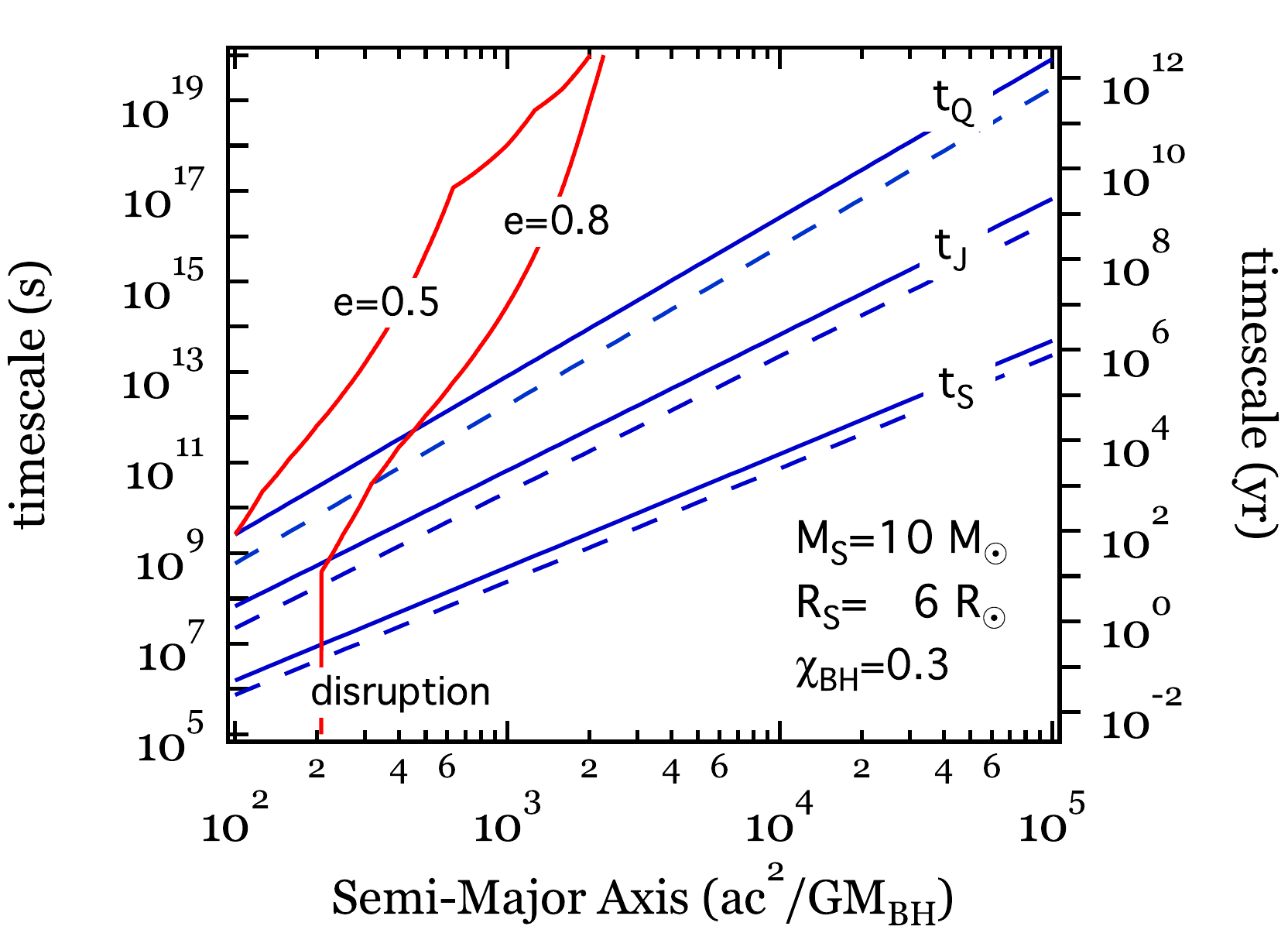,height=2.1in}
\end{center}
\caption{Characteristic timescales similar to the ones shown in the left panel of Fig.~\ref{fig:timescales} but assuming a black-hole spin $\chi=0.3$ and eccentricities $e=0.5$ (solid lines) and $e=0.8$ (dashed lines). The green line in the left panel shows the characteristic timescale for orbital evolution of a $10\,M_\odot$ star due to the presence of a stellar wind at a mass loss rate of $10^{-7}\,M_\odot~{\rm yr}^{-1}$. The red lines in the right panel show the characteristic timescale for orbital evolution due to the tidal dissipation of the orbital energy. For stars sufficiently close to Sgr~A$^\ast$, the effect of stellar winds is negligible, while the tidal dissipation of their orbital energies occurs at timescales comparable to the timescale of precession due to the quadrupole moment of Sgr~A$^\ast$. Taken from Ref.~\cite{Psaltiswinds13}.}
\label{fig:windtimescales}
\end{figure}

Figure~\ref{fig:windtimescales} shows the characteristic timescales for the orbital evolution of a $10\,M_\odot$ star including the effects of its stellar wind and the tidal dissipation of its orbital energy. For stars sufficiently close to Sgr~A$^\ast$, the former effect is negligible, while the latter effect occurs at timescales comparable to the timescale of orbital-plane precession induced by the quadrupole moment of Sgr~A$^\ast$. Thus, the tidal dissipation of stars on orbits close to Sgr~A$^\ast$ are a potential source of systematic uncertainty for tests of the no-hair theorem with the observations of such stars~\cite{Psaltiswinds13}.

\begin{figure}[ht]
\begin{center}
\psfig{figure=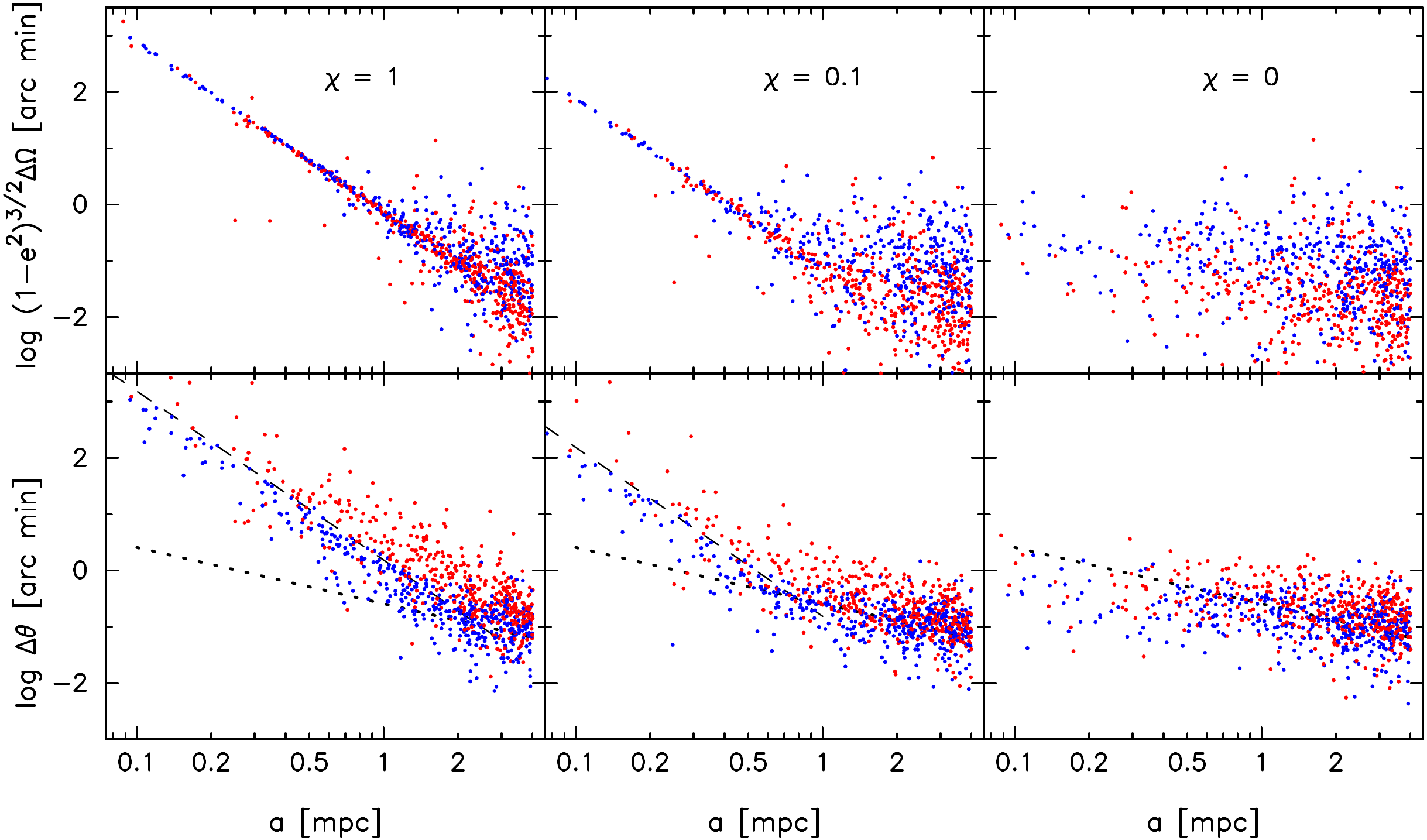,height=4.2in,angle=270}
\end{center}
\caption{Simulated 10 yr evolution of the orbital angular momenta of stars around Sgr~A$^\ast$ due to frame-dragging (dashed lines) and stellar perturbations (dotted lines) as measured by (top panels) the nodel angle $\Delta\Omega$ and (bottom panels) the angle $\Delta\theta$ between the inital and final orbital angular momentum as a function of the orbital semi-major axis $a$. The three panels correspond to a Kerr black hole with a dimensionless spin (a) $\chi=1$, (b) $\chi=0.1$, and (c) $\chi=0$. The blue and red dots correspond to $30M_\odot$ stars with initial orbital eccentricities $0\leq e \leq0.7$ and $0.7< e\leq1$, respectively. In the lower panels, the dashed and dotted lines correspond to the precessions induced by frame-dragging (setting $e=2/3$) and stellar perturbations, respectively. Taken from Ref.~\cite{Merritt10}.}
\label{fig:Merrittsim}
\end{figure}

Reference~\cite{Merritt10} performed extensive N-body simulations of a populations of stars orbiting around Sgr~A$^\ast$ and assessed their impact on the precessions of the orbital plane of a star caused by the spin and quadrupole moment of Sgr~A$^\ast$. Reference~\cite{Merritt10} showed that the effect of the quadrupole moment on the orbit of such a star is masked by the effect of the spin for the group of stars known to orbit Sgr~A$^\ast$. However, if a star can be detected within $\sim1000$ Schwarzschild radii of Sgr~A$^\ast$ and if it can be monitored over a sufficiently long period of time, this technique may also measure the spin and even the quadrupole moment of Sgr~A$^\ast$ (see Fig.~\ref{fig:Merrittsim}). Reference~\cite{Sadeghian11} found similar results for the effect of such perturbations using methods of orbital perturbation theory. References~\cite{Will14a,Will14b} derived ``cross-term'' expressions to be added to the right-hand-side of Eq.~(\ref{eq:EOM}) describing the coupling between the potential of the central black hole and the potential due to other stars at the first post-Newtonian order, which may need to be incorporated in long-term N-body simulations of stars orbiting around Sgr~A$^\ast$. Reference~\cite{Sadeghian13} calculated the perturbing effects of a distribution of dark matter around Sgr~A$^\ast$ and demonstrated that its effect is small compared to the effects of the spin and quadrupole moment of Sgr~A$^\ast$. References~\cite{Antonini10,Antonini13} analyzed tidal disruption events of S-stars in N-body simulations finding a $\lesssim1\%$ probability for such an event to take place over the full lifetime of an S-star.

\begin{figure}[ht]
\begin{center}
\psfig{figure=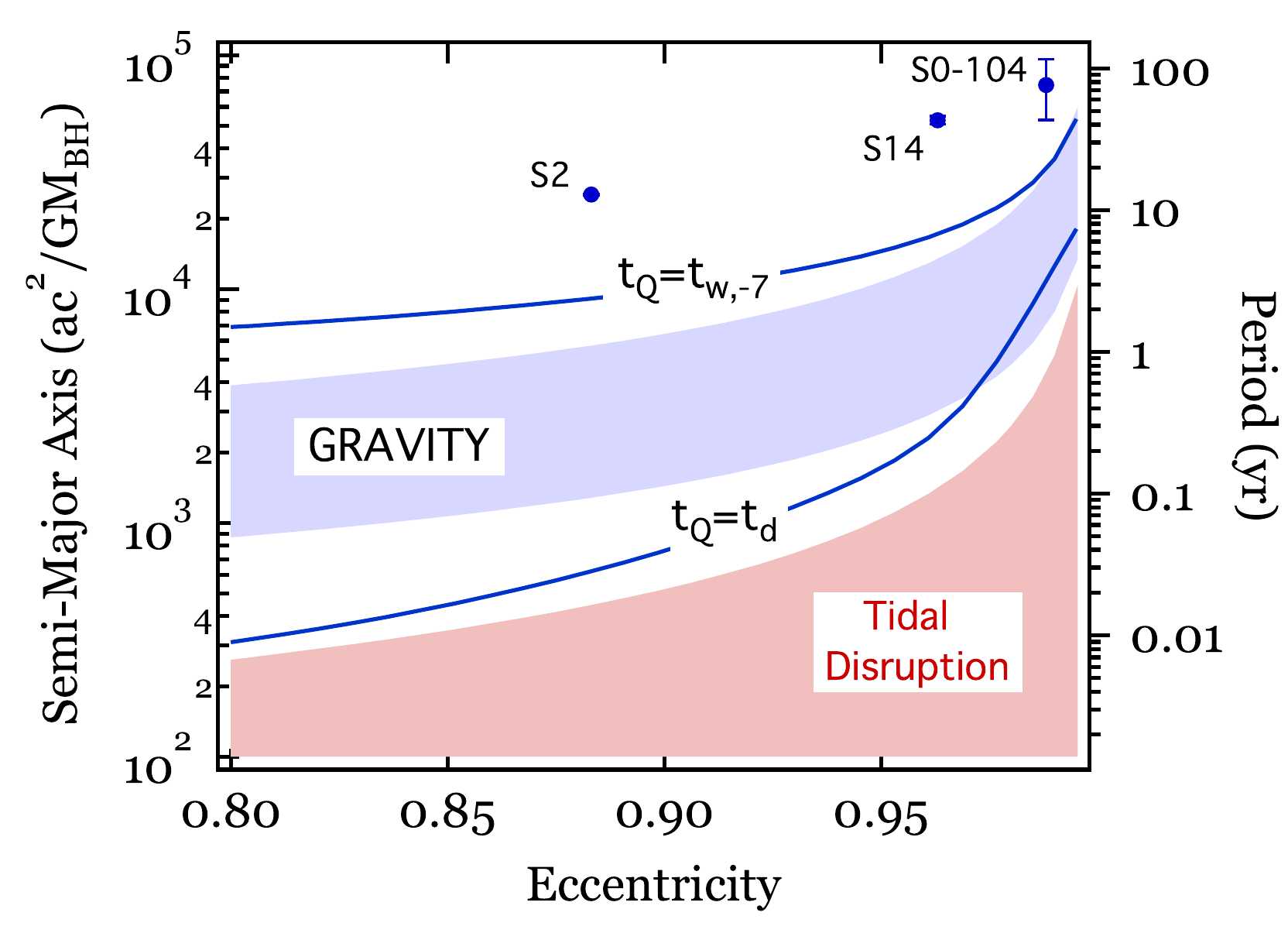,height=2.1in}
\psfig{figure=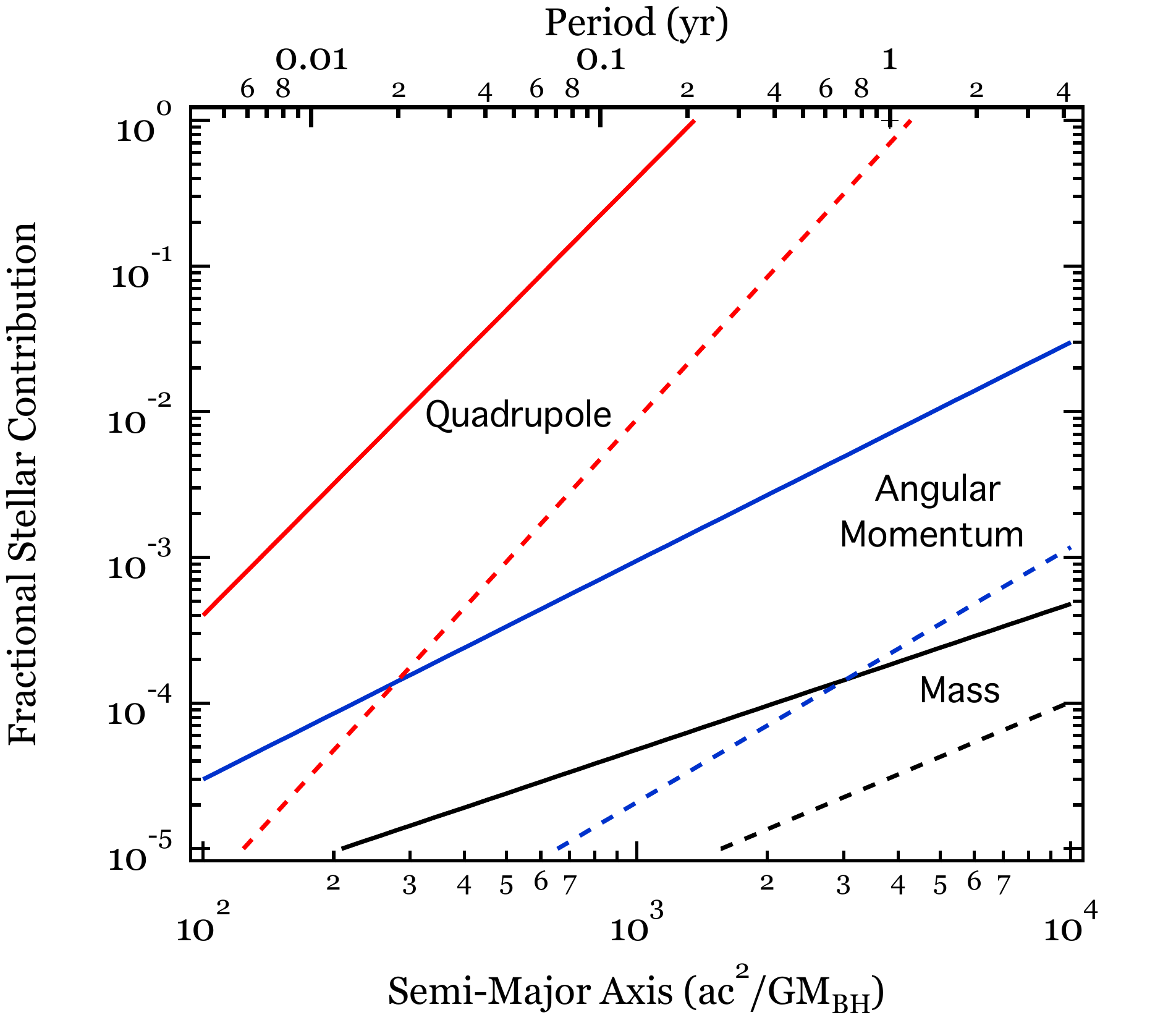,height=2.45in}
\end{center}
\caption{The left panel (taken from Ref.~\cite{Psaltiswinds13}) shows the range of the orbital parameters of stars on orbits around Sgr~A$^\ast$ for which the stars follow nearly test-particle trajectories, bounded by the two blue curves showing the loci of orbital parameters at which the timescale of orbital-plane precession due to the quadrupole moment of the black hole ($t_Q$) is equal to the orbital evolution timescale due to stellar winds ($t_{w,-7}$) and due to tidal dissipation ($t_d$), respectively. The blue shaded area shows the range of orbital parameters for which frame dragging will be detectable with GRAVITY at a signal-to-noise ratio of five, assuming a range of astrometric accuracies between $10-200~{\rm \mu as}$. In the red shaded area, such stars are tidally disrupted at pericenter. All curves are for a black-hole spin of $\chi=0.3$ and a $10\,M_\odot$ star. The three filled circles show the orbital parameters of the three stars nearest to Sgr~A$^\ast$ that are presently known. The right panel (taken from Ref.~\cite{PWK15}) shows the fractional contribution to the mass, spin, and quadrupole moment of Sgr~A$^\ast$ inside the orbit of a star due to the enclosed distribution of other stars and objects. These fractional contributions represent the limiting accuracies to which the corresponding properties of Sgr~A$^\ast$ can be inferred using observations of orbits of stars. The solid and dahsed lines correspond to stellar distributions with a density profile $n\propto \tilde{a}^{-2}$ and $n\propto \tilde{a}^{-7/4}$, respectively.}
\label{fig:windtimescales2}
\end{figure}

Figure~\ref{fig:windtimescales2} shows the range of the orbital parameters of S-stars for which the stars follow nearly test-particle orbits. This region is bounded by two curves along which the timescale of orbital-plane precession due to the quadrupole moment of the black hole ($t_Q$) is equal to the orbital evolution timescale due to stellar winds and due to tidal dissipation, respectively. Inside the blue shaded region, GRAVITY can detect the effect of frame-dragging assuming a signal-to-noise ratio of five and a range of astrometric accuracies between $10-200~{\rm \mu as}$~\cite{Psaltiswinds13}. Figure~\ref{fig:windtimescales2} also shows the fractional contribution to the mass, spin, and quadrupole moment of Sgr~A$^\ast$ inside the orbit of a star due to the enclosed distribution of other stars and objects as estimated by Ref.~\cite{PWK15}. These fractional contributions represent the limiting accuracies to which the corresponding properties of Sgr~A$^\ast$ can be inferred using observations of orbits of stars (and pulsars; see Sec.~\ref{sec:pulsars}).

\begin{figure}[ht]
\begin{center}
\psfig{figure=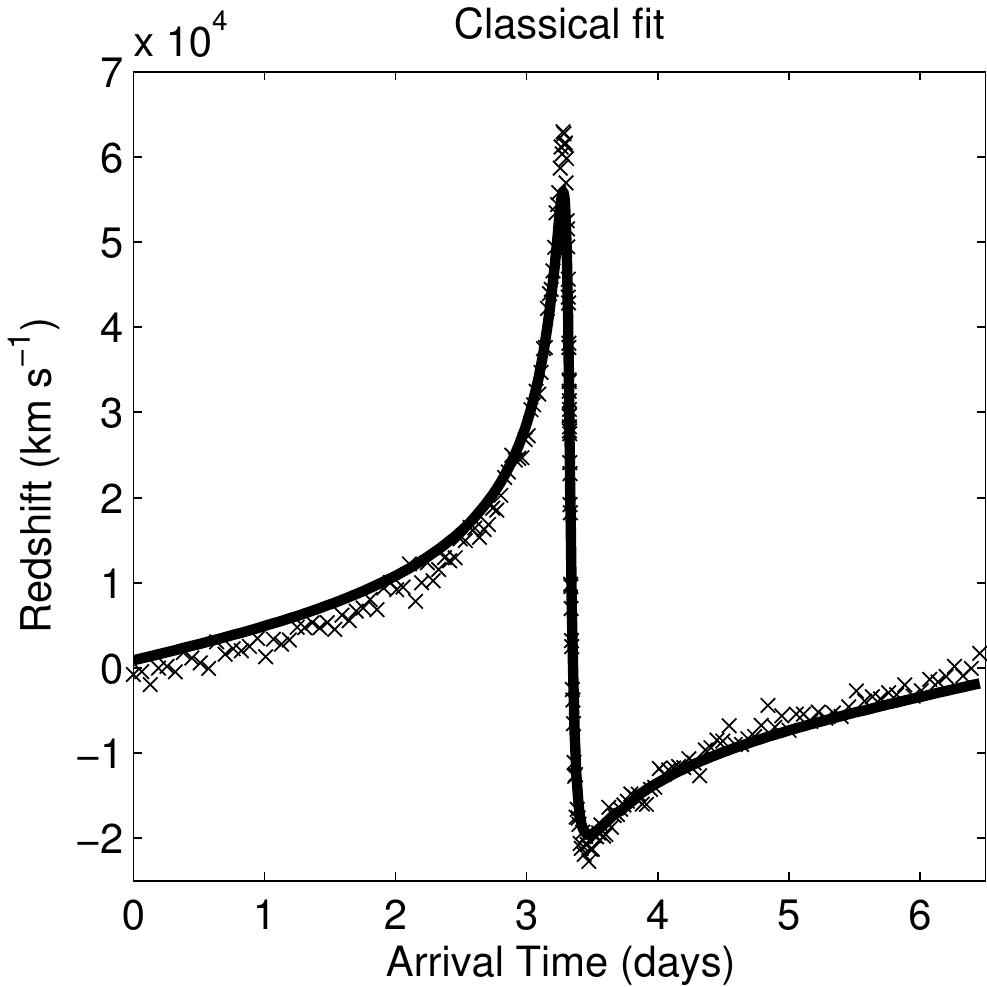,height=1.99in}
\psfig{figure=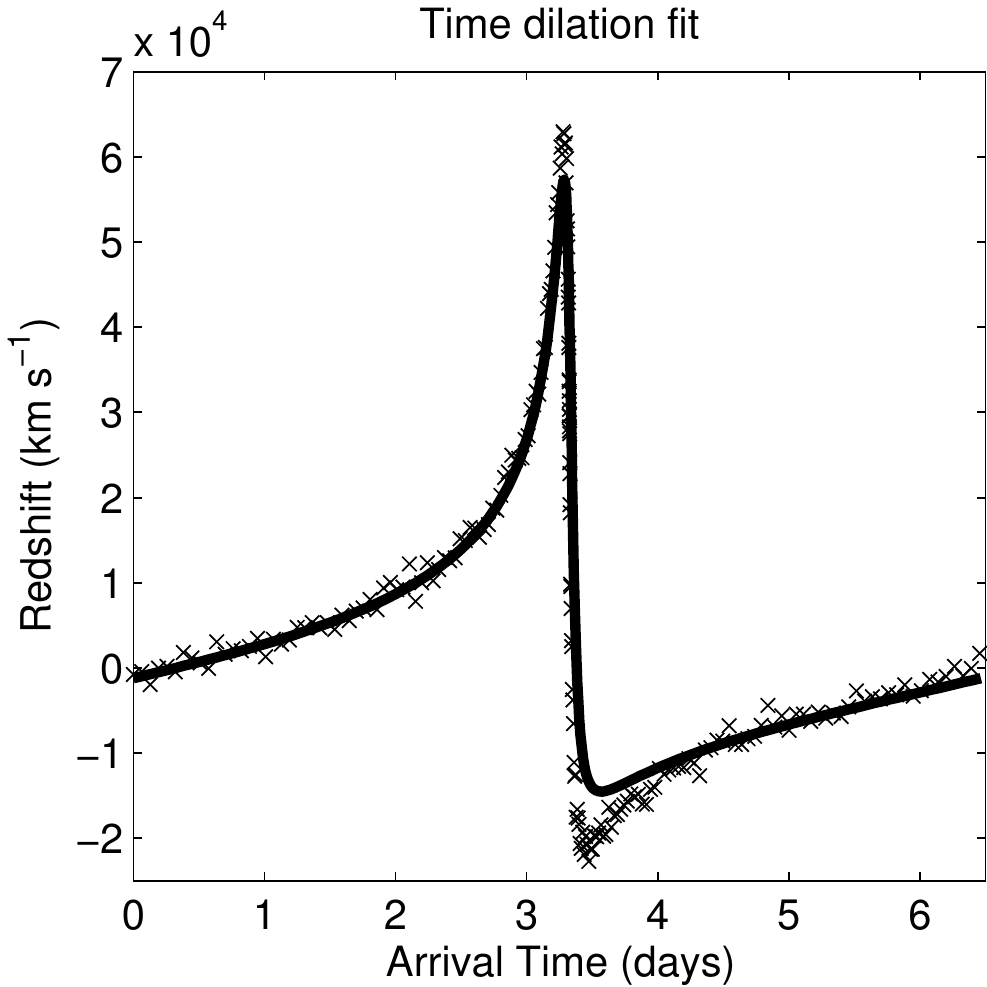,height=1.99in}
\psfig{figure=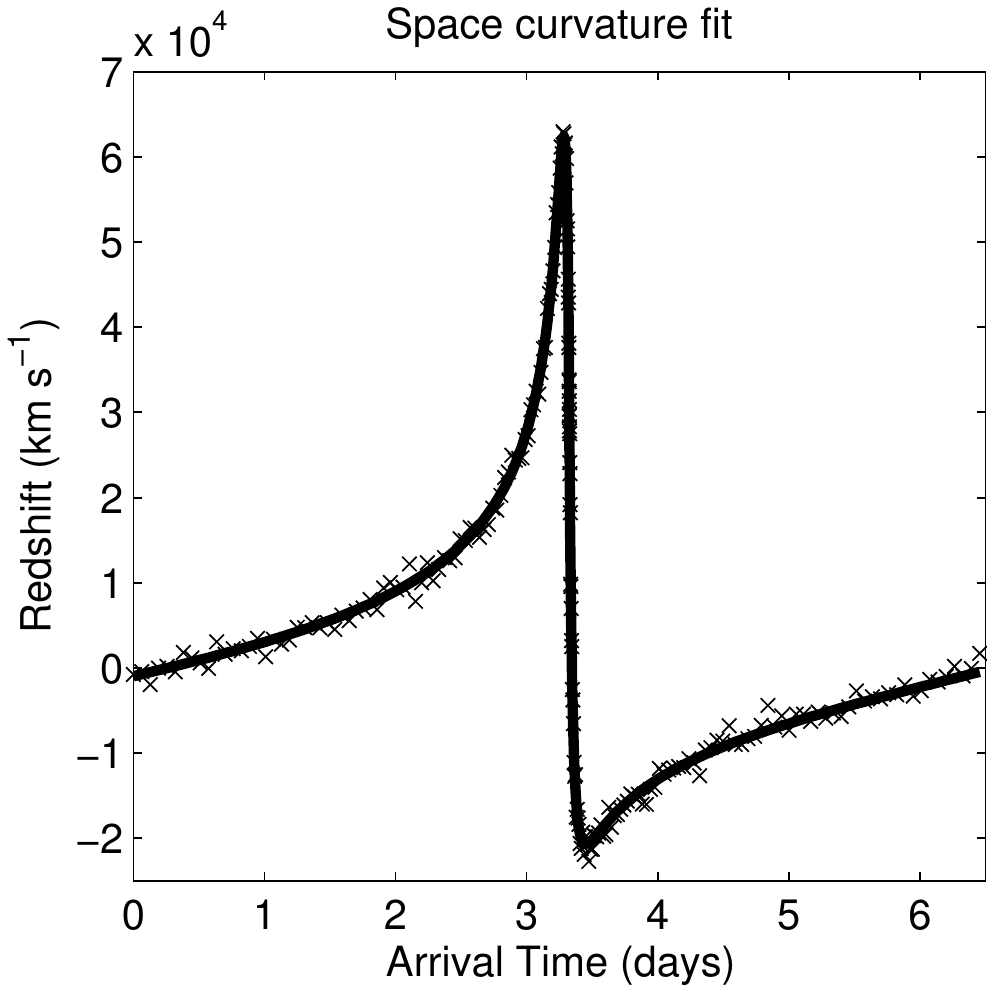,height=1.99in}
\end{center}
\caption{Simulated radial velocity data for a full orbit of an S2-like star with a semi-major axis 100 times shorter than the one of S2. The three panels show data fits which include different gravitational effects: Keplerian orbit only (left panel), gravitational time delay included (center panel), Schwarzschild spacetime curvature included (right panel). Taken from Ref.~\cite{AngelilProc11}.}
\label{fig:redshifts}
\end{figure}

In addition to the precession of the orbit of the star, photons emitted from the star may also be Doppler-shifted and gravitationally lensed by Sgr~A$^\ast$ and experience a corresponding time delay. Reference~\cite{AngelilProc11} simulated radial velocity curves for a full orbit of an S2-like star with a semi-major axis 100 times shorter than the one of S2 (see Fig.~\ref{fig:redshifts}). The fits of the simulated data shown in Fig.~\ref{fig:redshifts} include the contributions of different gravitational effects: the Keplerian orbit alone, the combined redshift of the special-relativistic Doppler effect and the gravitational redshift experienced by photons emitted from the star in the potential of Sgr~A$^\ast$, and the Schwarzschild spacetime curvature [c.f., Eq.~(\ref{eq:EOM})].

The relativistic effects, including those induced by the spin and the quadrupole moment of Sgr~A$^\ast$, are strongest near the pericenter of the stellar orbit (see Refs.~\cite{Rubilar01,Zucker06,Angelil10,Angeliletal10,Angelil11,Angelil14}). In the next few years, already existing instruments (e.g., the spectrograph SINFONI at the VLT~\cite{SINFONI1,SINFONI2}) will likely detect at least the redshift corrections due to the special-relativistic Doppler effect and the gravitational redshift~\cite{Zucker06} and the pericenter precession due to the Schwarzschild term in Eq.~(\ref{eq:EOM})~\cite{Rubilar01} for the star S2, in particular during its next pericenter passage in 2018~\cite{Ghez08,Gillessen09,Gillessen09b,Meyer12}. Reference~\cite{Iorio11} computed the time variations of S-stars averaged over one orbit up to the quadrupole order as well as in the presence of dark matter.

\begin{figure}[ht]
\begin{center}
\psfig{figure=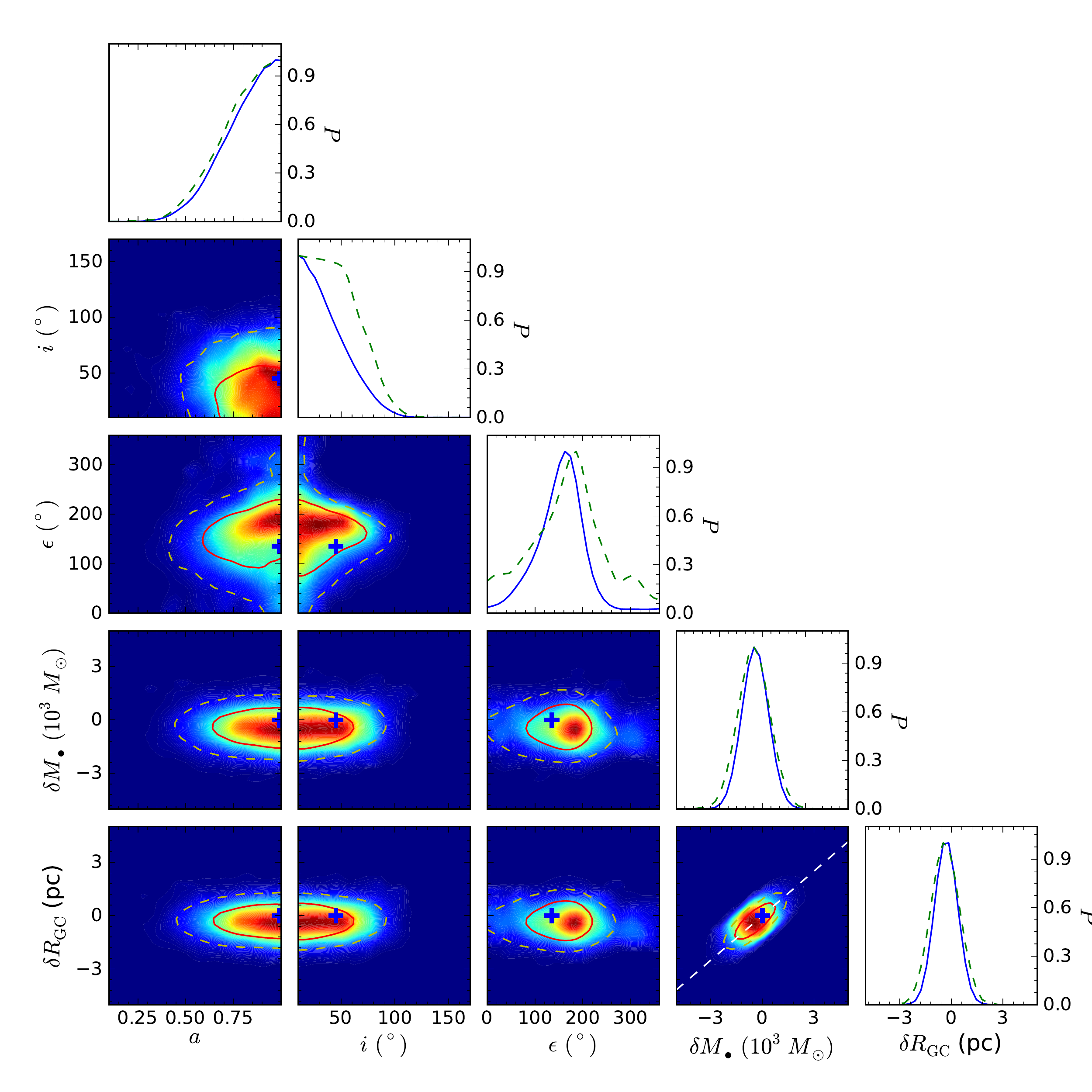,height=3.05in}
\psfig{figure=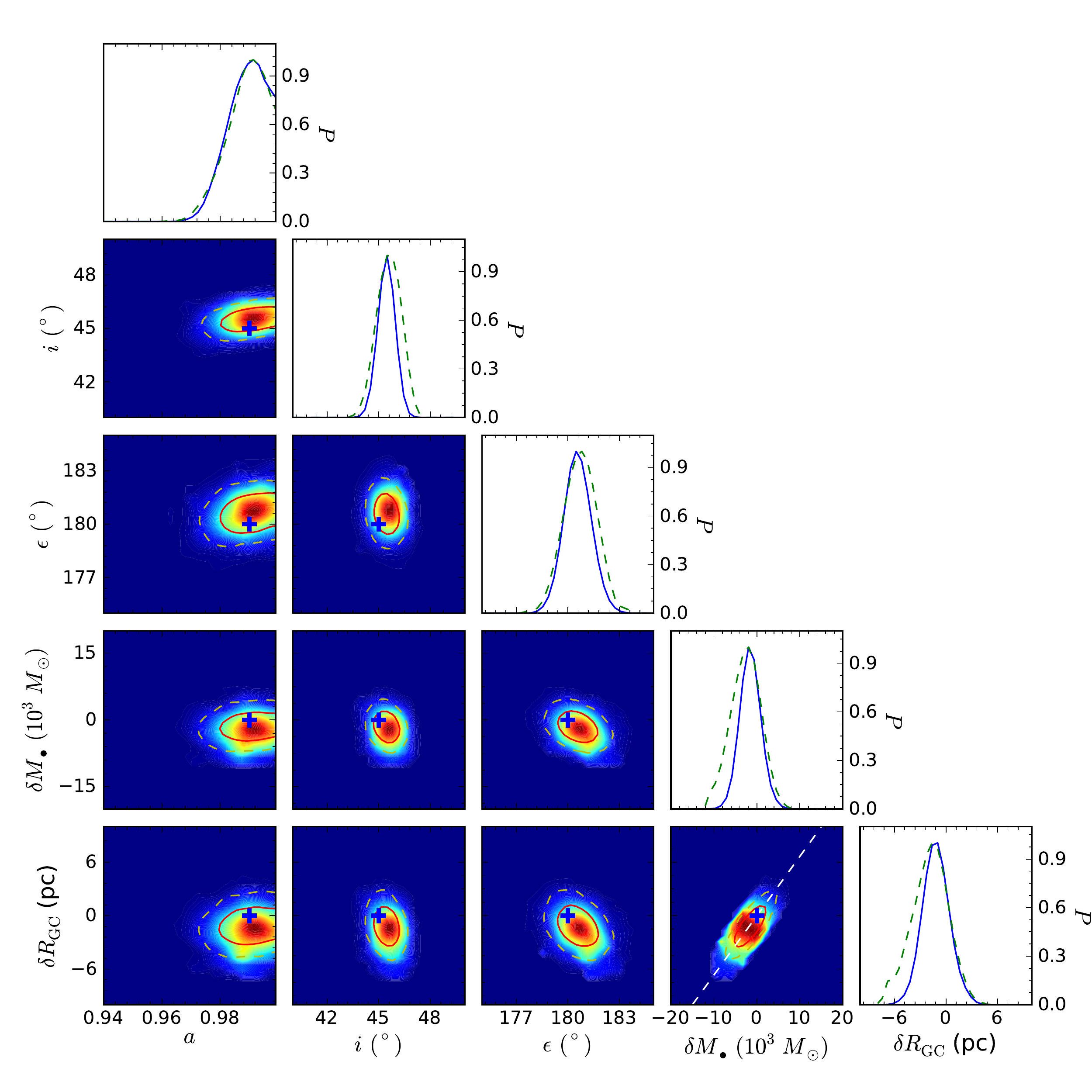,height=3.05in}
\end{center}
\caption{Simulated probability densities of the mass $\delta M$ (relative to a mass $M=4\times10^6\,M_\odot$), distance $\delta R_0$ (relative to a distance $R_0=8~{\rm kpc}$), and spin of Sgr~A$^\ast$, as well as of the orientation angles $i$ and $\varepsilon$ of the spin axis with respect to the plane of the sky and the line of sight, respectively, for (left panel) the star S2 and (right panel) a hypothetical star with a semi-major axis of $37.5~{\rm mas}$, an orbital eccentricity of $0.98$, and an inclination of $45^\circ$. The simulations consist of a set of 120 observations for each star over two to three full orbits ($\sim45~{\rm yrs}$) with astrometric and radial velocity precisions of $10~{\rm \mu as}$ and $1~{\rm km/s}$, respectively, and with a cadence $\propto r^{-1.5}$. Sgr~A$^\ast$ is assumed to have a spin value $\chi=0.99$ in each case. The uncertainties of the mass and distance are slightly lower for the star S2, while the uncertainty of the spin is slightly lower for the hypothetical star on a highly eccentric orbit. Taken from Ref.~\cite{Zhang15}.}
\label{fig:Zhangsim}
\end{figure}

Reference~\cite{Zhang15} calculated the positions of stars orbiting around Sgr~A$^\ast$ using a ray-tracing algorithm which includes all (general-)relativistic effects. Reference~\cite{Zhang15} simulated the precision with which the mass, distance, and spin of Sgr~A$^\ast$ as well as the orientation of the orbits of such stars can be determined assuming a set of 120 observations for each star over two to three full orbits ($\sim45~{\rm yrs}$) with astrometric and radial velocity precisions of $10~{\rm \mu as}$ and $1~{\rm km/s}$, respectively, and with a cadence $\propto r^{-1.5}$. Figure~\ref{fig:Zhangsim} shows two triangle plots of the probability densities of the mass $\delta M$ (relative to a mass $M=4\times10^6\,M_\odot$), distance $\delta R_0$ (relative to a distance $R_0=8~{\rm kpc}$), and spin of Sgr~A$^\ast$, as well as of the orientation angles $i$ and $\varepsilon$ of the spin axis with respect to the plane of the sky and the line of sight, respectively. These triangle plots correspond to the star S2 and a hypothetical star with a semi-major axis of $37.5~{\rm mas}$, an orbital eccentricity of $0.98$, and an inclination of $45^\circ$ for a black-hole spin $\chi=0.99$. In this simulation, Ref.~\cite{Zhang15} obtain the values $\chi=0.812^{+0.187}_{-0.257}$, $i=37^{\circ+39^\circ}_{~-27^\circ}$, $\varepsilon=156^{\circ+74^\circ}_{~-89^\circ}$, $\delta R_0=-0.30^{+0.98}_{-0.97}~{\rm pc}$, and $\delta M=-0.41^{+1.17}_{-1.15}\times10^3\,M_\odot$ for S2 and $\chi=0.989^{+0.010}_{-0.011}$, $i=45^{\circ+1^\circ}_{~-1^\circ}$, $\varepsilon=181^{\circ+1^\circ}_{~-1^\circ}$, $\delta R_0=-1.16^{+2.50}_{-2.34}~{\rm pc}$, and $\delta M=-1.53^{+3.55}_{-3.29}\times10^3\,M_\odot$ for the hypotheical star quoting $2\sigma$ uncertainties. In this scenario, the mass and distance of Sgr~A$^\ast$ can be measured very precisely with observations of either star. The precision of the spin measurement depends significantly on the eccentricity of the stellar orbit. Interestingly, the uncertainties of the mass and distance are slightly lower for the star S2 (eccentricity $e=0.88$~\cite{Ghez08,Gillessen09,Gillessen09b,Meyer12}), while the uncertainty of the spin is slightly lower for the hypothetical star on a highly eccentric orbit.

\begin{figure}[ht]
\begin{center}
\psfig{figure=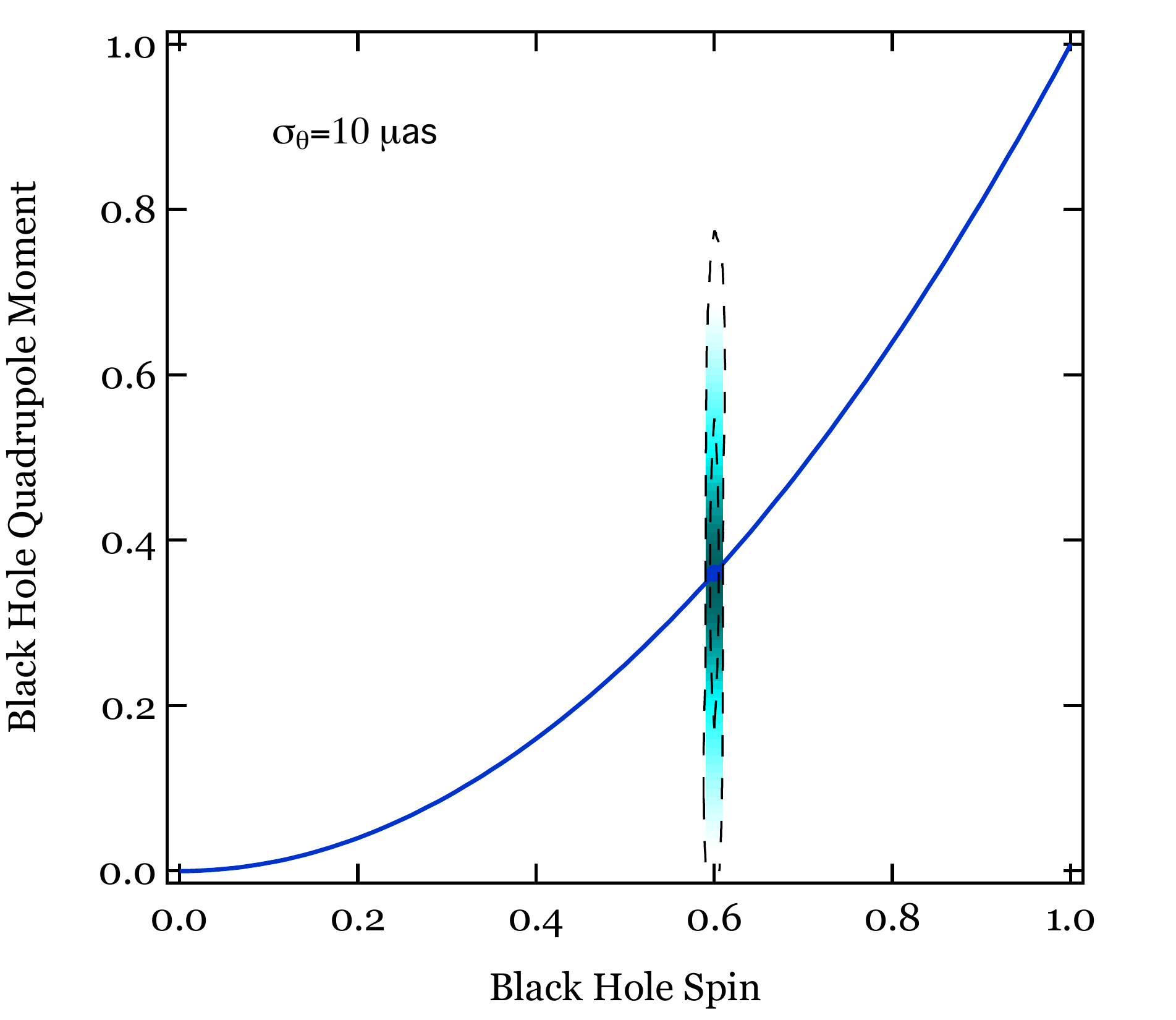,height=2.45in}
\psfig{figure=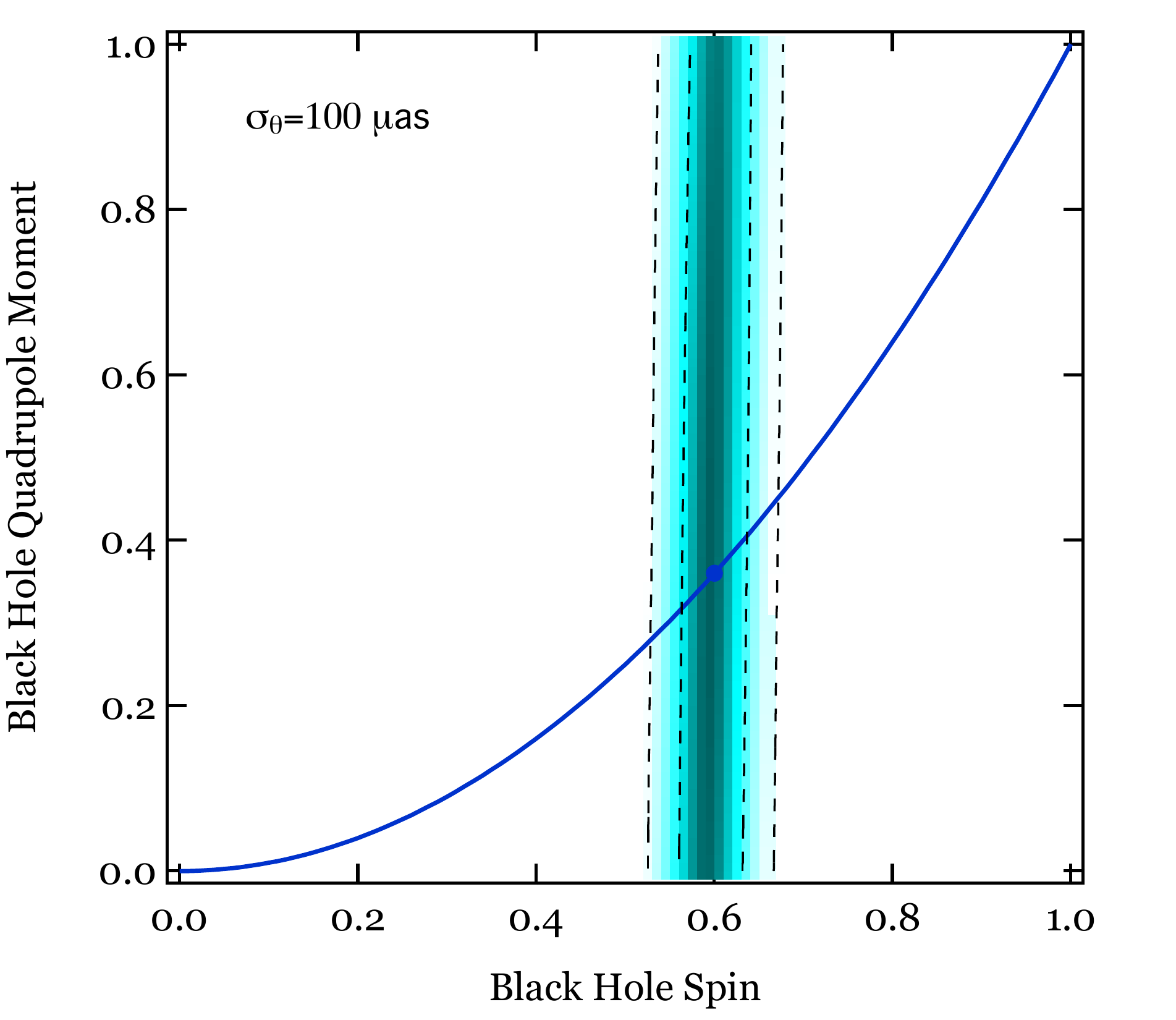,height=2.45in}
\end{center}
\caption{Posterior likelihood of measuring the spin and quadrupole moment of Sgr~A$^\ast$ by tracing $N=40$ orbits of two stars with GRAVITY, assuming an astrometric precision of (left panel) $10~{\rm \mu as}$ and (right panel) $100~{\rm \mu as}$. The dashed curves show the 68\% and 95\% confidence limits, while the solid curve shows the expected relation between these two quantities in the Kerr metric. The filled circle marks the assumed spin and quadrupole moment of a Kerr black hole with a value of the spin $a=0.6r_g$. The two stars are assumed to have orbital separations equal to $800r_g$ and $1000r_g$ and eccentricities of 0.9 and 0.8, respectively. Even at these relatively small orbital separations, tracing the orbits of stars primarily measures the spin of the black hole, unless a very high level of astrometric precision is achieved. Taken from Ref.~\cite{PWK15}.}
\label{fig:PWKstars}
\end{figure}

Reference~\cite{PWK15} estimated the precision with which the spin and quadrupole moment of Sgr~A$^\ast$ can be measured with GRAVITY observations of the nodal and apsidal precessions of two stars with semi-major axes of $800r_g$ and $1000r_g$ and eccentricities of 0.9 and 0.8, respectively. Figure~\ref{fig:PWKstars} shows the 68\% and 95\% confidence contours of the probability density of measuring the spin and quadrupole moment of Sgr~A$^\ast$ for GRAVITY observations of such stars over $N=40$ orbits with astrometric precisions of $10~{\rm \mu as}$ and $100~{\rm \mu as}$, respectively, assuming that Sgr~A$^\ast$ is a Kerr black hole with a value of the spin $a=0.6r_g$. Even at these relatively small orbital separations, tracing the orbits of stars primarily measures the spin of the black hole, unless a very high level of astrometric precision is achieved~\cite{PWK15}.

For one S-star, Ref.~\cite{PWK15} estimated that GRAVITY observations can measure its spin with a precision
\ba
  \sigma_\chi &\sim&  0.064\left(\frac{\sigma_\theta}{100\,\mu{\rm as}}\right)
  \left(\frac{N}{40}\right)^{-3/2}
  \left(\frac{\tilde{a}}{1000r_g}\right)^{1/2} 
  \left(\frac{r_g/D}{5.1\,\mu{as}}\right)^{-1} \nn \\ &&
  \left[\frac{(1-e)(1-e^2)^{1/2}}{0.12}\right]
  \left(\frac{\cos i}{0.5}\right)^{-1},
\label{eq:sigmachi}
\ea
where $\sigma_\theta$ is the astrometric precision of GRAVITY observations over $N$ orbits, $D$ is the distance of Sgr~A$^\ast$, and where $\tilde{a}$, $e$, and $i$ are the semi-major axis, eccentricity, and orbital inclination of the S-star.

Relativistic effects on the orbits of S-stars may also be imprinted on potential gravitational lensing events caused by the deflection of light rays by Sgr~A$^\ast$, which would result in the presence of two or more images of the same S-star. The position and magnification of images of gravitationally-lensed S-stars depend primarily on the mass and distance of Sgr~A$^\ast$, but may also be affected by the Schwarzschild part of the potential sourced by Sgr~A$^\ast$ or even its spin and quadrupole moment~\cite{Virbhadra98,Virbhadra99,Virbhadra02,DePaolis03,BozzaMancini04,DePaolis04,BozzaMancini05,Bozza06,Virbhadra08,BozzaMancini09,Virbhadra09,DePaolis11,WeiSW15,Linet15}. Gravitational lensing events of S-stars may be resolvable with instruments such as GRAVITY~\cite{BozzaGRAVITY} and could potentially reveal deviations from the Kerr metric~\cite{BinNun10,BinNun11,Bozza15}. See Ref.~\cite{BozzaReview} for a review. Stars such as S2 may also be used as a probe for intermediate-mass black holes~\cite{Gualandris10}. GRAVITY may also measure the spin and quadrupole moment of Sgr~A$^\ast$ by observing localized NIR flares in the accretion flow surrounding Sgr~A$^\ast$ over the course of several orbits (see Sec.~\ref{sec:variability}).

\section{Pulsar Timing}
\label{sec:pulsars}

Radio pulsars emit regular, steady beams of electromagnetic radiation, the periods of which can often be measured very precisely. Several binary pulsars have been discovered to date (see Ref.~\cite{Lorimer05} for a review) including the ``Hulse--Taylor'' pulsar PSR B1913+16~\cite{HulseTaylor75} and the double pulsar PSR J0737--3039~\cite{Burgay03,Lyne04}. Due to the compactness of such systems with semi-major axes $\sim 1\,R_\odot$ and orbital periods of only several hours, timing observations of double pulsars can be used for measurements of the parameters of the binary system including the masses of both stars and for consistency tests of general relativity with great precision~\cite{Kramer06,Breton08}.

As such, double pulsars can provide a very good testing ground for weak-field general relativity (see Refs.~\cite{Stairs03,KramerStairs08,KramerWex09} for reviews) and, in some cases, even for strong-field tests of particular theories of gravity such as Brans-Dicke gravity~\cite{BransDicke} and the second-order scalar-tensor theory by Damour and Esposito-Far\`ese~\cite{Damour93,Damour96} (see Ref.~\cite{DeDeo03}; \cite{Psaltis08,Antoniadisetal13,Zhu15}) as well as certain Lorentz-violating theories of gravity~\cite{Yagi14a,Yagi14b}. In these alternative theories of gravity, modifications in the strong-field regime such as the predicted existence of dipolar gravitational radiation lead to observable effects even in the weak-field regime. The orbital evolution of pulsars in binaries or triple systems may also be used to test the strong equivalence principle with high precision~\cite{Freire12,Ransom14}. A pulsar in a binary with a stellar-mass black hole could also reveal the presence of a ``large'' extra dimension in RS2-type braneworld gravity~\cite{RandallSundrum2} at the $\sim1~{\rm \mu m}$ level through its orbital evolution induced by mass loss of the black hole into the extra dimension~\cite{McWilliams10,Simonetti11} (c.f., Refs.~\cite{Emparan03,Tanaka03}).

Although a large number $(\sim200)$ of radio pulsars are thought to populate the stellar cluster at the Galactic center~\cite{Wharton12,Chennamangalam14}, as well as perhaps even pulsar--black hole binaries~\cite{Faucher11}, only five pulsars and one magnetar have been discovered within $15'$ of Sgr~A$^\ast$ to date, the closest of which has a distance of $\sim3''$ $(\sim1~{\rm pc})$~\cite{Morris02,Johnston06,Deneva09,Kennea13,Mori13,Rea13,Eatough13,Zadeh15}. The lack of additional pulsar detections near Sgr~A$^\ast$ is most likely caused by the strong scattering of radio waves in this region at typical observing frequencies $\sim1-2~{\rm GHz}$ (as well as, perhaps, free-free absorption~\cite{Rajwade15}), requiring pulsar searches at higher frequencies~\cite{CordesLazio02}. However, pulsar detections remain elusive even though several high-frequency surveys have already been carried out near the Galactic center~\cite{Kramer00,Johnston06,Deneva09,Macquart10,Bates11,Eatough13,Siemion13} and may require highly-sensitive instruments such as the Atacama Large Millimeter/submillimeter Array (ALMA) in Chile or the SKA in Australia and South Africa~\cite{Kramer15,Keane15,Shao15}.

\begin{figure}[ht]
\begin{center}
\psfig{figure=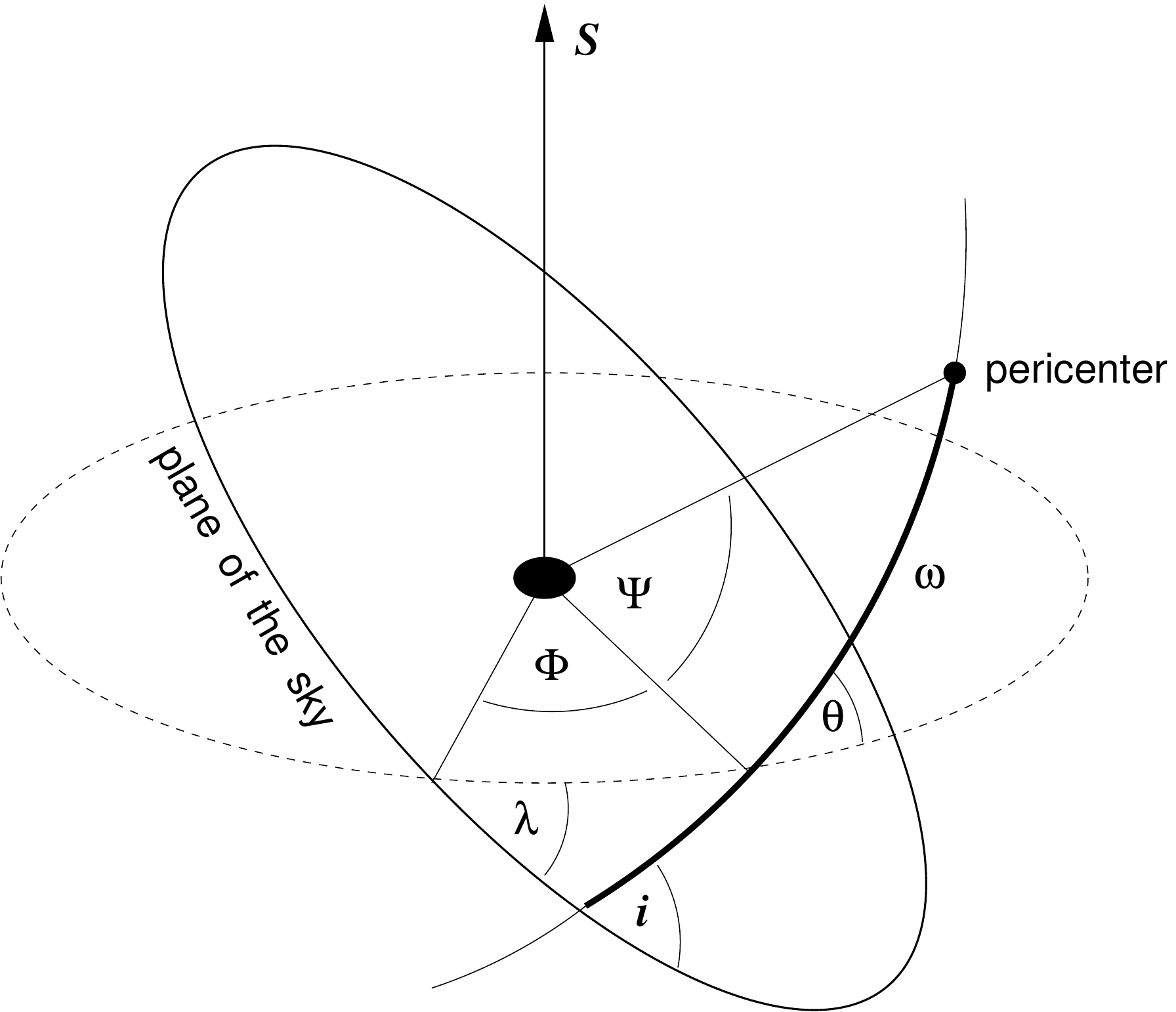,height=2.1in}
\psfig{figure=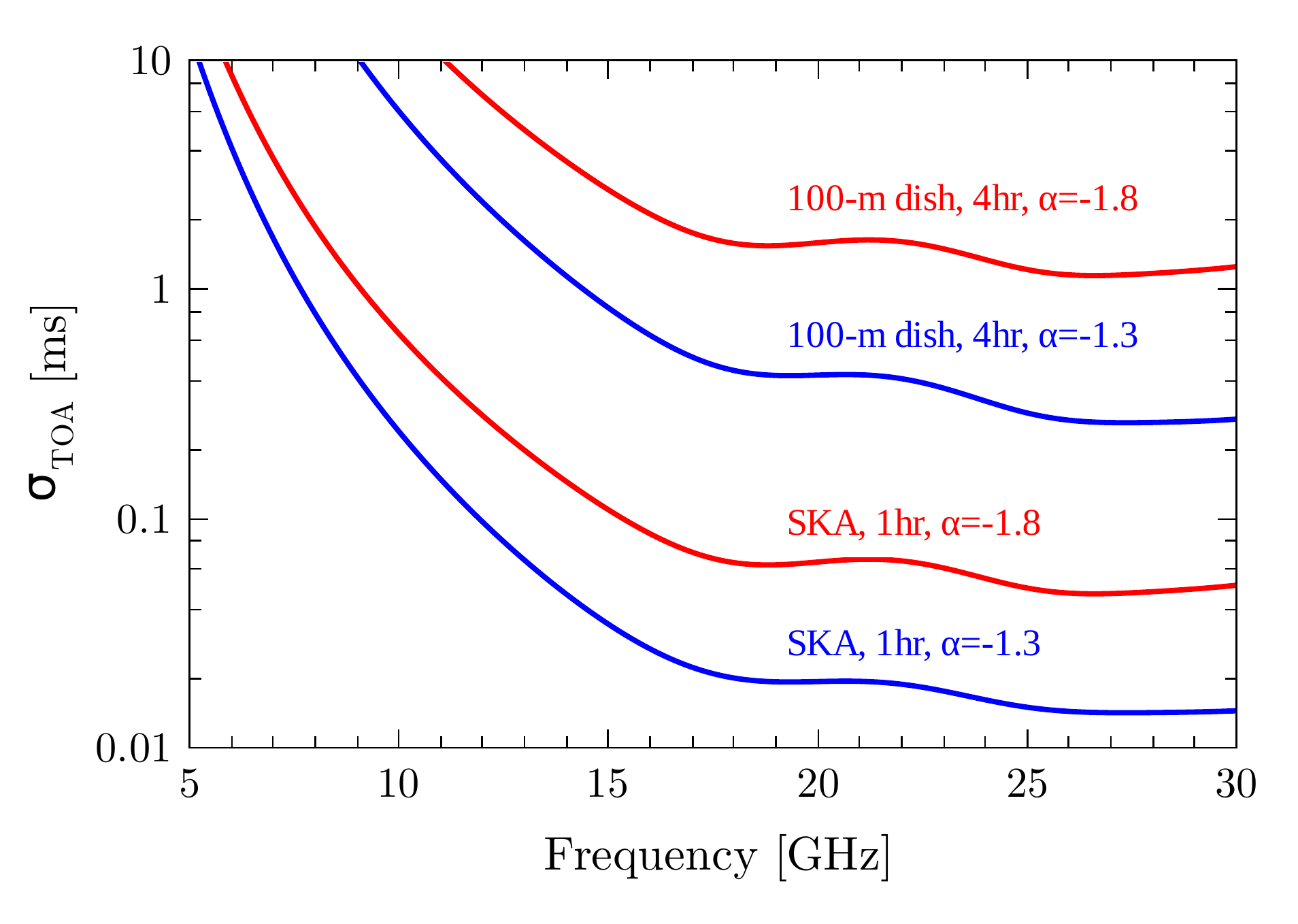,height=2.1in}
\end{center}
\caption{The left panel shows the geometry of a pulsar orbiting around Sgr~A$^\ast$ as characterized by a set of angles. The angles $i$ and $\omega$ are the orbital inclination and longitude of pericenter as measured from the ascending node in the plane of the sky. The pulsar orbit with respect to the equatorial plane of Sgr~A$^\ast$ is determined by the inclination $\theta$, the equatorial longitude of the ascending node $\Phi$, and the equatorial longitude of pericenter $\Psi$, while $\lambda$ denotes the angle between the line of sight and the spin axis $\vec{S}$ of Sgr~A$^\ast$ (c.f., Fig.~\ref{fig:Sgeometry}). The right panel shows the predicted uncertainty of the pulse arrival time for a pulsar arbiting around Sgr~A$^\ast$ for two different spectral indices $\alpha$, which includes the uncertainties from pulse phase jitter intrinsic to the pulsar, pulse broadening from scattering along the line of sight, and from interstellar scintillation. The curves assume a four-hour integration time for a 100~m radio telescope and a one-hour integration time for an SKA-like telescope, each with a bandwidth of 1~GHz. Observational frequencies above $\approx15~{\rm GHz}$ are favored and precisions of $\sim100~{\rm \mu s}$ seem achievable with an SKA-like telescope. Taken from Ref.~\cite{Liu12}.}
\label{fig:Pgeometry}
\end{figure}

\begin{figure}[ht]
\begin{center}
\psfig{figure=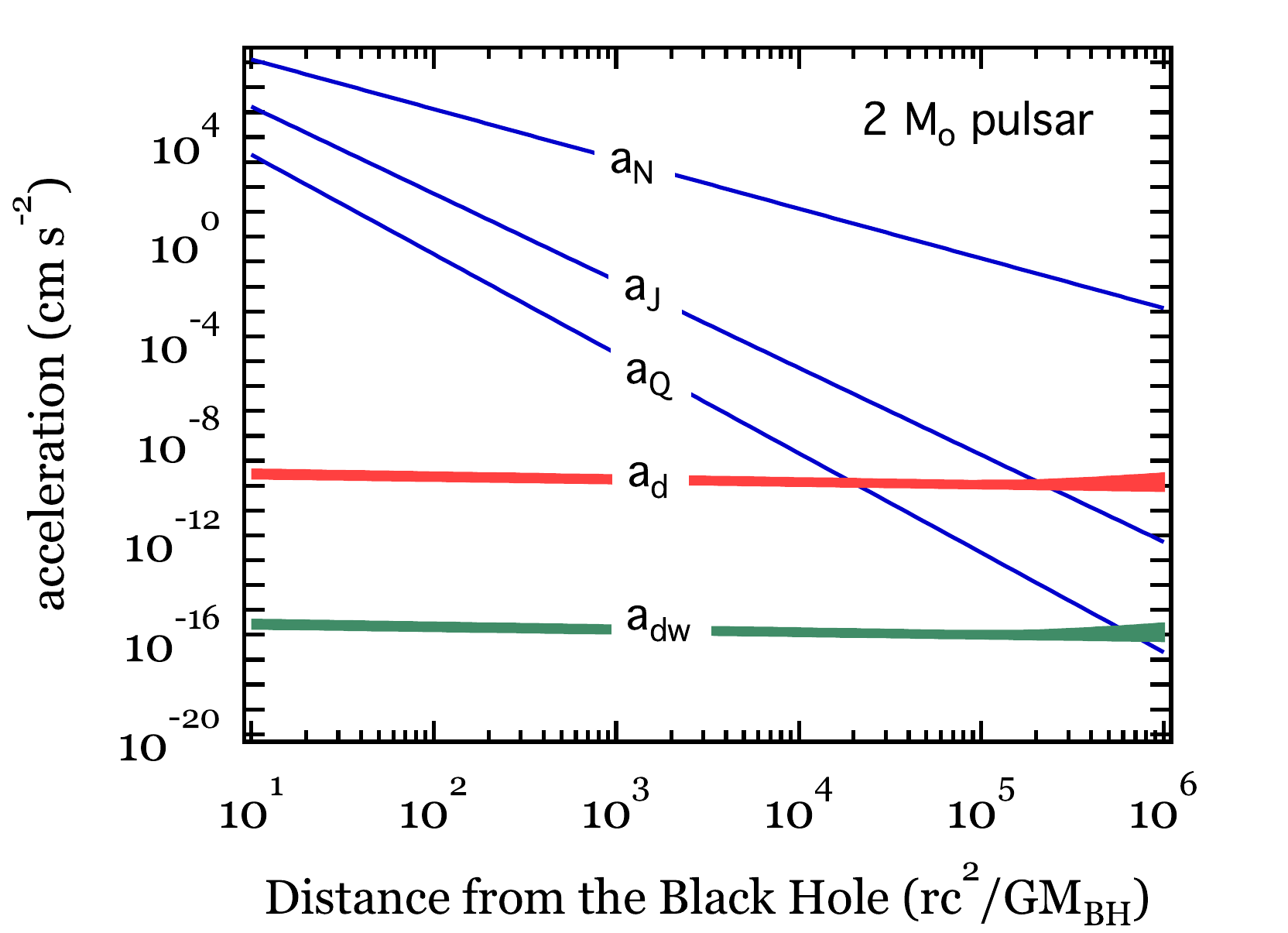,height=2.1in}
\psfig{figure=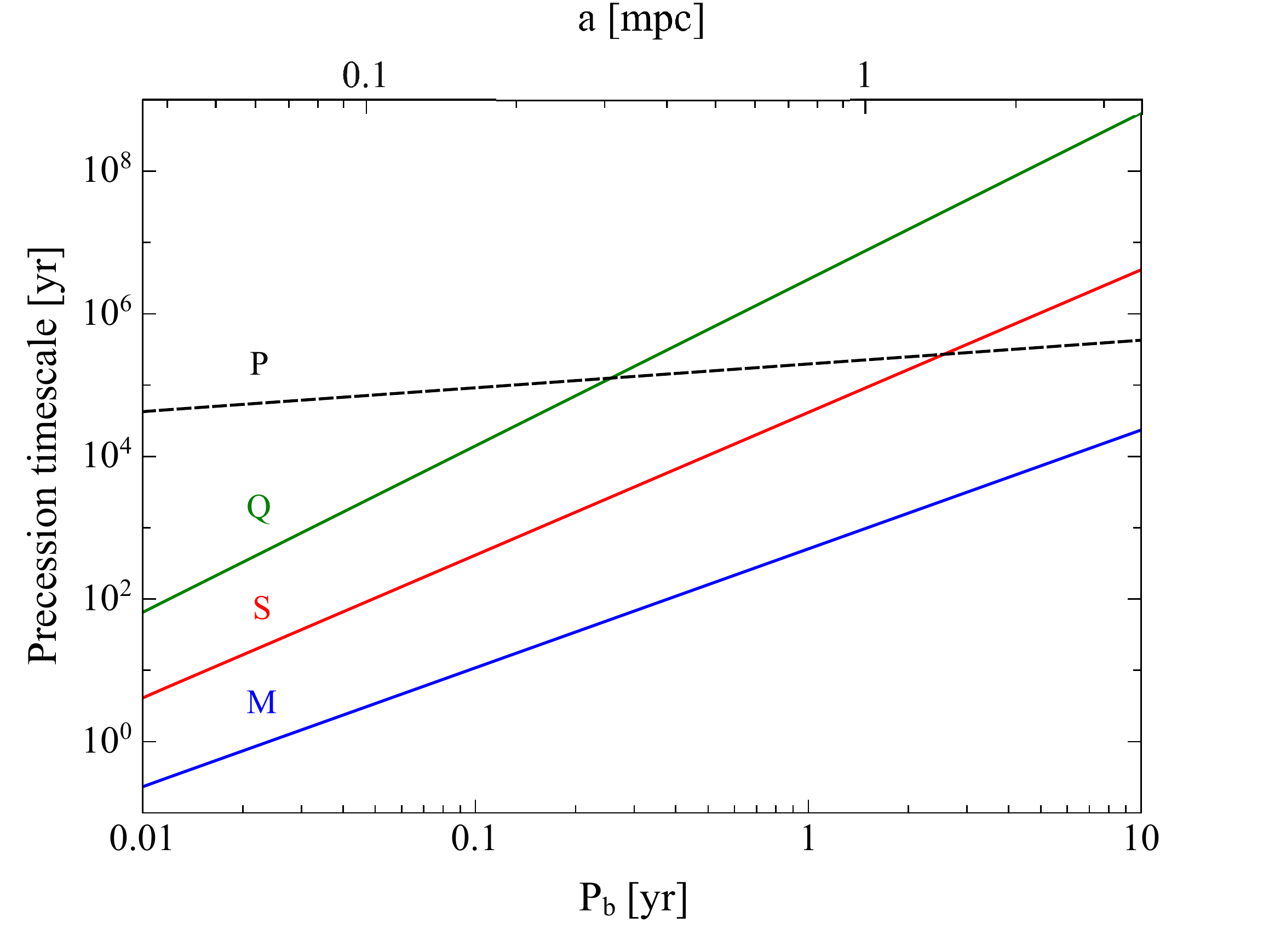,height=2.2in}
\end{center}
\caption{The left panel (taken from Ref.~\cite{Psaltisdrag12}) shows the accelerations of a $2\,M_\odot$ pulsar due to the Newtonian potential $(a_N)$ (without the Schwarzschild corrections), frame-dragging $(a_J)$, and quadrupolar torque $(a_Q)$, as well as the accelerations due to hydrodynamic drag ($a_d$; red line) and the gravitational interaction of the star with its wake ($a_{\rm dw}$; green line) assuming a spin of $\chi=0.1$ for Sgr~A$^\ast$. The right panel (taken from Ref.~\cite{Liu12}) shows the characteristic timescales of such a pulsar assuming an orbital eccentricity of 0.5 and $10^3$ $1\,M_\odot$ perturbing stars within 1~mpc of Sgr~A$^\ast$, where the letters M, S, Q and P correspond to the contributions from the mass monopole (including the Schwarzschild corrections), spin (frame dragging), quadrupole moment, and stellar perturbations, respectively.}
\label{fig:Ptimescales}
\end{figure}

Nonetheless, if a pulsar sufficiently close to Sgr~A$^\ast$ is discovered, pulsar timing could provide another means to measure the mass, spin, and quadrupole moment of Sgr~A$^\ast$ and, thereby, test the no-hair theorem. Such a pulsar experiences the same accelerations as an S-star; see Eq.~(\ref{eq:EOM}). Figure~\ref{fig:Pgeometry} shows the basic geometry of the binary with the corresponding definitions of angles which characterize the orbital motion of the pulsar around Sgr~A$^\ast$. Figure~\ref{fig:Pgeometry} also shows the uncertainties $\sigma_{\rm TOA}$ of the pulse arrival time (estimated by Ref.~\cite{Liu12}) for a pulsar with two different spectral indices as a function of observing frequency assuming a four-hour integration time for a 100~m radio telescope and a one-hour integration time for an SKA-like telescope, each with a bandwidth of 1~GHz. The TOA uncertainty includes three contributions according to the relation
\be
\sigma^2_{\rm TOA} \equiv \sigma^2_{\rm rn} + \sigma^2_{\rm j} + \sigma^2_{\rm scint},
\ee
where $\sigma_{\rm rn}$, $\sigma_{\rm j}$, and $\sigma_{\rm scint}$ correspond to the uncertainties due to radiometer noise, intrinsic pulse phase jitter, and interstellar scintillation (see Refs.~\cite{CordesShannon10,LiuScint11}), respectively. Timing observations are favored at observing frequencies above $\approx15~{\rm GHZ}$ and precisions of $\sim100~{\rm \mu s}$ seem achievable with an SKA-like telescope~\cite{Liu12}.

Figure~\ref{fig:Ptimescales} shows the accelerations of a $2\,M_\odot$ pulsar due to the Newtonian potential $(a_N)$, frame-dragging $(a_J)$, and quadrupolar torque $(a_Q)$ assuming a (dimensionless) spin $\chi=0.1$ of Sgr~A$^\ast$ [c.f., Eq.~(\ref{eq:defchi})]. Such a pulsar will also experience accelerations due to hydrodynamic drag ($a_d$) and the gravitational interaction of the star with its wake ($a_{\rm dw}$) as calculated by Ref.~\cite{Psaltisdrag12}; c.f., Fig.~\ref{fig:timescales}. Figure~\ref{fig:Ptimescales} likewise shows the timescales defined in Eqs.~(\ref{eq:tS})--(\ref{eq:tQ}) assuming an orbital eccentricity of 0.5, as well as the timescale of stellar perturbations for a distribution of $10^3$ $1\,M_\odot$ stars within 1~mpc of Sgr~A$^\ast$.

In addition to the spin and quadrupole moment of Sgr~A$^\ast$, such a system has three groups of parameters: non-orbital parameters such as the pulse period, the rates of change of this period, and their positions in the sky; five ``Keplerian'' parameters such as the eccentricity $e$, the orbital period $P_b$, and the semi-major axis; as well as five ``post-Keplerian'' parameters such as the mean rate of pericenter advance $\left< \dot{\omega} \right>$, the Einstein delay $\gamma_{\rm E}$ of the emitted radio pulse (a combination of the relativistic Doppler effect and the gravitational redshift), and the orbital period derivative $\dot{P}_b$. Assuming general relativity, the post-Keplerian parameters can be expressed in terms of the Keplerian parameters and the masses $m_1$ and $m_2$ of the pulsar and Sgr~A$^\ast$, respectively, according to the expressions~\cite{Robertson38,Blandford76,DamourTaylor92}
\ba
\left< \dot{\omega} \right> &=& \frac{6\pi}{P_b} \left(\frac{2\pi Gm}{c^3P_b}\right)^{2/3} (1-e^2)^{-1}, \\
\gamma_{\rm E} &=& e\left(\frac{2\pi}{P_b}\right)^{-1} \left(\frac{2\pi Gm}{c^3P_b}\right)^{2/3} \frac{m_2}{m} \left( 1 + \frac{m_2}{m} \right), \\
\dot{P}_b &=& -\frac{192\pi}{5} \left(\frac{2\pi \mathcal{M}}{P_b}\right)^{5/3} F(e), \\
\ea
where $m \equiv m_1 + m_2$ and
\ba
\mathcal{M} &\equiv& \frac{G M_\odot}{c^3} \frac{(m_1 m_2)^{3/5}}{(m_1+m_2)^{1/5}}, \\
F(e) &\equiv& \left( 1+\frac{73}{24}e^2 + \frac{37}{96}e^4 \right) (1-e^2)^{-7/2}.
\ea

Since $M_{\rm BH}\equiv m_2 \gg m_1$ and, therefore, $M_{\rm BH}\approx m$, the pulsar mass can be neglected in these equations. For Sgr~A$^\ast$, the equations for the mean pericenter advance and for the Einstein delay can then be written as (see Ref.~\cite{Liu12})
\ba
\left< \dot\omega \right> &\simeq& \frac{3}{1 - e^2} \left(\frac{2\pi}{P_{\rm b}}\right)^{5/3} \left(\frac{GM_{\rm BH}}{c^3}\right)^{2/3} \nn\\
    &\simeq& 0.269^{\circ}\,\frac{1}{1 - e^2} \left(\frac{P_{\rm b}}{1\,{\rm yr}}\right)^{-5/3} \left(\frac{M_{\rm BH}}{4\times 10^6\,M_\odot}\right)^{2/3}\,{\rm yr^{-1}}, \label{eq:Pomegadot} \\
\gamma_{\rm E} &\simeq& 2 e \left( \frac{P_{\rm b}}{2\pi}  \right)^{1/3} \left( \frac{GM_{\rm BH}}{c^3} \right)^{2/3} \nn\\
     &\simeq& 2500\,e \left(\frac{P_{\rm b}}{1\,{\rm yr}}\right)^{1/3} \left(\frac{M_{\rm BH}}{4\times 10^6\,M_\odot}\right)^{2/3}\,{\rm s} \label{eq:PgammaE}.
\ea
In addition, radio pulses experience a (Shapiro) time delay when passing through the gravitational potential of Sgr~A$^\ast$, which is given by the expression (\cite{Blandford76}; see Ref.~\cite{Liu12})
\ba
\Delta_{\rm S} &\simeq& \frac{2GM_{\rm BH}}{c^3}\, \ln \left( \frac{1 + e \cos\varphi} {1 - \sin i\sin(\omega + \varphi)} \right) \nn\\
    &\simeq& 39.4 \left(\frac{M_{\rm BH}}{4\times 10^6\,M_\odot}\right) \, \ln \left(\frac{1 + e \cos\varphi} {1 - \sin i\sin(\omega + \varphi)} \right)\,{\rm s}, \label{eq:Psdelay}
\ea
where $\omega$ and $\varphi$ are the angular distance of the pericenter in the orbital plane and the orbital phase of the pulsar, respectively, and $i$ is the inclination of the orbital plane with respect to the observer's line of sight.

Potential measurements of the mass, spin, and quadrupole moment of Sgr~A$^\ast$ with pulsar timing observations were discussed in detail by Ref.~\cite{Liu12}. The Keplerian parameters of the orbit of the pulsar could be measured relatively easily. Consequently, a measurement of either the pericenter advance, the Einstein delay, or the Shapiro delay would suffice to infer the mass of Sgr~A$^\ast$. However, the pericenter advance and the Shapiro delay are also affected by the spin of Sgr~A$^\ast$ and the Einstein delay cannot be separated from the Roemer delay which describes the contribution of the proper motion of the pulsar to the observed time delay (see the discussion in, e.g., Ref.~\cite{Liu12}). However, the mass of Sgr~A$^\ast$ and the inclination $i$ are also coupled via Kepler's third law, which defines the so-called mass function
\be
GM_{\rm BH} \simeq \left(\frac{cx}{\sin i}\right)^3 \left(\frac{2\pi}{P_{\rm b}}\right)^2,
\ee
where $x$ is the projected semi-major axis of the pulsar orbit (in light seconds), which is an observable Keplerian parameter.

\begin{figure}[ht]
\begin{center}
\psfig{figure=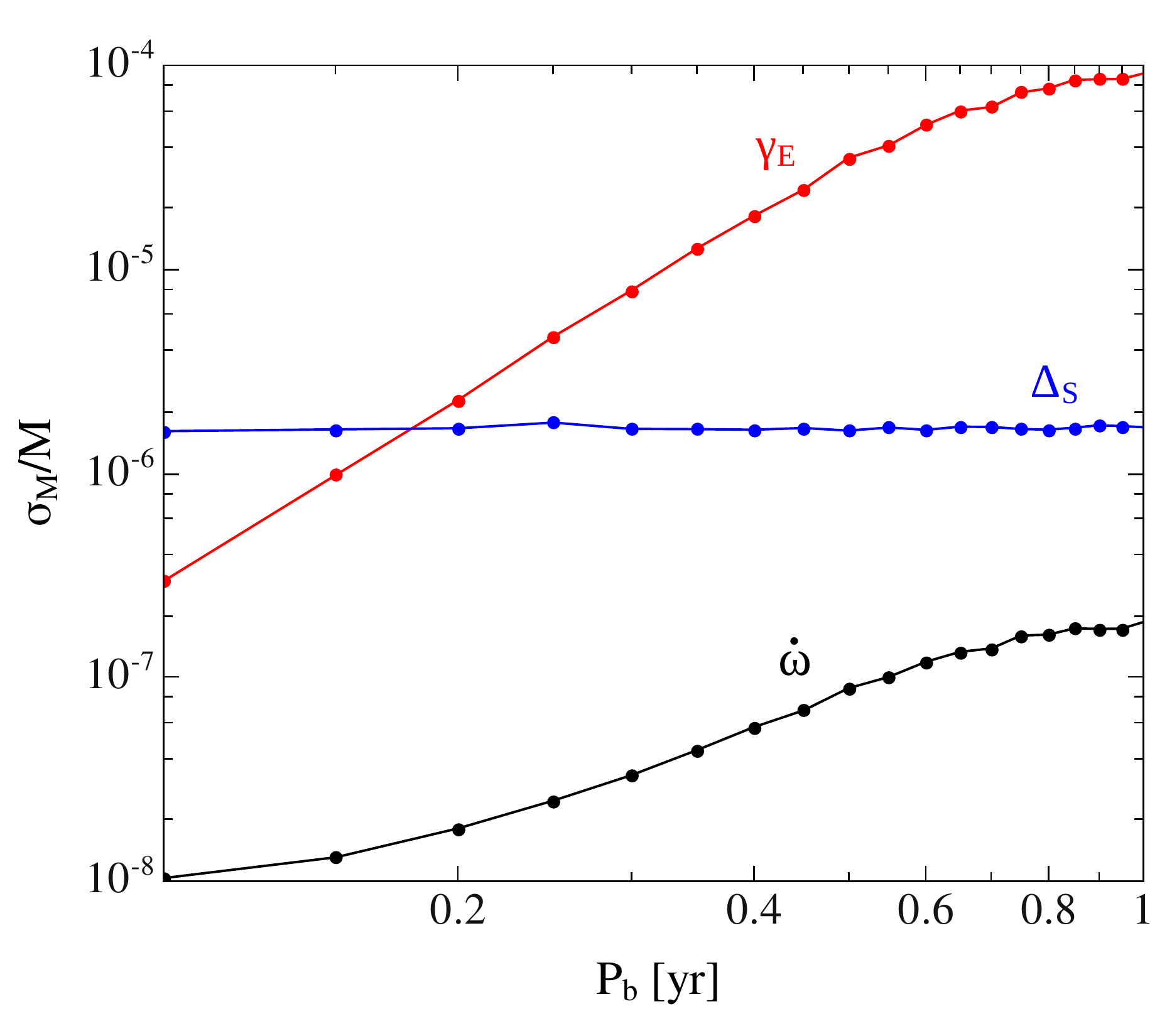,height=2.3in}
\psfig{figure=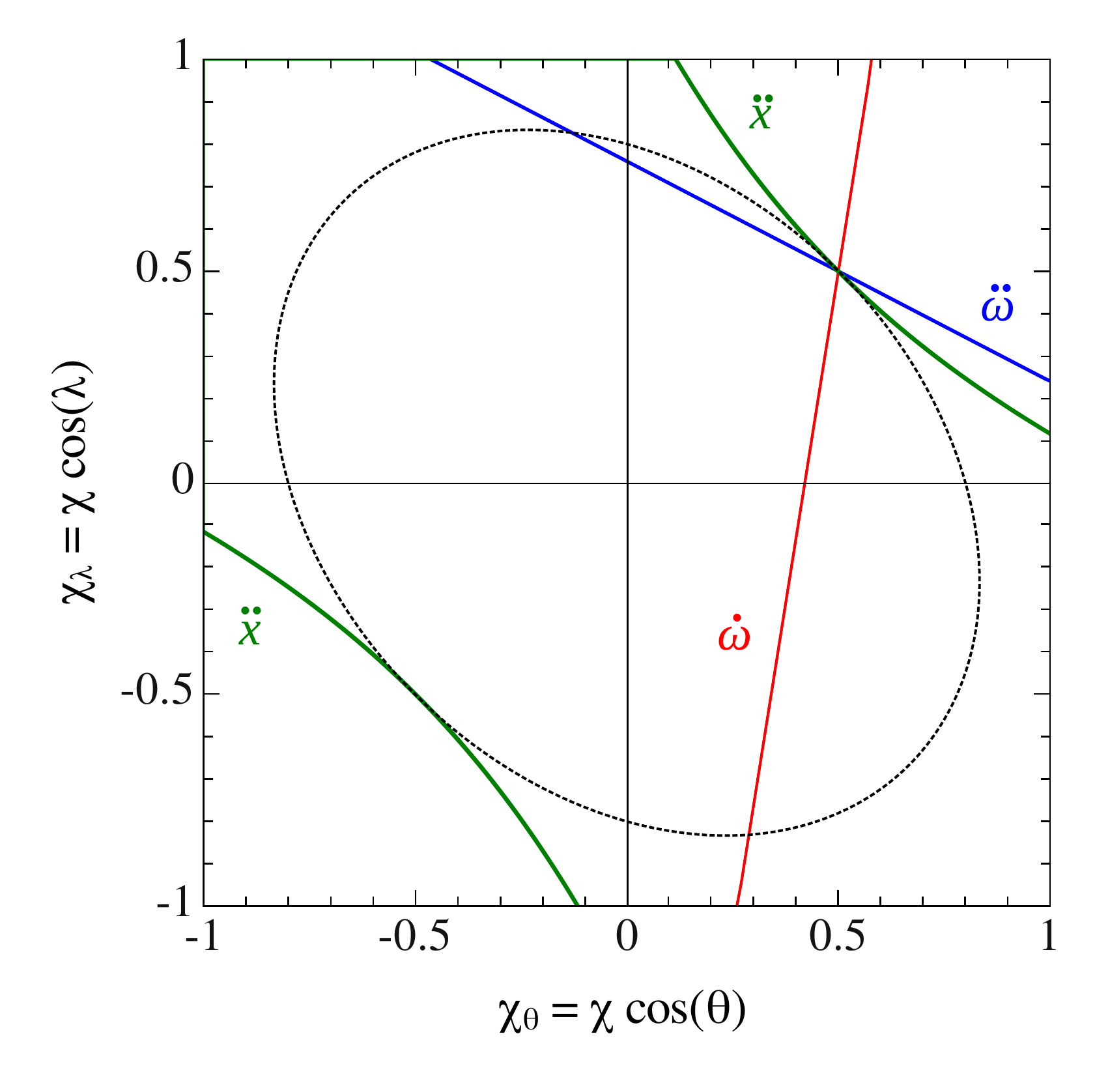,height=2.3in}
\end{center}
\caption{The left panel shows the simulated fractional precision of a mass measurement of Sgr~A$^\ast$ as a function of the orbital period $P_{\rm b}$ for the pericenter precession of the orbit of the pulsar $(\dot{\omega})$, Einstein delay $(\gamma_{\rm E})$, and Shapiro delay $(\Delta_{\rm S})$ assuming weekly measurements of the pulse arrival time with an uncertainty of $100~{\rm \mu s}$ over a time span of five years for a pulsar orbiting around a non-rotating black hole with an orbital eccentricity $e=0.5 $ and inclination $i=60^\circ$. Precision levels of $10^{-6}-10^{−7}$ seem realistic. The right panel illustrates the (simulated) determination of the spin orientation of Sgr~A$^\ast$ in the plane spanned by $\chi_\theta\equiv\chi\cos\theta$ and $\chi_\lambda\equiv\chi\cos\lambda$ (c.f., Fig.~\ref{fig:Pgeometry}) assuming an orbital period of 0.3~yr, an eccentricity $e=0.5$, a spin magnitude $\chi=1$, and angles $\Phi_0=\Psi_0=45^\circ$. The inferred spin magnitude of Sgr~A$^\ast$ in this simulation has a value $\chi=0.9997\pm0.0010$ at 95\% confidence. The dotted ellipse corresponds to the outer boundary of the allowed parameter space for Kerr black holes. Taken from Ref.~\cite{Liu12}.}
\label{fig:Pmassmeasurement}
\end{figure}

The contributions of the spin and the quadrupole moments can, then, be separated from the effect of the mass alone through a fit~\cite{WK99,Kopeikin97}. In practice, the mass is obtained within a model for the pulse arrival time, where the pericenter advance as well as the Shapiro, Roemer, and Einstein time delays are inferred simultaneously; see Ref.~\cite{Liu12}. Reference~\cite{Liu12} simulated the fractional precision of a mass measurement of Sgr~A$^\ast$ for the pericenter precession, Einstein delay, and Shapiro delay as a function of the orbital period of the pulsar. The left panel of Fig.~\ref{fig:Pmassmeasurement} shows the simulated fractional precision of such a mass measurement of Sgr~A$^\ast$ in the case that Sgr~A$^\ast$ does not rotate. This simulation assumes weekly measurements of the pulse arrival time with an uncertainty of $100~{\rm \mu s}$ over a time span of five years as well as an orbital eccentricity $e=0.5$ and inclination $i=60^\circ$ of the pulsar. Precision levels of $10^{-6}-10^{−7}$ seem achievable~\cite{Liu12}.

In order to infer the spin of Sgr~A$^\ast$, the precession of the orbital plane of the pulsar induced by frame-dragging has to be taken into account. For one orbit, the corresponding precession rates of the angles $\Phi$ and $\Psi$ (see Fig.~\ref{fig:Pgeometry}) are given by the expressions~\cite{Barker75} [c.f., Eqs.~(\ref{eq:dUpsilon})--(\ref{eq:dOmega}) and Ref.~\cite{Wex95}]
\ba
\dot\Phi &=& \Omega_{\rm LT} \label{eq:Phidot} \\
\dot\Psi &=& -3\,\Omega_{\rm LT}\,\cos\theta, \label{eq:Psidot}
\ea
where
\be
\Omega_{\rm LT} \equiv \frac{8\pi^2GM}{c^3P_{\rm b}^2} (1-e^2)^{-3/2} \chi
\label{eq:OmegaLT}
\ee
is the Lense-Thirring frequency. The longitude of pericenter $\omega$ and the projected semi-major axis $x$ can then be expressed in terms of a Taylor expansion,
\ba
\omega &=& \omega_0 + \dot\omega_0(t - t_0) + \frac{1}{2}\ddot{\omega}_0(t - t_0)^2 + \ldots, \label{eq:omegaseries}\\
x      &=& x_0      + \dot x_0    (t - t_0) + \frac{1}{2}\ddot{x}_0    (t - t_0)^2 + \ldots, \label{eq:xseries}
\ea
where the coefficients of these expansions and, thereby, the spin magnitude and orientation are obtained from a fit of the timing data~\cite{Wex98,WK99} (c.f., Ref.~\cite{Liu12}).

Reference~\cite{Liu12} also simulated the precision with which the spin and the quadrupole moment of Sgr~A$^\ast$ can be determined with timing observations of a pulsar orbiting around the Galactic center. The right panel of Fig.~\ref{fig:Pmassmeasurement} illustrates the determination of the spin orientation of Sgr~A$^\ast$ in the plane spanned by $\chi_\theta\equiv\chi\cos\theta$ and $\chi_\lambda\equiv\chi\cos\lambda$ as simulated for a pulsar with an orbital period of 0.3~yr, an eccentricity of $e=0.5$, angles $\Phi_0=\Psi_0=45^\circ$, and a spin magnitude $\chi=1$ of Sgr~A$^\ast$. The inferred spin magnitude of Sgr~A$^\ast$ in this simulation has a value $\chi=0.9997\pm0.0010$ at 95\% confidence. Note that the three curves corresponding to the coefficients $\dot{\omega}_0$, $\ddot{\omega}_0$, and $\ddot{x}_0$ of the expansions in Eqs.~(\ref{eq:omegaseries})--(\ref{eq:xseries}) have to intersect in one point if the orbit of a pulsar is unperturbed by other effects, which can serve as an independent test for the presence of such perturbations~\cite{Liu12}.

\begin{figure}[ht]
\begin{center}
\psfig{figure=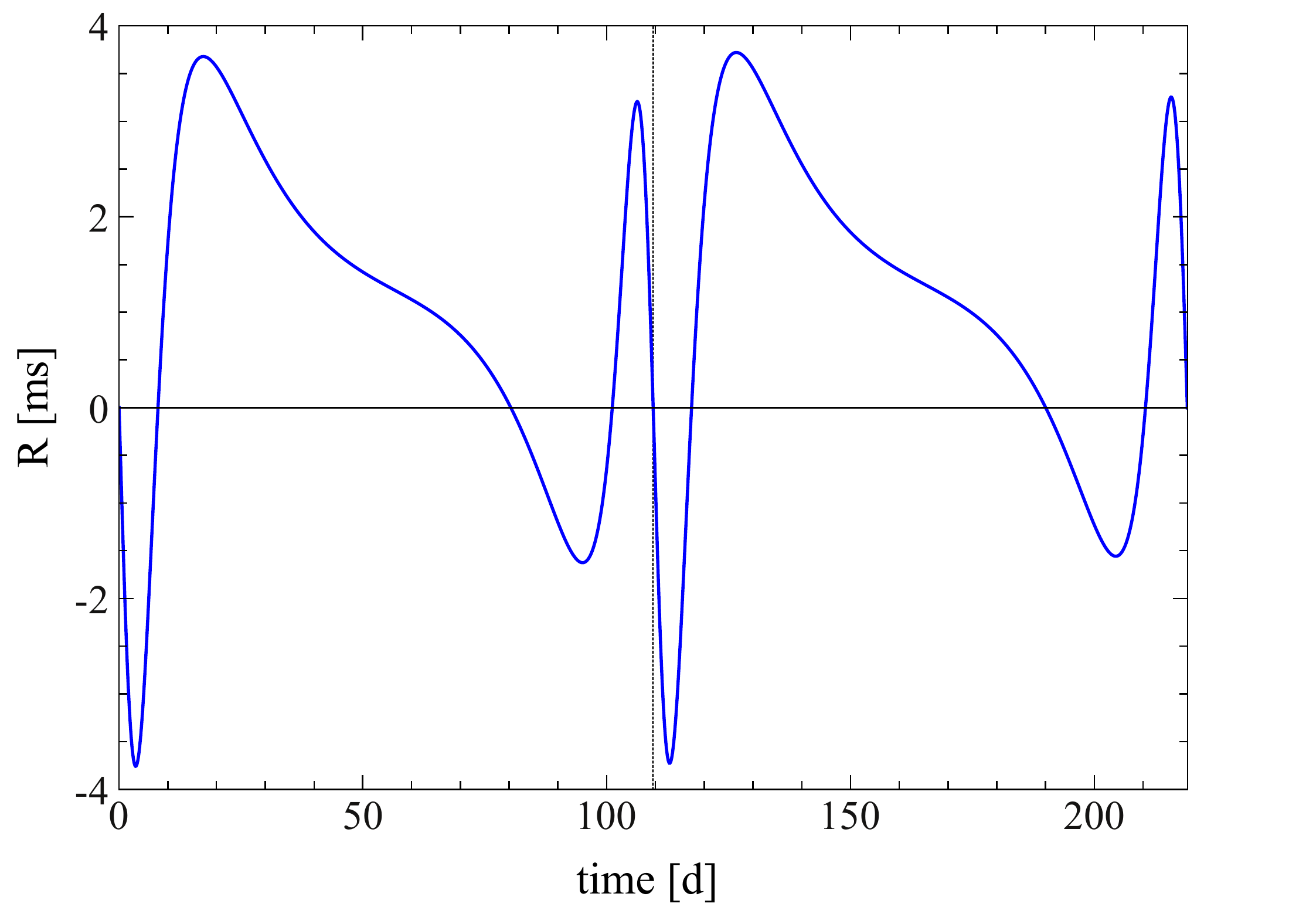,height=2.3in}
\psfig{figure=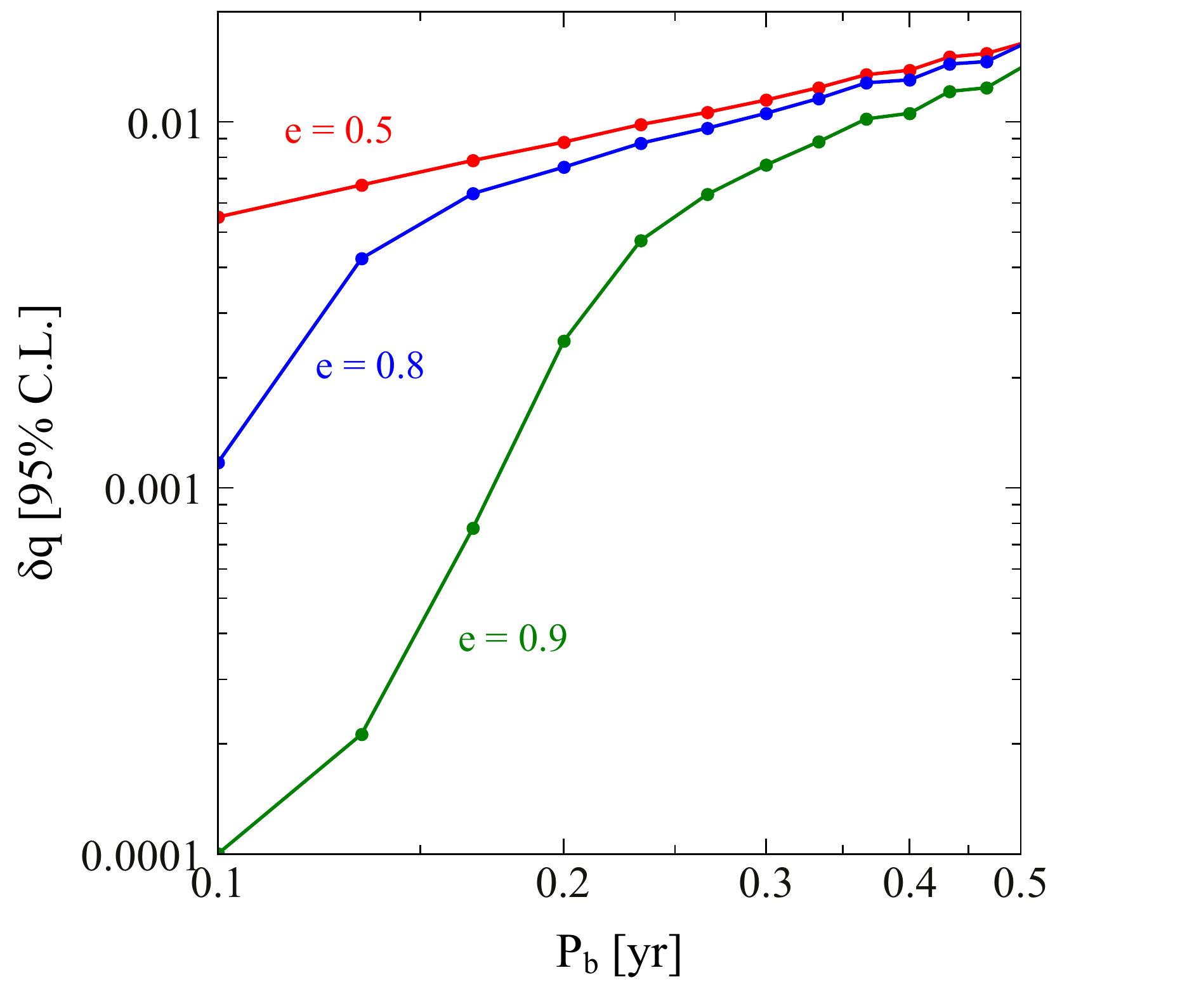,height=2.3in}
\end{center}
\caption{The left panel shows the simulated timing residuals due to the presence of the quadrupole moment of Sgr~A$^\ast$ assuming a maximally-rotating Kerr black hole. The other parameters of the simulation are the same as in the one for the determination of the spin; see the left panel of Fig.~\ref{fig:Pmassmeasurement}. The right panel shows the simulated precision of a measurement of the quadrupole moment for three different eccentricities of the orbit of the pulsar as a function of its orbital period obtainable over a five-year time span in the absence of perturbations. Taken from Ref.~\cite{Liu12}.}
\label{fig:PQ}
\end{figure}

The precession of the orbit of the pulsar induced by the quadrupole moment of Sgr~A$^\ast$ leads to a variation in the Roemer delay which can be identified from the timing residuals of the pulse arrival times once the effects of the mass monopole and frame-dragging have been subtracted~\cite{WK99}. Figure~\ref{fig:PQ} shows such characteristic timing residuals over the span of two orbits as simulated by Ref.~\cite{Liu12} assuming the same parameters as in the simulated determination of the spin of Sgr~A$^\ast$ (see the left panel of Fig.~\ref{fig:Pmassmeasurement}) and a spin value $\chi=1$ of Sgr~A$^\ast$. Figure~\ref{fig:PQ} also shows the precision of a measurement of the quadrupole moment as a function of the orbital period of the pulsar for different values of the eccentricity assuming (on average) weekly observations over a five-year baseline with a higher cadence around the time of pericenter passage. Thus, pulsar timing may test the no-hair theorem with high precision, especially if Sgr~A$^\ast$ has a high value of the spin~\cite{Liu12}.

\begin{figure}[ht]
\begin{center}
\psfig{figure=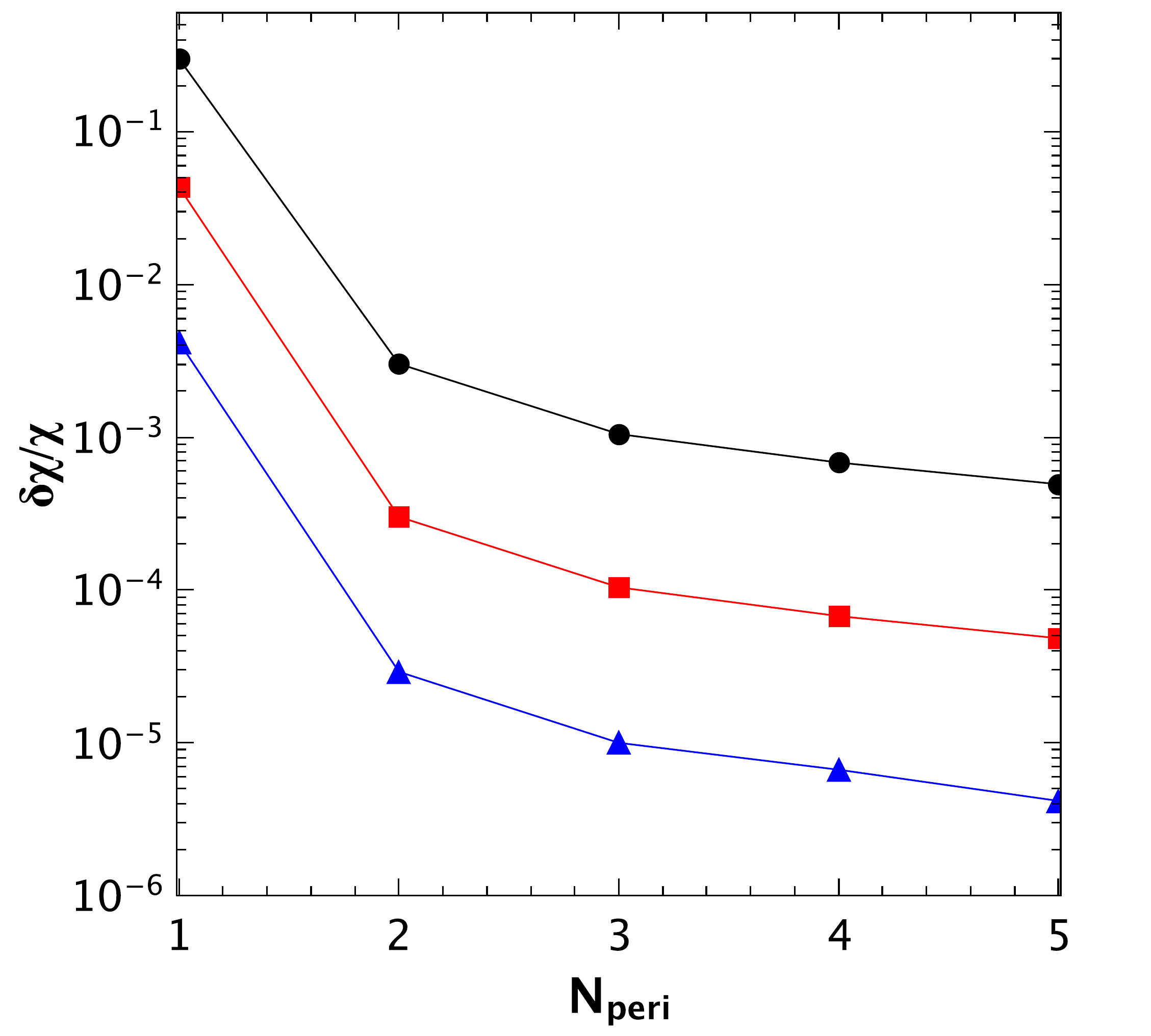,height=2.1in}
\psfig{figure=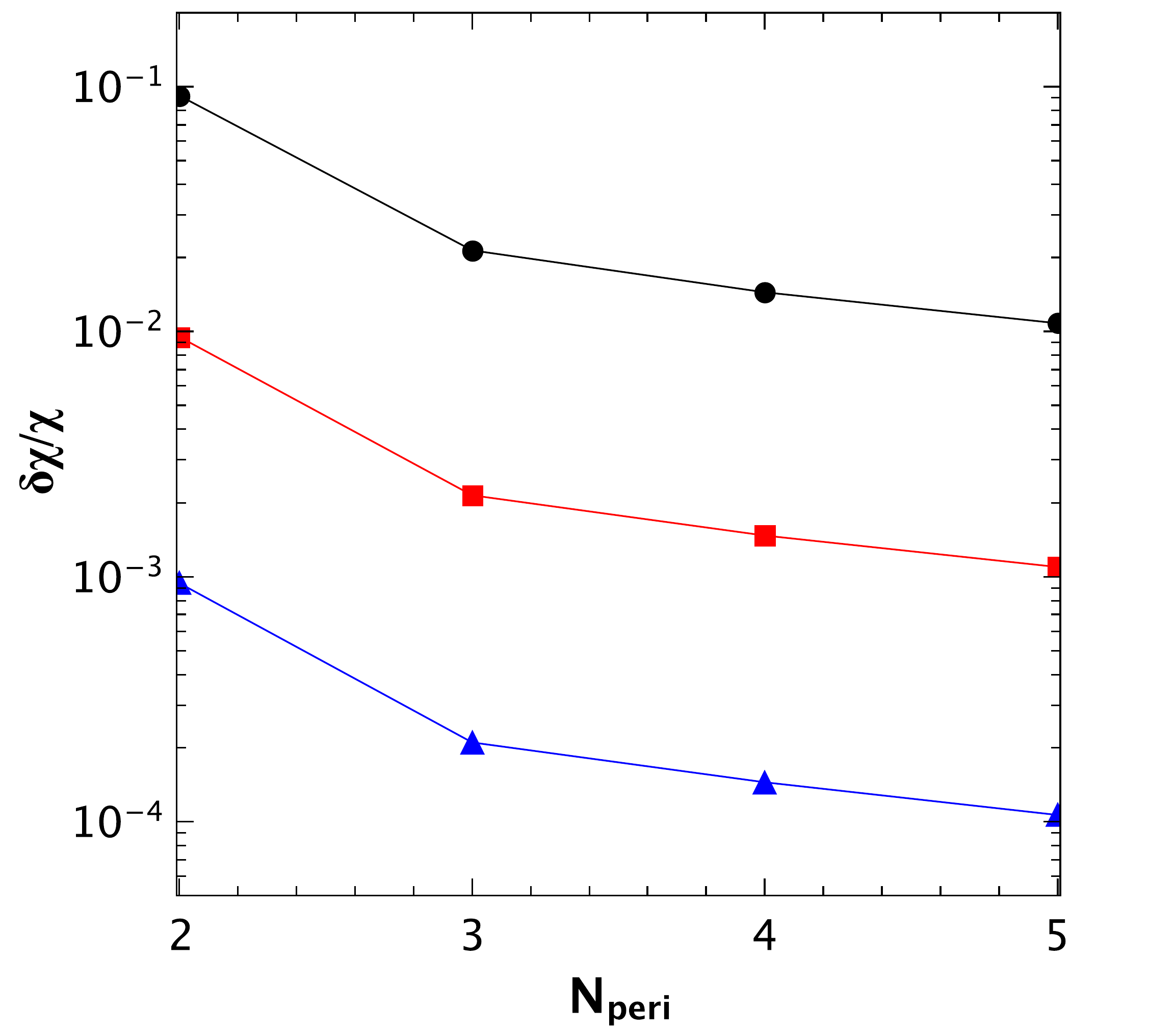,height=2.1in}
\end{center}
\caption{Fractional precision ($2\sigma$) for a measurement of the spin $\chi$ as a function of pericenter passages, based on a dense timing campaign, neglecting (left panel) and including (right panel) the effects of external perturbations. The pulsar has an assumed orbital period of 0.5~yr and an eccentricity of 0.8, while Sgr~A$^\ast$ has an assumed value of the spin $\chi=0.6$. The three curves correspond to timing precisions of $100~{\rm \mu s}$ (black), $10~{\rm \mu s}$ (red), and $1~{\rm \mu s}$ (blue), respectively. The orientation of the spin is the same as in Fig.~\ref{fig:Pmassmeasurement}. Taken from Ref.~\cite{PWK15}.}
\label{fig:PWK_spin}
\end{figure}

\begin{figure}[ht]
\begin{center}
\psfig{figure=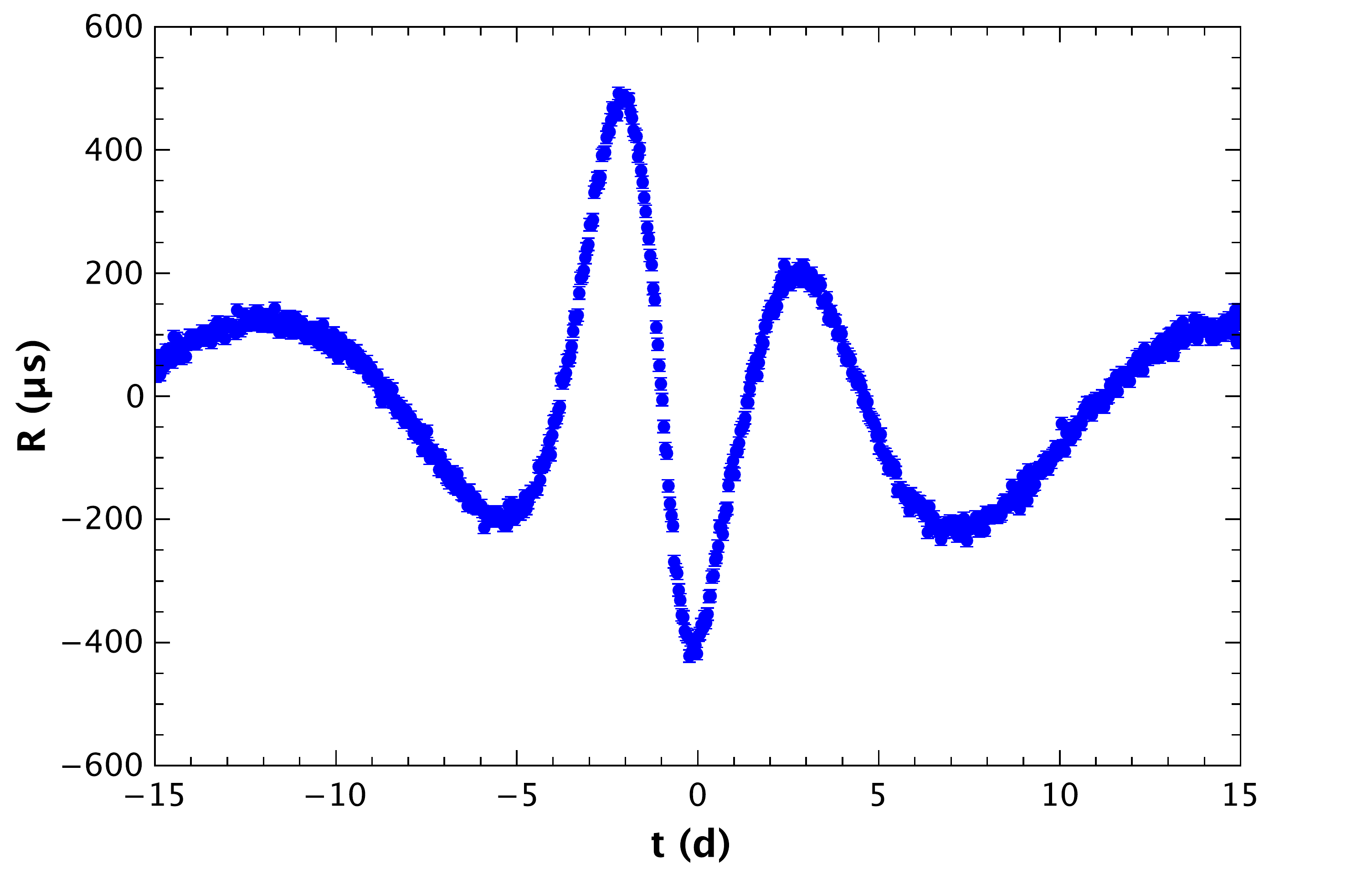,height=1.98in}
\psfig{figure=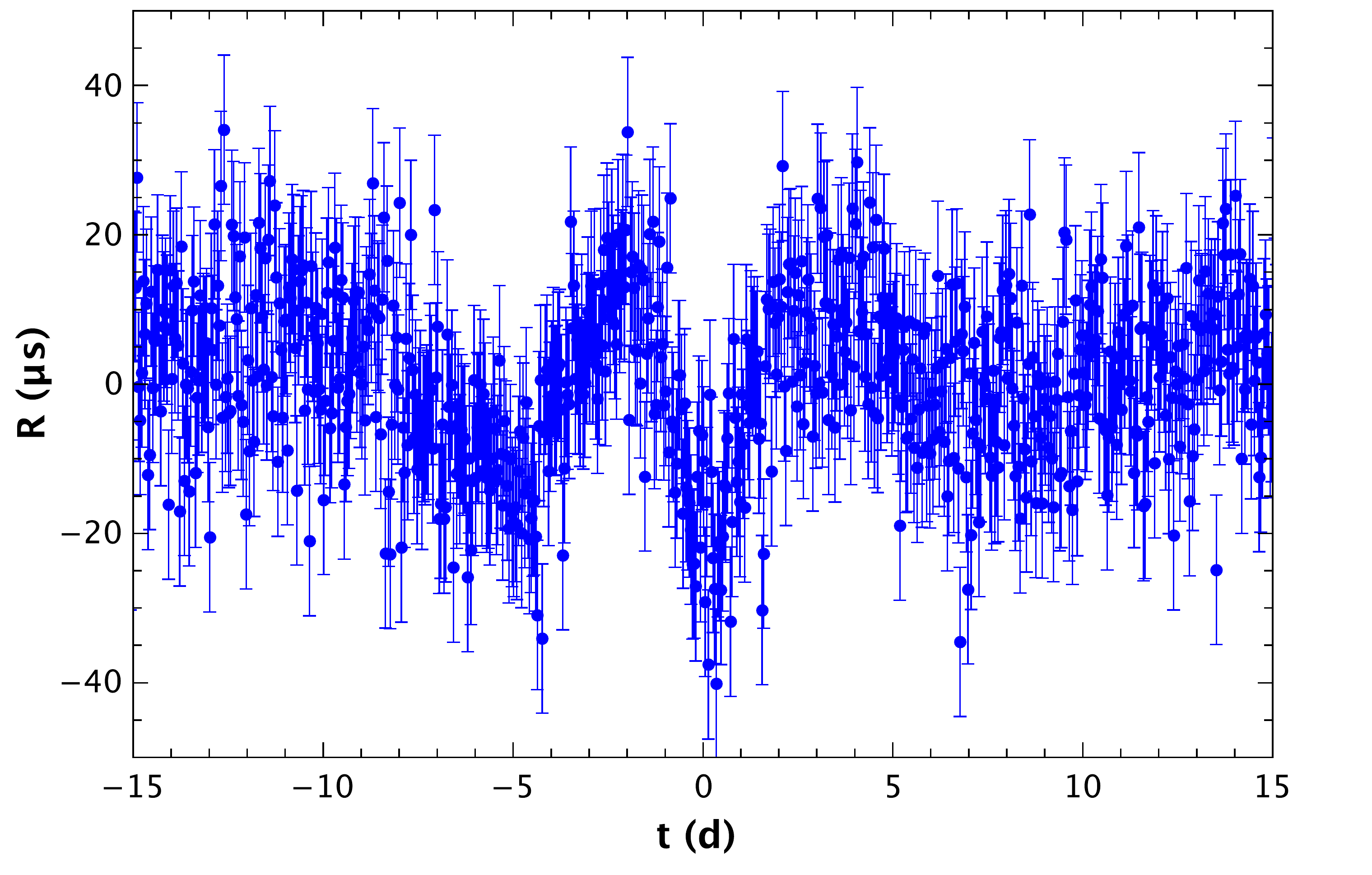,height=1.95in}
\end{center}
\caption{Quadrupolar timing residuals around pericenter passage observed over a few years for the case of a Kerr black hole with values of the spin $\chi=1$ (left panel) and $\chi=0.2$ (right panel), assuming the same pulsar and spin orientation as in Fig.~\ref{fig:PWK_spin} and a timing precision of $=10~{\rm \mu s}$. The quadrupole moment can be inferred with high precision for high values of the spin, but can still be measured accurately even for low values of the spin. Taken from Ref.~\cite{PWK15}.}
\label{fig:PWK_Q1}
\end{figure}

Reference~\cite{PWK15} refined the timing model used in Ref.~\cite{Liu12} by including higher-order post-Newtonian terms derived by Ref.~\cite{Wex95}. Figure~\ref{fig:PWK_spin} shows the fractional precision of a spin measurement as a function of pericenter passages for a pulsar with an orbital period of 0.5~yr and an eccentricity of 0.8 orbiting around a Kerr black hole with a value of the spin $\chi=0.6$, based on a dense timing campaign. The fractional precisions shown in the left and right panels neglect and include the effects of external perturbations, respectively. Figure~\ref{fig:PWK_Q1} shows the quadrupolar timing residuals around pericenter passage observed over a few years for different values of the black-hole spin, assuming the same pulsar and spin orientation as in Fig.~\ref{fig:PWK_spin} and a timing precision of $10~{\rm s}$. The quadrupole moment can be inferred with higher precision for higher values of the spin.

\begin{figure}[ht]
\begin{center}
\psfig{figure=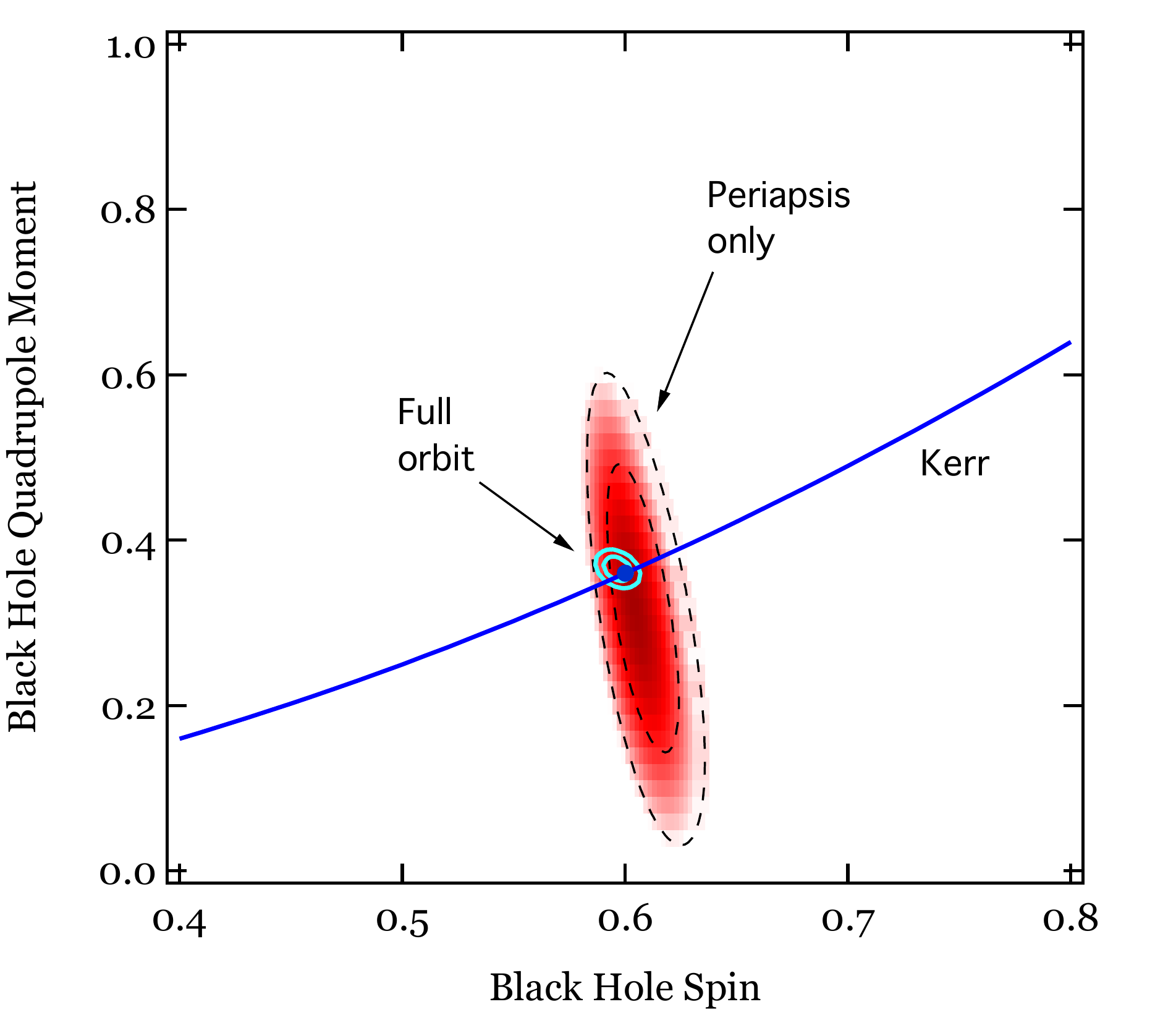,height=2.4in}
\psfig{figure=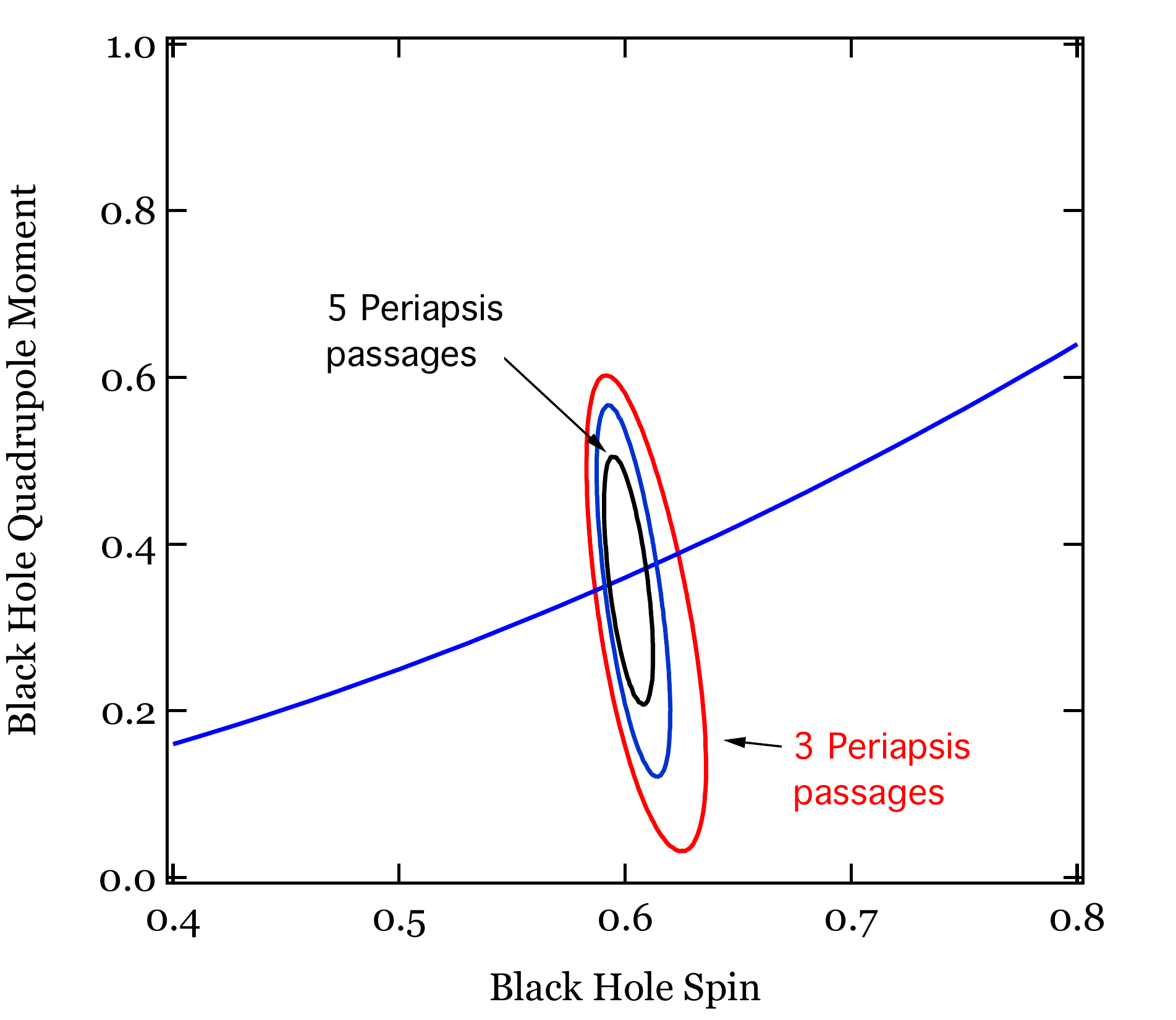,height=2.4in}
\end{center}
\caption{Simulated posterior likelihood of measuring the spin and quadrupole moment of Sgr~A$^\ast$ assuming a Kerr black hole with a value of the spin $\chi=0.6$. In the left panel, the dashed curves show the 68\% and 95\% confidence contours, while, in the right panel, the solid curves show the 95\% confidence contours. The solid curve shows the expected relation between the spin and quadrupole moment of a Kerr black hole. The pulsar is assumed to have an orbital period of 0.5~yr (corresponding to an orbital separation of $\approx2400r_g$) and an eccentricity of 0.8, while three time-of-arrival measurements per day with equal timing uncertainty of $100~{\rm \mu s}$ have been simulated. The left panel compares the uncertainties in the measurement when only three pericenter passages have been considered in the timing solution to those when the three full orbits are taken into account. The right panel shows the increase in the precision of the measurement when the number of pericenter passages is increased from three to five. Taken from Ref.~\cite{PWK15}.}
\label{fig:PWK_Q}
\end{figure}

Figure~\ref{fig:PWK_Q} shows the corresponding posterior likelihoods of measuring the spin and quadrupole moment of Sgr~A$^\ast$ for different observing campaigns assuming a timing precision of $100~{\rm \mu s}$ and a Kerr black hole with a value of the spin $\chi=0.6$. Even in the case of a comparably low timing precision of $100~{\rm \mu s}$ and the presence of external perturbations, a quantitative test of the no-hair theorem is possible after only a few pericenter passages and the spin and quadrupole moment of Sgr~A$^\ast$ can be measured with high precision after a few orbits~\cite{PWK15}.

\begin{figure}[ht]
\begin{center}
\psfig{figure=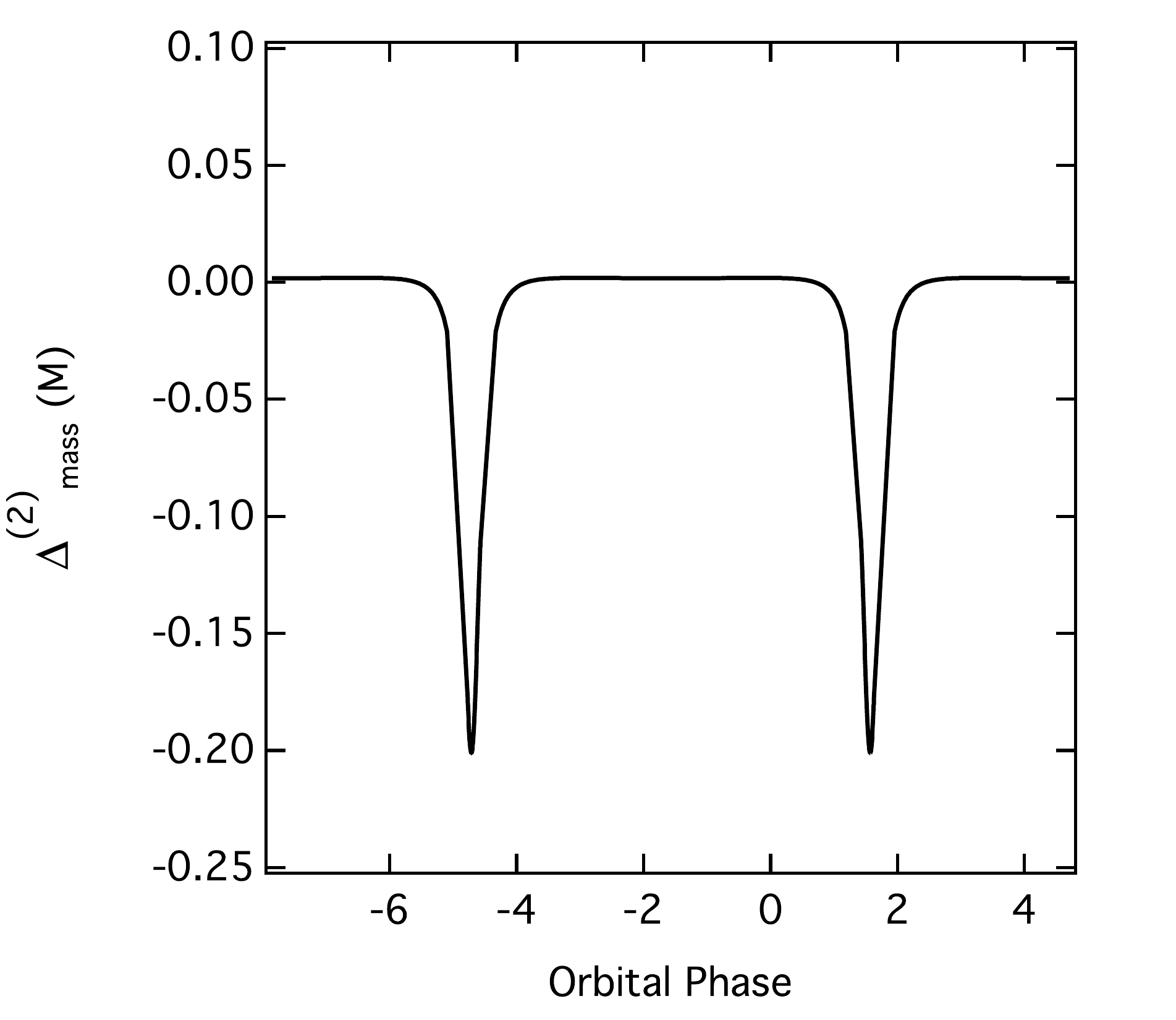,height=2.3in}
\psfig{figure=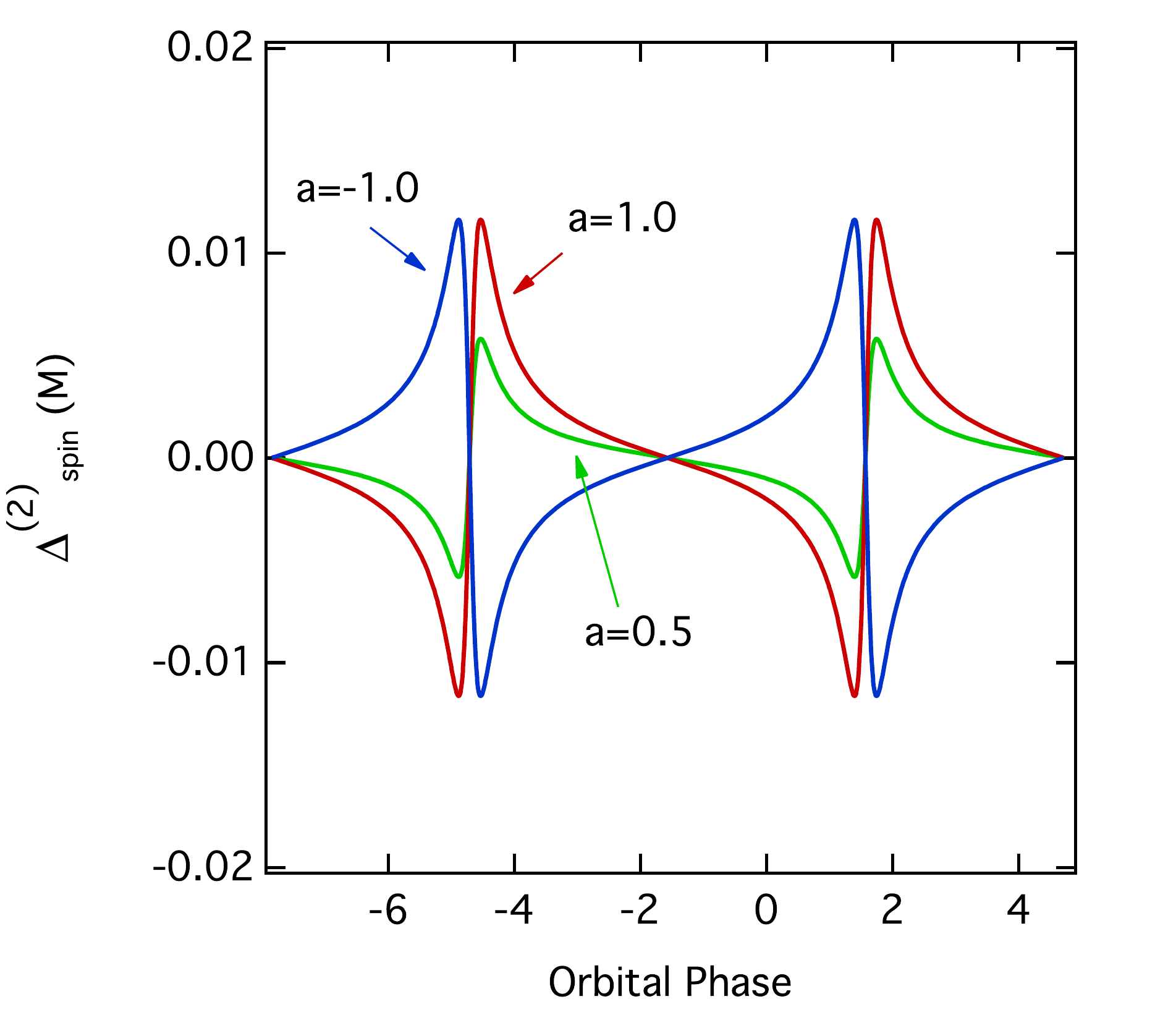,height=2.3in}
\psfig{figure=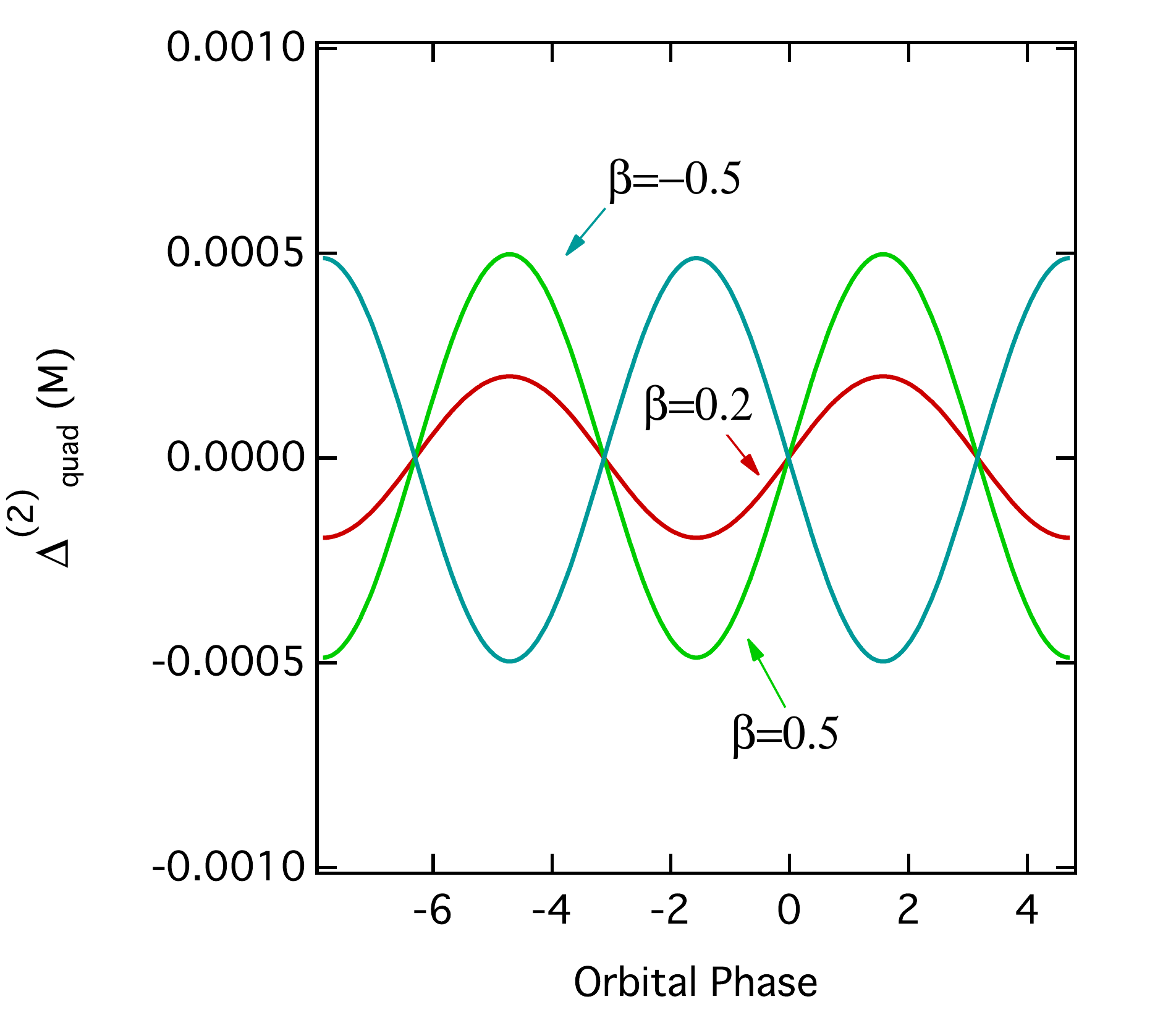,height=2.3in}
\psfig{figure=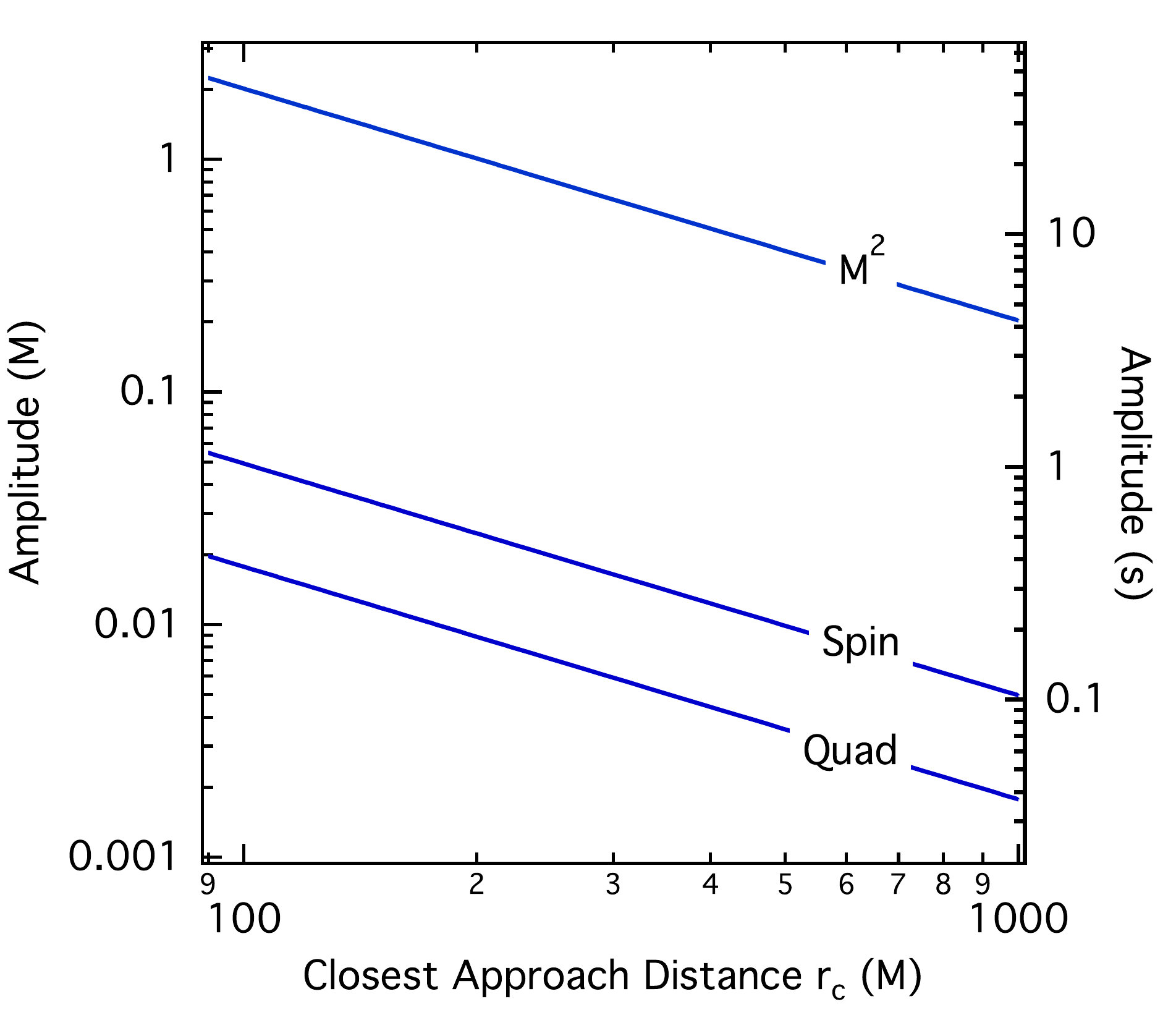,height=2.3in}
\end{center}
\caption{Second-order contributions to the (Shapiro) light travel time delay for a pulsar on a circular orbit around a spinning black hole as a function of orbital phase due to the mass (top left panel), spin (top right panel), and quadrupole moment (bottom left panel) of the black hole. The orbit of the pulsar has a radius of $1000r_g$ and an inclination of $80^\circ$, and the superior conjunction occurs at an orbital phase of $\pi/2$. The top right and bottom left panels show the second-order contributions for different values of the spin $a$ and the (dimensionless) quadrupole parameter $\beta$, respectively. The bottom right panel shows the corresponding amplitudes of the second-order Shapiro delay for a maximally spinning black hole but as a function of the closest approach distance $r_{\rm c}$. The axis on the right assumes a black hole mass $M=4.3\times10^6M_\odot$. Although all three second-order effects are much smaller than the (first-order) Shapiro time delay, they are much larger than the expected measurement uncertainties for observations of pulsars around Sgr~A$^\ast$ with 100m-class radio telescopes or the SKA. Taken from Ref.~\cite{CPL15}.}
\label{fig:CPL}
\end{figure}

Since the orbital parallax of the pulsar also makes a significant contribution to the observed timing signals, the distance of Sgr~A$^\ast$ can likewise be measured using pulsar timing. For $N$ equally distributed time-of-arrival measurements with an uncertainty $\sigma_{\rm TOA}$, the distance can be inferred with a fractional precision given by the equation~\cite{PWK15}
\ba
  \delta D &\sim& 2\,\frac{c\sigma_{\rm TOA}}{\sqrt{N}}\, \left(\frac{D}{a}\right)^2 \nn\\ 
           &\sim& 20\,{\rm pc}  \left(\frac{\sigma_{\rm TOA}}{10^2\,\mu{\rm s}}\right)  \left(\frac{N}{10^3}\right)^{-1/2}  \left(\frac{D}{8.3~{\rm kpc}}\right)^2
                  \left(\frac{a}{10^2\,{\rm au}}\right)^{-2} \;.
\ea

Reference~\cite{CPL15} calculated the Shapiro time delay experienced by photons emitted from a pulsar on an orbit around a black hole to second parameterized post-Newtonian order which also depends on the spin and quadrupole moment of the black hole (see, also, Refs.~\cite{Ashby10,Teyssandier14}). Reference~\cite{CPL15} used the metric of Butterworth and Ipser~\cite{num1,AlGendy14} as the underlying spacetime which describes a stationary and axisymmetric rotating fluid body in general relativity up to order $(GM/rc^2)^2$ in quasi-isotropic coordinates $(t,r,\theta,\phi)$ and depends on the mass $M$, spin $a$, and quadrupolar parameter $\beta$ of the fluid body.

Figure~\ref{fig:CPL} shows the second-order contributions to the Shapiro delay as a function of orbital phase due to the mass, spin, and quadrupole moment of the black hole, as well as the corresponding amplitudes as a function of the closest approach distance. Although all three second-order effects are much smaller than the first-order Shapiro time delay, they are much larger than the expected measurement uncertainties for observations of pulsars around Sgr~A$^\ast$ with 100m-class radio telescopes or the SKA. However, these effects will primarily introduce a small bias to the measurement of the quadrupole moment discussed in Ref.~\cite{PWK15}, because the quadrupole-order time-delay and orbital effects have very different signatures on the time-of-arrival measurements~\cite{CPL15}.

Reference~\cite{Pani16} showed that a binary pulsar orbiting around Sgr~A$^\ast$ could also be used as a probe of the distribution of dark matter at the Galactic center. Such a pulsar and its companion would experience a wind of dark-matter particles that can aﬀect the orbital motion through dynamical friction leading to a characteristic seasonal modulation of the orbit and a secular change of the orbital period~\cite{Pani16}. The strong gravitational lensing of a pulsar orbiting around Sgr~A$^\ast$ could potentially also be used as a probe of certain quantum gravity effects~\cite{Pen14}.

\section{A Framework for Strong-Field Tests}
\label{sec:metrics}

By defintion, strong-field tests of the no-hair theorem cannot rely on the parameterized post-Newtonian formalism and a careful modeling of the underlying spacetime is required instead. Constructing a suitable spacetime for this purpose is a highly nontrivial task. Since it is unclear at present whether general relativity is modified in the strong-field regime and, if so, in what manner, an efficient approach is to test the no-hair theorem using a model-independent framework. Such a phenomenological framework is provided by a parametrically deformed Kerr-like spacetime which encompasses many different theories of gravity at once. Kerr-like metrics generally do not derive from the action of any particular gravity theory. The underlying theory is usually unknown and insight into this theory is hoped to be gained through observations.

Kerr-like metrics are a class of so-called metric theories of gravity~\cite{MTW} typically obeying the full Einstein equivalence principle (EEP)~\cite{Will14}. The EEP is the foundation of a theory of gravity and is comprised of three fundamental principles, the weak equivalence principle (WEP), local Lorentz invariance (LLI), and local position invariance (LPI). The WEP postulates that the trajectory of a freely falling ``test'' body, i.e., a body that is not affected by forces such as electromagnetism or tidal gravitational forces, is independent of its internal structure and composition. In Newtonian gravity, this statement is equivalent to the equality of the inertial and gravitational mass of such a test body. The LLI states that the outcome of any local non-gravitational experiment is independent of the velocity of the freely-falling reference frame in which it is performed. The LPI postulates that the outcome of such an experiment is independent of its position and the time of its performance. It then follows from the EEP that gravitation can be described by the curvature of a spacetime (e.g.,~\cite{Will93}).

The only theories of gravity that are consistent with the EEP are metric theories of gravity. In these metric theories, the spacetime is endowed with a symmetric metric, the trajectories of freely falling test bodies are geodesics of that metric, and in local freely falling reference frames, the non-gravitational laws of physics are those of special relativity. This setup, then, allows for the calculation and prediction of possible observable signatures of the theory. The three components of the EEP have been thoroughly tested by many different experiments, at least in the weak-field regime~\cite{Will14}. Other requirements of the EEP, however, can be relaxed, such as the LLI for black holes in Lorentz-violating theories~\cite{YP09,Pani11,YYT12}.

In this section, I describe Kerr-like metrics and some of their properties, focusing primarily on three particular Kerr-like metrics. Throughout this section, I use geometric units, where $G=c=1$.

\subsection{Kerr-like Metrics}
\label{subsec:metrics}

Following the discovery of the Schwarzschild~\cite{Schwarzschild16} and Kerr~\cite{Kerr63} metrics in 1916 and 1963, respectively, Hartle and Thorne~\cite{HT1,HT2} constructed a metric for slowly rotating neutron stars with arbitrary (but small) quadrupole moments in the late 1960s (see Ref.~\cite{Boshkayev15a} and references therein for alternative forms of this metric). Tomimatsu and Sato~\cite{TS1,TS2} found a discrete family of spacetimes in 1972 that contains the Kerr metric as a special case. In 1985, Quevedo and Mashhoon~\cite{Quevedo85} constructed a metric of a rotating mass with an arbitrary quadrupole moment building on the static metric found by Erez and Rosen~\cite{Erez59,Dorosh65,Young69} in 1959. After a full decade of research, Manko and Novikov~\cite{MN92} found two classes of metrics in 1992 that are characterized by an arbitrary set of multipole moments.

Many exact solutions of the Einstein field equations are now known~\cite{Stephani03}. Of particular interest is the subclass of stationary, axisymmetric, vacuum (SAV) solutions of the Einstein equations, and especially those metrics within this class that are also asymptotically flat. Once an explicit SAV has been found, all SAVs can in principle be generated by a series of HKX-transformations (\cite{HKX1,HKX2} and references therein), which form an infinite-dimensional Lie group~\cite{GerochGroup1,GerochGroup2}. Each SAV is fully and uniquely specified by a set of scalar multipole moments~\cite{BeigSimon80,BeigSimon81} and can also be generated from a given set of multipole moments~\cite{Sibgatullin91,MankoSibgatullin93}. These solutions, however, are generally very complicated and often unphysical. For some astrophysical applications, such as the study of neutron stars, it is oftentimes more convenient to resort to a numerical solution of the field equations~\cite{num1,num2,num3,num4,num5}.

Kerr-like metrics focus on parameteric deviations from the Kerr metric and need not be vacuum solutions in general relativity. Several Kerr-like metrics have been constructed thus far (e.g., \cite{MN92,CH04,GB06,VH10,JPmetric,VYS11,Jmetric,CPR14,Lin16}), which depend on one or more free parameters that measure potential deviations from the Kerr metric and which include the Kerr metric as the special case when all deviations vanish. Observations can then be used to measure these deviations, should they exist, and, thereby, infer properties of the underlying theory of gravity. If no deviations are detected, the compact object is verified to be a Kerr black hole. If, on the other hand, nonzero deviations are measured, there are two possible interpretations. If general relativity still holds, the object is not a black hole but, instead, another stable stellar configuration or, perhaps, an exotic object~\cite{Hughes06}. Otherwise, the no-hair theorem would be falsified. Alternatively, within general relativity, the deviation parameters may also be interpreted as a measure of the systematic uncertainties affecting the measurement so that their effects can be treated in a quantitative manner.

The Kerr metric is the only stationary, axisymmetric, asymptotically flat, vacuum solution to the Einstein equations that possesses an event horizon and is free of timelike curves outside of the horizon. Hence, it uniqely describes black holes in general relativity~\cite{Israel67,Israel68,Penrose69,Carter71,Hawking72,Price72a,Price72b,Carter73,Robinson75}. Due to Hawking's rigidity theorem~\cite{Hawking72}, stationary (asymptotically flat, vacuum) black holes are automatically axisymmetric, and, thus, axisymmetry is not a requirement in general relativity. This, however, need not be the case outside of general relativity. In addition, the Kerr metric possesses a third constant of motion, the Carter constant, making geodesic motion integrable in this spacetime~\cite{Carter68}.

A Kerr-like metric necessarily has to differ from the Kerr metric in at least one of above properties (whether or not it admits the existence of a Carter-like constant). The many proposed metrics in the literature can be divided into two subclasses: those that are Ricci flat, i.e., $R_{\mu \nu} = 0$, and those that are not. In the former case, the metric in the far field satisfies the Laplace equation, and thus, when in asymptotically Cartesian and mass-centered coordinates, it can be expressed as a sum of mass and current multipole moments (see, e.g., Ref.~\cite{Thorne80}). For small deviations from the Kerr metric, one can relate these moments to each other via~\cite{CH04,VH10,Vigeland10}
\be
M_{\ell} + {\rm i}S_{\ell} = M({\rm i}a)^{{\ell}} + \delta M_{\ell} + {\rm i}\delta S_{\ell}\,,
\label{eq:mult}
\ee
where $\delta M_{\ell}$ and $\delta S_{\ell}$ are mass and current multipole deformations.

When the metric is not Ricci flat, the above parameterization of the metric in the far field (as a sum over mass and current multipole moments that depend only on the $\ell$ harmonic number) is not valid. Such metrics generically arise from explicit or implicit modifications to the Einstein-Hilbert action. In these cases, it is not clear what the general structure of a modification of Eq.~(\ref{eq:kerrmult}) would look like.

A second important distinction between different Kerr-like metrics is the degree of nonlinearity of their deviations from the Kerr metric. Some Kerr-like metrics have been defined as small (and, therefore, linear) perturbations away from the Kerr metric, while other metrics are ``exact,'' i.e., they are considered exact (and often nonlinear) solutions to (usually unknown) sets of field equations. This is an important difference, because deviations from the Kerr metric, should they exist, could be large and still satisfy the current observational constraints. Thus, there is no need to a priori limit Kerr-like spacetimes to the description of only small deviations.

In addition, it is sometimes useful in practice to compute the properties of Kerr-like metrics with linear deviations from the Kerr metric to all orders in the deviation parameters, i.e., without expanding the results of such computations to linear order in the deviation parameters. While an expansion in small deviation parameters can always be performed in analytic calculations, it is a lot more difficult and, in some cases, even impossible to enforce in other settings such as the ones involving magnetohydrodynamic simulations, which numerically solve the (nonlinear) geodesic equations. In this interpretation, Kerr-like metrics containing small perturbations from the Kerr metric also have to be considered exact. Similarly, it can be instructive to study nonlinear Kerr-like metrics also in the limit of small deviations, expanding these metrics to first order in the deviation parameters and treating the resulting metrics as perturbative. 

Examples of Kerr-like metrics defined as linear deviations from the Kerr metric include the bumpy Kerr metric~\cite{CH04,VH10}, the quasi-Kerr metric~\cite{GB06}, and the modified-gravity bumpy Kerr metric~\cite{VYS11}. The Manko-Novikov metric~\cite{MN92} and the metrics of Refs.~\cite{JPmetric,Jmetric} are examples of nonlinear Kerr-like metrics. The Manko-Novikov metric is Ricci flat, the quasi-Kerr metric is Ricci flat up to terms containing the quadrupole moment, and the bumpy Kerr metric is a vacuum solution of the linearized Einstein equations if the spin vanishes. The modified-gravity bumpy Kerr metric and the metrics of Refs.~\cite{JPmetric,Jmetric} are not Ricci flat.

On the other hand, the quasi-Kerr, the bumpy Kerr, and the Manko-Novikov metrics, as well as the metric of Ref.~\cite{JPmetric} are stationary, axisymmetric, and do not possess a Carter-like constant, while the modified-gravity bumpy Kerr metric also admits an approximate Carter-like constant. The metric of Ref.~\cite{Jmetric} possess an exact Carter-like constant. All of these metrics are asymptotically flat. The quasi-Kerr, the bumpy Kerr, and the Manko-Novikov metrics harbor naked singularities, while the modified gravity bumpy Kerr metric and the metric of Ref.~\cite{Jmetric} describe black holes. The metric of Ref.~\cite{JPmetric} generally harbors a naked singularity, which is located at the Killing horizon and can have either spherical or disjoint topology, but describes a black hole for small values of the deviation parameter when it is linearized in that parameter. See Ref.~\cite{pathologies} and references therein for a detailed discussion. When linearized in its deviation parameters, the metric of Ref.~\cite{Jmetric} can be mapped to the modified-gravity bumpy Kerr metric in at least certain cases~\cite{Jmetric}. 

\begin{figure}[ht]
\begin{center}
\psfig{figure=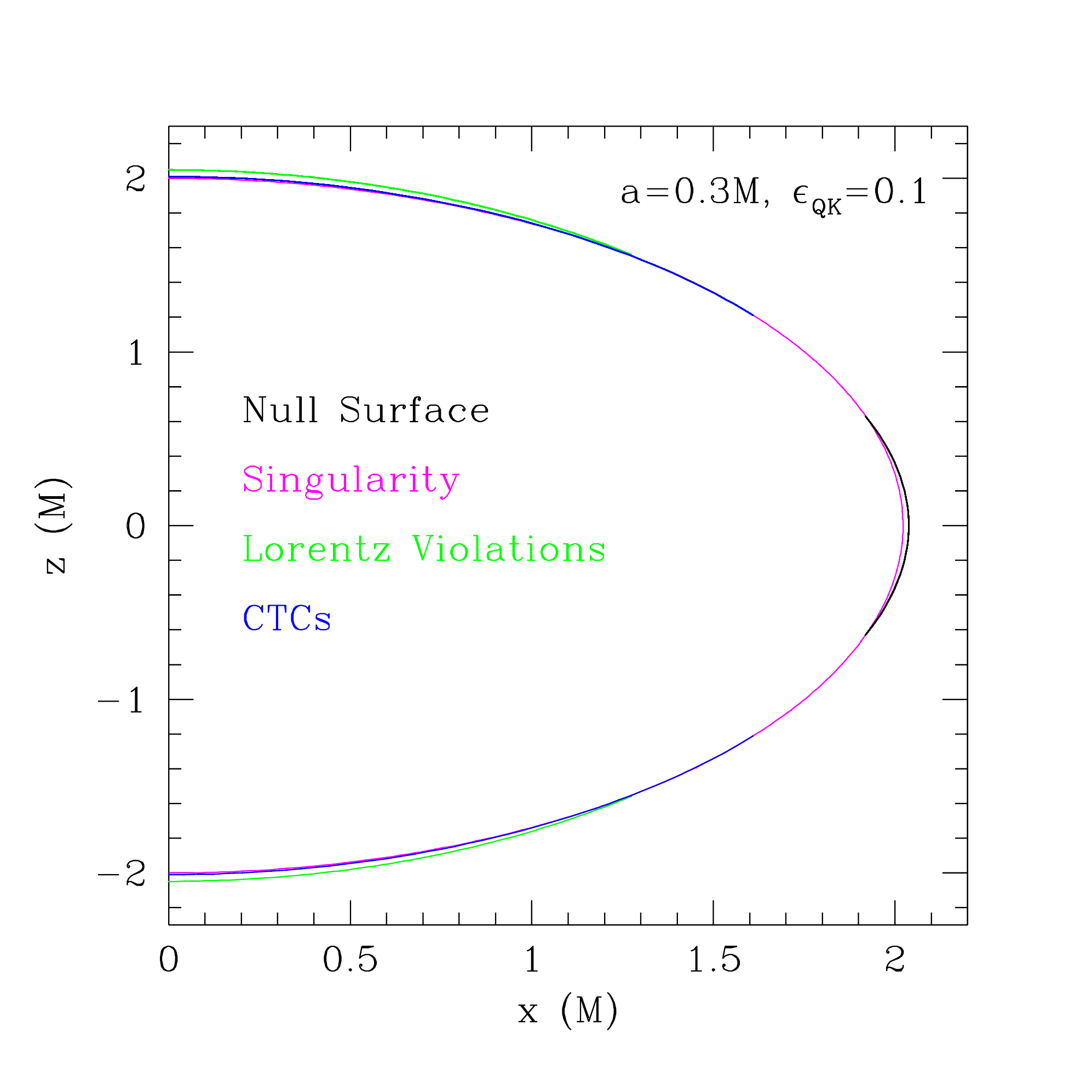,height=2.03in}
\psfig{figure=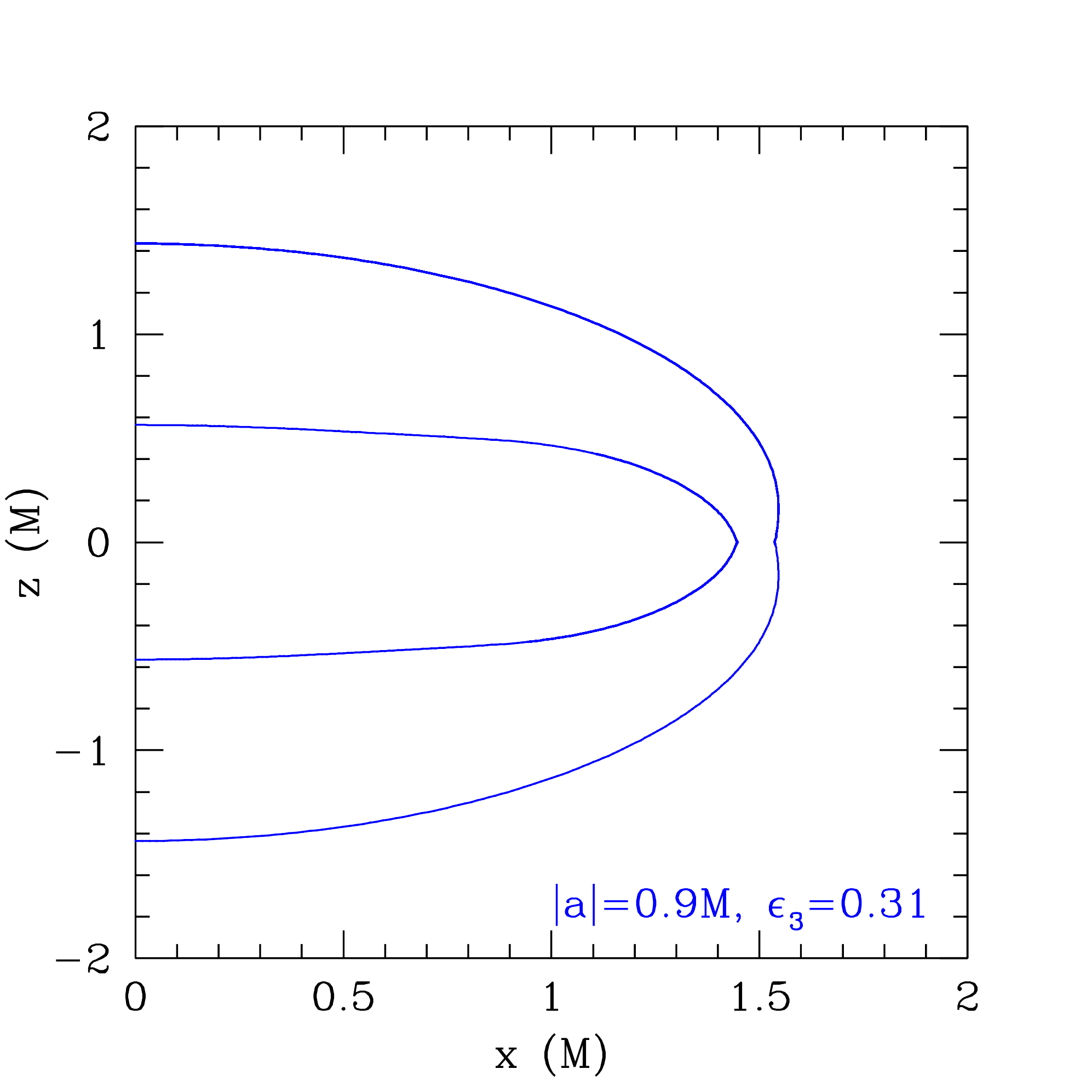,height=2.03in}
\psfig{figure=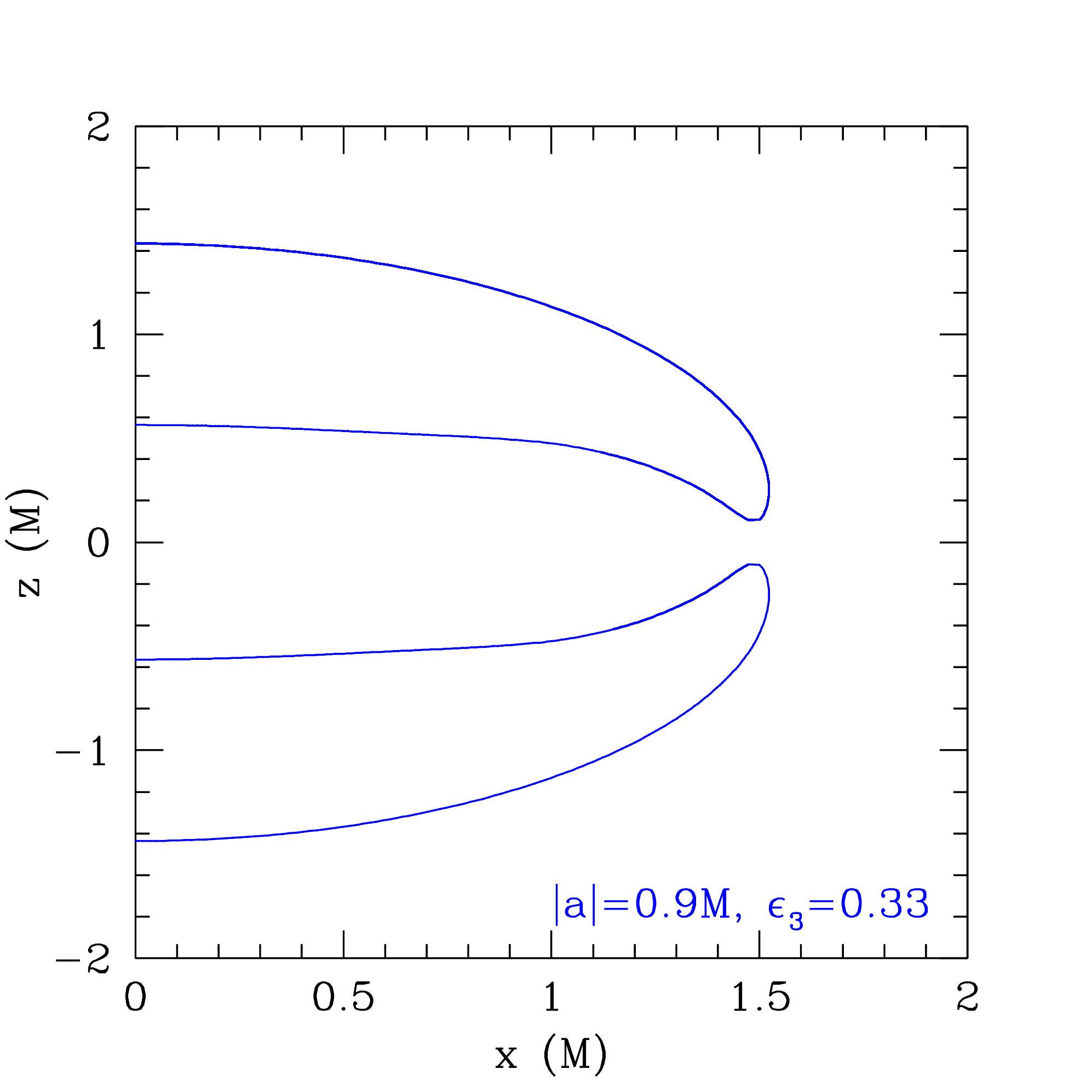,height=2.03in}
\end{center}
\caption{The left panel shows the null surface and regions with Lorentz violations and closed timelike curves (denoted ``CTCs") in the quasi-Kerr metric~\cite{GB06} for values of the spin $a=0.3M$ and the deviation parameter $\epsilon=0.1$. This metric harbors a naked singularity located at the null surface and Lorentz violations as well as closed timelike curves exist around the poles. Such pathological regions have to be excluded by the introduction of a suitable cutoff radius which shields outside observers from their adverse effects. The center and right panels show the location of the naked singularity harbored by the metric of Ref.~\cite{JPmetric} for values of the spin $|a|=0.9M$ and different values of the deviation parameter $\epsilon_3$. At this value of the spin, the naked singularity is of spherical topology if $\epsilon_3\lesssim0.32$ and of disjoint topology otherwise. Taken from Ref.~\cite{pathologies}.}
\label{fig:QKvalidity}
\end{figure}

By construction, however, Kerr-like metrics often contain pathological regions of space where singularities, closed timelike curves, or violations of Lorentz symmetry exist, such as outside of the central objects of the quasi-Kerr, bumpy Kerr, and Manko-Novikov metrics (see Ref.~\cite{pathologies}). The left panel of Fig.~\ref{fig:QKvalidity} shows an example of such regions in the quasi-Kerr metric. These regions are unphysical and have to be excised by introducing a cutoff radius, which acts as an artificial event horizon. All photons and matter particles that pass through this horizon are considered ``captured'' and are excluded from the domain outside of the horizon. the presence of a cutoff radius, therefore, limits the ability of these metrics to serve as a framework for observational tests of the no-hair theorem. They impact both EMRI observations in the gravitational-wave spectrum, as well as electromagnetic observations of accretion flows, since both depend sensitively on the behavior of the metric near the innermost stable circular orbit (ISCO); see the discussion in Ref.~\cite{JPmetric}. Note that causality is violated everywhere in the Kerr metric if the spin exceeds the Kerr bound in Eq.~(\ref{eq:kerrbound}), because, in that case, any event in that spacetime can be connected to any other event by both a future and a past directed timelike curve~\cite{Carter68,Carter73} (see, also, Ref.~\cite{Clarke82}). This property usually also restricts the applicability of Kerr-like metrics to values of the spin for which the Kerr metric harbors a black hole.

The modified-gravity bumpy Kerr metric and the metrics of Refs.~\cite{JPmetric,Jmetric} are free of such pathologies exterior to the central object making them particularly suited for tests of the no-hair theorem. In the case of the metric of Ref.~\cite{JPmetric}, a cutoff radius has to be introduced just outside of the central naked singularity. This, however, does not limit the applicability of this metric in practice, because the cutoff radius can always be chosen so that the ISCO still lies in the domain exterior to the cutoff~\cite{JPmetric}. This metric was later generalized by Ref.~\cite{CPR14} to include two independent types of deviations.

Note, however, that the metrics of Refs.~\cite{VH10,JPmetric,CPR14} have been constructed by the use of the Newman-Janis algorithm~\cite{NewmanJanis1,NewmanJanis2} to generate rotating solutions from static seeds and it is not guaranteed that this procedure can be applied consistently to general metrics which are not solutions of the Einstein field equations. This is not a surprise, because it is still not fully clear why the Newman-Janis algorithm works even in general relativity and what the necessary conditions are so that a static metric can be used as a seed in this method~\cite{Talbot69,Newman72,Schiffer73,Finkelstein75,Gurses75,Quevedo92}. However, there are at least several known examples of black hole solutions other than the Kerr solution for which this is indeed the case~\cite{Dianyon88,Kim99,Yazadjiev00,Whisker06,Adamo14}. Recently, Ref.~\cite{Visser16} constructed a much simpler form of the Newman-Janis algorithm in Kerr-Schild coordinates.

While the existence of a Carter-like constant in the modified-gravity bumpy Kerr metric and the metric of Ref.~\cite{Jmetric} necessarily restricts the scope of these metrics to include only black hole metrics that admit a Carter-like constant, it allows for the separation of the geodesic equations, which can greatly facilitate the study of observables in these spacetimes. Nonetheless, it would be desirable to employ an even more general Kerr-like metric which contains all Kerr-like metrics with at least two constants of motion. At present, however, no such metric is known. In its most general form, the modified-gravity bumpy Kerr metric covers the entire class of stationary, axisymmetric black hole metrics with small deviations from the Kerr metric which admit three constants of motion and have a Carter-like constant that is quadratic in the momentum~\cite{VYS11}. Whether or not the corresponding property holds for the nonlinear metric of Ref.~\cite{Jmetric} is unclear; see the discussion in Ref.~\cite{Jmetric}.

Recently, Ref.~\cite{Lin16} proposed a Kerr-like metric in the form of a Kerr metric for which the mass $M$ is replaced by two deviation functions $m_1(r)$ and $m_2(r)$ which reduce to the mass $M$ if all deviations from the Kerr metric vanish. This metric harbors a black hole and is free of curvature singularities outside of the event horizon~\cite{Lin16}. As can easily be seen from the form of the metric elements, for certain ranges of the deviation parameters the exterior domain in this metric is also free of pathological regions.

This metric can be mapped to the static black hole solution found by Bardeen~\cite{BardeenBH} (which describes black holes in general relativity with a magnetic monopole coupled to a nonlinear electromagnetic field~\cite{BeatoGarcia00}) and a corresponding rotating solution constructed by Ref.~\cite{BambiModesto13} based on the Newman-Janis algorithm~\cite{NewmanJanis1,NewmanJanis2}. Whether or not this stationary solution belongs to the same theory and physical setup as the static solution by Bardeen is unclear (c.f., the discussion on the applicability of the Newman-Janis algorithm above). The spacetime of Ref.~\cite{Lin16} can also be mapped to several metrics in certain quantum-gravity inspired scenarios~\cite{Nicolini06,Smailagic10,Modesto10,Zhangetal15}. See Ref.~\cite{Lin16} for these mappings. I discuss a generalization of this metric together with a mapping to the metric of Ref.~\cite{Jmetric} in \ref{appLin16}. References~\cite{Abdujabbarov15,Zhidenko16} expressed generic deviations from the Kerr metric in terms of a continued-fraction expansion.

Here, I focus on three particular Kerr-like metrics, which have been used frequently in the context of no-hair tests with electromagnetic observations of Sgr~A$^\ast$ and of black holes in general: the quasi-Kerr metric~\cite{GB06}, the metric of Ref.~\cite{JPmetric}, and the metric of Ref.~\cite{Jmetric}.

The quasi-Kerr metric derives from the Hartle-Thorne metric~\cite{HT1,HT2} and contains an independent quadrupole moment which is not assumed to depend on mass and spin through Eq.~(\ref{eq:kerrmult}). The quasi-Kerr metric modifies the quadrupole moment of the Kerr metric by the amount
\begin{equation}
\delta M_2 = -\epsilon M^3,
\end{equation}
where the parameter $\epsilon$ measures deviations from the Kerr metric. The full quadrupole moment is then
\begin{equation}
M_2 = -M\left(a^2+\epsilon M^2\right).
\label{eq:qradmoment}
\end{equation}
In Boyer-Lindquist-like coordinates, i.e., in spherical-like coordinates that reduce to Boyer-Lindquist coordinates in the Kerr limit, the quasi-Kerr metric $g_{\rm \mu\nu}$ is given by the expression
\be
g_{\rm \mu\nu} = g_{\rm \mu\nu}^{\rm K} + h_{\rm \mu\nu},
\label{eq:QKmetric}
\ee
where the correction $h_{\rm \mu\nu}$ to the Kerr metric $g_{\rm \mu\nu}^{\rm K}$ in Eq.~(\ref{eq:kerr}) is diagonal with the components (of the contravariant metric)
\ba
h^{tt}&=&\left(1-\frac{2M}{r}\right)^{-1}\left[\left(1-3\cos^2\theta\right)\mathcal{F}_1(r)\right], \nonumber \\
h^{rr}&=&\left(1-\frac{2M}{r}\right)\left[\left(1-3\cos^2\theta\right)\mathcal{F}_1(r)\right], \nonumber \\
h^{\theta\theta}&=&-\frac{1}{r^2}\left[\left(1-3\cos^2\theta\right)\mathcal{F}_2(r)\right], \nonumber \\
h^{\phi\phi}&=&-\frac{1}{r^2\sin^2\theta}\left[\left(1-3\cos^2\theta\right)\mathcal{F}_2(r)\right].
\label{eq:h}
\ea
The functions $\mathcal{F}_{1,2}(r)$ are given in Appendix~A of Ref.~\cite{GB06}. Recently, the Hartle-Thorne metric was extended to include terms that are of higher order in the quadrupole moment~\cite{Frutos15,Frutos15b}.

While the quasi-Kerr metric has the advantage of being of a relativily simple form, it depends on only one deviation parameter and, strictly speaking, can only be applied to slowly to moderately spinning compact objects. The metric of Ref.~\cite{JPmetric} depends on one infinite set of deviation parameters, which are the coefficients of a series expansion of a deviation function $h(r,\theta)$ (which could also be of a more general form~\cite{JPmetric}). The nonvanishing components of this metric can be written as
\ba
g_{tt}&=&-[1+h(r,\theta)] \left(1-\frac{2Mr}{\Sigma}\right), \nn \\
g_{rr}&=&\frac{ \Sigma[1+h(r,\theta)] }{ \Delta + a^2\sin^2\theta h(r,\theta) }, \nn \\
g_{\theta\theta}&=&\Sigma, \nn \\
g_{\phi\phi}&=&\left[ \sin^2\theta \left( r^2 + a^2 + \frac{ 2a^2 Mr\sin^2\theta }{\Sigma} \right) + h(r,\theta) \frac{a^2(\Sigma + 2Mr)\sin^4\theta }{\Sigma} \right], \nn \\
g_{t\phi}&=&-\frac{ 2aMr\sin^2\theta }{ \Sigma }[1+h(r,\theta)] \label{eq:JPmetric},
\ea
where
\be
h(r,\theta) \equiv \sum_{k=1}^\infty \left( \epsilon_{2k} + \epsilon_{2k+1}\frac{Mr}{\Sigma} \right) \left( \frac{M^2}{\Sigma} \right)^{k}
\label{h(r,theta)}
\ee
and where $\epsilon_2$ is usually set to zero in order to be consistent with the parameterized post-Newtonian constraints on deviations from general relativity (c.f., Ref.~\cite{Will14}). The lower-order coefficients $\epsilon_0$ and $\epsilon_1$ are neglected here. The coefficient $\epsilon_0$ vanishes due to the requirement that the metric be asymptotically flat, and the coefficient $\epsilon_1$ is likewise strongly constrained by weak-field tests of gravity. The latter coefficient can also be absorbed into a trivial rescaling of the mass and, thus, plays no role~\cite{Jmetric}. The generalization of this metric by Ref.~\cite{CPR14} decouples the set of coefficients $\epsilon_k$ into two independent infinite sets of deviation parameters $\epsilon_k^t$, $\epsilon_k^r$ and reduces to the metric of Ref.~\cite{JPmetric} when $\epsilon_k^t=\epsilon_k^r$ for all $k$~\cite{CPR14}.

The metric of Ref.~\cite{JPmetric} harbors a naked singularity unless it is expanded to linear order in the deviation parameters in which case it describes a black hole for small deviations from the Kerr metric. Focusing only on the lowest-order nonvanishing deviation parameter $\epsilon_3$, the event horizon is then locted at the radius
\be
r_H = r_+^{\rm K} \left[ 1 - \frac{ \epsilon_3 a^2 M^3 \sin^2\theta }{ 2\sqrt{M^2-a^2} \left( 2M r_+^{\rm K} - a^2\sin^2\theta \right)^2 } \right],
\ee
where $r_+^{\rm K}$ is the event horizon of a Kerr black hole, see Eq.~(\ref{eq:kerrhor}). If no event horizon exists, the naked singularity is located at the Killing horizon which can be of either spherical or disjoint topology. Again, setting all deviation parameters other than $\epsilon_3$ to zero, the Killing horizon is of spherical topology for values of the deviation parameter $\epsilon_3\leq \epsilon_3^{\rm bound}$ and of disjoint topology otherwise, where
\ba
\epsilon_3^{\rm bound} &\equiv& \frac{1}{3125(a/M)^2} \bigg[ 1024 \left( 4 + \sqrt{16-15(a/M)^2} \right) \nn \\
&&- 160(a/M)^2 \left(40+7\sqrt{16-15(a/M)^2} \right) \nn \\
&& + 150 (a/M)^4 \left(15+\sqrt{16-15(a/M)^2} \right) \bigg]
\label{ep3bound}
\ea
assuming $|a|\leq M$~\cite{pathologies}. The center and left panels of Fig.~\ref{fig:QKvalidity} show the location of the naked singularity with spherical and disjoint topology, respectively, for a value of the spin $|a|=0.9M$. The metric of Ref.~\cite{CPR14} has similar properties~\cite{CPR14}.

Still, even with the expanded scope of this metric compared to the one of the quasi-Kerr metric, it is possible to introduce additional degrees of freedom with suitably chosen deviation functions in more general Kerr-like metrics. In general relativity, stationary, axisymmetric, and asymptotically flat metrics that admit the existence of integrable two-dimensional hypersurfaces generally depend on only four functions. As a consequence of Frobenius's theorem, such hypersurfaces are automatically guaranteed to exist if the spacetime is also vacuum. Such metrics can be written in the form of the Papapetrou line element where the metric is expressed with respect to the Weyl-Papapetrou coordinates and has only three metric functions. Out of these functions only two are independent, while the third one can be derived from the other two.

This happens for three reasons. First, the symmetries imposed, the assumption of asymptotic flatness and the vanishing of the Ricci tensor allow for the spacetime to have integrable two-dimensional hypersurfaces that are orthogonal to the two Killing fields and on which one can define coordinates that can be carried along integral curves of these Killing fields to the rest of the spacetime. Thus the metric can be written in a $2\times2$ block form. If one chooses as one of the coordinates the determinant of the $(t,\phi)$ part of the metric, then the metric can be written in a form that has only four independent functions. Second, the field equations imply, by the vanishing of the Ricci tensor (i.e., the vacuum assumption), that the coordinate $\rho$ which is defined by the determinant of the $(t,\phi)$ part of the metric is a harmonic function and thus one can define the second coordinate on the two-dimensional hypersurfaces as the harmonic conjugate of $\rho$ and absorb one of the functions in the process, reducing the independent functions to three. Finally the vacuum field equations imply that the third of the functions is related to the other two and can be determined up to the addition of a constant \cite{Wald84}.

In alternative theories of gravity, however, the vacuum assumption does not necessarily imply the existence of two-dimensional integrable hypersurfaces. Therefore, one would expect that metrics which describe black holes in alternative theories of gravity depend on at least four independent functions. This motivated the construction of the Kerr-like metric of Ref.~\cite{Jmetric} (and, earlier, of the modified-gravity bumpy Kerr metric~\cite{VYS11}), which has the nonvanishing components
\ba
g_{tt} &=& -\frac{\tilde{\Sigma}[\bar{\Delta}-a^2A_2(r)^2\sin^2\theta]}{[(r^2+a^2)A_1(r)-a^2A_2(r)\sin^2\theta]^2}, \nn \\
g_{t\phi} &=& -\frac{a[(r^2+a^2)A_1(r)A_2(r)-\bar{\Delta}]\tilde{\Sigma}\sin^2\theta}{[(r^2+a^2)A_1(r)-a^2A_2(r)\sin^2\theta]^2}, \nn \\
g_{rr} &=& \frac{\tilde{\Sigma}}{\bar{\Delta} A_5(r)}, \nn \\
g_{\theta \theta} &=& \tilde{\Sigma}, \nn \\
g_{\phi \phi} &=& \frac{\tilde{\Sigma} \sin^2 \theta \left[(r^2 + a^2)^2 A_1(r)^2 - a^2 \bar{\Delta} \sin^2 \theta \right]}{[(r^2+a^2)A_1(r)-a^2A_2(r)\sin^2\theta]^2},
\label{eq:Jmetric}
\ea
where
\ba
\bar{\Delta} &\equiv& \Delta + \beta M^2,
\label{eq:beta}\\
A_1(r) &=& 1 + \sum_{n=3}^\infty \alpha_{1n} \left( \frac{M}{r} \right)^n, 
\label{eq:A1}\\
A_2(r) &=& 1 + \sum_{n=2}^\infty \alpha_{2n} \left( \frac{M}{r} \right)^n, 
\label{eq:A2}\\
A_5(r) &=& 1 + \sum_{n=2}^\infty \alpha_{5n} \left( \frac{M}{r} \right)^n, 
\label{eq:A5}\\
\tilde{\Sigma} &=& \Sigma + f(r), 
\label{eq:Sigmatilde}\\
f(r) &=& \sum_{n=3}^\infty\epsilon_n \frac{M^n}{r^{n-2}}.
\label{eq:f}
\ea

The metric of Ref.~\cite{Jmetric} contains the four free functions $f(r)$, $A_1(r)$, $A_2(r)$, and $A_5(r)$ that depend on four sets of parameters which measure potential deviations from the Kerr metric. In addition, this metric depends on the deviation parameter $\beta$. In the case when all deviation parameters vanish, i.e., when $f(r)=0$, $A_1(r)=A_2(r)=A_5(r)=1$, $\beta=0$, this metric reduces to the Kerr metric in Eq.~(\ref{eq:kerr}). Formally, the parametrization also includes the Kerr-Newman metric and potential deviations from it if $\beta=Q^2/M^2$, where $Q$ is the electric charge of the black hole. However, astrophysical black holes should be electrically neutral, because any residual electric charge is expected to neutralize quickly. Therefore, I will treat the parameter $\beta$ primarily as a pure deviation from the Kerr metric.

The deviation functions in Eqs.~(\ref{eq:A1})--(\ref{eq:f}) are written as power series in $M/r$ (but could also be of a more general form~\cite{Jmetric}). The lowest-order coefficients of these series vanish so that the deviations from the Kerr metric are consistent with all current weak-field tests of general relativity (c.f., Ref.~\cite{Will14}) as in the case of the metric of Ref.~\cite{JPmetric} discussed above. However, certain restrictions on these functions and on the deviation parameter $\beta$ exist which are determined by the properties of the event horizon. 

The event horizon itself is independent of all deviation parameters except for the parameter $\beta$ and is located at the radius
\be
r_+ \equiv M+\sqrt{M^2-a^2-\beta M^2}.
\label{eq:KNhor}
\ee
Thus, the event horizon coincides with the event horizon of a Kerr black hole if the parameter $\beta$ vanishes.

\begin{figure}[ht]
\begin{center}
\psfig{figure=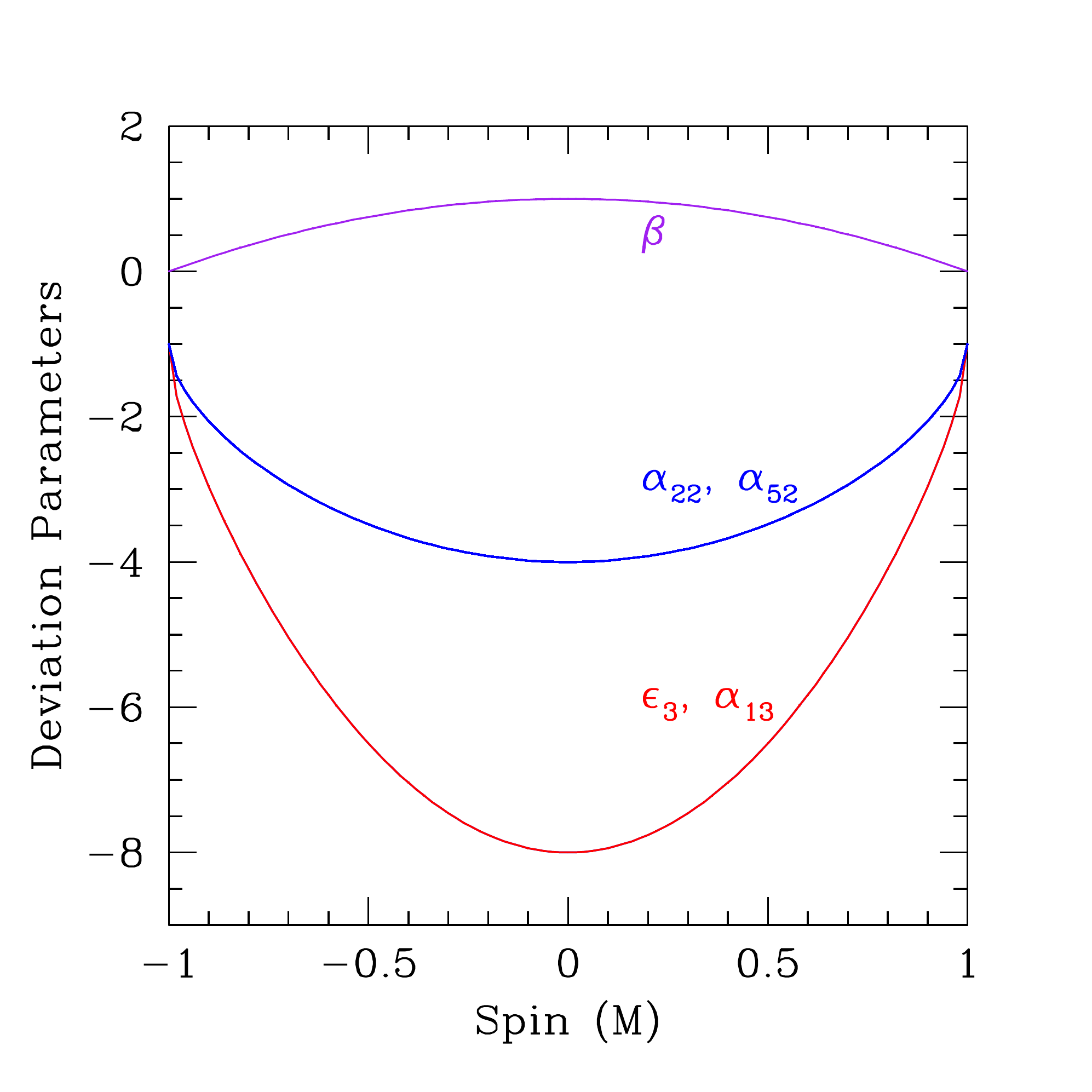,height=2.4in}
\end{center}
\caption{Allowed values of the deviation parameters $\epsilon_3$, $\alpha_{13}$, $\alpha_{22}$, $\alpha_{52}$, and $\beta$ as a function of the spin neglecting higher-order terms in the deviation functions. The purple line shows the maximum values of the parameter $\beta$, while the blue and red lines show the minimum values of the parameters $\alpha_{22}$, $\alpha_{52}$ and $\epsilon_3$, $\alpha_{13}$, respectively.}
\label{fig:allowedvalues}
\end{figure}

In order for an event horizon exist, the parameter $\beta$ must obey the usual relation
\be
\beta M^2 \leq M^2-a^2
\label{eq:lowerbounds1}
\ee
and the functions $\tilde{\Sigma}$, $A_1(r)$, $A_2(r)$, and $A_5(r)$ have to be positive everywhere on and outside of the event horizon. In the case of the lowest-order metric, i.e., when this metric is truncated at the lowest nonvanishing order in the deviation parameters, the latter requirement can be rewritten as the relations
\ba
\epsilon_3 &>& B_3,~~~~~\alpha_{13} > B_3, \nn \\
\alpha_{22} &>& B_2,~~~~~\alpha_{52} > B_2,
\ea
where
\ba
B_2 &\equiv& - \frac{\left(M+\sqrt{M^2-a^2}\right)^2}{M^2}, \nn \\
B_3 &\equiv& - \frac{\left(M+\sqrt{M^2-a^2}\right)^3}{M^3}.
\label{eq:lowerbounds2}
\ea
Otherwise, this metric harbors a naked singularity instead of a black hole~\cite{Jmetric}. Figure~\ref{fig:allowedvalues} shows the allowed values of the five deviation parameters in the lowest-order metric as a function of the spin.

Reference~\cite{Suvorov15} defined and computed multipole moments of the Kerr-Newman metric as a vacuum solution in $f(R)$ gravity theories finding that the relation of the Kerr multipole moments in Eq.~(\ref{eq:kerrmult}) is preserved in a modified form with the simple substitution $M\rightarrow-\sqrt{M^2-Q^2}$. Consequently, the multipole moments of the metric of Ref.~\cite{Jmetric} in the case when $\beta$ is the only nonvanishing deviation parameter are given by the relation
\begin{equation}
M_{l}+{\rm i}S_{l}=M\sqrt{1-\beta}({\rm i}a)^{l},
\label{eq:betamult}
\end{equation}
at least as long as this metric is interpreted as a vacuum solution in $f(R)$ gravity. In particular, the first three multipole moments are: $M_0=M\sqrt{1-\beta}$, $S_1=M\sqrt{1-\beta}a$, and $M_2=-M\sqrt{1-\beta}a^2$. Note that I use a different sign convention for the multipole moments in Eq.~(\ref{eq:betamult}) compared to the one used by Ref.~\cite{Suvorov15} so that the relation in Eq.~(\ref{eq:kerrmult}) is recovered in the limit $\beta\rightarrow0$.

The metric of Ref.~\cite{Jmetric} can be mapped to known black hole solutions of specific alternative theories of gravity. These include the black hole solutions of RS2 braneworld gravity~\cite{RS2BH} and MOG~\cite{MOG}, where black holes are effectively described by the Kerr-Newman metric. Here, the parameter $\beta$ can be mapped to the tidal charge $\beta_{\rm tidal}$ in the RS2 model and to the coupling constant $\alpha$ in MOG via the equations $\beta=\beta_{\rm tidal}/M^2$ and $\beta=\alpha(1+\alpha)^2$, respectively. In both cases, the parameter $\beta$ can be either positive or negative as long as Eq.~(\ref{eq:lowerbounds1}) is fulfilled. The metric of Ref.~\cite{Jmetric} can also be mapped to the black hole solutions of EdGB gravity~\cite{Mignemi93,Kanti96,YS11,Pani11,AY14,M15} and dCS gravity~\cite{YP09,Pani11,YYT12} up to linear order in the spin, the Bardeen metric~\cite{BardeenBH,BambiModesto13}, as well as to other Kerr-like metrics; see \ref{appendix} for the detailed mappings. Table~\ref{tab:mappings} summarizes the known mappings of the metric of Ref.~\cite{Jmetric} to specific black-hole solutions.

\begin{table}[h]
\begin{center}
\footnotesize
\begin{tabular}{lll}
\multicolumn{3}{c}{}\\
Black Hole Metric   & Nonvanishing Deviation Parameters  & Validity  \\
\hline
Kerr    & --- & \\
Kerr-Newman	& $\beta=Q^2_{\rm el}/M^2$	& \\
RS2	& $\beta=\beta_{\rm tidal}/M^2$	& \\
MOG	& $\beta=\alpha(1+\alpha)^2$	& \\
EdGB	& $\alpha_{13} = -\frac{1}{6}\zeta_{\rm EdGB},~\alpha_{14} = -\frac{14}{3}\zeta_{\rm EdGB},~\alpha_{15} = -\frac{173}{15}\zeta_{\rm EdGB},~\ldots$ & ${\cal O}(a)$, ${\cal O}(\zeta_{\rm EdGB})$ \\
        & $\alpha_{23} = -\frac{13}{30}\zeta_{\rm EdGB},~\alpha_{24} = -\frac{16}{3}\zeta_{\rm EdGB},~\alpha_{25} = -\frac{197}{15}\zeta_{\rm EdGB},~\ldots$ & \\
        & $\alpha_{52} = \zeta_{\rm EdGB},~\alpha_{53} = 3\zeta_{\rm EdGB},~\alpha_{54} = \frac{70}{3}\zeta_{\rm EdGB}$,~\ldots	& \\ 
dCS	& $\alpha_{24} = \frac{5}{8}\zeta_{\rm dCS},~\alpha_{25}=\frac{15}{14}\zeta_{\rm dCS},~\alpha_{26}=\frac{27}{16}\zeta_{\rm dCS}$	& ${\cal O}(a)$, ${\cal O}(\zeta_{\rm dCS})$ \\
Bardeen & $\alpha_{13} = -\frac{3}{2}\frac{g^2}{M^2},~\alpha_{14}=2\alpha_{13},~\alpha_{15}=\frac{3g^2\left(4a^2+5g^2-16M^2\right)}{8M^4},~\ldots$ & at least ${\cal O}(g^4)$\\
        & $\alpha_{23} = \alpha_{13},~\alpha_{24}=\alpha_{14},~\alpha_{25}=\alpha_{15},~\ldots$ & \\
        & $\alpha_{53} = -2\alpha_{13},~\alpha_{54}=-2\alpha_{14},~\alpha_{55}=-2\alpha_{15},~\ldots$ & \\
\hline
\label{tab:mappings}
\end{tabular}
\caption{Mappings of the metric of Ref.~\cite{Jmetric} to known black hole solutions. The mappings to the EdGB and dCS metrics are only valid up to linear order in the spin and for small values of the respective deviation parameters, while the mapping to the Bardeen metric is valid for small values of the parameter $g^2$ at least up to ${\cal O}(g^4)$. There are no such restrictions on the other mappings. Note that Bardeen metric for nonzero values of the spin constructed by Ref.~\cite{BambiModesto13} may not belong to the same theory and physical setup as the static solution by Bardeen~\cite{BardeenBH}.}
\end{center}
\end{table}

\subsection{Astrophysical Properties}
\label{subsec:properties}

The properties of the spacetime of a black hole play an important role in the characteristics of astrophysical observables such as the electromagnetic radiation emitted from a surrounding accretion flow. Kerr-like spacetimes can have properties that differ significantly from the properties of the Kerr spacetime leading to modified observed fluxes and spectra. These signals, then, encode properties of the underlying spacetime which may be inferred by observations.

Many of these effects are somewhat generic to Kerr-like spacetimes but can vary in magnitude. Here, I demonstrate some of these effects for nonzero values of the deviation parameter $\alpha_{13}$ in the metric of Ref.~\cite{Jmetric}; see Ref.~\cite{Jmetric} and Table~\ref{tab:devparams} regarding the other parameters of this metric. A description of the corresponding effects in the quasi-Kerr metric and the metric of Ref.~\cite{JPmetric} can be found in Refs.~\cite{GB06,PaperI,PaperIII} and Refs.~\cite{JPmetric,Johannsenreview12,BambiDiskCode}, respectively. Properties of other Kerr-like metrics were analyzed in Refs.~\cite{CH04,VH10,Gair08,Brink08,Apostolatos09,Lukes2,Lukes3,Lukes4,VYS11,pathologies,pcGR1,Flathmann15,Boshkayev15,Kleihaus15,Shoom16,Maselli16}. Reference~\cite{CardosoQueimada15} studied the possibility of spinning up certain classes of Kerr-like metrics past extremality with point particles or accretion disks, which would, thereby, violate the cosmic censorship conjecture~\cite{Penrose69}.

\begin{figure*}[ht]
\begin{center}
\psfig{figure=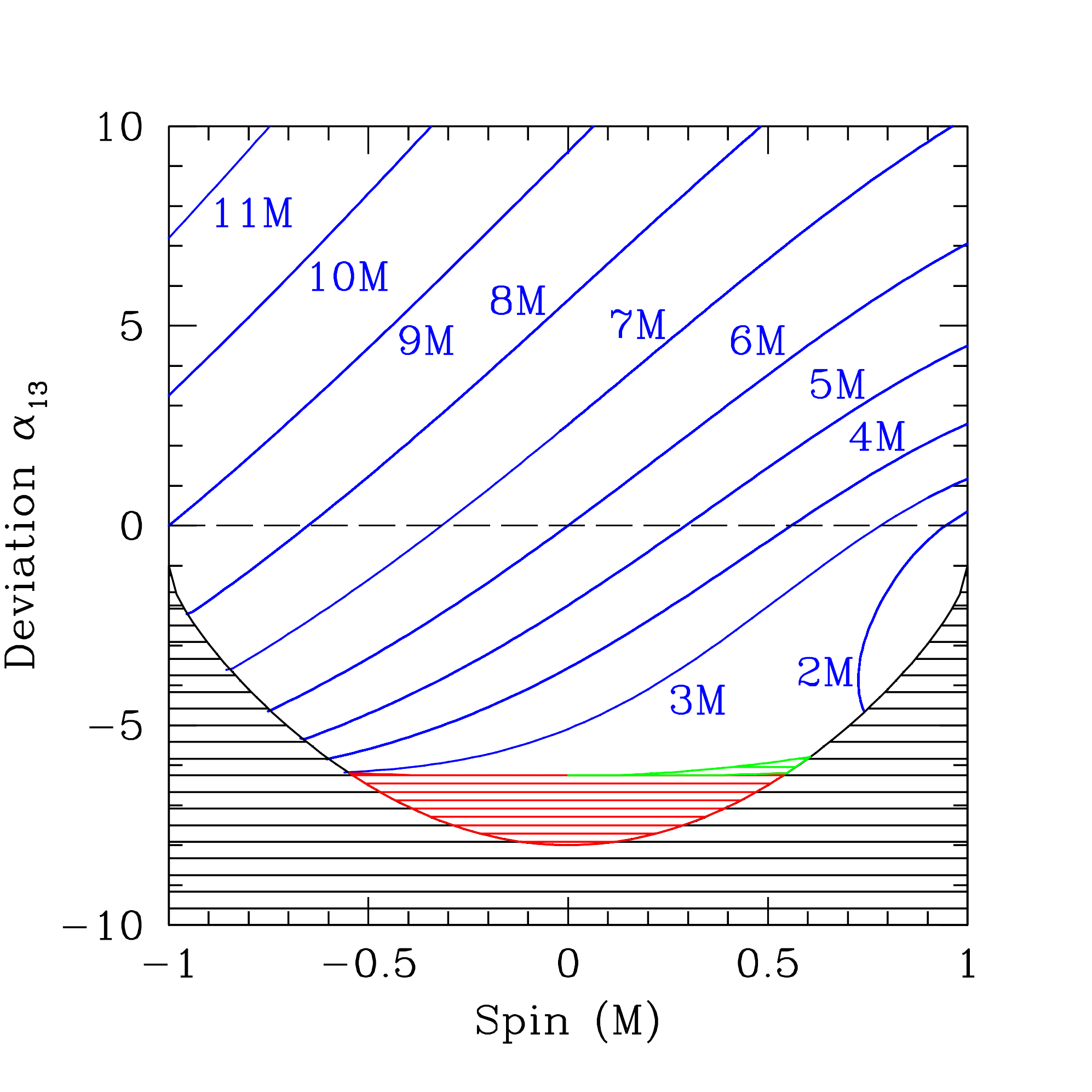,height=2.03in}
\psfig{figure=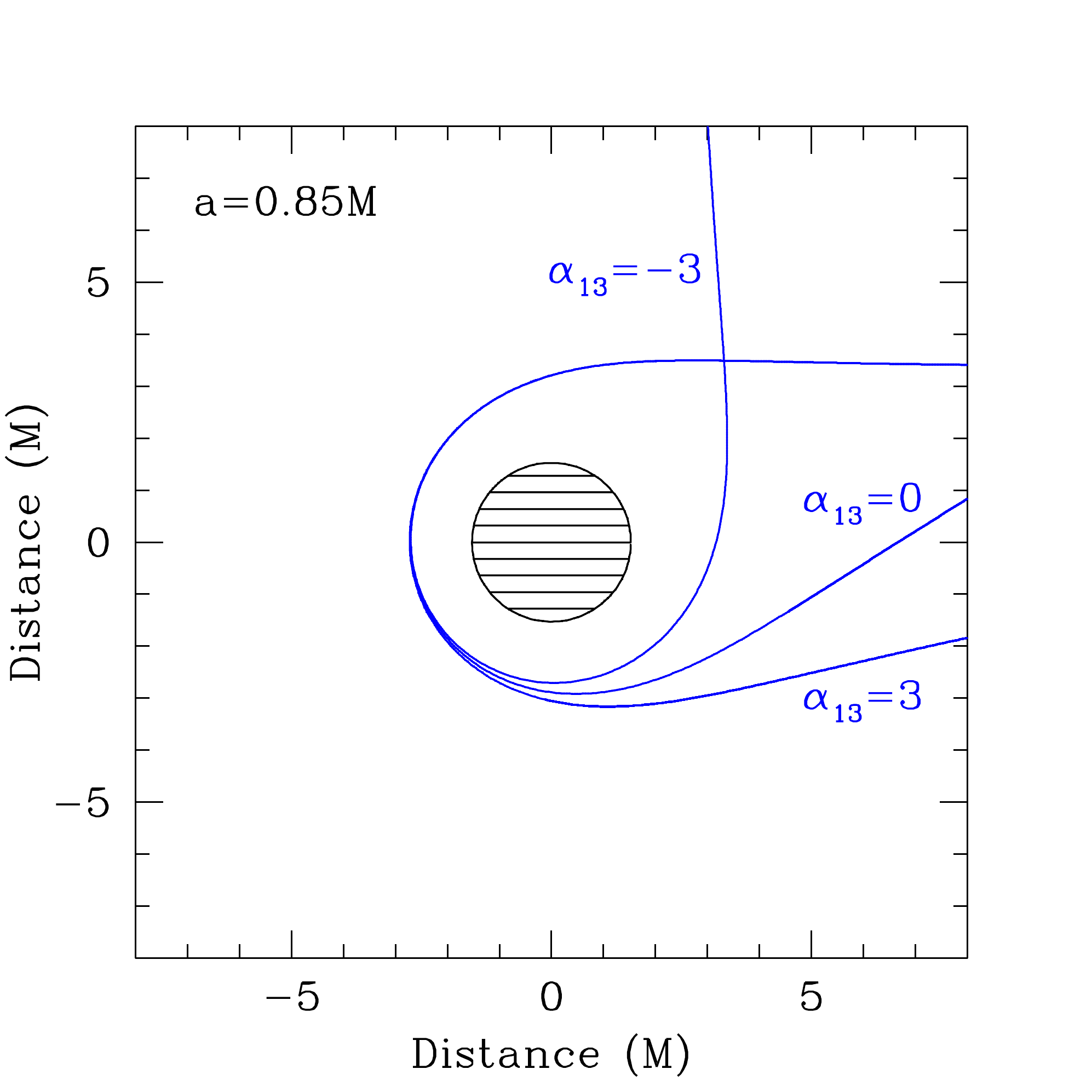,height=2.03in}
\psfig{figure=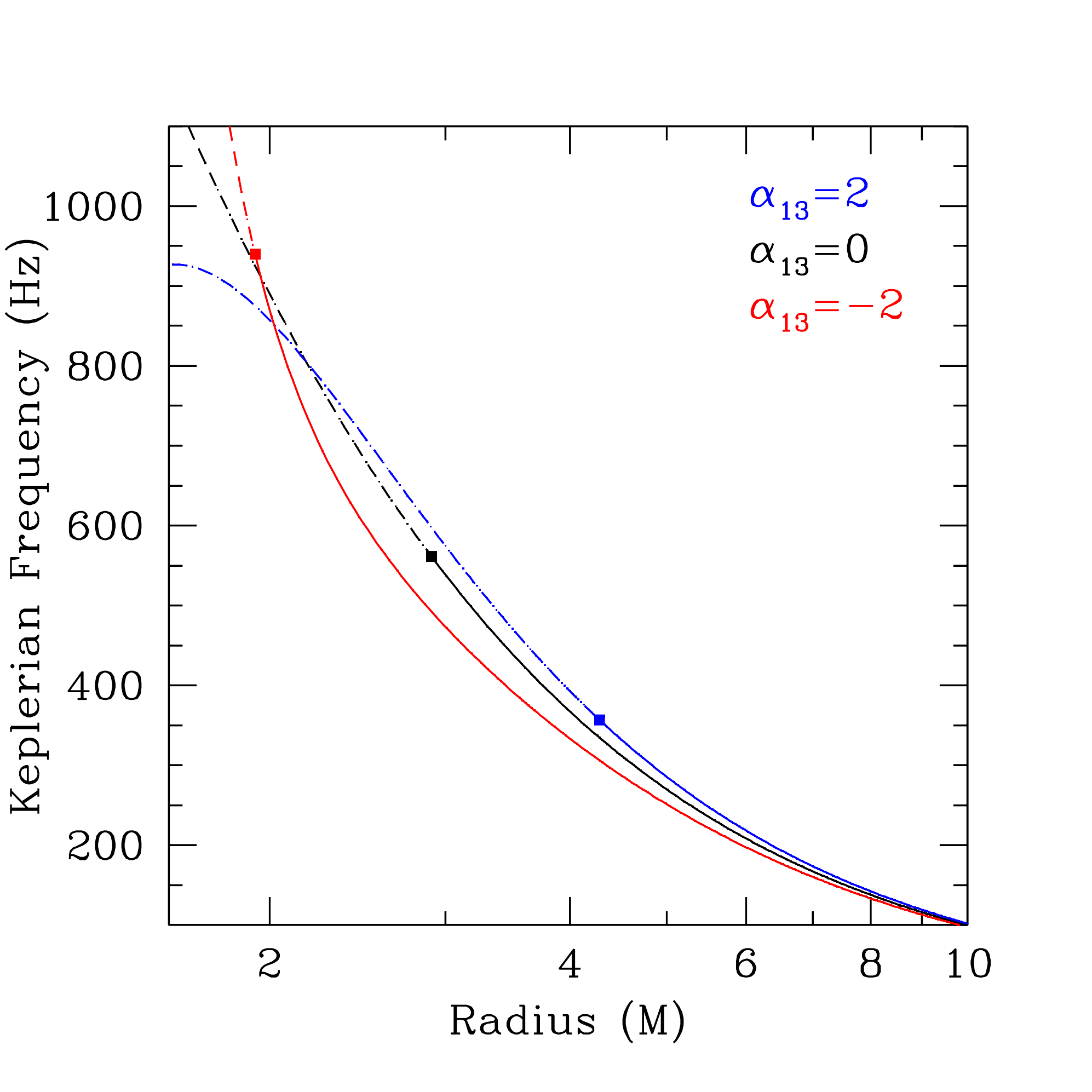,height=2.03in}
\end{center}
\caption{Effects of the deviation parameter $\alpha_{13}$ on the ISCO radius, the amount of lightbending, and the Keplerian frequency of a particle on a circular euqatorial orbit. The left panel (taken from Ref.~\cite{Jmetric}) shows contours of constant ISCO radius as a function of the spin and the deviation parameter. At a fixed value of the spin, the location of the ISCO increases for increasing values of the deviation parameter. In the green shaded region, the energy has two local minima and the ISCO is located at the outer radius where these minima occur. In the red shaded region, circular equatorial orbits do not exist at radii $r\sim2.5M$ and the ISCO is located at the outer boundary of this radial interval. The black shaded region marks the excluded part of the parameter space. The center panel shows trajectories of photons lensed by a black hole with a (counterclockwise) spin $a=0.85M$ for several values of the parameter $\alpha_{13}$. The shaded region corresponds to the event horizon. The right panel (taken from Ref.~\cite{Jmetric}) shows the Keplerian frequency $\nu_\phi= c^3 \Omega_\phi/2\pi GM$ as a function of radius for a black hole with mass $M=10\,M_\odot$ and spin $a=0.8M$ for different values of the parameter $\alpha_{13}$. At a given radius, the Keplerian frequency increases for increasing values of the parameter $\alpha_{13}$. The dot denotes the location of the ISCO.}
\label{fig:isco}
\end{figure*}

For nonzero values of the parameter $\alpha_{13}$ the coordinate locations of the circular photon orbit and of the ISCO are shifted compared to their coordinate locations in the Kerr metric. Similarly, photons experience either stronger or weaker lightbending near the black hole and the orbital frequencies of test particles are altered. Figure~\ref{fig:isco} shows the dependence of the ISCO radius on the spin and the deviation parameter $\alpha_{13}$. Figure~\ref{fig:isco} also shows the modified lightbending of photons around a black hole with a spin $a=0.85M$ and the Keplerian frequency of a particle on a circular equatorial orbit around a black hole with spin $a=0.8M$ as a function of radius for different values of the deviation parameter $\alpha_{13}$. 

\begin{figure*}[ht]
\begin{center}
\psfig{figure=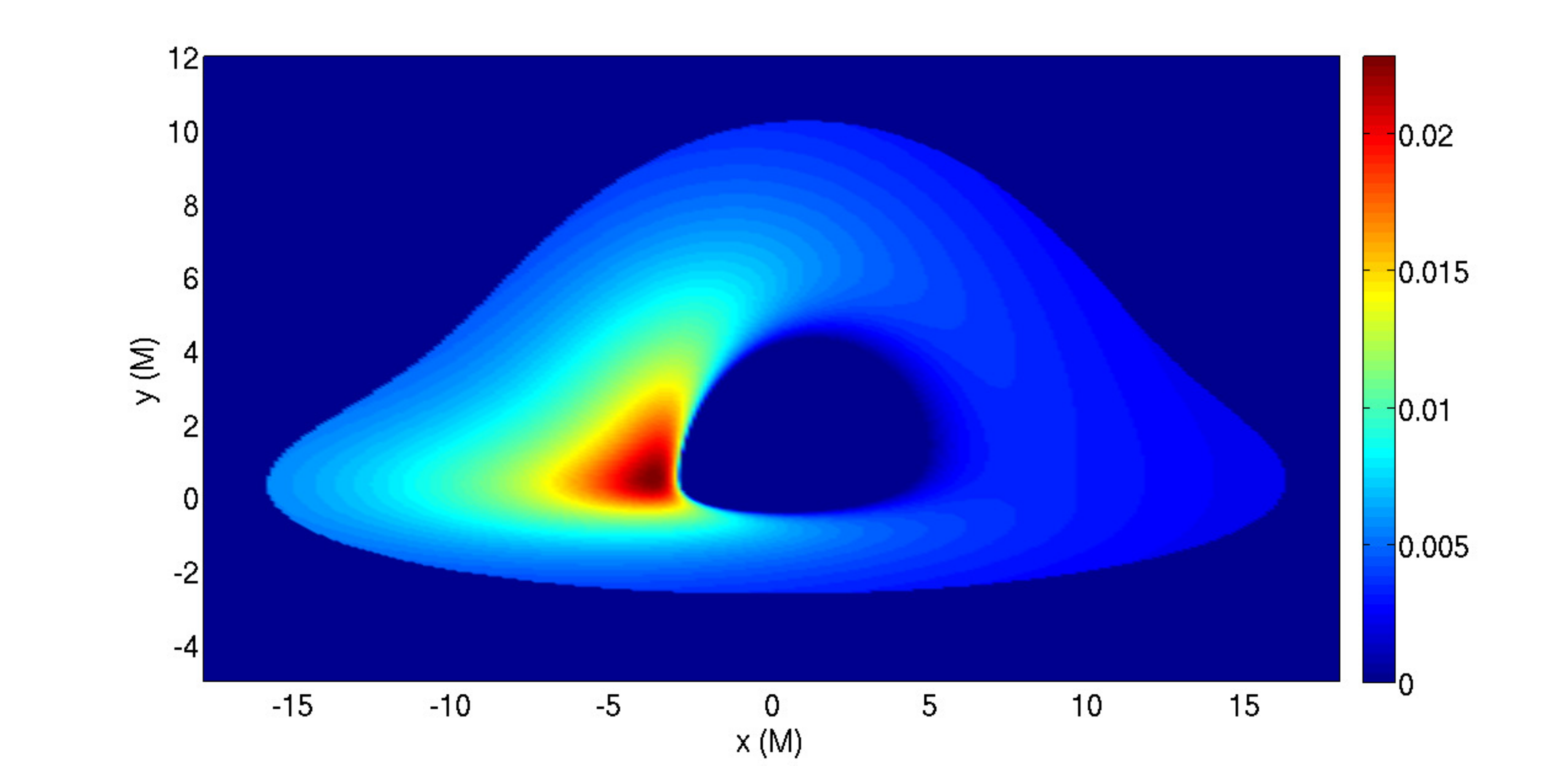,height=1.5in}
\psfig{figure=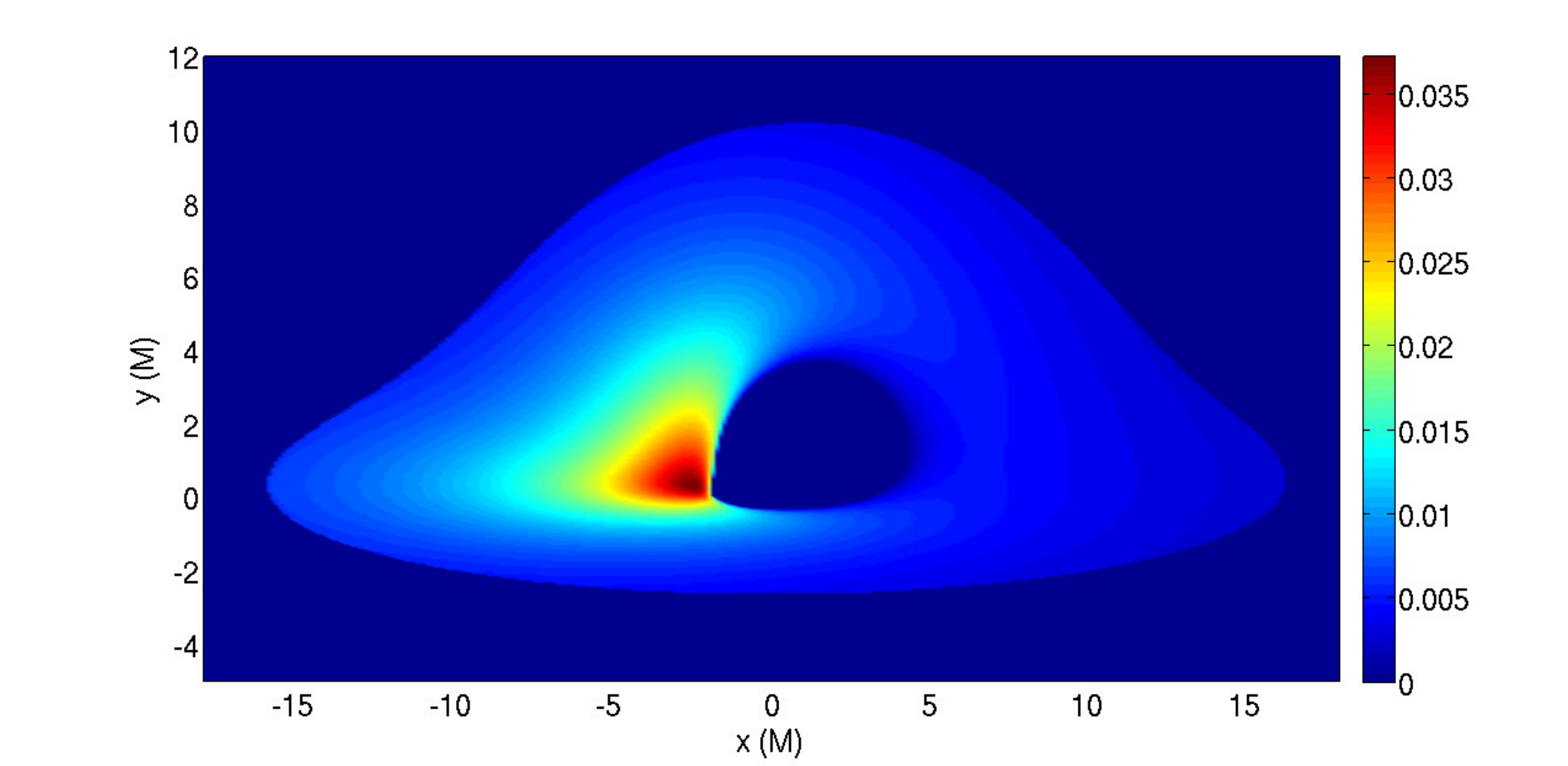,height=1.5in}
\end{center}
\caption{Direct images of geometrically thin accretion disks around (left panel) a Kerr black hole with spin $a=0.95M$ and (right panel) a Kerr-like black hole with the same spin and a value of the deviation parameter $\alpha_{13}=-1$. Both panels show the observed number flux density of (radially-symmetric) thermal disk emission at $1~{\rm keV}$ in units of ${\rm keV^{-1}~cm^{-2}~s^{-1}}$. The highest emission originates from a strongly localized region near the ISCO on the side of the black hole that is approaching the observer where Doppler boosting and beaming are particularly high. This region shifts toward the black hole and emits a significantly higher flux for negative values of the parameter $\alpha_{13}$. Taken from Ref.~\cite{Xrayprobes}.}
\label{fig:redshift}
\end{figure*}

Relativistic boosting and beaming as well as the gravitational redshift of photons propagating through such a black-hole spacetime likewise are important factors in the observed radiation. Figure~\ref{fig:redshift} illustrates the combination of these effects (as well as of the ISCO shift and the modified lightbending) using direct images of geometrically thin Novikov-Thorne~\cite{NTdisk} accretion disks around black holes with values of the spin $a=0.95M$ and the parameter $\alpha_{13}=0$ and $\alpha_{13}=-1$, respectively. The thermal radiation emitted by the disk is radially symmetric, but the observed disk flux has a much more complicated structure.

Since all of these effects typically depend on both the spin and deviation parameters of a given Kerr-like metric, there is an inherent degeneracy between the spin and the deviation parameters which can complicate the detection of a potential deviation from the Kerr metric. This is the case especially for the ISCO radius (see, e.g., the left panel of Fig.~\ref{fig:isco}), the location of which allows for spin measurements of Kerr black holes based on, e.g., their continuum x-ray emission~\cite{McClintock14} or relativistically broadened iron lines~\cite{Reynolds14}. For non-Kerr black holes, however, the dependence of the ISCO radius on the spin and the deviation parameters leads to a strong correlation between all of these parameters if the location of the ISCO is the primary quantity being measured, either directly or indirectly~\cite{PJ11,Krawcz12,PaperIII,PaperIV,BambiDiskCode,Xrayprobes}.

A property of the metric of Ref.~\cite{Jmetric} which is not generic to Kerr-like metrics is the fact that it can be written in a Kerr-Schild-like form which removes all coordinate singularities at the location of the event horizon~\cite{Jmetric}. This allows for a consistent treatment of accretion flows in fully relativistic magnetohydrodynamic simulations; c.f., e.g., Refs.~\cite{McKinney1,McKinney2}.

Table~\ref{tab:devparams} summarizes several important properties of the quasi-Kerr metric~\cite{GB06} and the metrics of Refs.~\cite{JPmetric,Jmetric}. Table~\ref{tab:devparams} lists the type of the compact object harbored by each of these metrics, their respective deviation parameters together with the metric elements affected by them in the Boyer-Lindquist-like forms given in Eqs.~(\ref{eq:QKmetric}), (\ref{eq:JPmetric}), and (\ref{eq:Jmetric}), respectively, and the magnitude of the effect of the (lowest-order) deviation parameters on the location of the ISCO, the lightbending, and the orbital frequency $\nu_\phi$ of a particle on a circular equatorial orbit around the compact object.

\begin{table*}[h]
\begin{center}
\footnotesize
\begin{tabular}{lllllll}
\multicolumn{6}{c}{}\\
Metric   & Object  & Parameters   & Modified Metric Elements   & ISCO   & Lightbending   & Frequency $\nu_\phi$ \\
\hline 
Ref.~\cite{GB06}  & NS~\cite{pathologies} & $\epsilon$       & $(t,t)$, $(r,r)$, $(\theta,\theta)$, $(\phi,\phi)$ & strong~\cite{PaperI} & strong~\cite{PaperI} & weak~\cite{PaperIII} \\
Ref.~\cite{JPmetric} & NS$^{\rm a}$~\cite{pathologies} & $\epsilon_3$, $\epsilon_4$,~$\ldots$	& $(t,t)$, $(r,r)$, $(\phi,\phi)$, $(t,\phi)$ & strong~\cite{JPmetric} & strong~\cite{Johannsenreview12} & weak~\cite{PaperIV} \\
Ref.~\cite{Jmetric}& BH~\cite{Jmetric} & $\epsilon_3$, $\epsilon_4$,~$\ldots$  &   $(t,t)$, $(r,r)$, $(\theta,\theta)$, $(\phi,\phi)$, $(t,\phi)$ & weak~\cite{Jmetric} & strong & weak~\cite{Jmetric}\\
  && $\alpha_{13}$, $\alpha_{14}$,~$\ldots$  &   $(t,t)$, $(\phi,\phi)$, $(t,\phi)$ & strong~\cite{Jmetric} & strong & weak~\cite{Jmetric}\\
  && $\alpha_{22}$, $\alpha_{23}$,~$\ldots$  &   $(t,t)$, $(\phi,\phi)$, $(t,\phi)$ & strong~\cite{Jmetric} & strong & weak~\cite{Jmetric}\\
  && $\alpha_{52}$, $\alpha_{53}$,~$\ldots$  &   $(r,r)$ & none~\cite{Jmetric} & strong & none~\cite{Jmetric}\\
  && $\beta$  &   $(t,t)$, $(r,r)$, $(\phi,\phi)$, $(t,\phi)$ & strong & strong & weak \\
\hline
\end{tabular}
\footnotetext{}{$^{\rm a}$BH for small deviations at linear order.  }
\caption{Properties of the Kerr-like metrics of Refs.~\cite{GB06,JPmetric,Jmetric}. The table lists the type of the compact object (NS -- naked singularity, BH -- black hole), nonvanishing deviation parameters, modified metric elements, and the effects of these parameters on the location of the ISCO, lightbending, and the Keplerian frequency $\nu_\phi$ of matter particles on circular equatorial orbits around the compact object.}
\label{tab:devparams}
\end{center}
\end{table*}

\section{Very-Long Baseline Interferometric Observations of the Accretion Flow}
\label{sec:EHT}

Sgr~A$^\ast$ is a prime target of high-resolution VLBI observations with the EHT~\cite{Doele09a,Doele09b,Fish09}. Initial VLBI observations of Sgr~A$^\ast$ in 2007--2009 at 230~GHz with a three-station array comprised by the James Clerk Maxwell Telescope (JCMT) and Sub-Millimeter Array (SMA) in Hawaii, the Submillimeter Telescope Observatory (SMTO) in Arizona, and several dishes of the Combined Array for Research in Millimeter-wave Astronomy (CARMA) in California resolved structures on scales of only $4r_S$~\cite{Doele08}, where $r_S\equiv2r_g$ is the Schwarzschild radius of Sgr~A$^\ast$. Similar observations also detected time variability on these scales in Sgr~A$^\ast$ and measured a closure phase along the Hawaii--SMA--SMTO triangle~\cite{Fish11}. In 2009--2013, follow-up observations with the same telescope array [also including the Caltech Submillimeter Observatory (CSO) in Hawaii] have led to an increased data set including numerous closure phase measurements~\cite{Fish15} and the detection of polarized emission originating from within a few Schwarzschild radii~\cite{JohnsonScience15}. Such measurements have demonstrated the feasibility of VLBI imaging of Sgr~A$^\ast$ with the EHT on event horizon scales. 

In 2015, the existing three-station EHT array has been expanded to include ALMA in Chile, the Large Millimeter Telescope (LMT) in Mexico, the South Pole Telescope (SPT), the Plateau de Bure Interferometer (PdB) in France, and the Pico Veleta Observatory (PV) in Spain; see Ref.~\cite{FishALMA} for a recent description of the EHT. Simulations based on such enlarged telescope arrays support the possibility of probing the accretion flow of Sgr~A$^\ast$ in greater detail and of directly imaging the shadow of Sgr~A$^\ast$~\cite{Doele09a}. At around 230~GHz, the emission from Sgr~A$^\ast$ becomes optically thin (see the right panel of Fig.~\ref{fig:ehtsources}) and the shadow will be (at least partially) unobscured by the surrounding accretion flow (see Ref.~\cite{Bro09} and references therein). The sensitivity and resolution of this enlarged array will be greatly increased, caused primarily by ALMA which will have a sensitivity that is about 50 times greater than the sensitivity of the other stations and the long baselines from the stations in the Northern hemisphere to the SPT. In addition, this array allows for the measurement of closure phases along many different telescope triangles, some of which depend very sensitively on the parameters of Sgr~A$^\ast$~\cite{CP}, as well as of closure amplitudes along telescope quadrangles.

In this section, I first review the importance of the shadow for tests of the no-hair theorem. Second, I focus on studies of the accretion flow of Sgr~A$^\ast$ as a means for such tests. Third, I discuss other potential tests of the no-hair theorem based on orbiting hot spots in the accretion flow.

\subsection{The Shadow of Sgr~A$^\ast$}
\label{subsec:shadows}

The shadow is a prominent feature of resolved accretion flow images of supermassive black holes. Such a shadow is the projection of the circular photon orbit onto the sky along null geodesics. For a Kerr black hole, the circular photon orbit is located at a coordinate radius that ranges from $9r_g$ in the case of a maximally counterrotating black hole to $1r_g$ in the case of a maximally corotating black hole~\cite{Bardeen73}. The shadow is expected to be surrounded by a bright ring corresponding to photon trajectories that wind around the black hole many times. Thanks to the long path length through the emitting medium of the photons that comprise the ring, these photons can make a significantly larger contribution to the observed flux than individual photons outside of the ring.

Images of shadows of Kerr black holes, either with or without optically and geometrically thin accretion disks, have been calculated by a number of different authors~\cite{Bardeen73,CunninghamBardeen73,Luminet79,Sikora79,FukueYokoyama88,Jaroszynski92,Viergutz93,Falcke00,Takahashi04,BeckwithDone05,Bozza08,PaperII,Chan13,Interstellar,Moore15,merger15,
Pu16}. Images of shadows and accretion flows around non-Kerr black holes and exotic objects in general relativity or other theories of gravity were analyzed by Refs.~\cite{PaperII,BambiYoshida10,Amarilla10,Bambietal12,AE12,Bambi13b,J13rings,Abdujabbarov13,Atamurotov13,ZilongBambi14,Vincent14,Grenzebach14,MOG,Ghasemi15,Atamurotov15,Wei15,Herdeiro15PRL,
Yang15,Tinchev15,Abdujabbarov16,Krawcz16}. References~\cite{ChenJing12,Liu12a} studied the strong gravitational lensing near Kerr-like compact objects. Black hole shadows are also clearly visible in several (three-dimensional) general-relativistic magnetohydrodynamic simulations (GRMHD) reported to date~\cite{Moscibrodzka09,Dexter09,ShcherbakovPenna11,ChanPsaltis15a}.

\begin{figure*}[ht]
\begin{center}
\psfig{figure=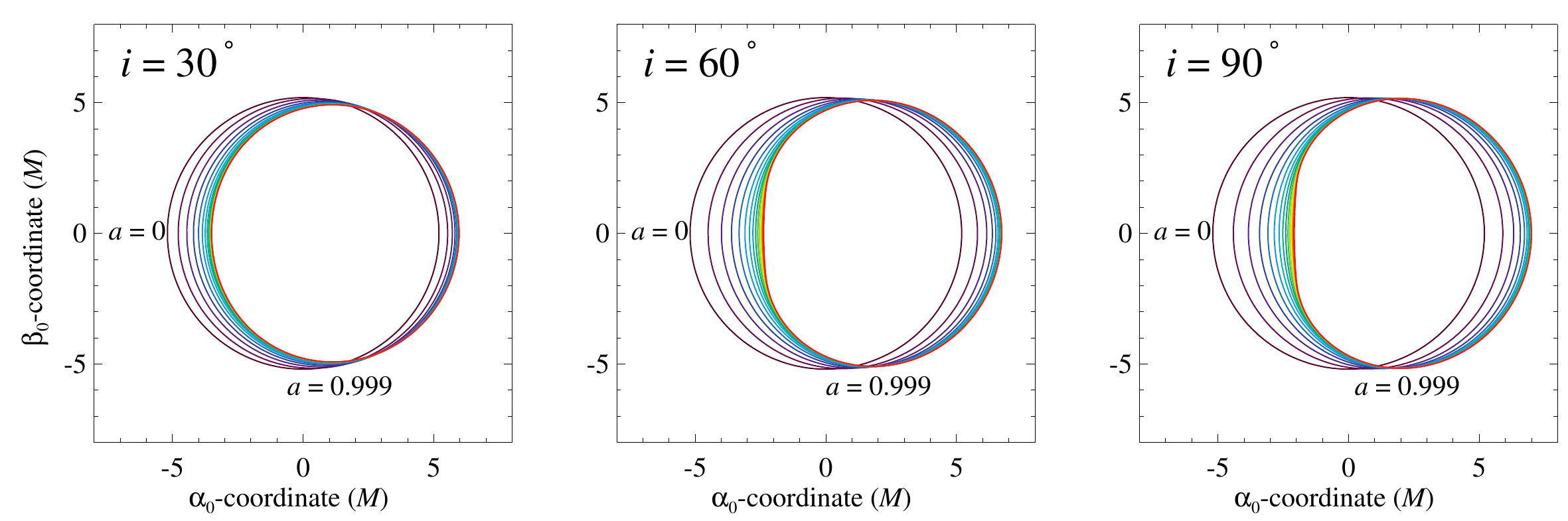,height=1.85in}
\psfig{figure=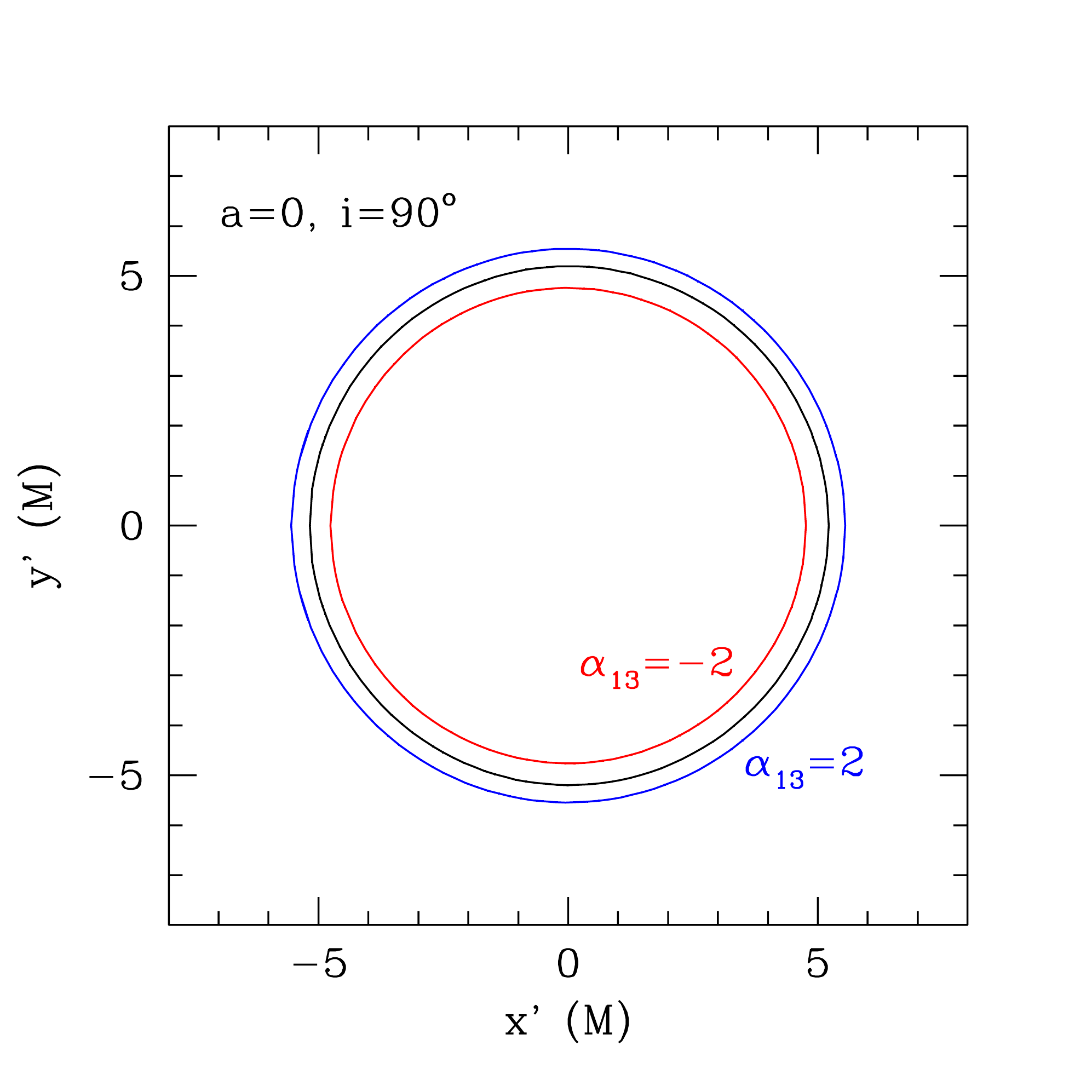,height=2.02in}
\psfig{figure=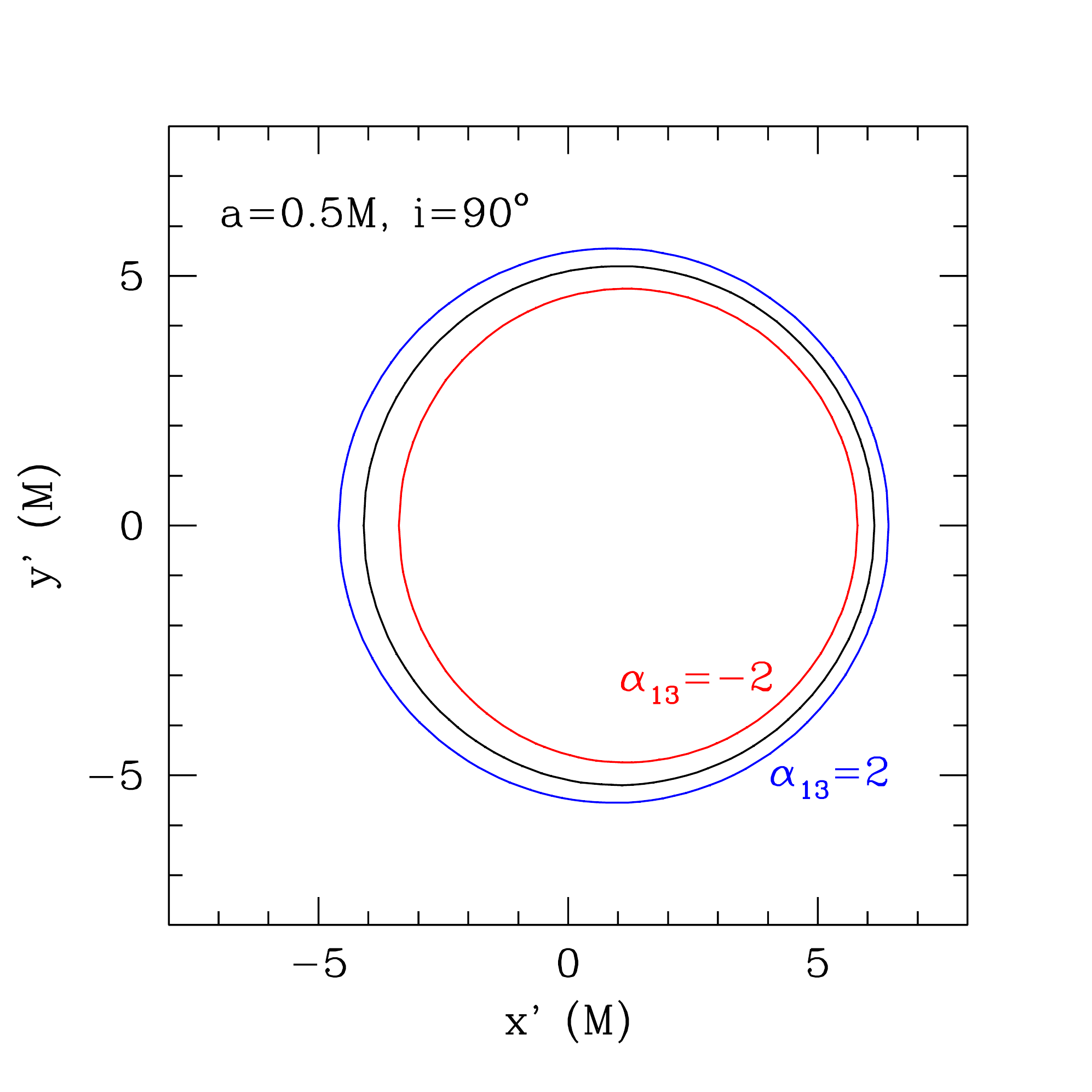,height=2.02in}
\psfig{figure=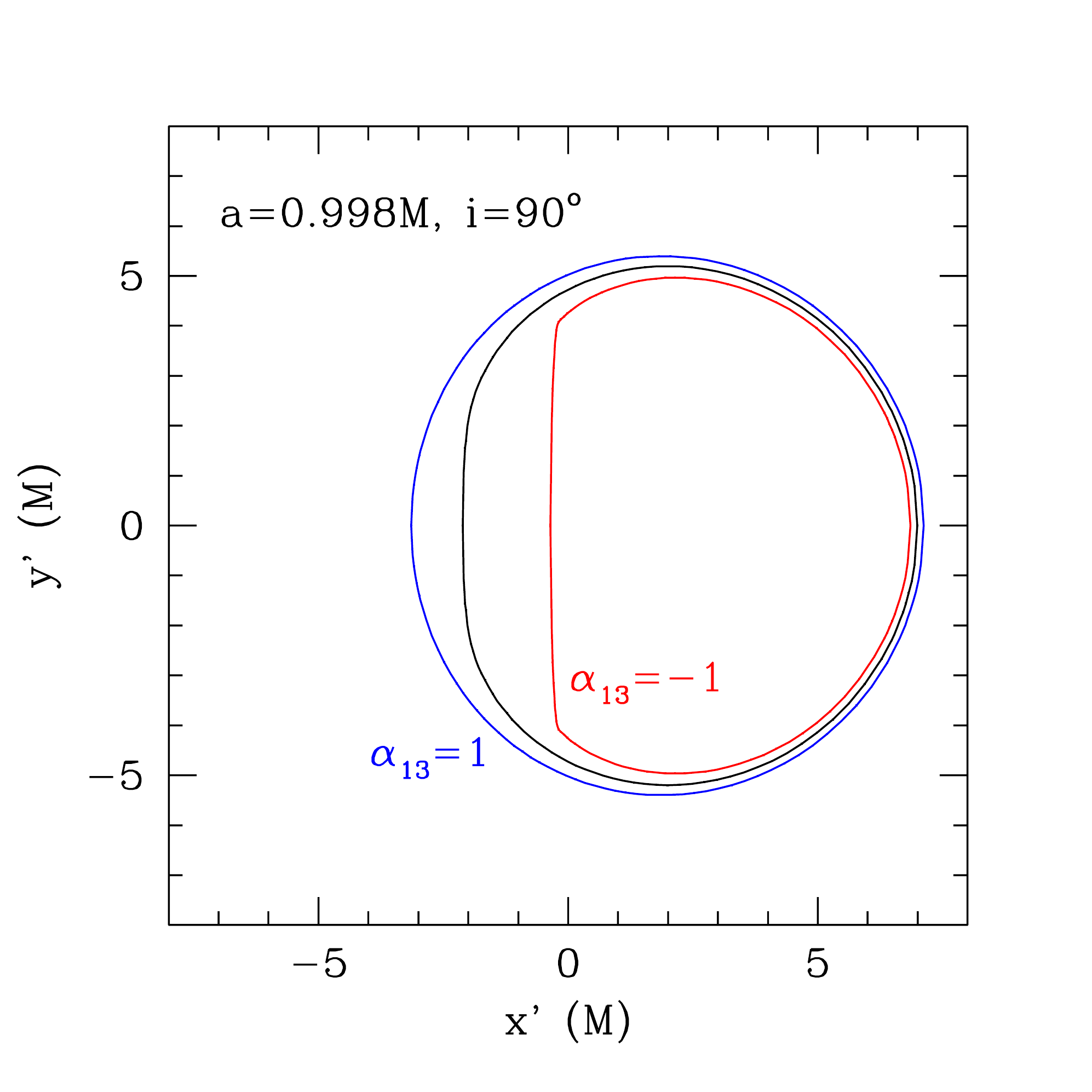,height=2.02in}
\end{center}
\caption{Shadows of Kerr (top row; taken from Ref.~\cite{Chan13}) and non-Kerr (bottom row; taken from Ref.~\cite{J13rings}) black holes with different spins $a$, inclinations $i$, and deviations from the Kerr metric $\alpha_{13}$. In the top panels, different colors represent different spins ranging from $a=0$ (black) to $a=0.999r_g$ (red). For each inclination, the size of the shadow is determined primarily by the black hole mass $M$ (in gravitational units) and depends only weakly on the black hole spin and the inclination. The displacement of the shadow along the $x$-axis is a function of the spin and the inclination. Moreover, the shadow retains its nearly circular shape unless the black hole spin is very high and the inclination is large. For nonzero values of the parameter $\alpha_{13}$, however, the size of the shadow is altered and its shape can be significantly aymmetric.}
\label{fig:shadowshapes}
\end{figure*}

Since the shape of the shadow of a black hole is determined only by the geometry of the underlying spacetime, it is independent of the complicated structure of the accretion flow making it an excellent target of imaging observations with the EHT. For a Kerr black hole, the shape of the shadow depends uniquely on the mass, spin, and inclination of the black hole (e.g., \cite{Falcke00}). For a Schwarzschild black hole, the shadow is exactly circular and centered on the black hole. For Kerr black holes with nonzero values of the spin and the inclination, the shadow is displaced off center and retains a nearly circular shape~\cite{Takahashi04,BeckwithDone05,PaperII}, except for extremely high spin values $a\gtrsim0.9r_g$ and large inclinations, in which case the shape of the shadow becomes asymmetric~\cite{PaperII,Chan13}.

However, images of black hole shadows can be significantly altered if the no-hair theorem is violated. For black holes that are described by a Kerr-like metric, the shape of the shadow can become asymmetric~\cite{PaperII, Abdujabbarov13,Atamurotov13,Grenzebach14,Ghasemi15,Atamurotov15,Wei15,Herdeiro15PRL,Tinchev15,Abdujabbarov16} and its size can vary significantly~\cite{AE12,J13rings,Abdujabbarov13,Grenzebach14,Ghasemi15,Atamurotov15,Herdeiro15PRL,Tinchev15,Abdujabbarov16}. Figure~\ref{fig:shadowshapes} shows several examples of shadows of Kerr and Kerr-like black holes for different values of the spin $a$, the inclination $i$, and the deviation parameter $\alpha_{13}$ in the metric of Ref.~\cite{Jmetric}. Kerr naked singularities, i.e., compact objects described by the Kerr metric with values of the spin exceeding the Kerr bound in Eq.~(\ref{eq:kerrbound}), do not have a well-defined shadow making them easily distinguishable from the shadows of black holes~\cite{BambiFreese09,HiokiMaeda09}.

\begin{figure*}[ht]
\begin{center}
\psfig{figure=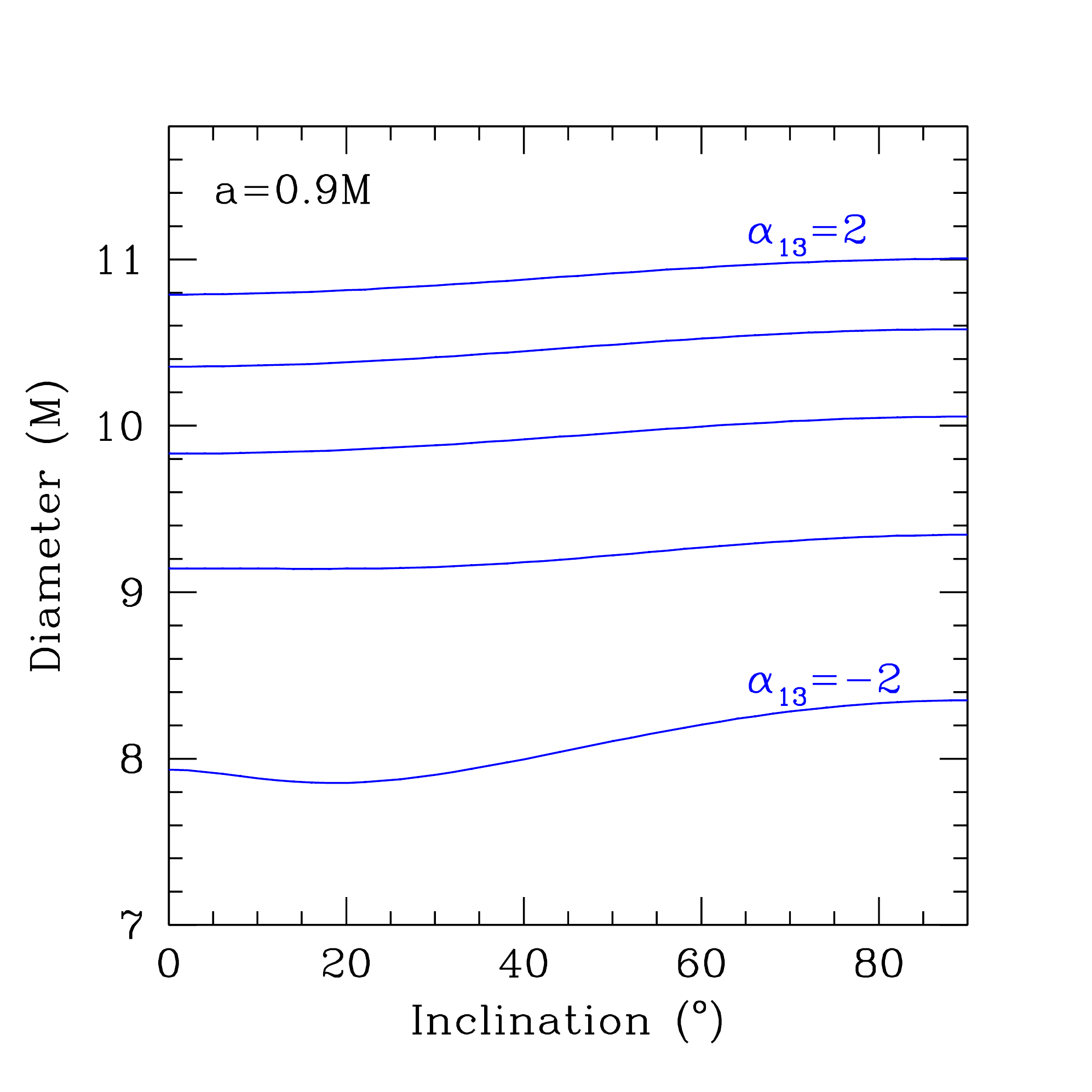,height=2.02in}
\psfig{figure=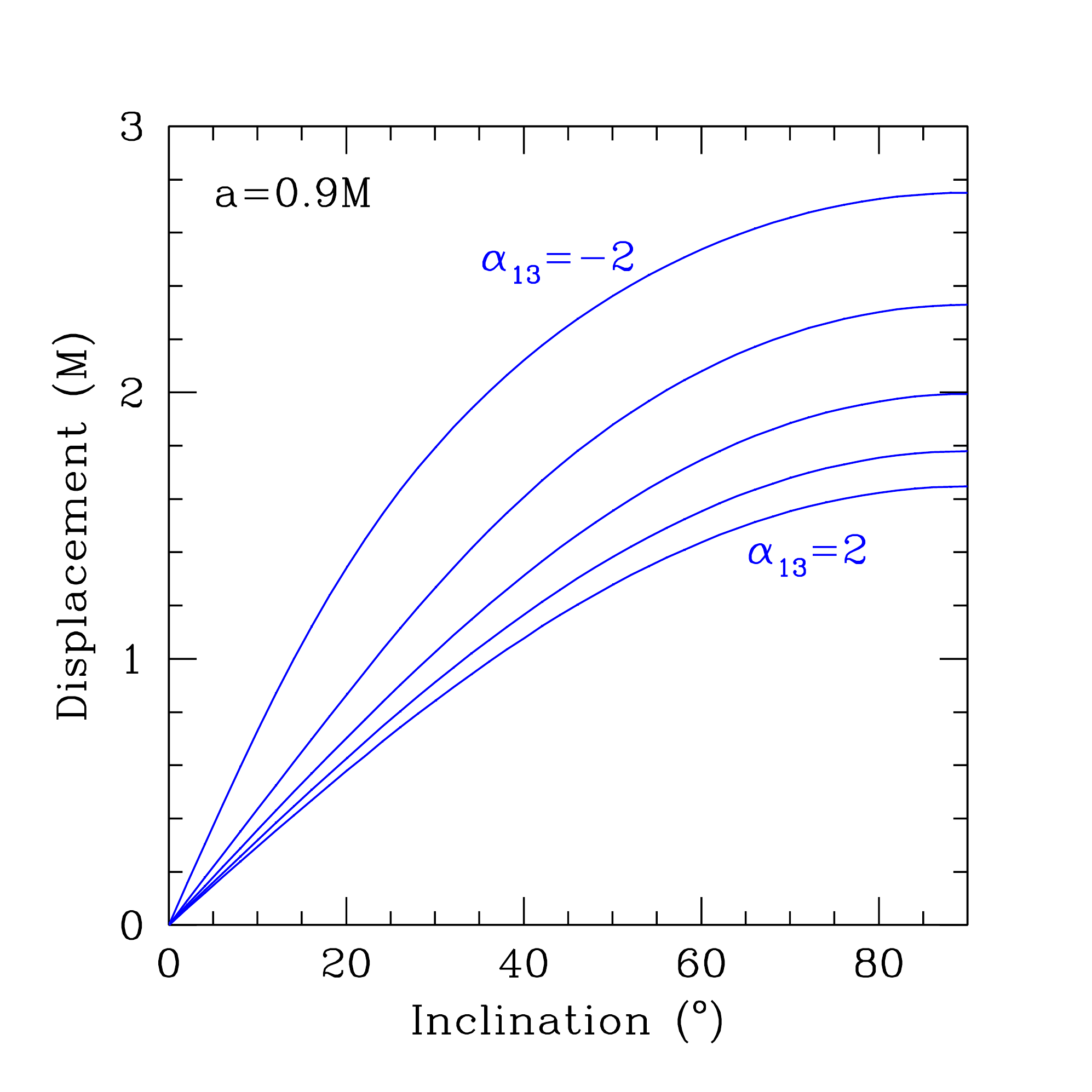,height=2.02in}
\psfig{figure=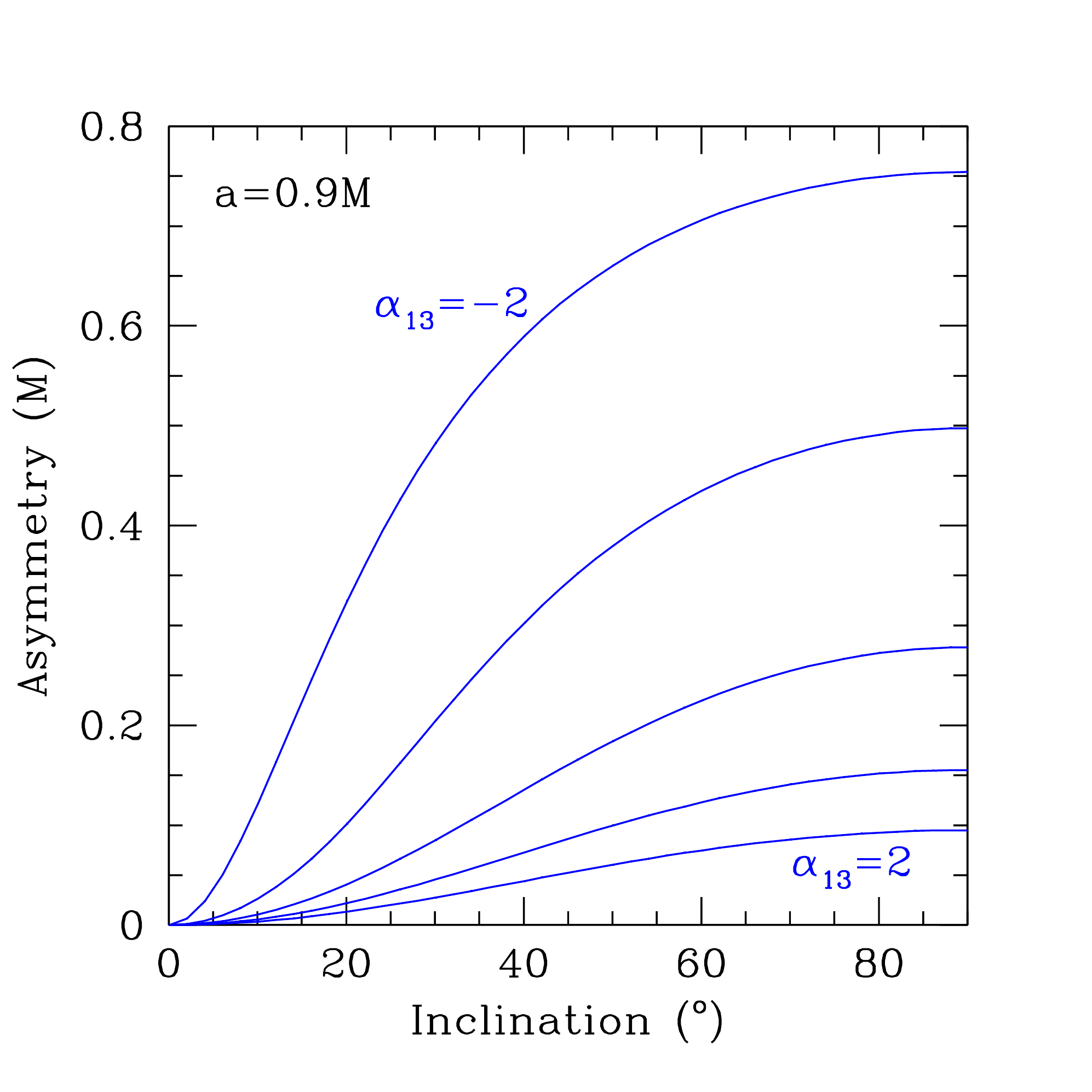,height=2.02in}
\psfig{figure=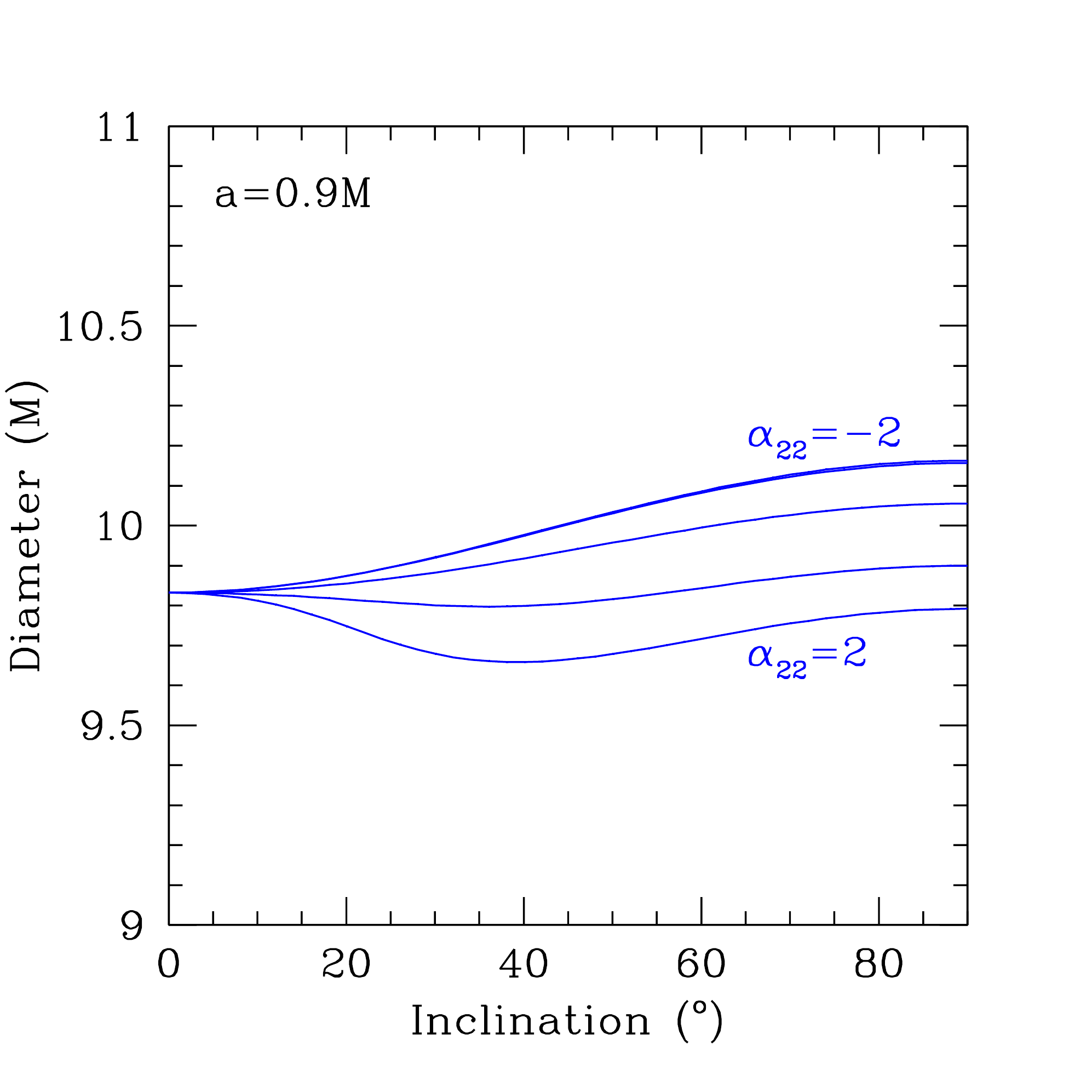,height=2.02in}
\psfig{figure=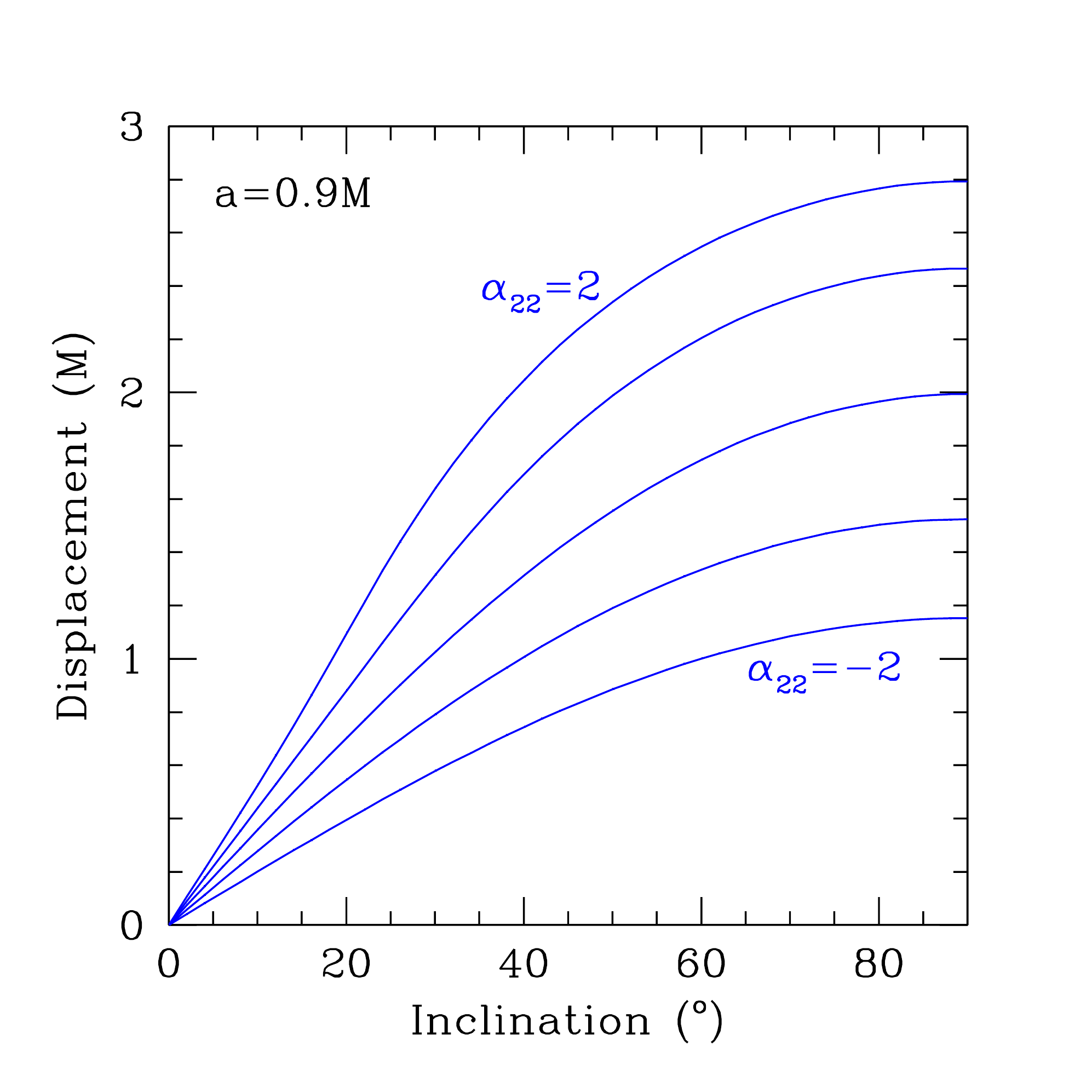,height=2.02in}
\psfig{figure=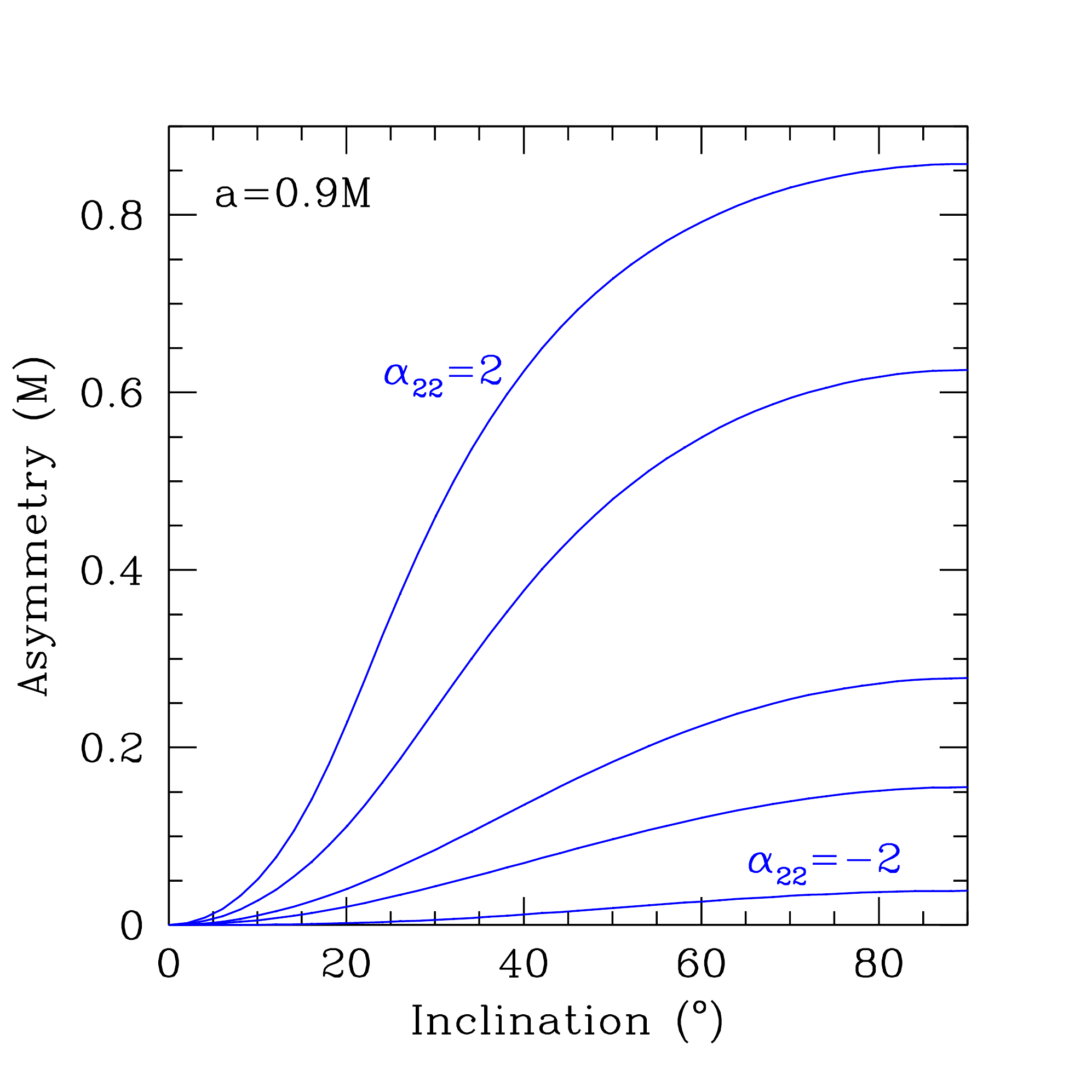,height=2.02in}
\psfig{figure=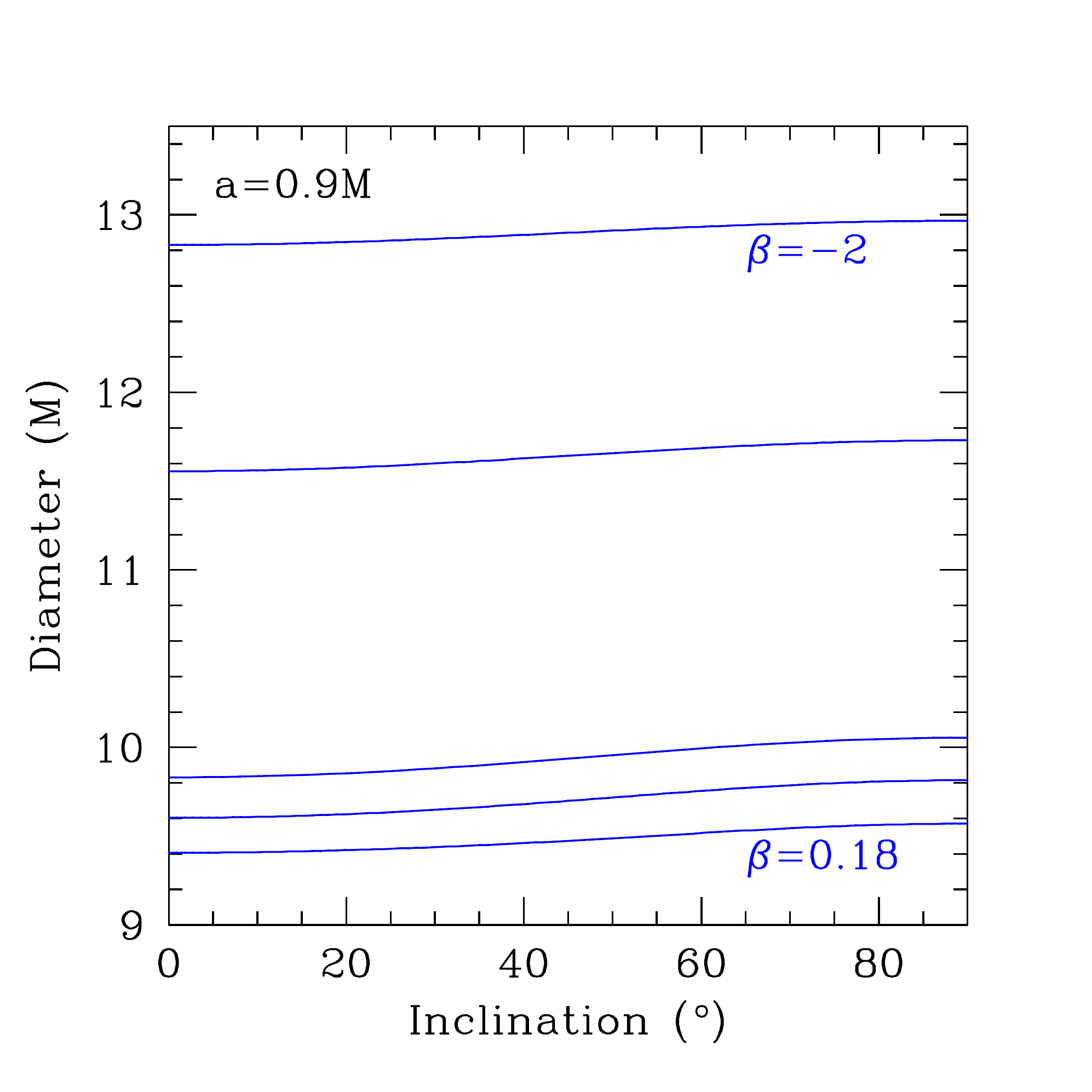,height=2.02in}
\psfig{figure=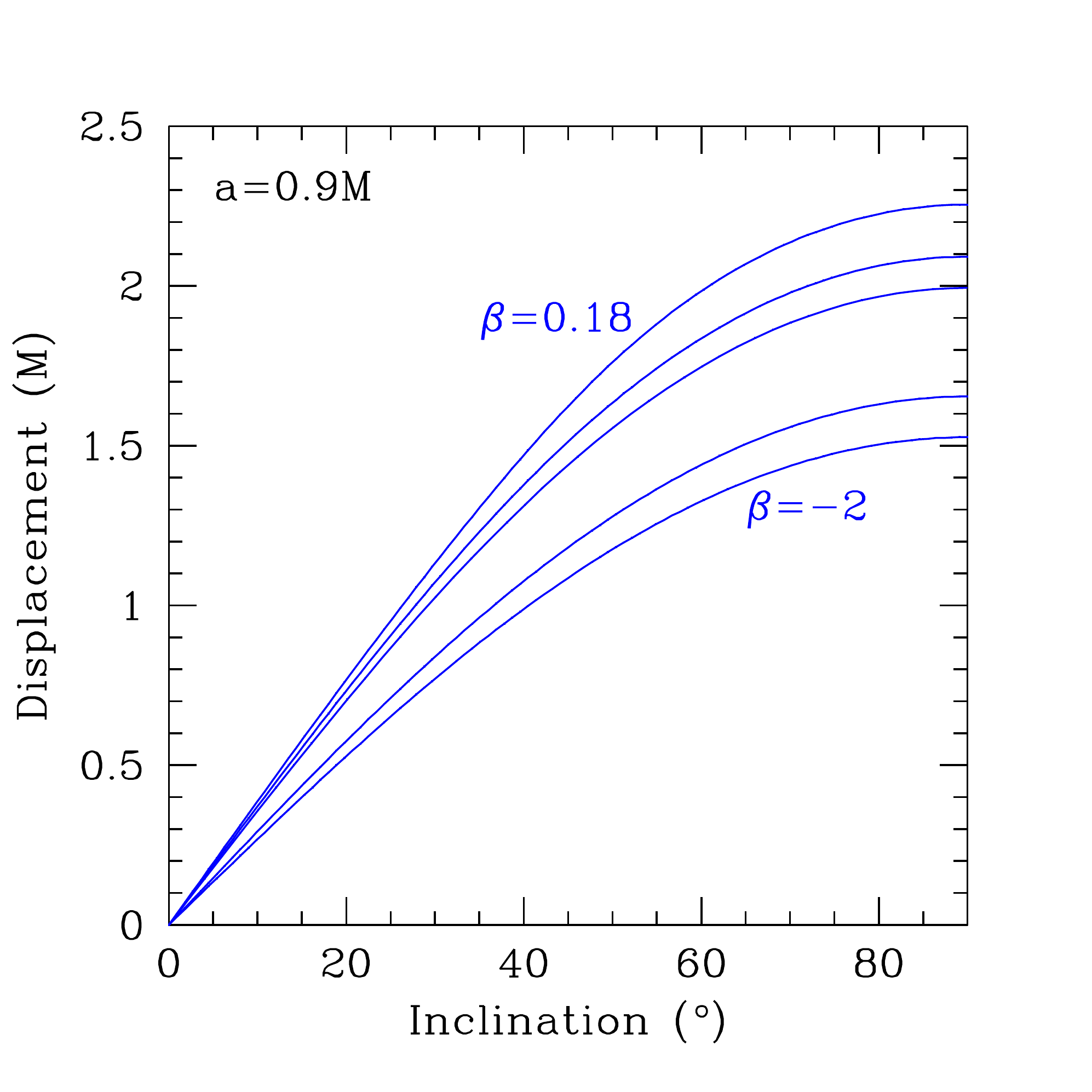,height=2.02in}
\psfig{figure=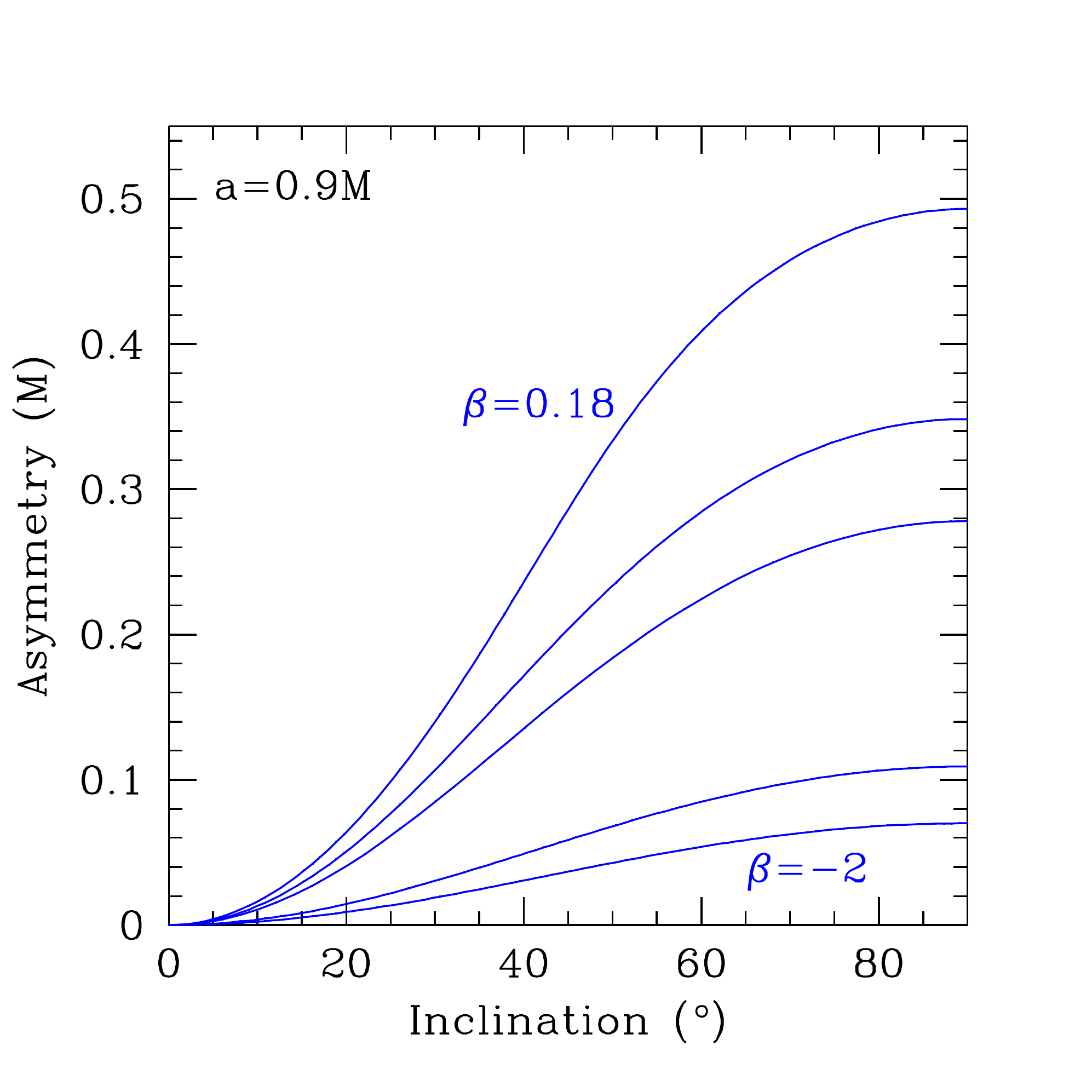,height=2.02in}
\end{center}
\caption{Diameter (left column), displacement (center column), and asymmetry (right column) of the shadows of Kerr-like black holes with a spin $a=0.9M$ as a function of the inclination for values of the deviation parameters $\alpha_{13}=-2,-1,0,1,2$ (top row), $\alpha_{22}=-2,-1,0,1,2$ (center row), and $\beta=-2,-1,0,0.1,0.18$ (bottom row) in the metric of Ref.~\cite{Jmetric}. The shadow diameter depends only weakly on the parameter $\alpha_{22}$, while it is practically constant for fixed values of the parameters $\alpha_{13}$ and $\beta$. The shadow displacement is affected by all three deviation parameters but depends primarily on the spin. Negative values of the parameter $\alpha_{13}$ and positive values of the parameters $\alpha_{22}$ and $\beta$ can cause the shadow shape to be significantly more asymmetric than the shadow of a Kerr black hole with the same spin. The size and asymmetry of the shadow are direct measures of the degree to which the no-hair theorem is violated.}
\label{fig:rings09}
\end{figure*}

Several authors have quantified the effects of the spin and inclination as well as of potential deviations from the Kerr metric on the position and shape of the shadow. Reference~\cite{Takahashi04} showed that the displacement of the shadow occurs in the direction perpendicular to the spin axis of the black hole with an approximately linear dependence on the spin and characterized the shape of the shadow in terms of the maximum and minimum width of the shadow. Reference~\cite{PaperII} defined the displacement $D$ as the mean of the maximum and minimum abscissae $x'_{\rm max}$ and $x'_{\rm min}$ on the axis perpendicular to the spin $(y')$ axis, as well as the diameter $L$ and asymmetry $A$ of the shadow as an angular average of its radius and the root mean square of its radius, respectively, which is easier to measure in practice. These expressions are given by the equations
\ba
D \equiv \frac{ \left| x'_{\rm max} + x'_{\rm min} \right| }{ 2 }, \\
L \equiv \frac{1}{\pi} \int_0^{2\pi} \bar{R} d\vartheta, \\
A \equiv 2 \sqrt{ \frac{ \int_0^{2\pi} \left(\bar{R} - \left< \bar{R} \right> \right)^2 d\vartheta }{ 2\pi } },
\label{eq:DLA}
\ea
where
\be
\bar{R} \equiv \sqrt{ (x'-D)^2+y'^2 }
\ee
is the average radius and
\be
\tan \vartheta \equiv \frac{y'}{x'}.
\label{eq:vartheta}
\ee

The displacement of the shadow around Kerr black holes is reminiscent of the location of the caustics in the Kerr spacetime~\cite{RauchBlandford94,Bozza08}. References~\cite{PaperII,Chan13,J13rings} computed approximate expressions of the displacement, size, and asymmetry of the shadow of a Kerr black hole. Reference~\cite{PaperII} and Ref.~\cite{J13rings} found approximate expressions of the displacement, diameter, and asymmetry for the shadow of the compact object described by the quasi-Kerr metric~\cite{GB06} and of the Kerr-like black hole described by the metric of Ref.~\cite{Jmetric} (in terms of the parameters $\alpha_{13}$ and $\alpha_{22}$), respectively. See \ref{approx} for a list of these expressions, including approximate expressions of the displacement, diameter, and asymmetry of the shadow in terms of the parameter $\beta$ in the metric of Ref.~\cite{Jmetric}. Figure~\ref{fig:rings09} shows the displacement, diameter, and asymmetry of the shadow of Kerr-like black holes with spin $a=0.9r_g$ as a function of the inclination for different values of the deviation parameters $\alpha_{13}$, $\alpha_{22}$, and $\beta$ in the metric of Ref.~\cite{Jmetric}. Reference~\cite{Abdujabbarov15} expressed the polar curve $R(\vartheta$) as an expansion in Legendre polynomials.

\begin{figure*}[h]
\begin{center}
\psfig{figure=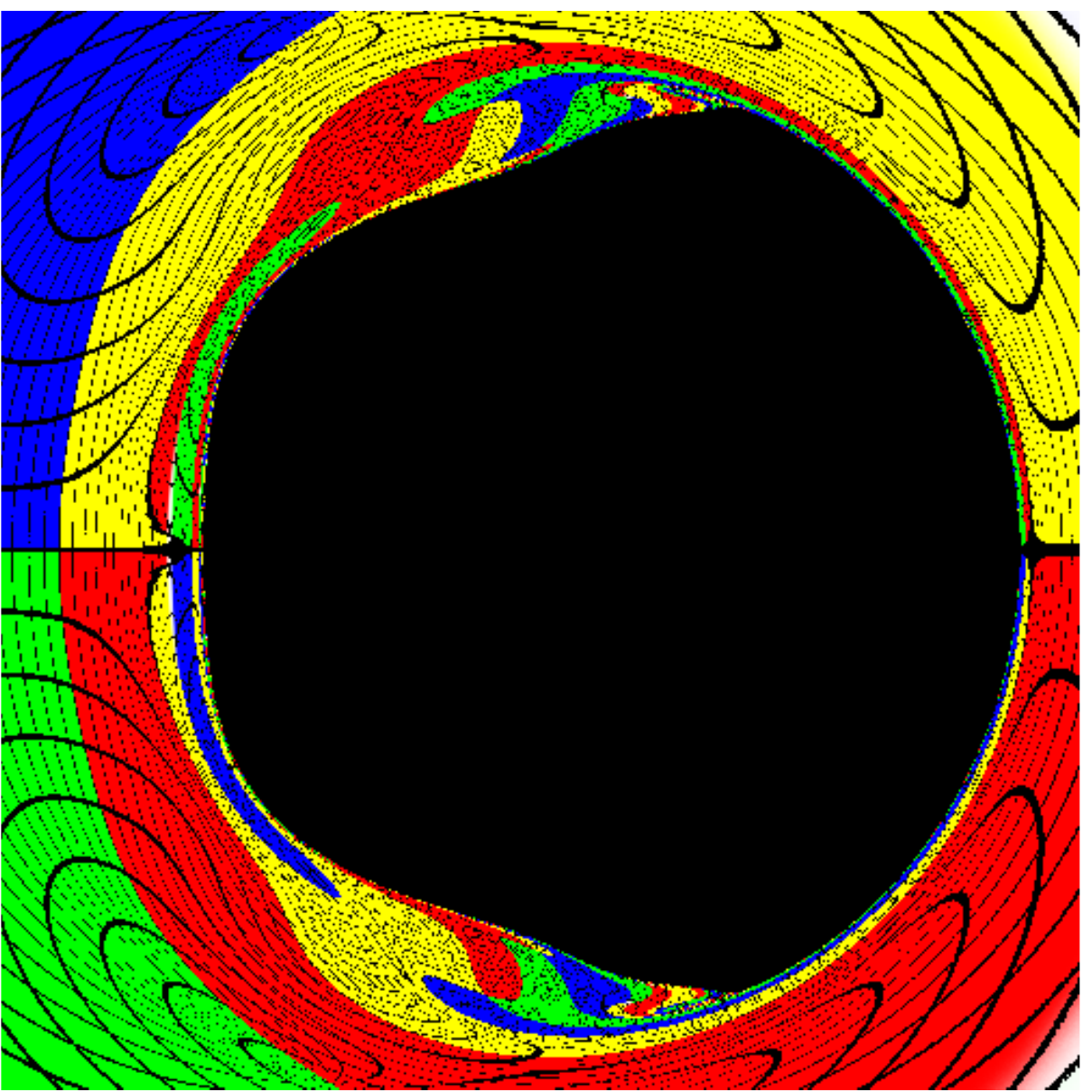,height=2.02in}
\psfig{figure=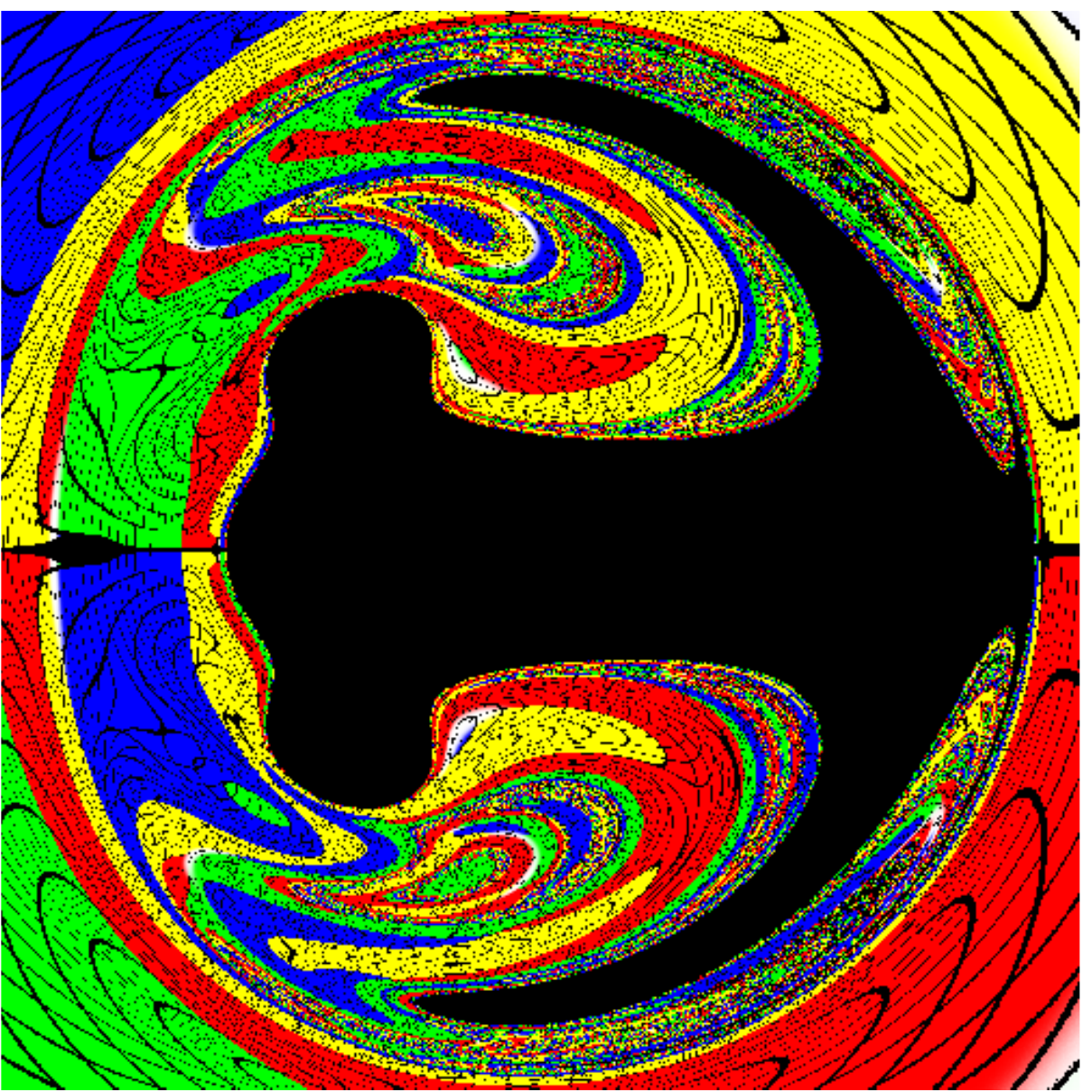,height=2.02in}
\psfig{figure=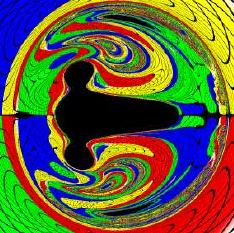,height=2.0in}
\end{center}
\caption{Images of shadows of Kerr black holes surrounded by a toroidal-shaped scalar field which can carry a significant fraction of the total mass and angular momentum. The images correspond to objects with values (in units of $1/\mu$ where $\mu$ is the mass of the boson particle and $G=c=1$) of the Arnowitt--Deser--Misner (ADM) mass $M_{\rm ADM}\approx0.965,0.966,0.975$, black-hole mass $M_{\rm H}\approx0.157,0.020,0.018$, ADM angular momentum $J_{\rm ADM}\approx0.789,0.834,0.85$, and black-hole angular momentum $J_{\rm H}\approx0.0713,0.0030,0.002$, respectively, viewed at an inclination of $90^\circ$~\cite{Herdeiro15PRL}. The deformations of the shadows deviate greatly from a nearly circular shape (not shown) for values of the fraction $J_{\rm H}/M_{\rm H}^2>1$. In this case, the shadow can even be disconnected as shown in the right panel (corresponding to the fraction $J_{\rm H}/M_{\rm H}^2\approx6.2\times10^3$) and include two ``eyebrow`` shadows above and below the central ``hammer-like`` shadow. The different colors (red, blue, yellow, green) illustrate the gravitational lensing by the object. The extreme deformations of the shadow cannot be modeled by any presently known Kerr-like metric, but should be easily distinguishable from the deformed shadows shown in Fig.~\ref{fig:shadowshapes} with EHT observations. Taken from Ref.~\cite{Herdeiro15PRL}.}
\label{fig:KBHSHshadows}
\end{figure*}

Recently, Ref.~\cite{Herdeiro15PRL} analyzed the shapes of shadows of Kerr black holes surrounded by a stable toroidal-shaped scalar field which can carry a significant fraction of the total mass and angular momentum of the system~\cite{Herdeiro14PRL} (c.f., Refs.~\cite{Arvanitaki11,Brito15}). For the case in which the angular momentum of the black hole exceeds the Kerr bound in Eq.~(\ref{eq:kerrbound}) (i.e., $J_{\rm H}/M_{\rm H}^2>1$), Ref.~\cite{Herdeiro15PRL} showed that the deformations of the shadows of such objects can be extreme and that disjoint shadows can appear in the image. Figure~\ref{fig:KBHSHshadows} shows three images of the shadow of such an object with increasingly extreme deformations for values of the dimensionless angular momentum of the black hole $J_{\rm H}/M_{\rm H}^2\approx2.9,7.5,6.2\times10^3$, respectively, which cannot be modeled by any presently known (vacuum) Kerr-like metric satisfying Eq.~(\ref{eq:kerrbound}). However, such extreme deformations of the shadow should be easily distinguishable from the deformed shadows shown in Fig.~\ref{fig:shadowshapes} with EHT observations (c.f., Fig.~\ref{fig:VincentImages}). Note that the (numerical) analysis of Ref.~\cite{Herdeiro15PRL} finds no singularities or pathological regions on or outside of the event horizon implying that the central object in the presence of the scalar field remains a black hole even if the Kerr bound is violated. Reference~\cite{Abdolrahimi15} showed that the exterior domain of a Kerr black hole coupled to other matter sources with a more general configuration can likewise be free of singularities in that case.

\subsection{The Accretion Flow of Sgr~A$^\ast$}
\label{subsec:accretionflow}

In addition to the shadow, the accretion flow that surrounds Sgr~A$^\ast$ probes the innermost region near the event horizon outside of the black hole and can reveal important characteristics of the underlying spacetime. Unlike active galactic nuclei, Sgr~A$^\ast$ is underluminous, with a bolometric luminosity of $10^{−9}$ in Eddington units~\cite{loweff}. While the detailed morphology of the emitting region remains uncertain, the existing spectral and polarization data across the electromagnetic spectrum have provided insight into some of the properties of the accretion flow which include a peaked (often approximated by a Maxwellian) electron distribution function with a power-law high-energy tail (e.g., \cite{Serabyn97,FalckeGoss98,Hornstein02,ZhaoYoung03,Genzeletal03,GhezWright04,Marrone06,Bremer11,Dodds-Eden11,Schodel11}; see the left panel of Fig.~\ref{fig:ehtsources}), nearly equipartition magnetic fields~\cite{Eatough13}, and variability (see Sec.~\ref{sec:variability}). See Ref.~\cite{Falcke13} for a review.

Several plausible models for the accretion flow of Sgr~A$^\ast$ exist~(e.g., \cite{Narayan98,Blandford99,Falcke99,Ozel00,Yuan03,Loeb07,Straub12,VincentTorus15}), many of which are categorized as radiatively inefficient accretion flows (RIAFs). Assuming that Sgr~A$^\ast$ is a Kerr black hole, Refs.~\cite{Bro09a,Bro11a} combined the early EHT observations of Sgr~A$^\ast$ in 2007--2009~\cite{Doele08,Fish11} with measurements of its spectral energy distribution and obtained values of the spin magnitude and direction employing the RIAF model of Refs.~\cite{Yuan03,Bro06a}. Likewise, Ref.~\cite{Huang09} analyzed the early EHT data of Sgr~A$^\ast$ using an accretion flow model with plasma wave heating for several different values of the spin and disk inclination. References~\cite{Moscibrodzka09,Dexter09,Dexter10,Ricarte15} fitted the early EHT and spectral data to sets of images obtained from GRMHD simulations~\cite{Gammie03,Fragile07,Fragile09,McKinney09}. In the future, the determination of the spin and orientation of the black hole can be complemented with a multiwavelength study of polarization~\cite{Bro06b} (c.f., Refs.~\cite{Schnittman09,Schnittman10,Schnittman16}). See Refs.~\cite{Falcke13,Yuan14} for reviews on accretion flow models of Sgr~A$^\ast$. The outer extent of the accretion flow could be constrained by the observation (or lack thereof) of X-ray flares originating from shock waves caused by the interaction of S-stars and their winds with the accretion flow~\cite{Christie16}.

References~\cite{ChanPsaltis15a,ChanPsaltis15b} computed images and spectra for a set of six GRMHD simulations with different magnetic field configurations, black-hole spins, and thermodynamic properties and showed that the combination of current spectral and early EHT observations rules out all models with strong funnel emission. Reference~\cite{ChanPsaltis15c} showed that GRMHD simulations for disk-dominated models produce short timescale variability in accordance with current observationations, while GRMHD simulations for jet-dominated models generate only slow variability, at lower flux levels. Neither set of models show any X-ray flares, which most likely indicate that additional physics, such as particle acceleration mechanisms, need to be incorporated into the GRMHD simulations. A similar analysis by Ref.~\cite{Gold16} showed that current observations favor models with ordered magnetic fields near the black hole event horizon, although both disk- and jet-dominated emission can satisfactorily explain most of the current EHT data. Reference~\cite{Gold16} also showed that stronger model constraints should be possible with upcoming circular polarization and higher frequency ($349~{\rm GHz}$) measurements.

Reference~\cite{PsaltisAlign15} argued that the angular momentum vector of the accretion flow (and perhaps of the black hole itself) is aligned with the angular momentum vector of the inner disk of stars within $\sim3''$ of Sgr~A$^\ast$. Reference~\cite{Kim16} inferred the spin orientation of Sgr~A$^\ast$ from the 2007--09 EHT data using a Bayesian estimator based on different GRMHD simulations reported in Refs.~\cite{ChanPsaltis15a,ChanPsaltis15b}. These results are broadly consistent with the spin orientation obtained by Refs.~\cite{Bro09a,Bro11a} but have a larger overall uncertainty~\cite{Kim16}. Reference~\cite{Encinas16} showed that different disk and jet models in GRMHD simulations based on those by Ref.~\cite{Moscibrodzka09} are consistent with the closure phase measurement by Ref.~\cite{Fish11} and tend to favor higher inclinations, while the spin magnitude and orientation are only poorly constrained by the same measurement.

The follow-up observations with the EHT in 2009--2013 measured a number of closure phases along the SMTO--CARMA--Hawaii triangle which have a nonzero (positive) mean~\cite{Fish15}. This implies that the millimeter emission from Sgr~A$^\ast$ is asymmetric on scales of a few Schwarzschild radii and can be used to break the $180^\circ$ rotational degeneracy of amplitude data alone. Since the sign of these closure phase measurements remained stable over most observing nights, the implied asymmetry in the image of Sgr A$^\ast$ is likely persistent and unobscured by refraction due to interstellar electrons along the line of sight~\cite{Fish15}.

Ref.~\cite{Brodericketal15} updated the RIAF analysis of Refs.~\cite{Bro09a,Bro11a} including the EHT data of Ref.~\cite{Fish15} finding an improvement of the constraints on the spin magnitude and orientation as well as on the inclination by about a factor of two. While the $180^\circ$ degeneracy of the spin orientation angle is now removed, a reflection degeneracy in the inclination remains. One of these inclinations is in remarkable agreement with the orbital angular momentum of the infrared gas cloud G2~\cite{Gillessen12,Pfuhl15} and the clockwise disk of young stars surrounding Sgr~A$^\ast$~\cite{Luetal09,Genzel10}, possibly suggesting a relationship between the accretion flow of Sgr~A$^\ast$ and these features~\cite{Brodericketal15}.

\begin{figure*}[ht]
\begin{center}
\psfig{figure=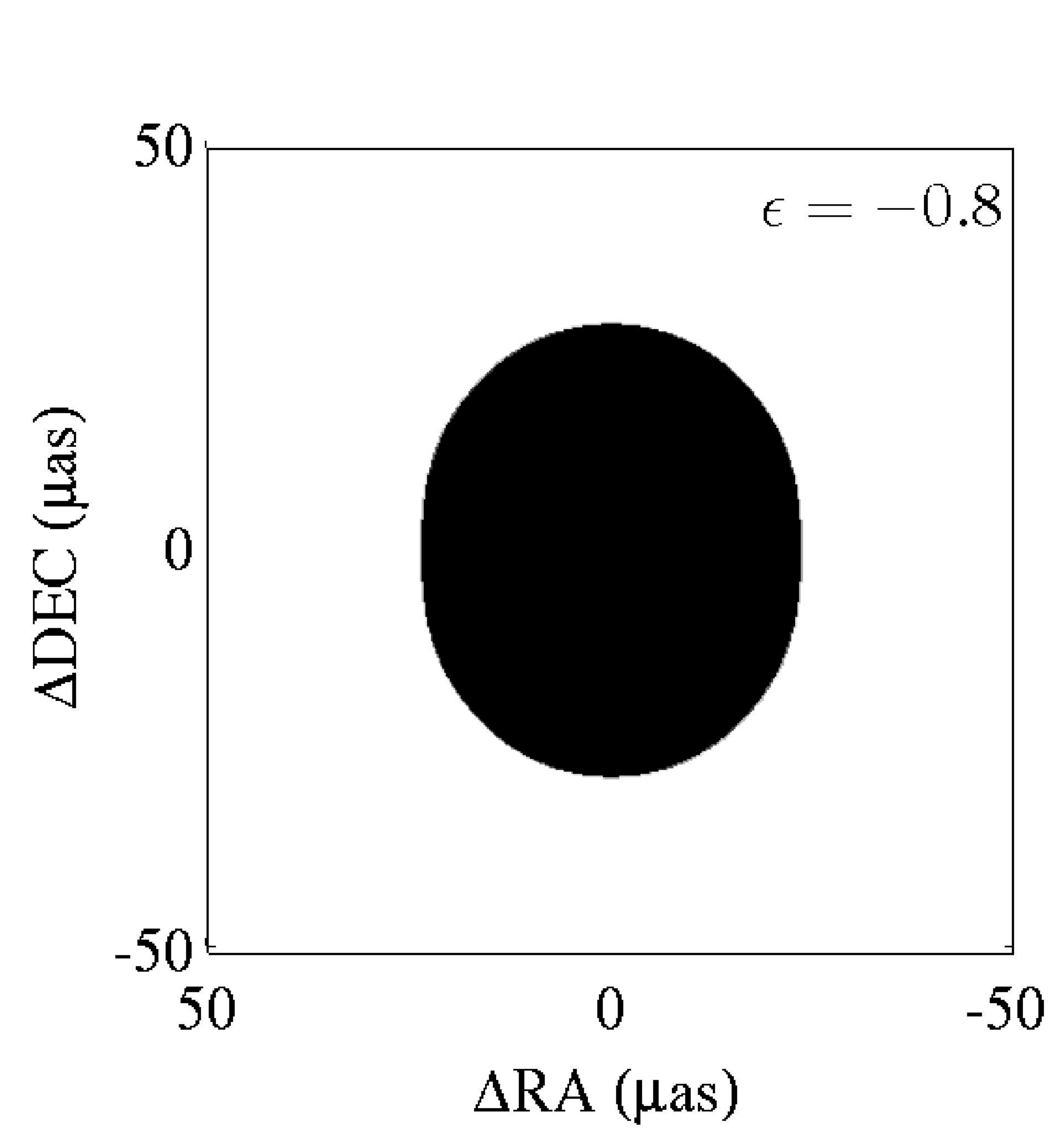,height=2.07in}
\psfig{figure=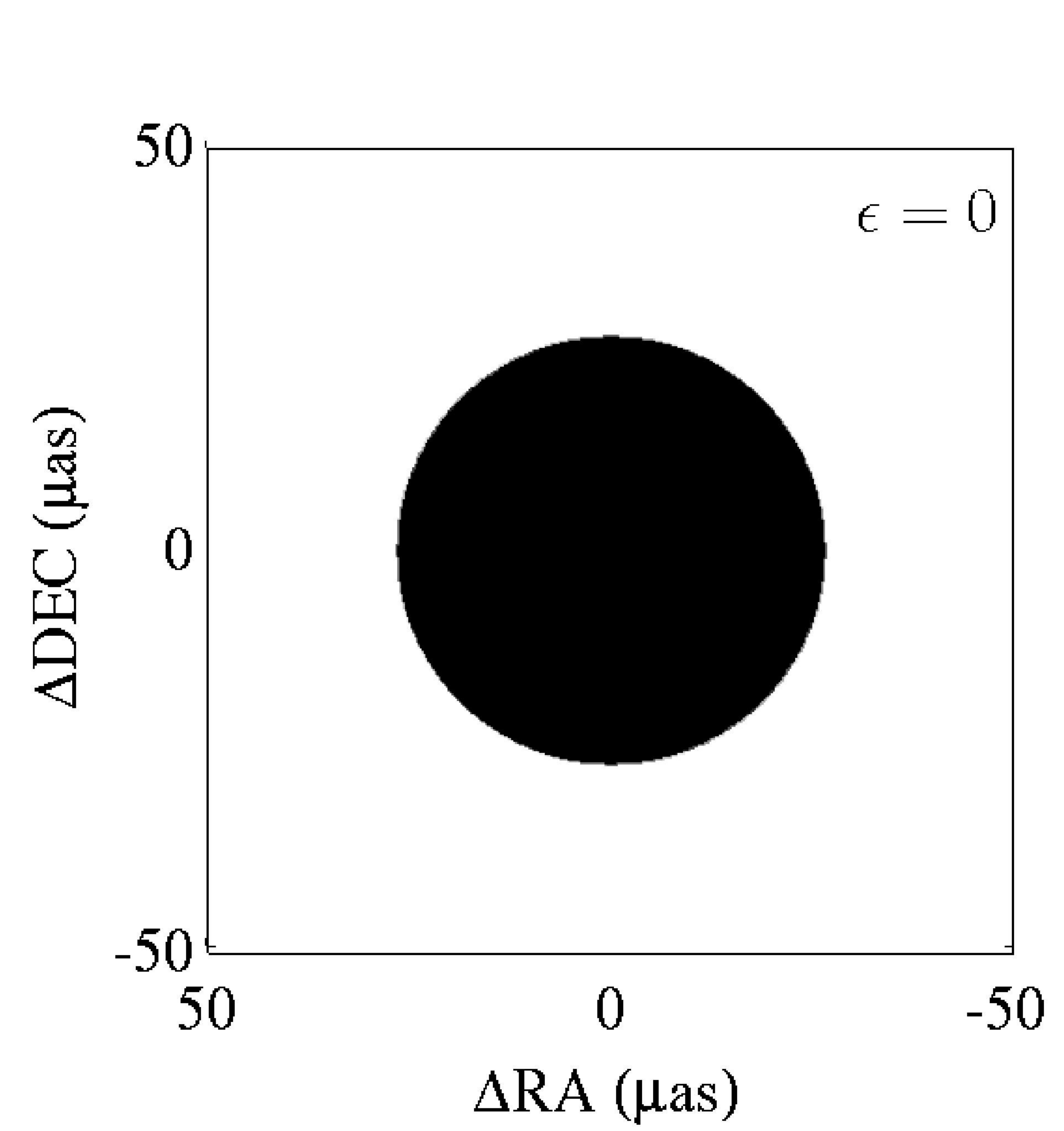,height=2.07in}
\psfig{figure=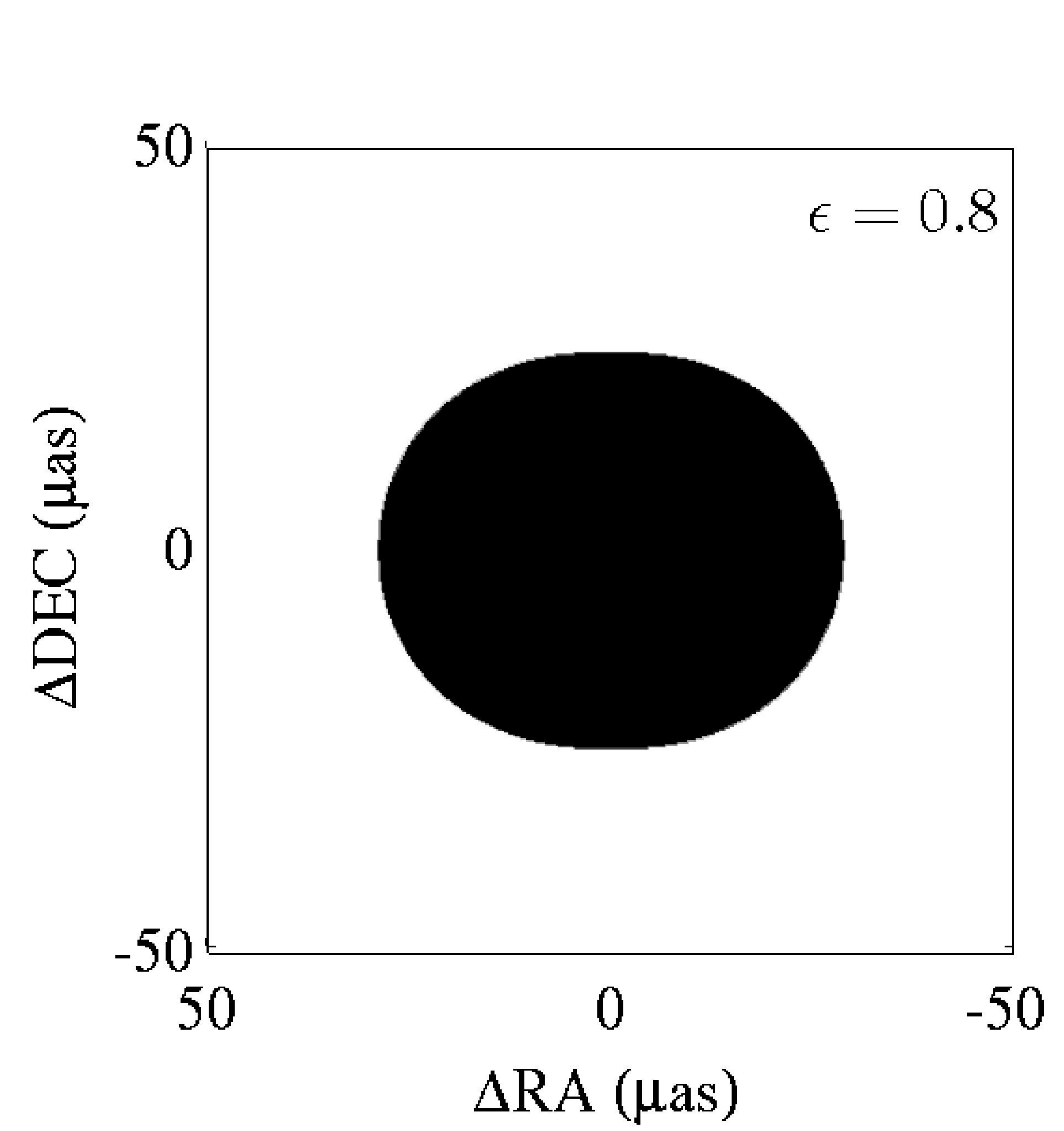,height=2.07in}
\psfig{figure=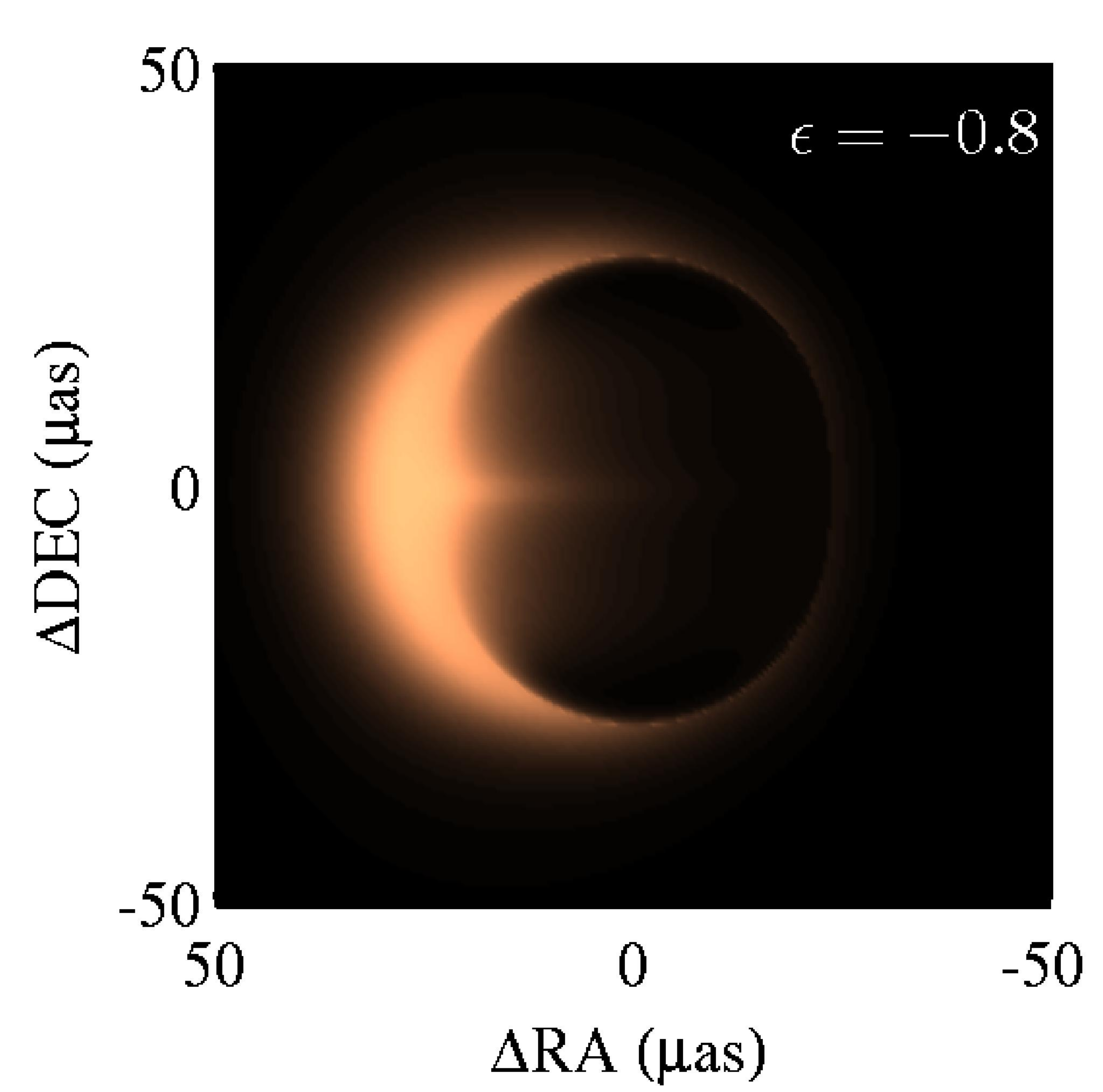,width=2.02in}
\psfig{figure=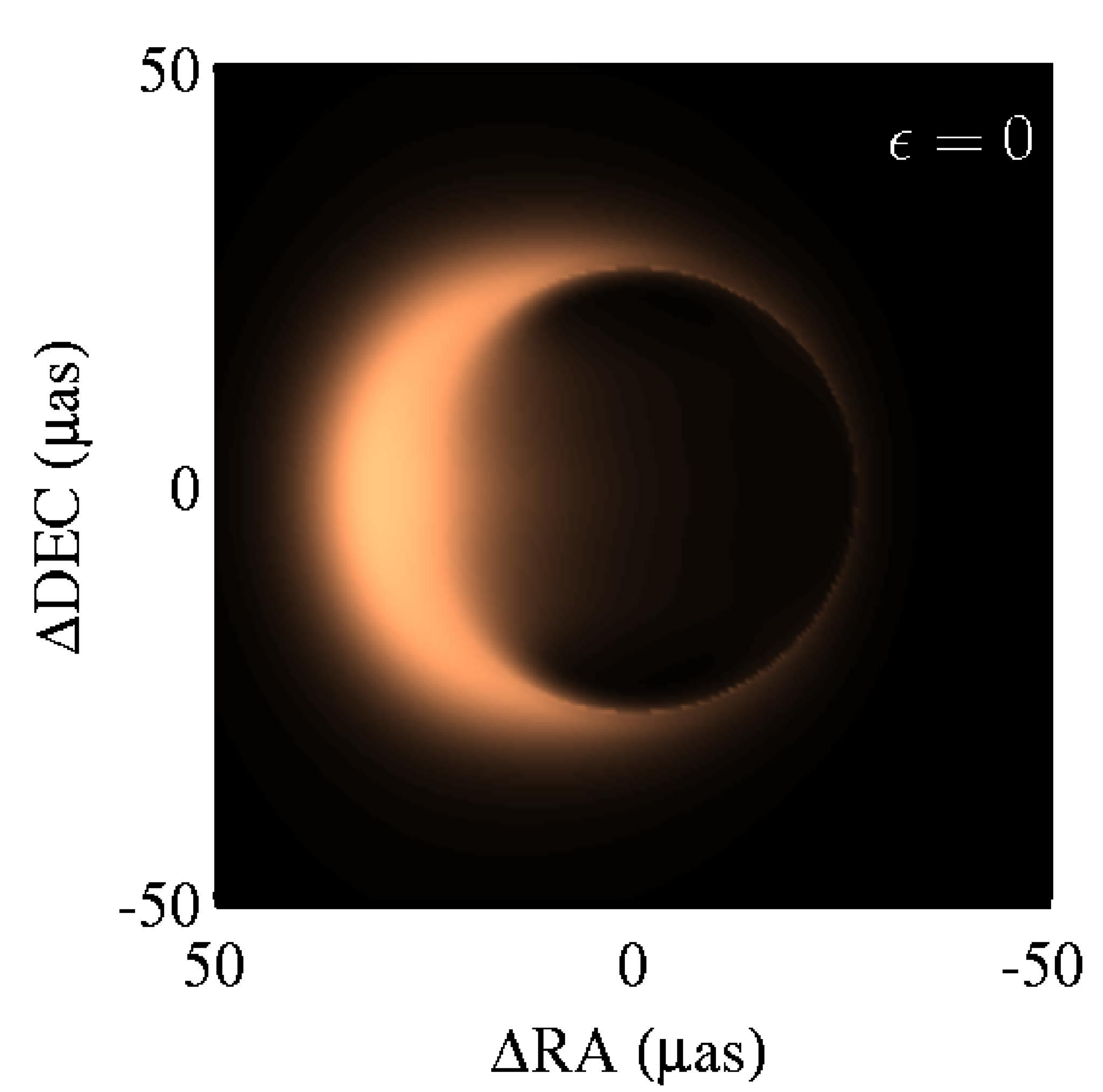,width=2.02in}
\psfig{figure=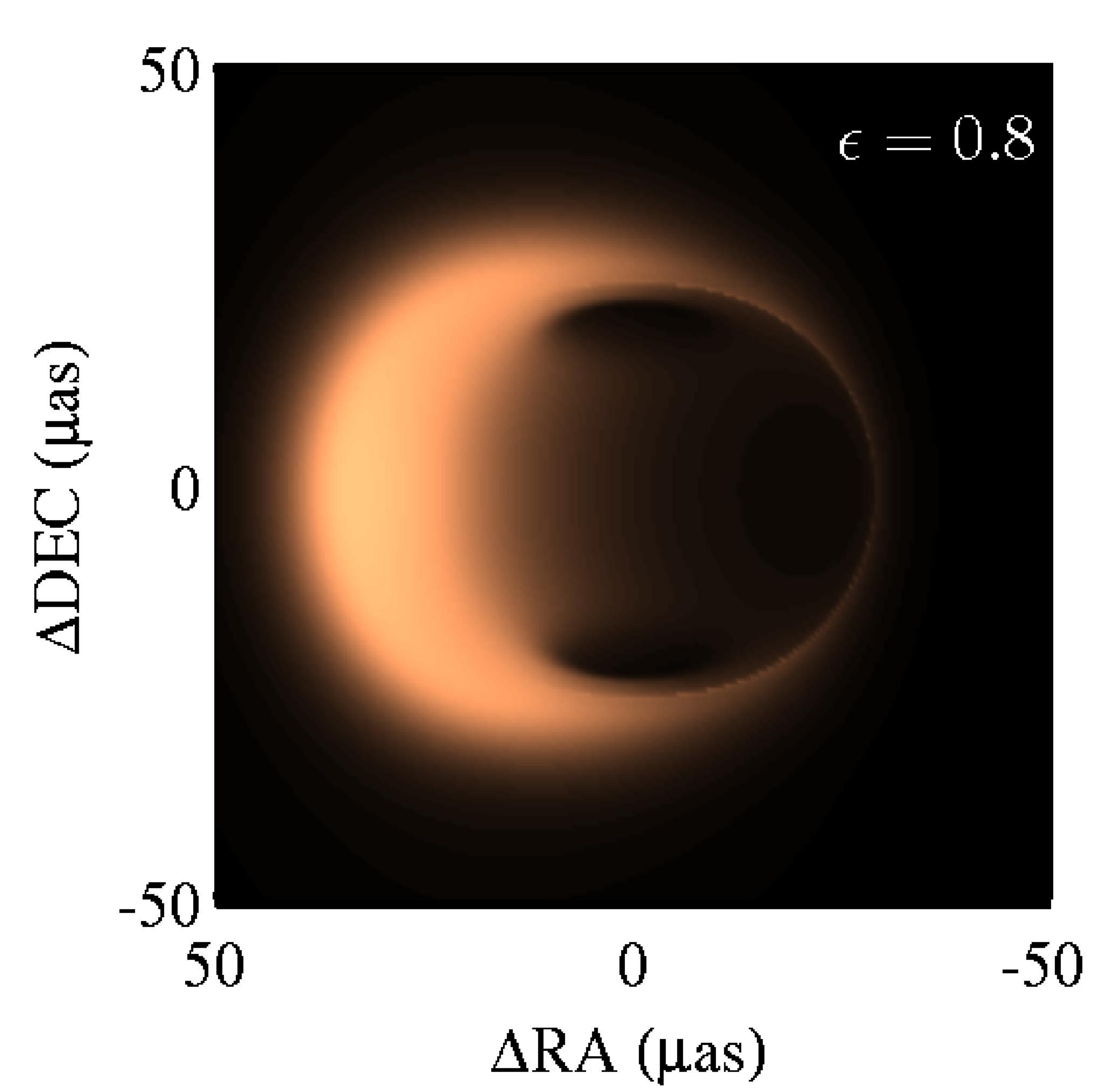,width=2.02in}
\end{center}
\caption{Examples of (top row) black hole shadows for non-rotating black holes viewed at an inclination $\theta=90^\circ$ with illustrative values of the quadrupolar parameter $\epsilon$ and (bottom row) spectrally fit RIAF model images at 230~GHz for the same parameter values. For negative values of the deviation parameter $\epsilon$ (left panels), the shadow has a more prolate shape than the shadow of a Kerr black hole with the same spin (central panels), while for positive values of the deviation parameter (right panels), the shadow has a more oblate shape. In all cases, the shadow is clearly visible in the model images even though it is partially obscured by the accretion flow on the left side of the shadow due to relativistic boosting and beaming. Nonzero values of the deviation parameter $\epsilon$ modify the morphology and measured intensity of the crescent and, for sufficiently negative values of the deviation parameter, the crescent acquires a more pronounced tongue-like flux feature in the equatorial plane of the black hole. Taken from Ref.~\cite{Bro14}.}
\label{fig:RIAF1}
\end{figure*}

\begin{figure*}[ht]
\begin{center}
\psfig{figure=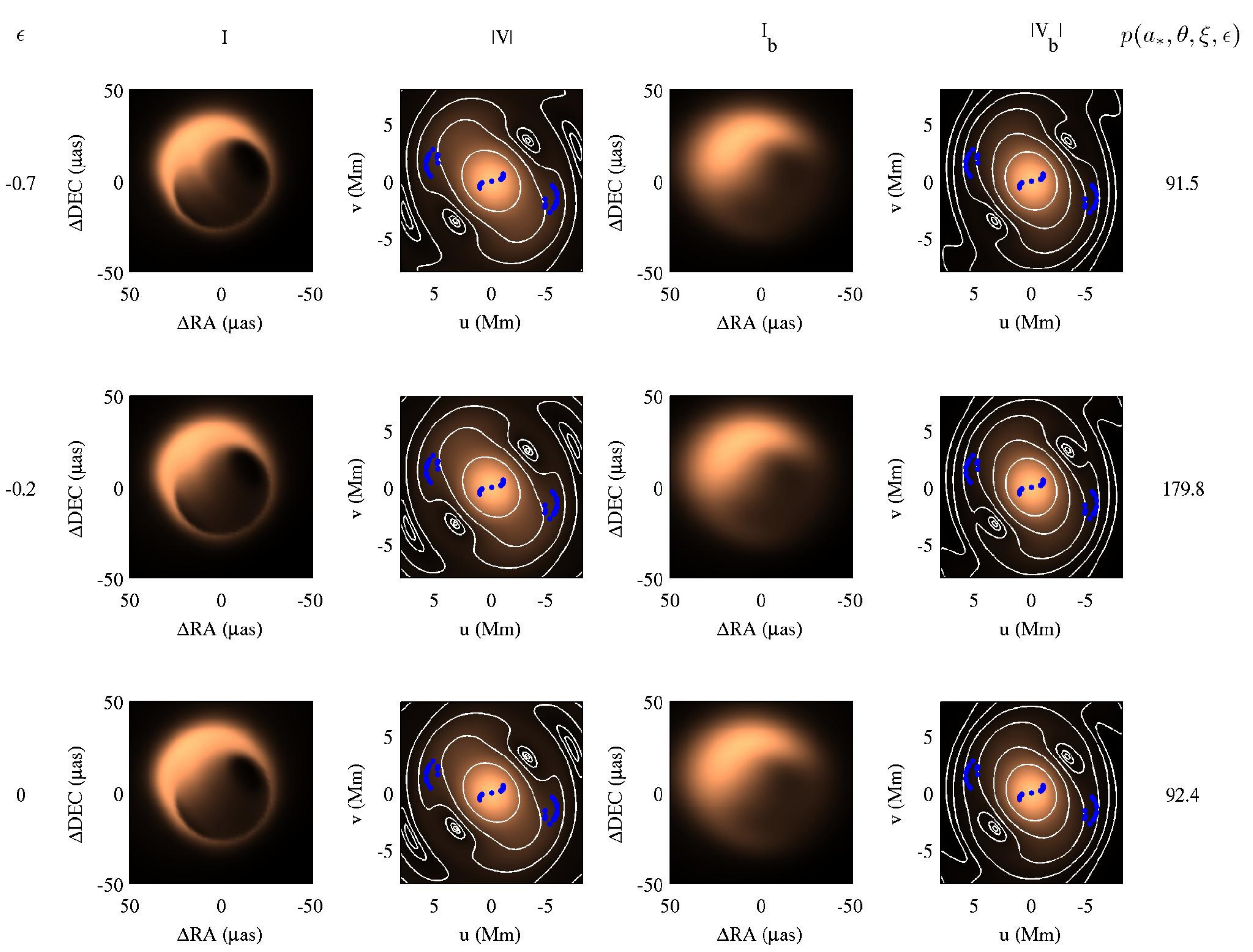,height=4.5in}
\end{center}
\caption{RIAF images and visibility magnitudes of Sgr~A$^\ast$ assuming values of the spin magnitude $a=0r_g$, spin orientation $\xi=127^\circ$, and inclination $\theta=65^\circ$ for (top to bottom rows) different values of the parameter $\epsilon$. The panels in columns 1-2 and 3-4, respectively, show the corresponding intrinsic and scatter-broadened images. Even though these images look similar, existing EHT data (blue points in the visibility magnitude plots) can already distinguish them assigning each image at different likelihood $p(a,\theta,\xi,\epsilon)$ of being consistent with the data. Taken from Ref.~\cite{Bro14}.}
\label{fig:RIAF2}
\end{figure*}

Reference~\cite{Bro14} performed an analysis of RIAF images of Sgr~A$^\ast$ similar to the ones of Refs.~\cite{Bro09a,Bro11a}, but using the quasi-Kerr metric as the underlying spacetime. For nonzero values of the deviation parameter $\epsilon$, the shadow becomes asymmetric~\cite{PaperII} (see the discussion in Sec.~\ref{subsec:shadows}). Reference~\cite{Bro14} showed that images of accretion flows in the quasi-Kerr spacetime can be significantly different from images of accretion flows around Kerr black holes revealing the asymmetric distortions of the shadow. Figure~\ref{fig:RIAF1} shows a set of black hole shadows and the corresponding RIAF images for values of the spin $a=0r_g$, inclination $\theta=90^\circ$, and different values of the deviation parameter $\epsilon$.

Reference~\cite{Bro14} also showed that such differences in the RIAF images can be distinguished already by early EHT data~\cite{Doele08,Fish11}. Figure~\ref{fig:RIAF2} shows RIAF images and visibility magnitudes of Sgr~A$^\ast$ with and without the effect of the observed scatter-broadening of such images assuming a Schwarzschild black hole with inclination $\theta=65^\circ$ and orientation $\xi=127^\circ$ for different values of the parameter $\epsilon$. Even though the images look similar, their corresponding likelihoods of being consistent with the EHT data vary by about a factor of two.

Reference~\cite{Bro14} only considered values of the spin and the deviation parameter $\epsilon$ for which the ISCO lies at a radius $r\geq4r_g$ and neglected all radiation passing through a cutoff radius located at $r=3r_g$ in order to avert the adverse impact of the naked singularity harbored by this metric. The cutoff radius acts as an artificial event horizon and effectively turns the compact object into a black hole for the purposes of the simulated images. Thereby, Ref.~\cite{Bro14} actually underestimate the effects of the spin and the deviation parameter on the images, which are strongest near the compact object. However, with this choice the quasi-Kerr metric can also be applied to ``rapidly'' spinning black holes, although it is, strictly speaking, only valid for slowly to moderately spinning black holes (see the discussion in Ref.~\cite{pathologies} and Sec.~\ref{subsec:metrics}).

\begin{figure*}[ht]
\begin{center}
\psfig{figure=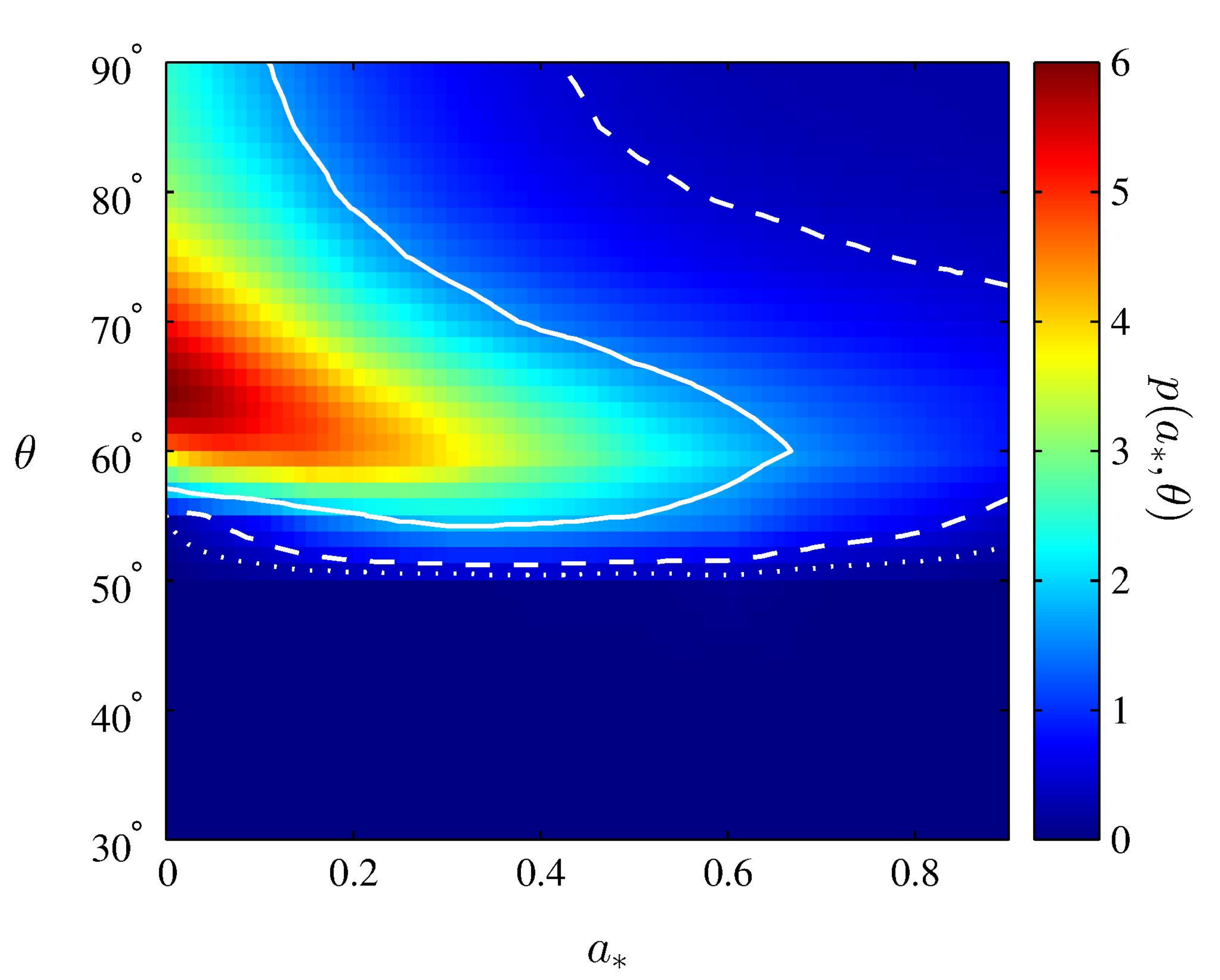,height=2.4in}
\psfig{figure=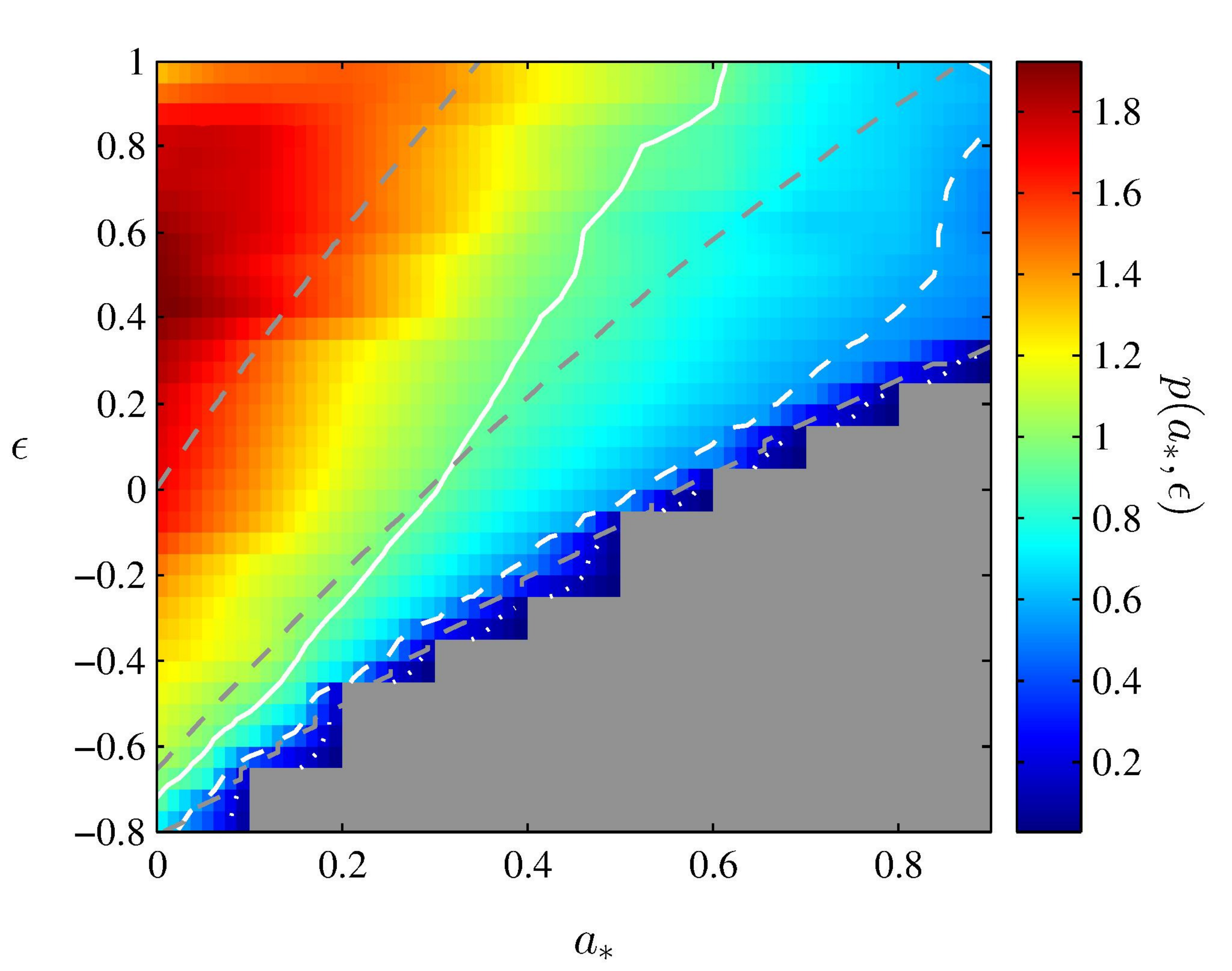,height=2.4in}
\psfig{figure=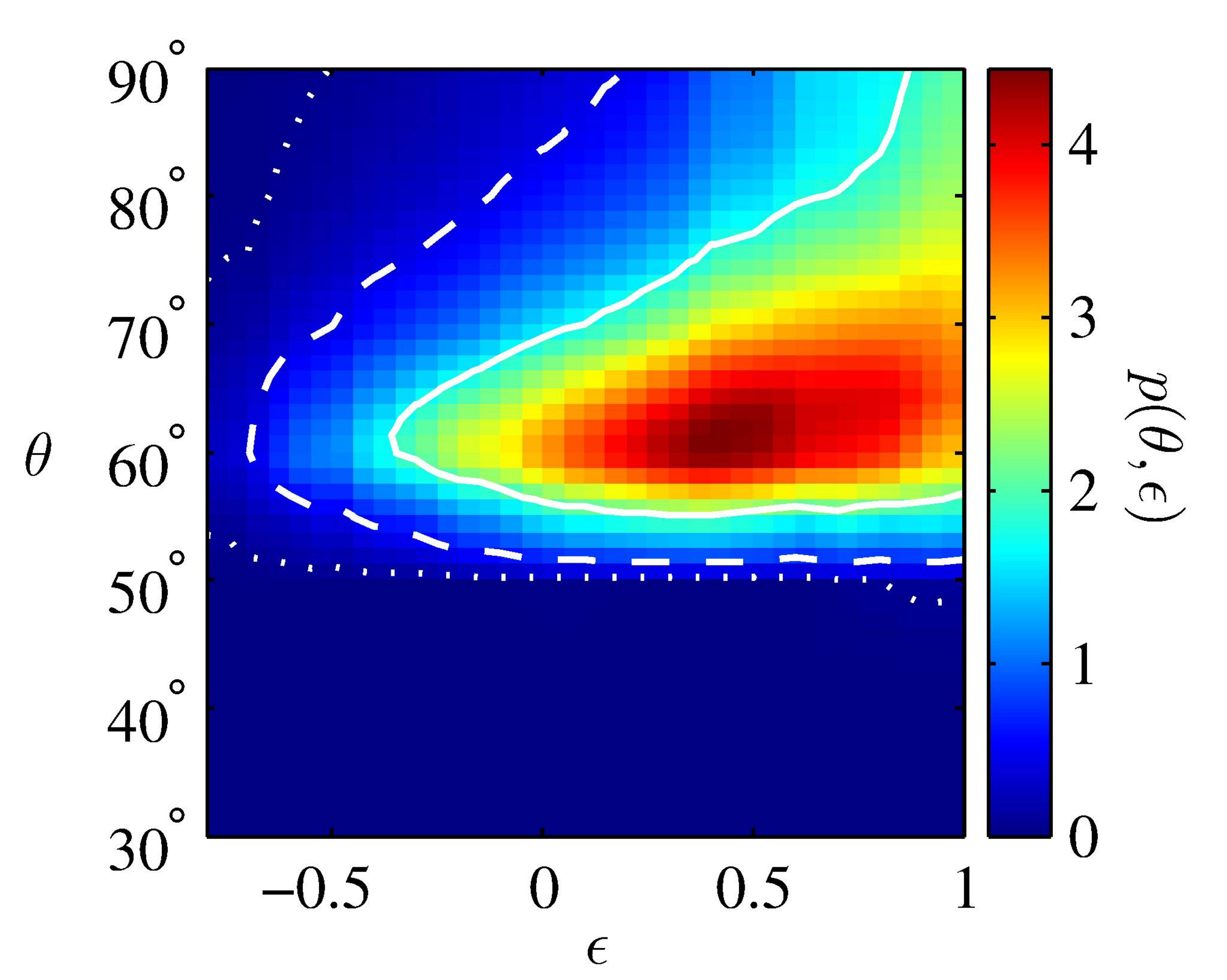,height=1.6in}
\psfig{figure=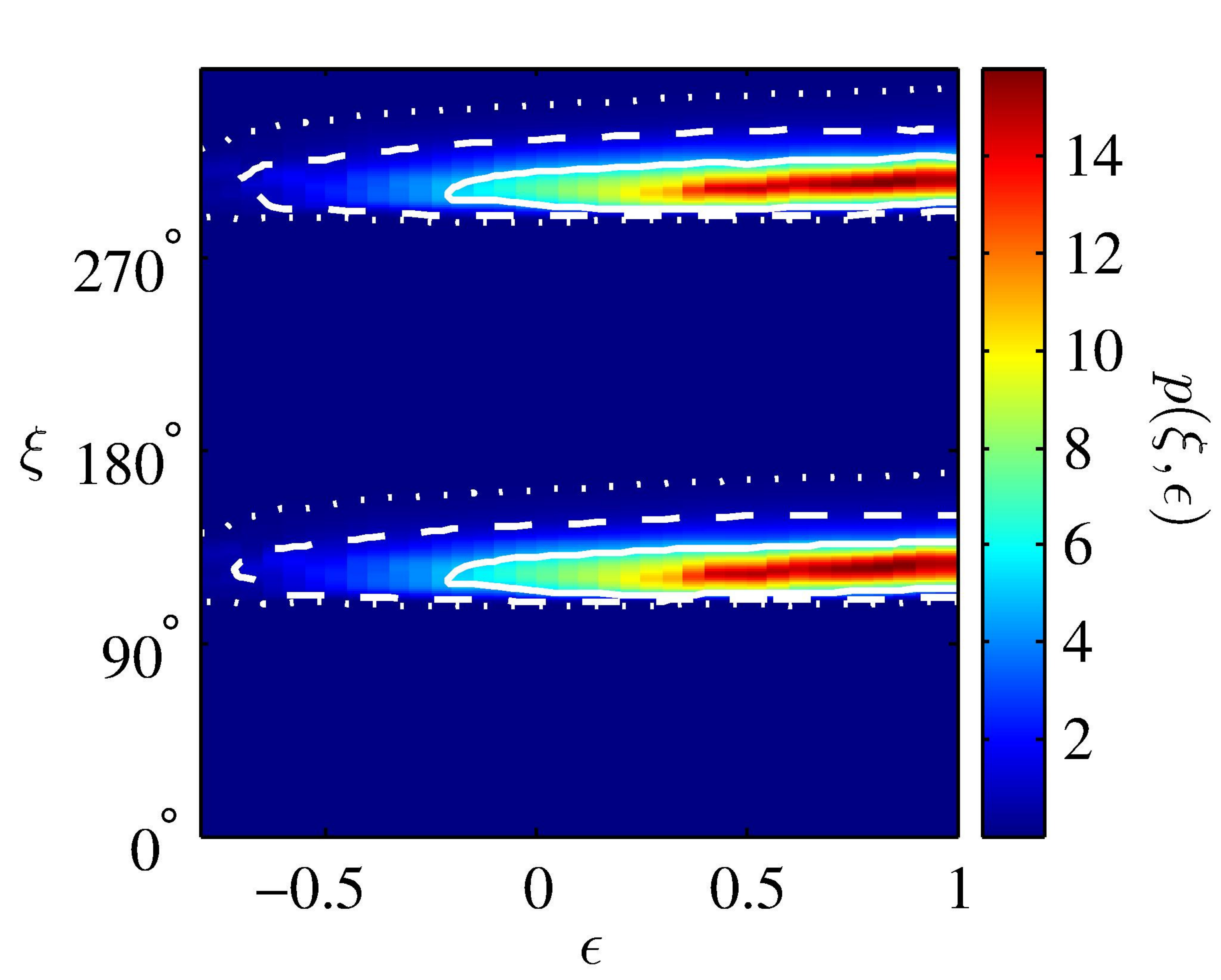,height=1.6in}
\psfig{figure=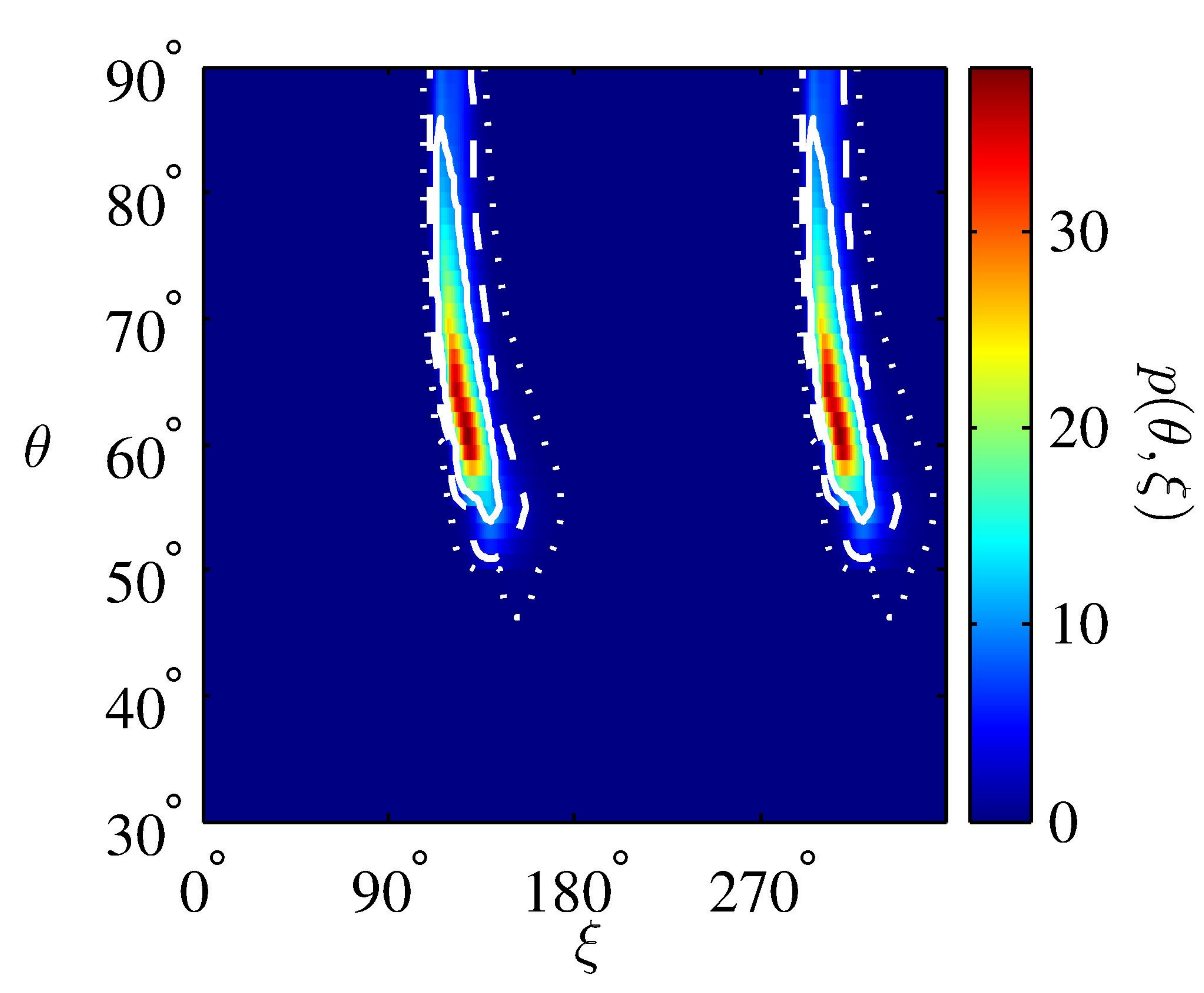,height=1.6in}
\end{center}
\caption{2D posterior probability densities as a function of (top row, left panel) dimensionless spin magnitude $a_*$ and inclination $\theta$, (top row, right panel) spin magnitude and quadrupolar deviation $\epsilon$, (bottom row, left panel) inclination and quadrupolar deviation, (bottom row, center panel) spin orientation and quadrupolar deviation, and (bottom row, right panel) inclination and spin orientation, respectively marginalized over all other quantities. In each panel, the solid, dashed, and dotted lines show the $1\sigma$, $2\sigma$, and $3\sigma$ confidence regions, respectively. In the top right panel, lines of constant ISCO radius are shown as dashed gray lines, corresponding to $6r_g$, $5r_g$, and $4r_g$ from top to bottom, while the gray region in the lower right is excluded. Taken from Ref.~\cite{Bro14}.}
\label{fig:RIAF3}
\end{figure*}

Fitting the early EHT data to a library of RIAF images, Ref.~\cite{Bro14} showed that previous measurements of the inclination and spin position angle in the same RIAF model~\cite{Bro09a,Bro11a} are robust to the inclusion of a quadrupolar deviation from the Kerr metric. Figure~\ref{fig:RIAF3} shows the 2D posterior probability densities of various combinations of the spin magnitude, spin orientation, inclination, and quadrupolar deviation, each marginalized over the remaining two parameters not shown. The spin magnitude and the quadrupolar deviation are strongly correlated, roughly along lines of constant ISCO radius as shown in Fig.~\ref{fig:RIAF3}, while the spin and the inclination are only modestly correlated. The spin orientation could be determined only up to a $180^\circ$ degeneracy. Reference~\cite{Bro14} obtained constraints (with $1\sigma$ errors) on the spin magnitude $a_*=0^{+0.7}$, spin orientation $\xi=127^{\circ+17^\circ}_{~-14^\circ}$ (up to a $180^\circ$ degeneracy), and inclination $\theta=65^{\circ+21^\circ}_{~-11^\circ}$, while constraints on the deviation parameter $\epsilon$ remained weak. However, such constraints within a specific RIAF model will improve dramatically with EHT observations using larger telescope arrays~\cite{Johannsenetal15}.

\begin{figure*}[h]
\centering{
\includegraphics[width=7.2cm]{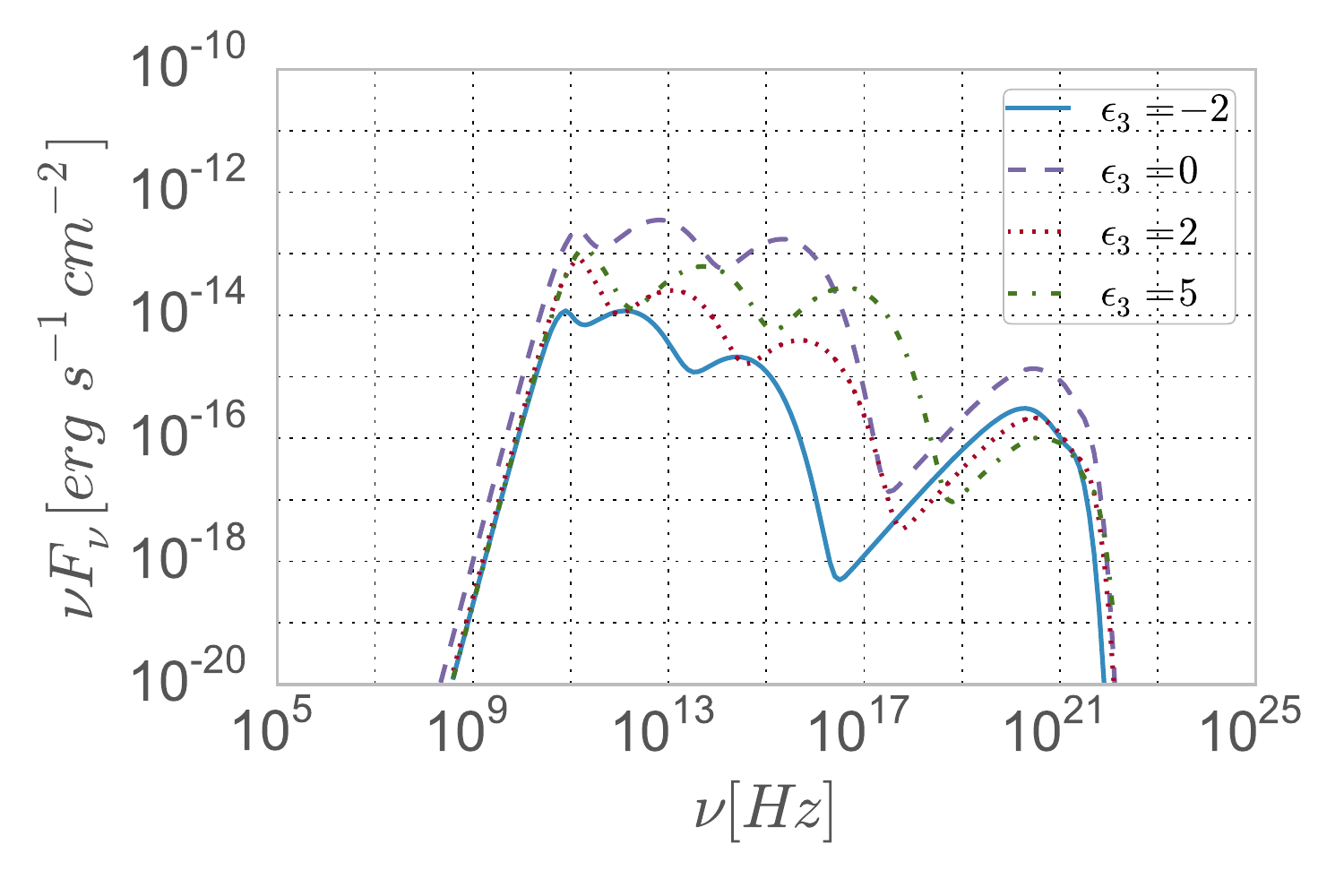}
\includegraphics[width=7.2cm]{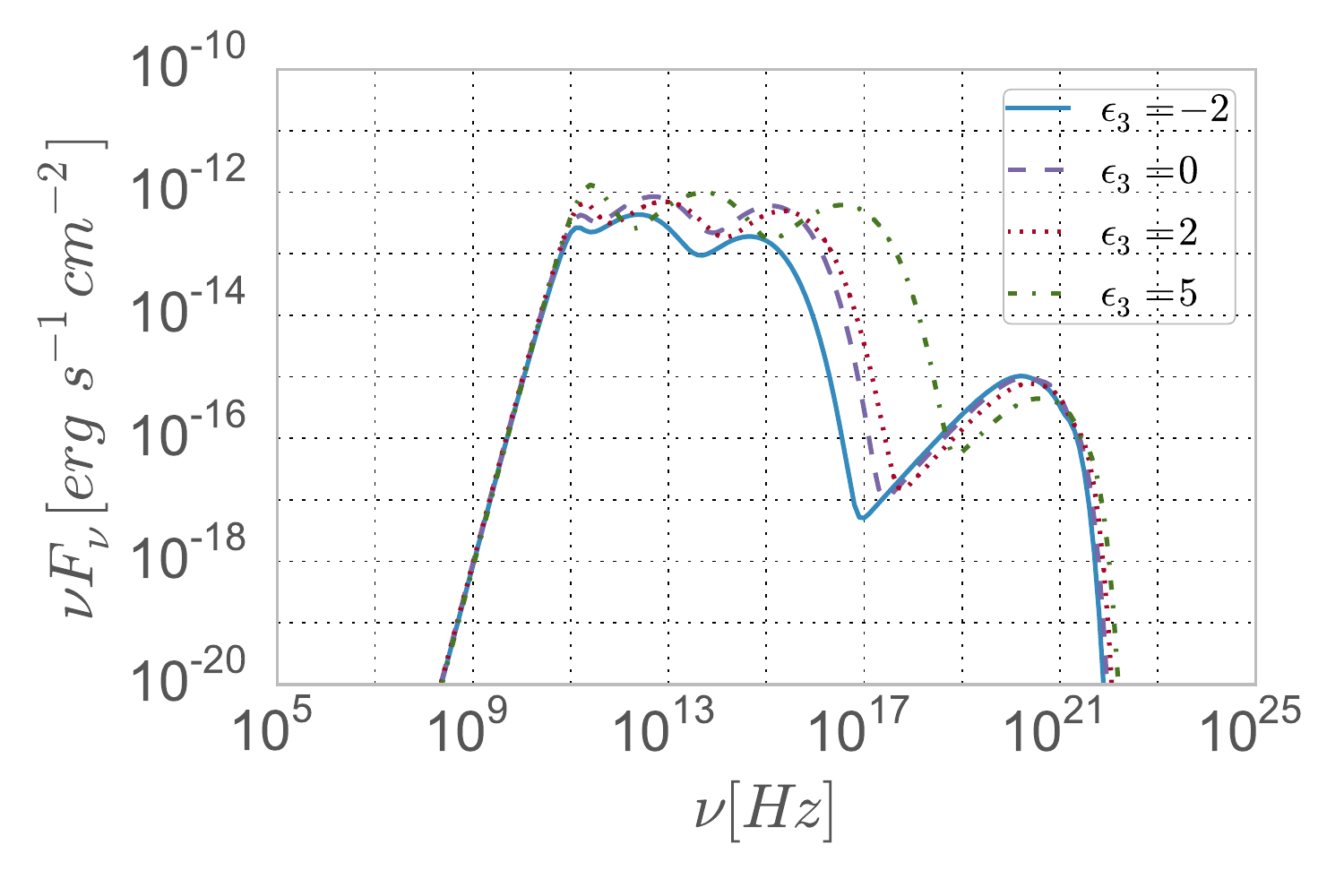}
}
\caption{Simulated spectra emitted by a toroidal accretion flow surrounding Sgr~A$^\ast$ in the model of Refs.~\cite{Abramowicz78,Jaroszynski80,Komissarov06,Straub12,VincentTorus15} for different values of the deviation parameter $\epsilon_3$ in the metric of Ref.~\cite{JPmetric}. All spectra were computed for fixed values of the spin $a=0.5r_g$, inclination $\theta=60^\circ$, specific angular momentum of the fluid particles (left panel) $\lambda=0.3$ and (right panel) $\lambda=0.6$, magnetic to total pressure ratio $\beta=0.1$, polytropic index $n=3/2$, central energy density $\rho_c = 10^{-17}$~g/cm$^3$, and central electron temperature $T_c = 0.02\,T_v$ where $T_v$ is the virial temperature. The spectra show a significent dependence on the deviation parameter. Taken from Ref.~\cite{BambiTorus15}.}
\label{fig:BambiTorus}
\end{figure*}

Building on the work of Refs.~\cite{Straub12,VincentTorus15} who calculated images and spectra for a toroidal accretion flow around a Kerr black hole in the model of Refs.~\cite{Abramowicz78,Jaroszynski80,Komissarov06}, Reference~\cite{BambiTorus15} calculated spectra for the same torus model in the background of the Kerr-like metric of Ref.~\cite{JPmetric} and showed that such spectra can depend significantly on deviations from the Kerr metric. Figure~\ref{fig:BambiTorus} shows two sets of spectra for different values of the deviation parameter, where the black hole has fixed values of the spin $a=0.5r_g$ and inclination $\theta=60^\circ$ and the torus has fixed values of the specific angular momentum of the fluid particles $\lambda=0.3$ or $\lambda=0.6$ (shown in the left and right panels, respectively), magnetic to total pressure ratio $\beta=0.1$, electron to ion temperature ratio $\xi=0.1$, polytropic index $n=3/2$, central energy density $\rho_c = 10^{-17}$~g/cm$^3$, and central electron temperature $T_c = 0.02\,T_v$, where $T_v$ is the virial temperature.

\begin{figure*}[h]
\centering{
\includegraphics[height=5.3cm]{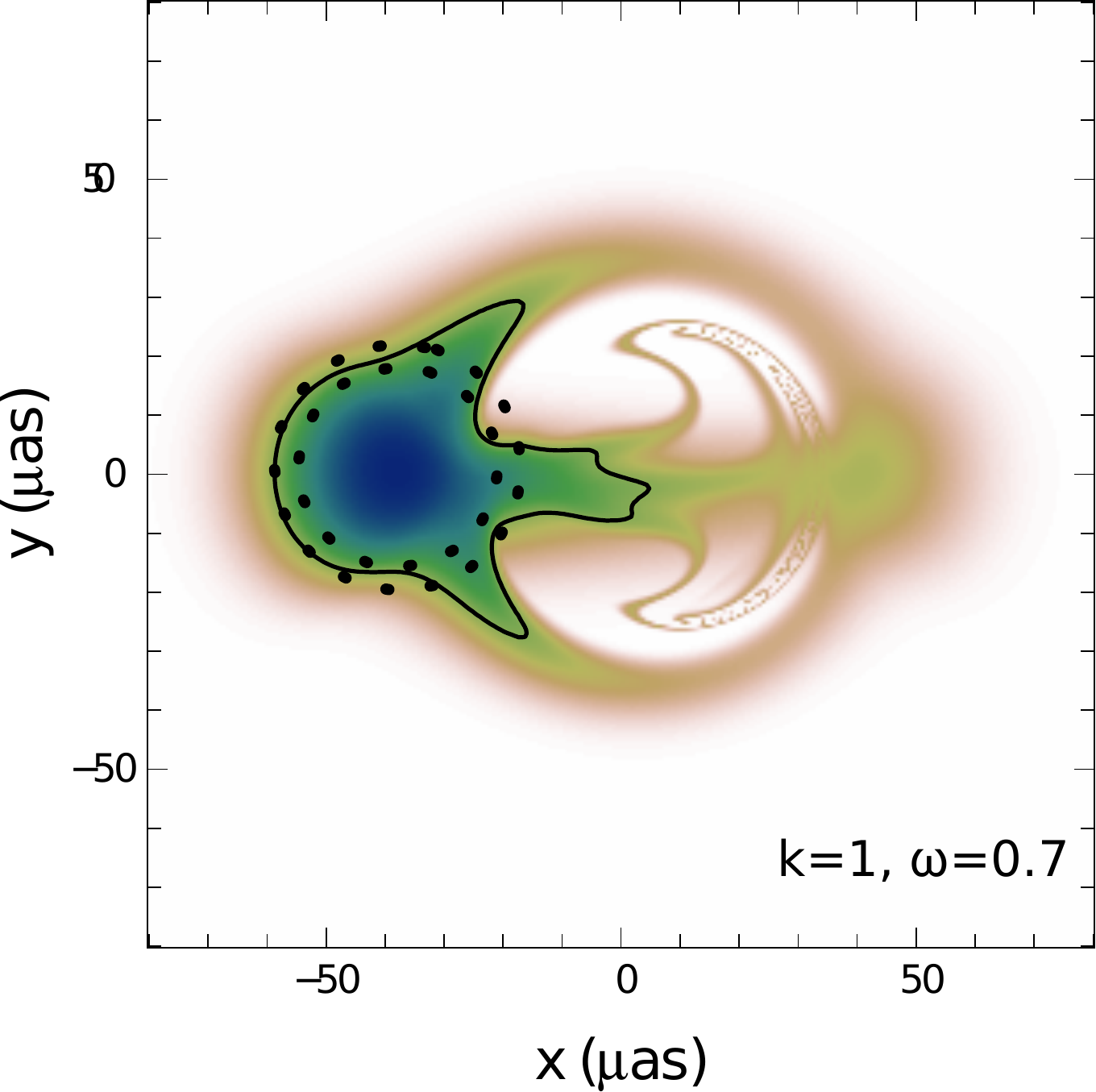}
\includegraphics[height=5.42cm]{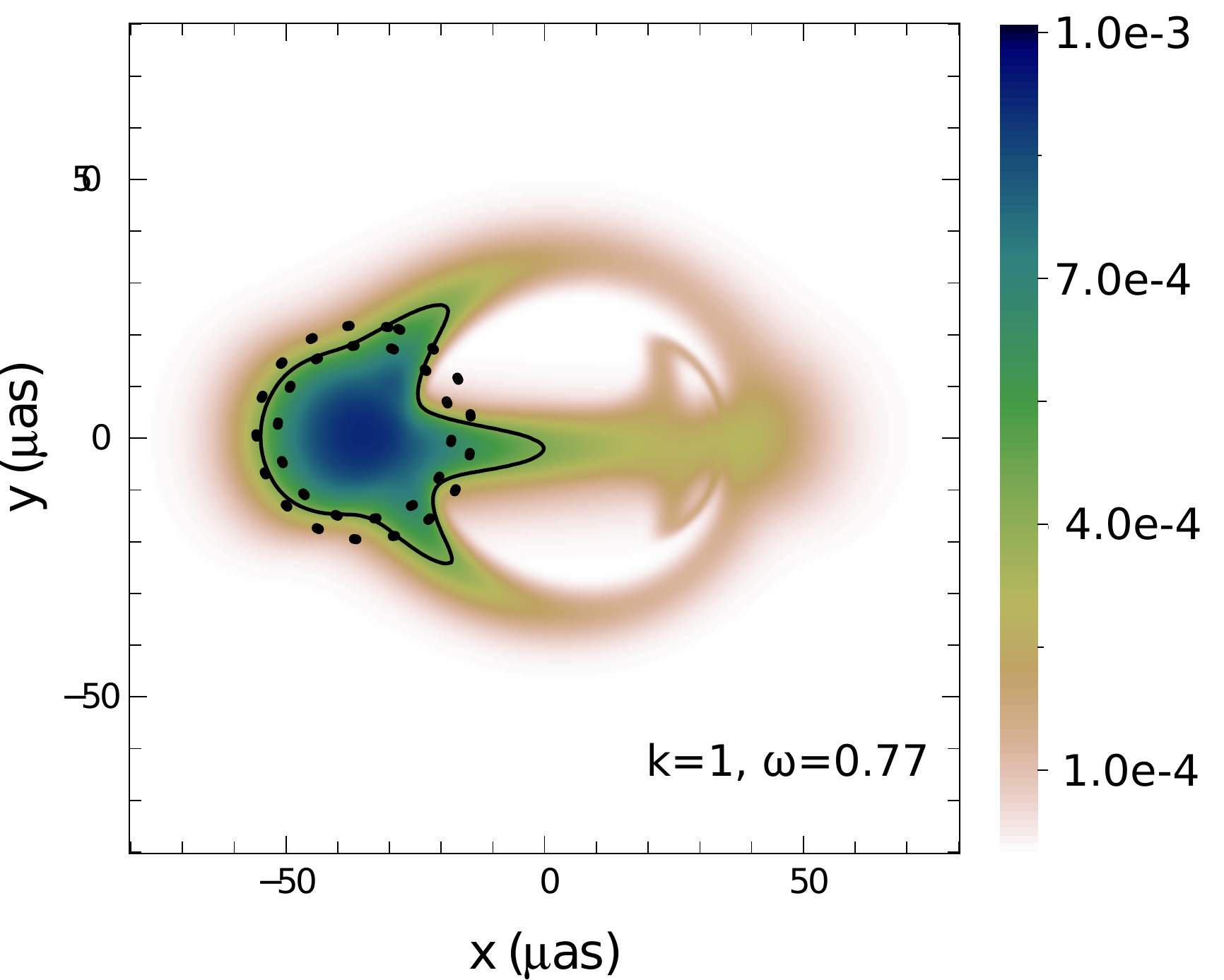}
}
\caption{Images of boson stars with (left panel) an ADM mass $M=1.26m_{\rm p}^2/m$ and ADM spin $a=0.82r_g$ and (right panel) an ADM mass $M\approx1.27m_{\rm p}^2/m$ (corresponding to the maximum of the curve $M(\omega)$, where $\omega$ is the frequency of the scalar field; see Ref.~\cite{Vincent15images}) and ADM spin $a=0.80r_g$ using the accretion flow model of Refs.~\cite{Straub12,VincentTorus15} with a fixed inner radius. These images are computed at a wavelength of 1.3~mm and an inclination of $85^\circ$. Although the size of the shadow in each image is similar to the size of the shadow of a Kerr black hole with the same mass and spin, the images show hammer-like features (c.f., Fig.~\ref{fig:KBHSHshadows}) which should allow for these images to be distinguished from the corresponding images of Kerr black holes. In each image, the dotted circles show the $1\sigma$ confidence limits on the angular size of the emitting region imposed by the EHT measurement of Ref.~\cite{Doele08}, centered on the maximum of the intensity distribution. The solid black contour encompasses the region emitting 50\% of the total flux. Taken from Ref.~\cite{Vincent15images}.}
\label{fig:VincentImages}
\end{figure*}

Reference~\cite{Vincent15images} simulated 1.3~mm images of boson stars~\cite{Feinblum68,Kaup68,Ruffini69,Arvanitaki11,Brito15} surrounded by such a toroidal accretion flow with a fixed inner radius (motivated by the accretion flows surrounding Kerr black holes) producing an ``effective'' shadow. Reference~\cite{Vincent15images} pointed out that the apparent sizes of the shadows in the latter setup are very similar to the sizes of the shadows of Kerr black holes with the same mass and spin leading to a potential confusion problem (see, also, Ref.~\cite{Herdeiro15PRL}). Figure~\ref{fig:VincentImages} shows simulated images for two such configurations with different values of the ADM mass (measured in units of $m_{\rm p}^2/m$, where $m_{\rm p}$ is the Planck mass and $m$ is the mass of one boson with a typical value corresponding to $\sim10^{-16}~{\rm eV}$) and spin. However, at least for extreme modifications of the shadow which is comprised by multiple shadows, these hammer-like features reveal clearly visible symmetric structures across the equatorial plane of the object which should be easily detectable with the EHT. References~\cite{Torres00,Troitsky15} argued that certain properties of boson stars are also consistent other observed characteristics of Sgr~A$^\ast$ such as its low accretion rate.

\subsection{Detecting the Shadow of Sgr~A$^\ast$}

Given the complexities of the accretion flow, a key question is how accurately the shadow can be detected with the EHT. Since the shape of the shadow can reveal potential deviations from the Kerr metric directly (see Sec.~\ref{subsec:shadows}), such a measurement can, at least in principle, evade the systematic uncertainties that arise from the unknown details of the accretion flow.

\begin{figure}[ht]
\begin{center}
\psfig{figure=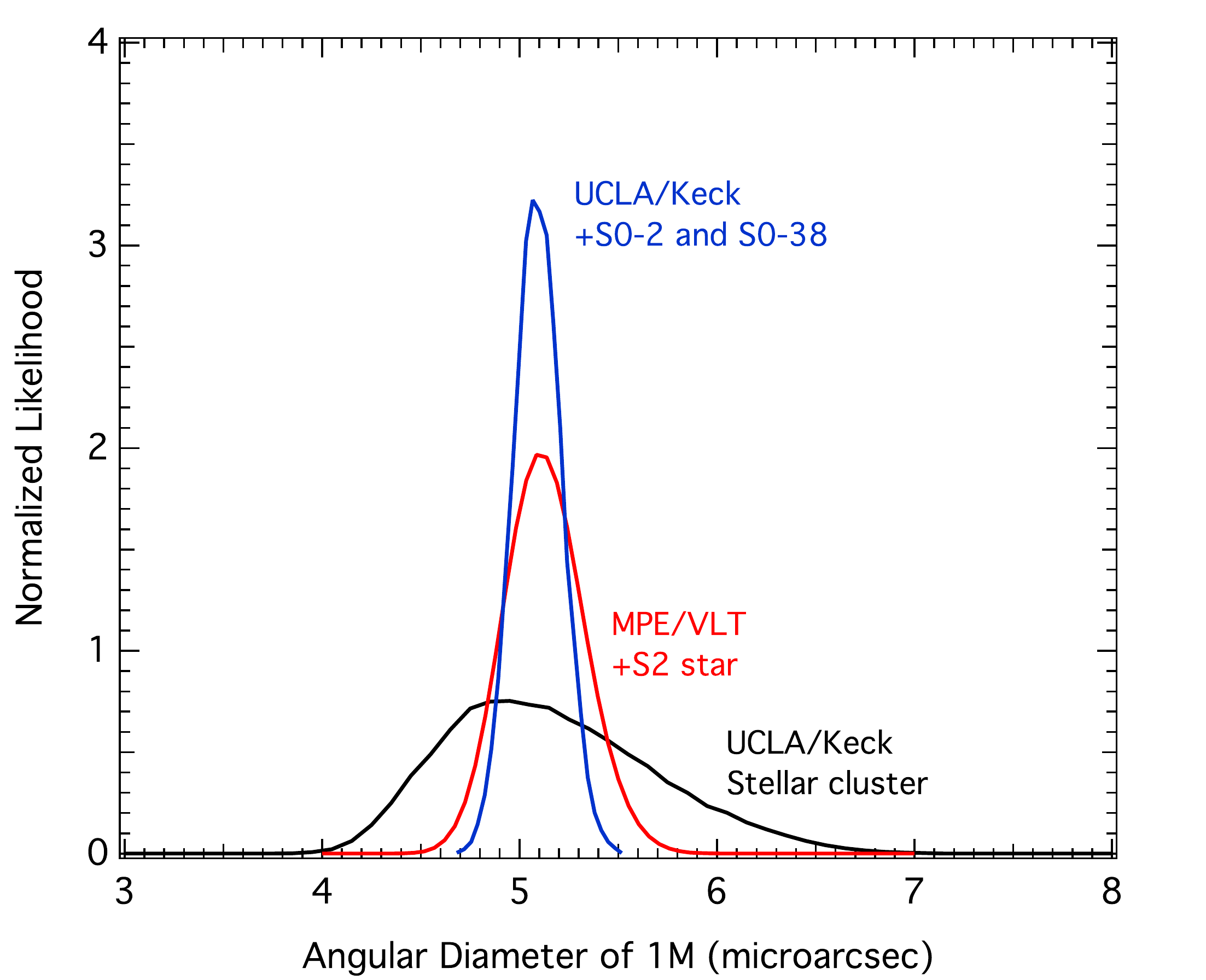,height=2.4in}
\end{center}
\caption{Posterior likelihood of the angular size of one gravitational radius (i.e., $GM/c^2D$) for Sgr~A$^\ast$, as inferred from fitting Keplerian orbits to astrometric observations of S-stars~\cite{Ghez08,Gillessen09,Boehle15}. The posterior likelihood in the analysis of Ref.~\cite{Boehle15} corresponds to an angular size of one gravitational radius of $5.09\pm0.17~{\rm \mu arcsec}$. Taken from Ref.~\cite{Psaltis14}.
}
\label{fig:size1M}
\end{figure}

Since the size of the shadow is determined primarily by the mass-distance ratio $M/D$, the existing mass and distance measurements, for which mass and distance are correlated either roughly as $M\sim D^2$ in the case of observations of stellar orbits~\cite{Ghez08,Gillessen09} or as $D\sim M^0$ in the case of the maser observations by Ref.~\cite{Reid14}, can be improved by measurements of this ratio with the EHT~\cite{SMBHmasses}. If Sgr~A$^\ast$ is indeed a Kerr black hole, then its angular radius measured by upcoming EHT observations has to coincide with the angular radius inferred from existing measurements of the mass and distance of Sgr~A$^\ast$ which constitutes a null test of general relativity~\cite{Psaltis14}. Figure~\ref{fig:size1M} shows the posterior likelihoods of the angular size of one gravitational radius for Sgr~A$^\ast$ obtained from two different sets of observations of the S-stars orbiting around the Galactic center~\cite{Ghez08,Gillessen09}. The posterior likelihood in the analysis of Ref.~\cite{Gillessen09} corresponds to an angular size of one gravitational radius of $5.12\pm0.29~{\rm \mu arcsec}$.

Reference~\cite{SMBHmasses} used simple scaling arguments to estimate the precision for a measurement of the size of the shadow at a wavelength $\lambda$ with an EHT array comprised of five to six stations. Assuming that the photon ring surrounding the shadow contributes $\sim1/15$ to the total flux and that the signal-to-noise ratio of such a measurement scales linearly with the uncertainty reported in early EHT observations~\cite{Doele08,Fish09}, Ref.~\cite{SMBHmasses} found an uncertainty of
\be
\sigma \simeq 4.3 \times \left( \frac{\lambda}{{\rm 1~mm}} \right)^{-2} \left[ \frac{53}{21} \left(\frac{\lambda}{{\rm 1~mm}}\right)^{-1} -1 \right]^{-1}\, {\rm \mu as}.
\ee

\begin{figure*}[h]
\centering{
\includegraphics[width=7.5cm]{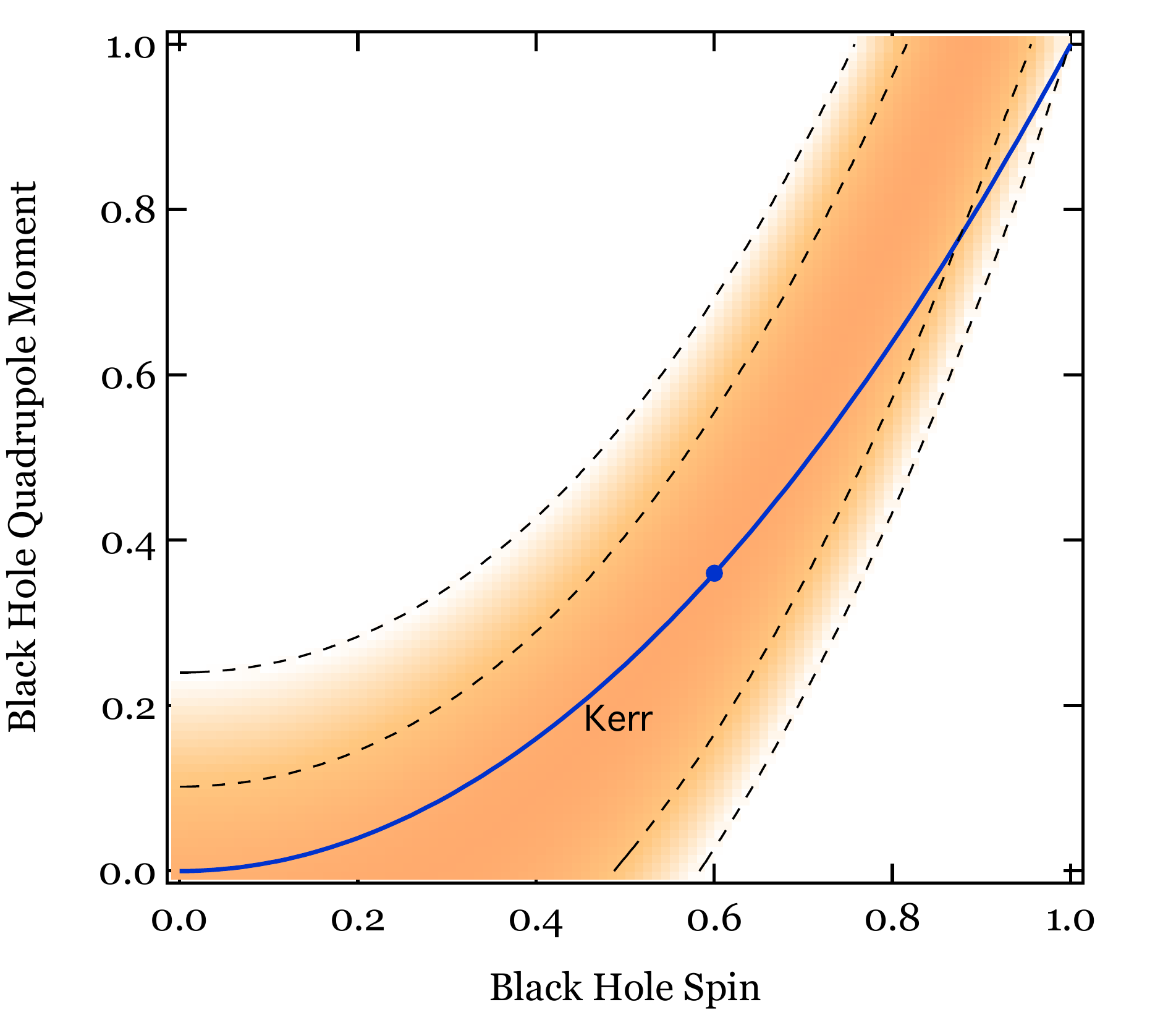}
\includegraphics[width=7.5cm]{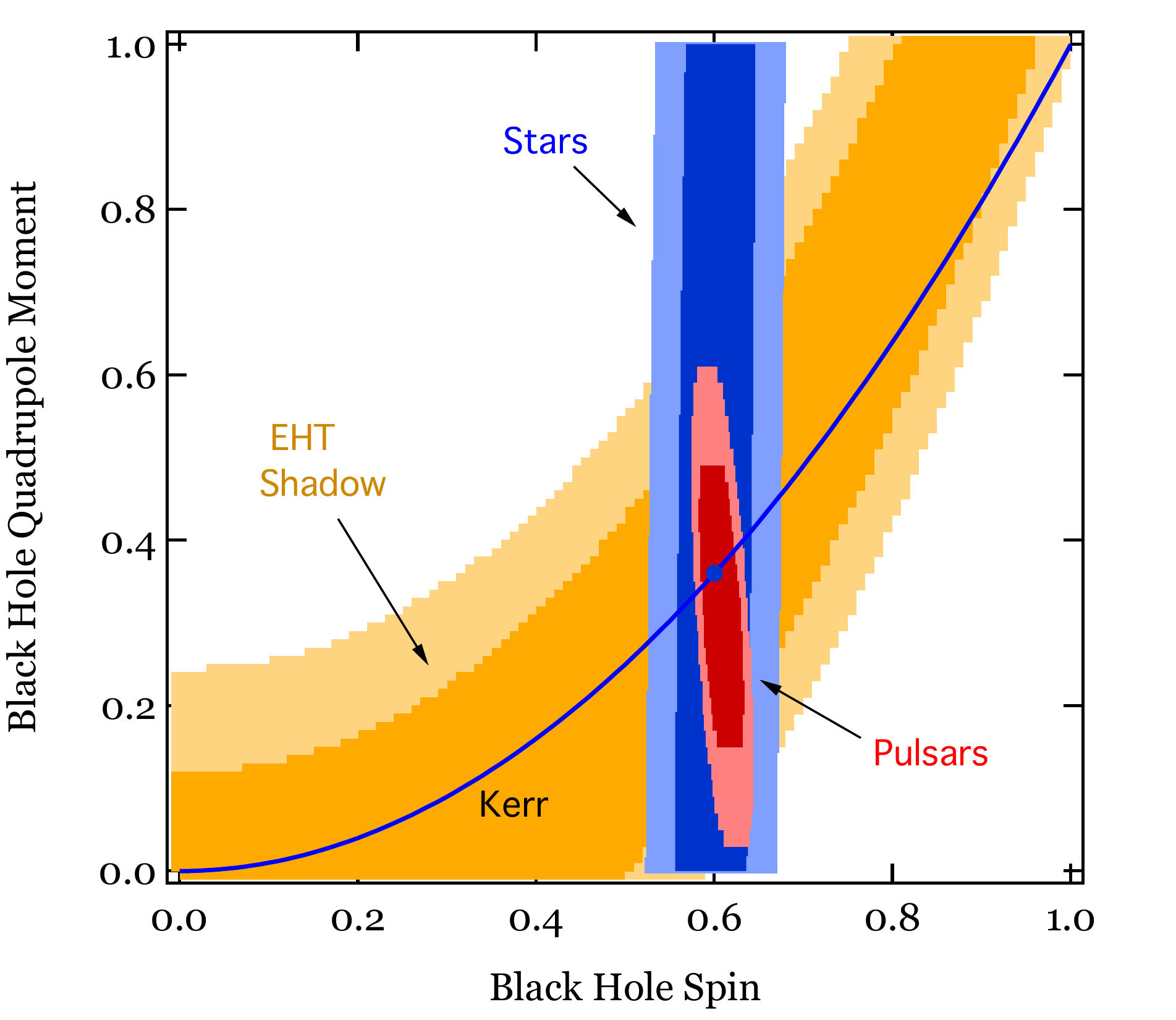}
}
\caption{The left panel shows the 68\% and 95\% confidence contours of the posterior likelihood for an EHT measurement of the asymmetry of the shadow as a function of the spin $a$ and the (dimensionless) quadrupole moment $q$ of Sgr~A$^\ast$. The solid curve shows the expected relation between spin and quadrupole moment for a Kerr black hole, while the filled circle marks the assumed spin and quadrupole moment ($a=0.6r_g$, $q=0.36$). As expected, the contours of maximum likelihood closely follow the Kerr relation, because any violation of the no-hair theorem would have caused a measurable asymmetry in the shadow shape. The right panel shows the 68\% and 95\% confidence contours of the posterior likelihood of simulated GRAVITY and pulsar-timing observations (see Fig.~\ref{fig:PWK_Q}) together with the contours for the EHT measurement. The contours of the GRAVITY and pulsar-timing observations are nearly orthogonal to the contours of the EHT measurement reducing the uncertainty of a combined measurement significantly. Taken from Ref.~\cite{PWK15}.}
\label{fig:PWK15EHT}
\end{figure*}

Based on this estimate, Ref.~\cite{PWK15} argued that the EHT can measure the asymmetry of the shadow as defined in Eq.~(\ref{eq:DLA}) with a precision of $\sigma_{\rm A}=0.9~{\rm \mu as}$. Assuming a Gaussian distribution of the asymmetry with a width $\sigma_{\rm A}$ and a dependence of the asymmetry on the spin and quadrupole moment of Sgr~A$^\ast$ as found in Ref.~\cite{PaperII} [see Eq.~(\ref{eq:QKasym})], Ref.~\cite{PWK15} obtained a Bayesian likelihood of such a measurement. Figure~\ref{fig:PWK15EHT} shows this likelihood as a function of the spin and the quadrupole moment for a Kerr black hole with a value of the spin $a=0.6r_g$. Figure~\ref{fig:PWK15EHT} also shows the corresponding likelihoods of their simulated measurements of the spin and quadrupole moment using GRAVITY observations of two stars and pulsar-timing observations of three periase passages of a low-precision pulsar (see Fig.~\ref{fig:PWK_Q}). The contours of the GRAVITY and pulsar-timing observations are nearly orthogonal to the contours of the EHT measurement reducing the uncertainty of a combined measurement significantly~\cite{PWK15}.

\begin{figure}[ht]
\begin{center}
\psfig{figure=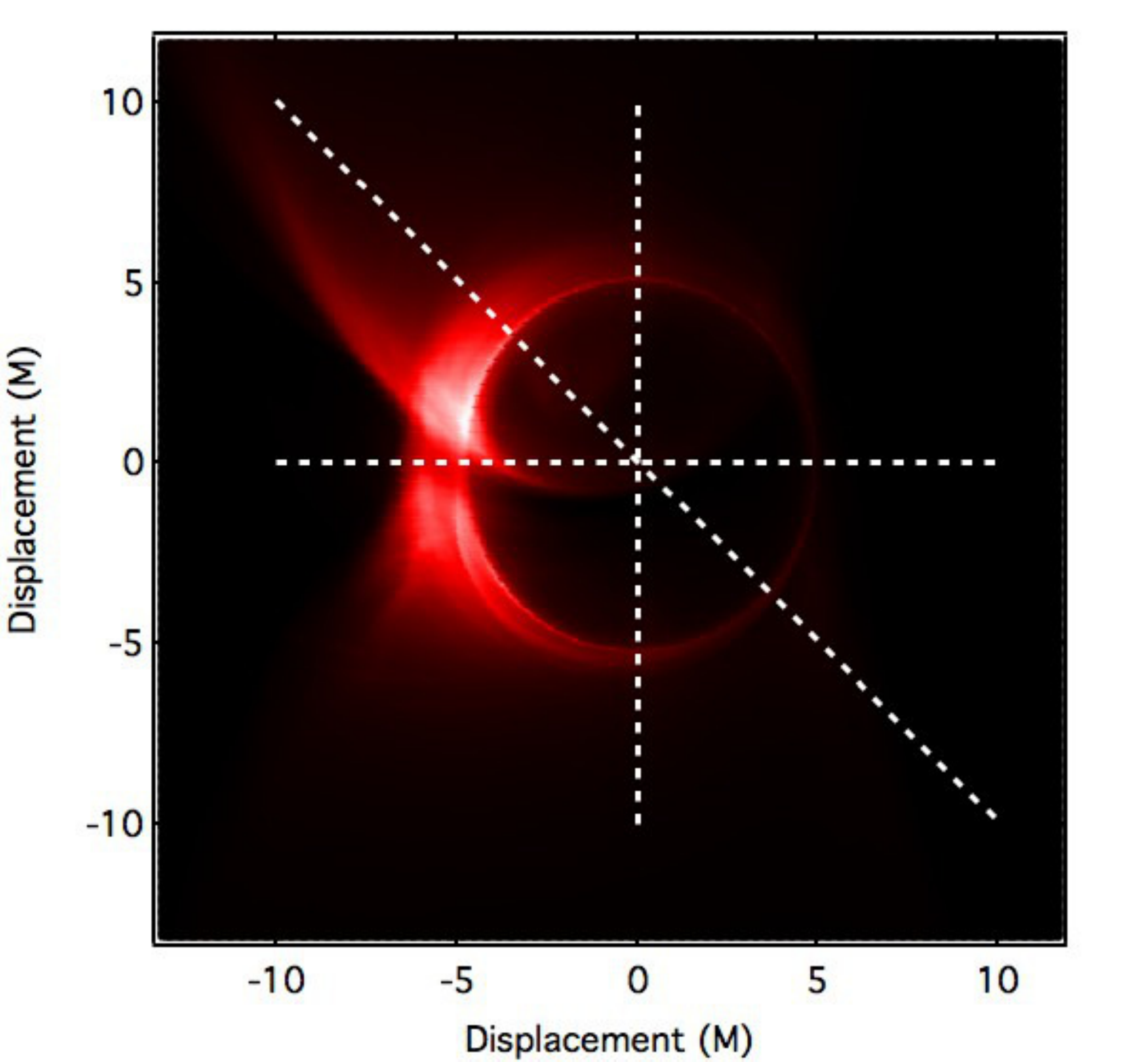,height=2.4in}
\psfig{figure=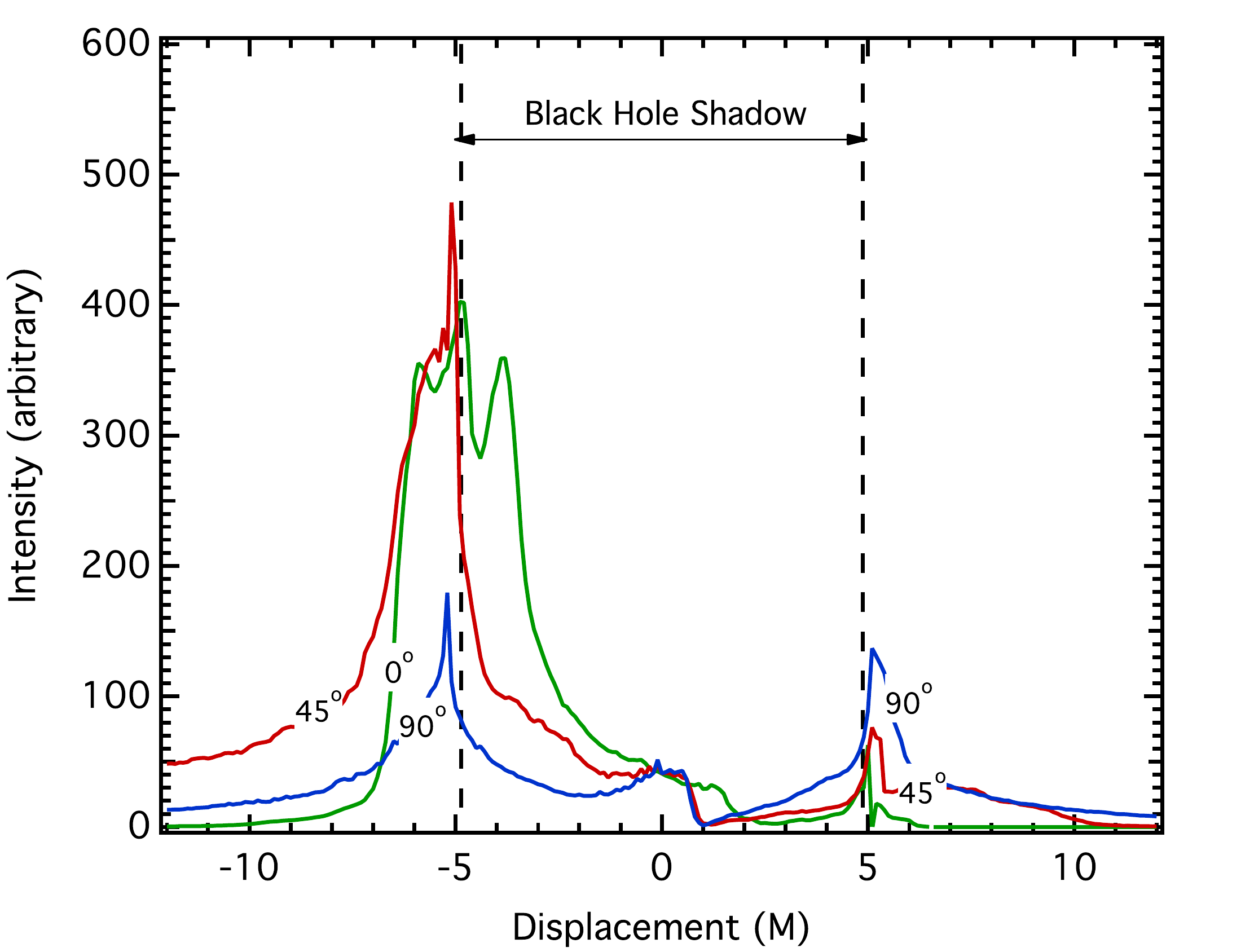,height=2.4in}
\psfig{figure=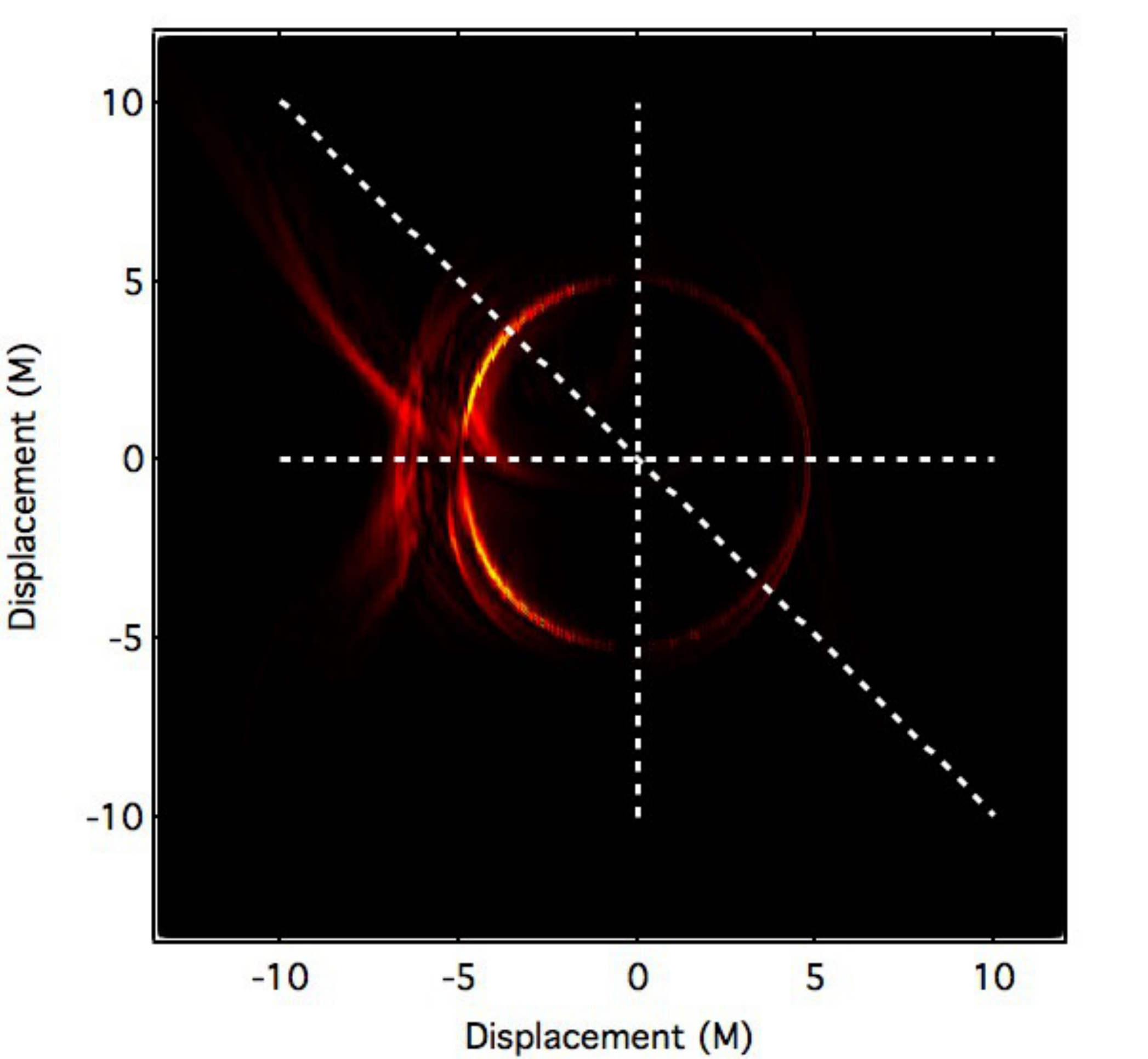,height=2.4in}
\psfig{figure=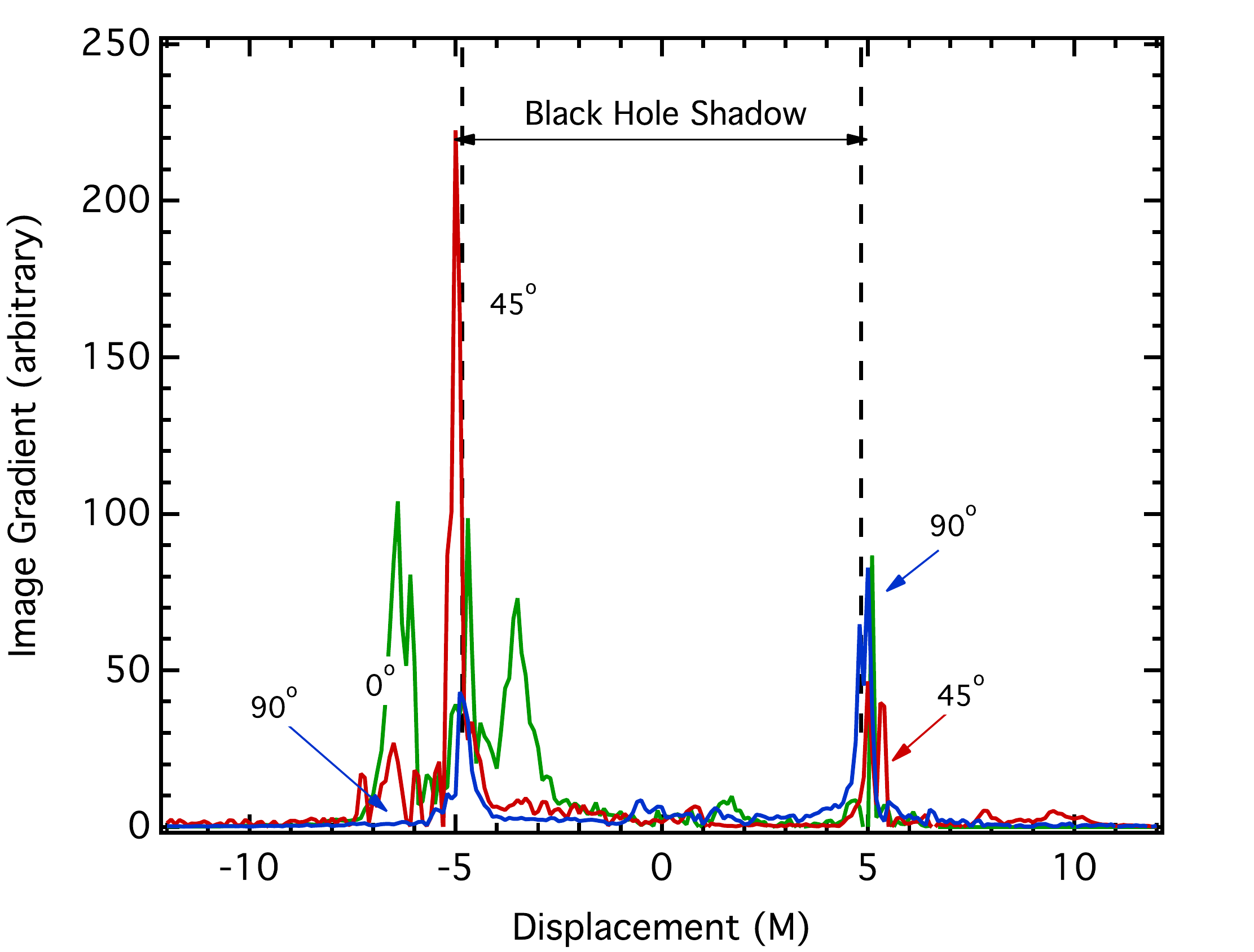,height=2.4in}
\end{center}
\caption{The top row shows (left panel) a simulated 1.3~mm image of Sgr~A$^\ast$, as calculated from a GRMHD simulation of accretion onto a Schwarzschild black hole with an inclination $\theta=60^\circ$~\cite{ChanPsaltis15a}, and (right panel) the brightness profiles of the image along the three indicated cross sections at $0^\circ$, $45^\circ$, and $90^\circ$ with respect to the equatorial plane shown on the left panel. In all cases, the rim of the black-hole shadow corresponds to a sharp drop in the brightness which is consistent to within $\sim0.5r_g$. The bottom row shows (left panel) a map of the magnitude of the gradient of the image brightness and (right panel) profiles of the magnitude of the gradient along the same cross sections as above. The bright rim along the boundary of the black-hole shadow is clearly visible showing prominent peaks within $\sim0.25r_g$ of the location of the shadow. Taken from Ref.~\cite{Psaltis14}.}
\label{fig:PsaltisSize1}
\end{figure}

\begin{figure}[ht]
\begin{center}
\psfig{figure=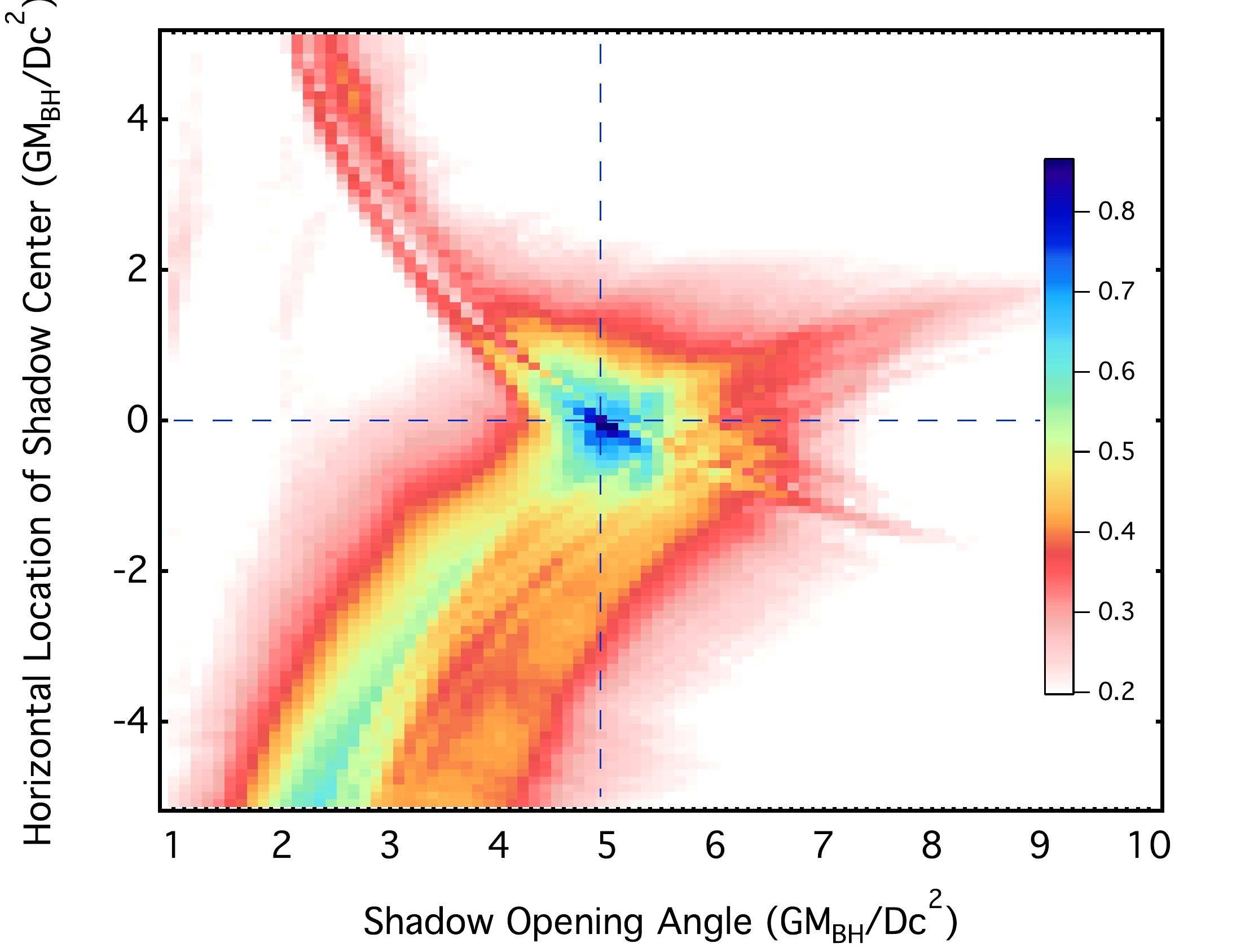,height=2.3in}
\psfig{figure=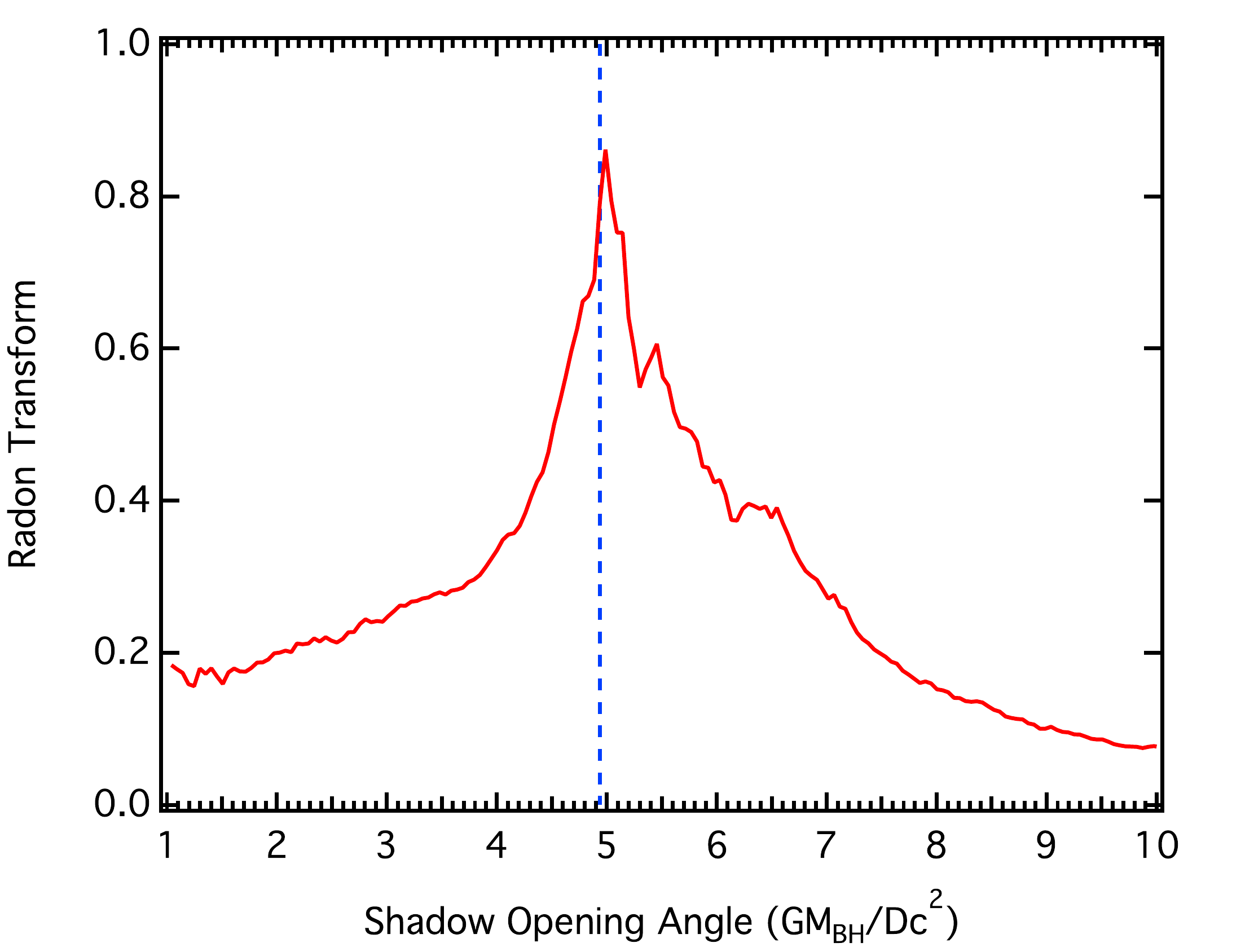,height=2.3in}
\end{center}
\caption{Left: two-dimensional cross section of the Radon transform of the top left panel in Fig.~\ref{fig:PsaltisSize1} as a function of the opening angle of the shadow and the horizontal location of the center of the black-hole shadow when the vertical location of the center of the shadow is set to zero. Right: cross section of the Radon transform for black-hole shadows centered at the known location of the black hole. The peak of the cross section is centered at the expected opening angle for the simulated black hole (vertical dashed line) and has a fractional HWHM of 9\%. Taken from Ref.~\cite{Psaltis14}.}
\label{fig:PsaltisSize2}
\end{figure}

Reference~\cite{Psaltis14} estimated the accuracy with which the size of the shadow can be determined with EHT observations at 1.3~mm employing an image of a Schwarzschild black hole from GRMHD simulations of the accretion flow around Sgr~A$^\ast$~\cite{ChanPsaltis15a}. Figure~\ref{fig:PsaltisSize1} shows the simulated image together with the brightness profiles along three chords across the image which are sharply peaked near the rim of the shadow and consistent with the location of the shadow to within $0.5r_g$. Figure~\ref{fig:PsaltisSize1} also shows a map of the magnitude of the gradient of the image brightness together with profiles of the magnitude of the gradient along the same cross sections as in the simulated image. The bright rim along the boundary of the black-hole shadow is clearly visible showing prominent peaks within $\sim0.25r_g$ of the location of the shadow.

Reference~\cite{Psaltis14} then used an edge detection scheme for interferometric data and a pattern matching algorithm based on the Hough/Radon transform to demonstrate that the shadow of the black hole in this image can be localized to within $\sim9\%$. Figure~\ref{fig:PsaltisSize2} shows the two-dimensional cross section of the Radon transform of the top left panel in Fig.~\ref{fig:PsaltisSize1} as a function of the opening angle of the shadow and the horizontal location of the center of the black-hole shadow when the vertical location of the center of the shadow is set to zero. Figure~\ref{fig:PsaltisSize2} also shows the cross section of the Radon transform for black-hole shadows centered at the known location of the black hole.

In practice, such an image will have to be reconstructed from observed EHT data which will be affected by other uncertainties such as electron scatter broadening, atmospheric fluctuations, and instrumental noise. Reference~\cite{deblurring} performed a reconstruction algorithm for a simulated scatter-broadened RIAF image of Ref.~\cite{Bro09a} based on a simulated one-day observing run of a seven-station EHT array assuming realistic measurement conditions. This deblurring algorithm corrects the distortions of the simulated visibilities by interstellar scattering so that the resolution of the image is predominantly determined by the instrumental beam.

\begin{figure}[ht]
\begin{center}
\psfig{figure=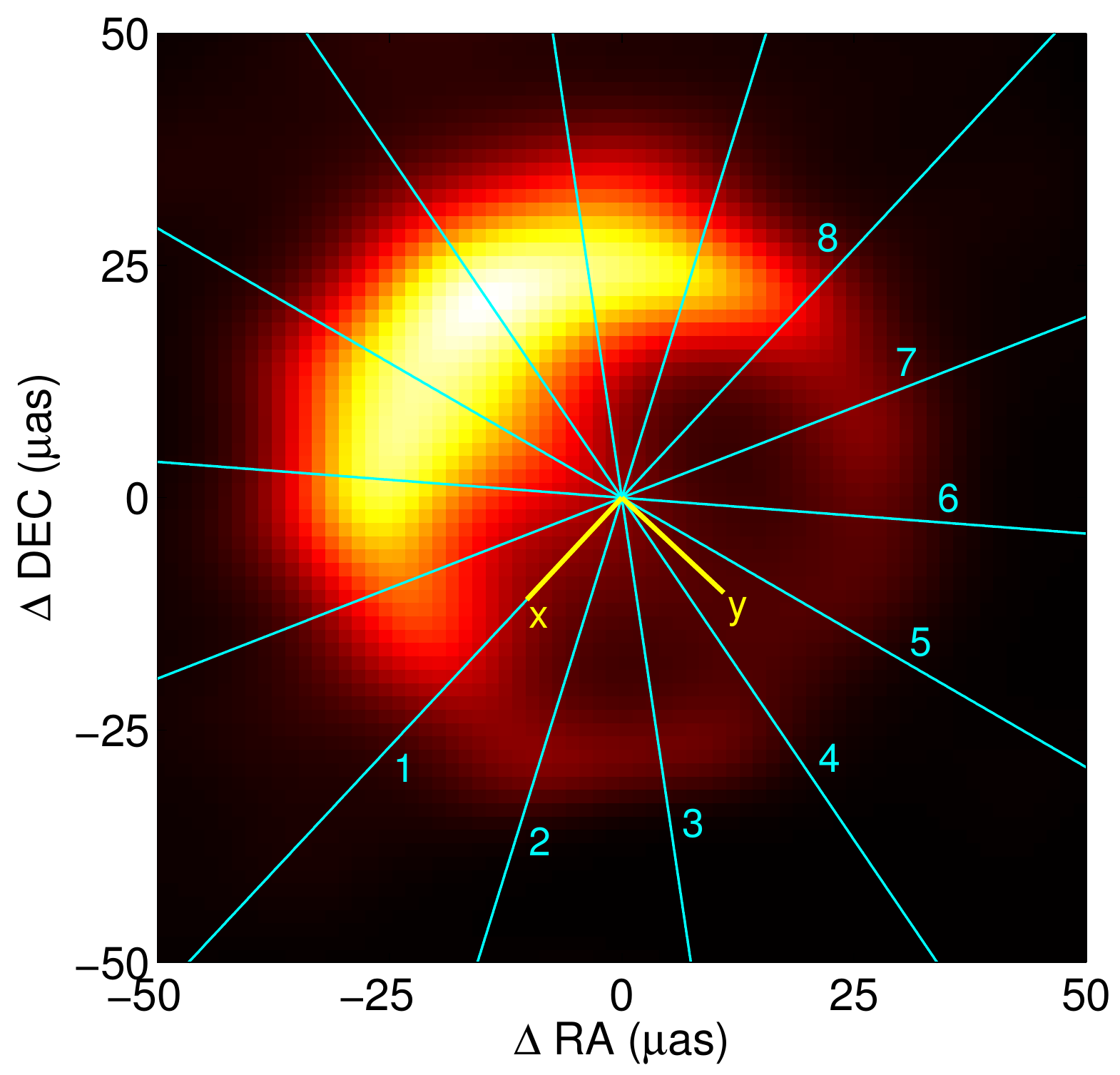,height=2.7in}
\psfig{figure=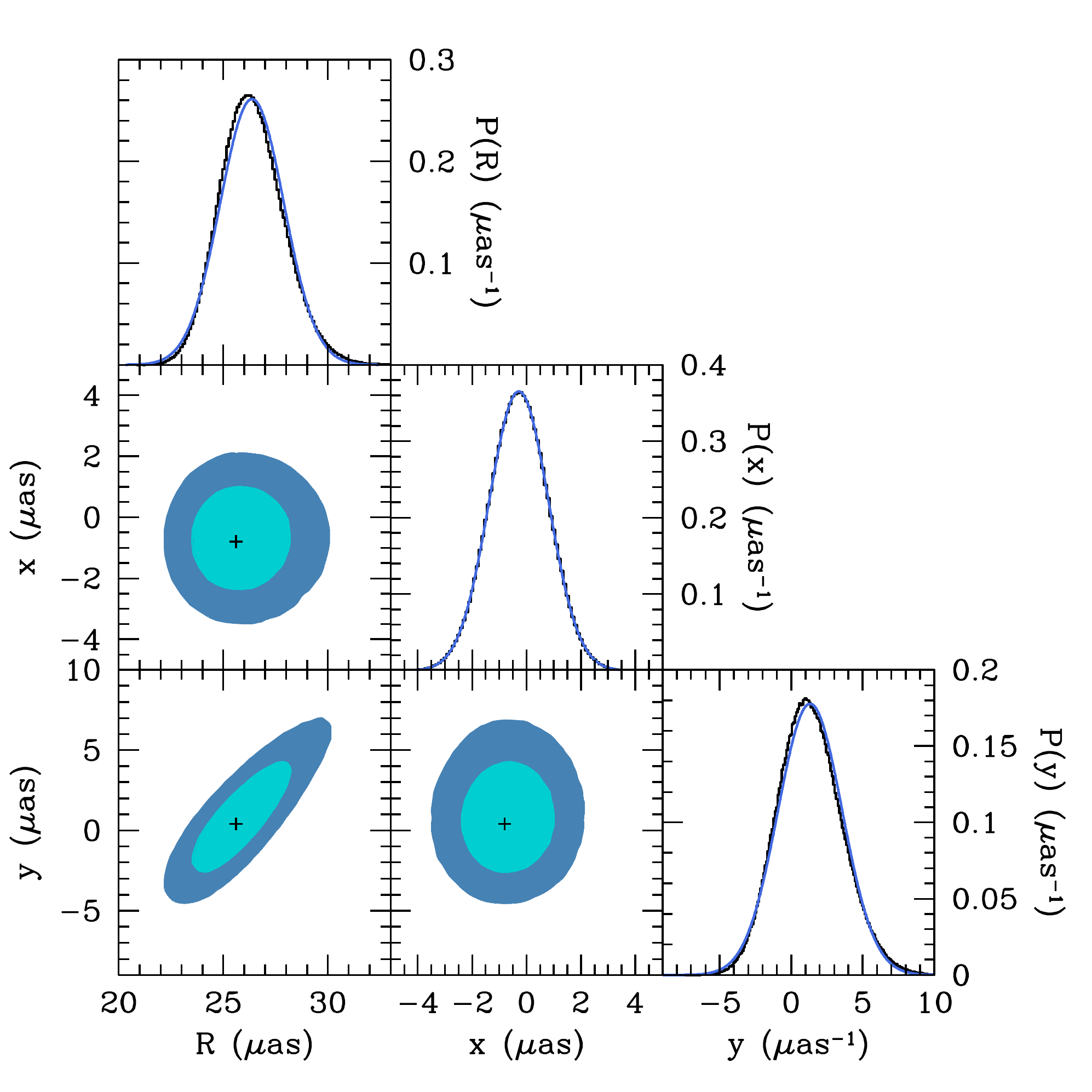,height=2.9in}
\end{center}
\caption{The left panel shows a reconstructed image of Sgr~A$^\ast$ for a simulated EHT observation at 230~GHz with a seven-station array taken from Ref.~\cite{deblurring}. The image shows seven chords for which the respective angular radii are determined from Gaussian fits of the brightness profile along the chord sections labeled ``1'',$\ldots$,``8.'' The right panel shows the resulting distributions of the angular radius $R$ of the shadow and the offset $(x,y)$ of the corresponding image center relative to the center of the chords using a Markov chain Monte Carlo sampling of a small region around the center of the chords. The inferred angular radius of $\approx1.5~{\rm \mu as}$ corresponds to a precision of 6\% and a length of $\approx0.16r_g$. Taken from Ref.~\cite{JohannsenPRL}.}
\label{fig:ringimages}
\end{figure}

Figure~\ref{fig:ringimages} shows the reconstructed image of Ref.~\cite{deblurring} with seven chords across the image. Reference~\cite{JohannsenPRL} employed a Markov chain Monte Carlo algorithm to infer the angular radius $R$ and a potential offset $(x,y)$ from the chosen center of the shadow in this image from Gaussian fits of the brightness profile along the chord sections labeled ``1'',$\ldots$,``8'' finding $R=(26.4\pm1.5)~{\rm \mu as}$, $x=(-0.3\pm1.1)~{\rm \mu as}$, $y=(1.3\pm2.2)~{\rm \mu as}$. This estimate of the angular radius is consistent with the actual angular radius of the shadow $R\approx27.6~{\rm \mu as}$ at the $1\sigma$ level corresponding to the values of the mass $M=4.3\times10^6\,M_\odot$, distance $D=8~{\rm kpc}$, and spin $a=0$ used in the simulated image shown in Fig.~\ref{fig:ringimages}; there is no significant offset $(x,y)$ of the image center. Figure~\ref{fig:ringimages} also shows a triangle plot of the $1\sigma$ and $2\sigma$ confidence contours of the resulting marginalized 2D probabilitiy densities and the corresponding marginalized 1D probability densities with a Gaussian fit. Since the radius estimate would be exact for a true image for which the specific intensity peaks at the shadow corresponding to the longest optical path length of photons in the accretion flow, the method of Ref.~\cite{JohannsenPRL} seems to have no significant bias.

\begin{figure*}[ht]
\begin{center}
\psfig{figure=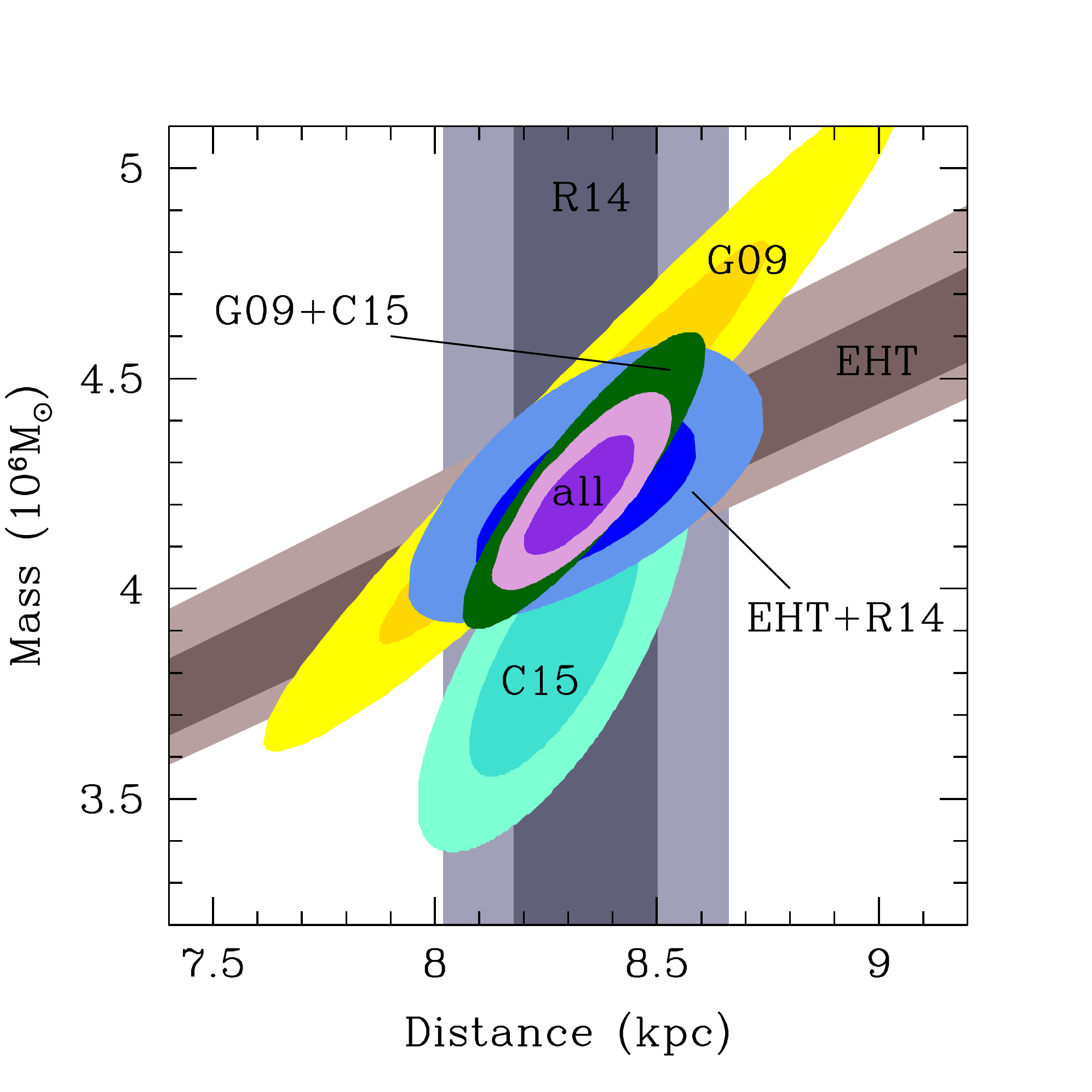,width=2.03in}
\psfig{figure=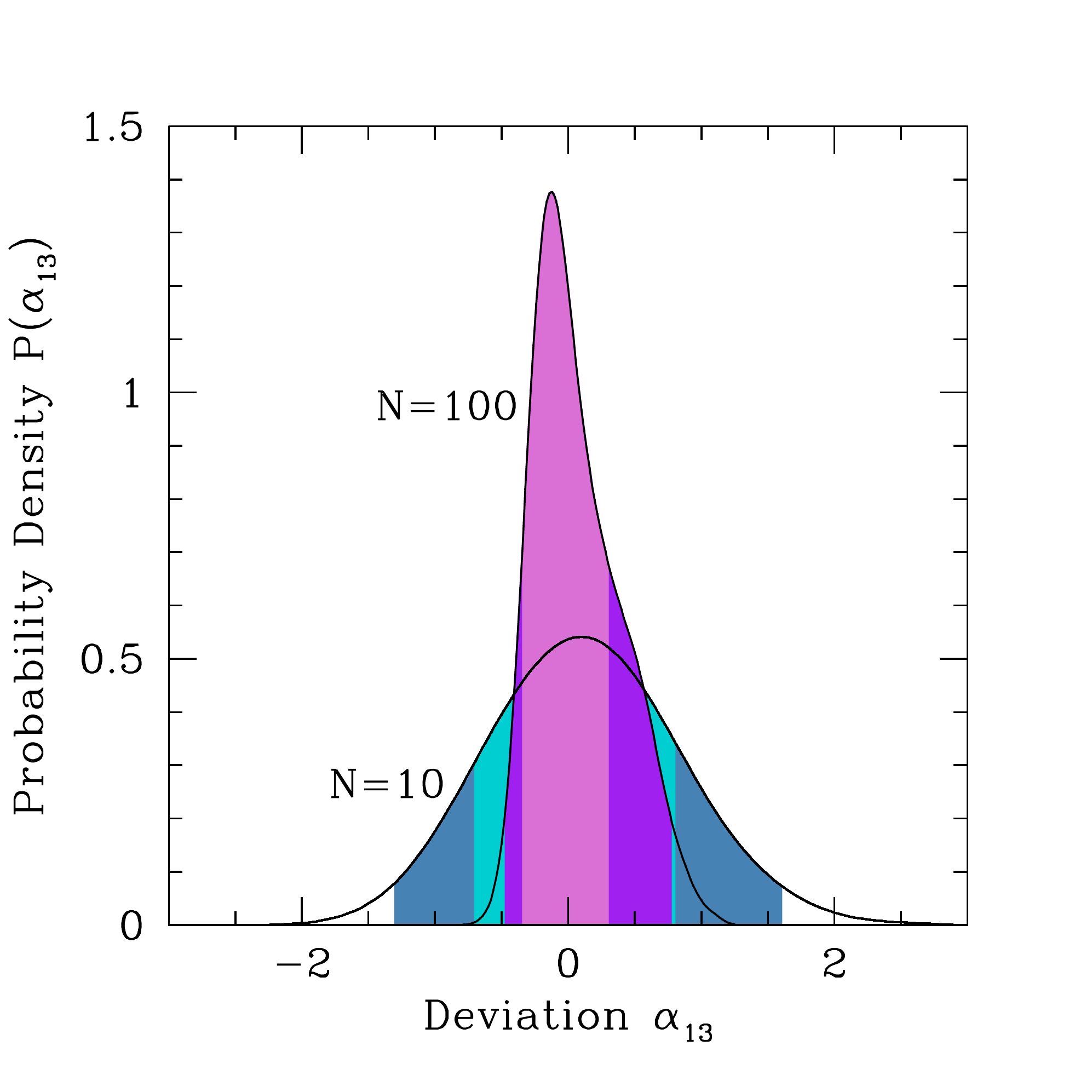,width=2.03in}
\psfig{figure=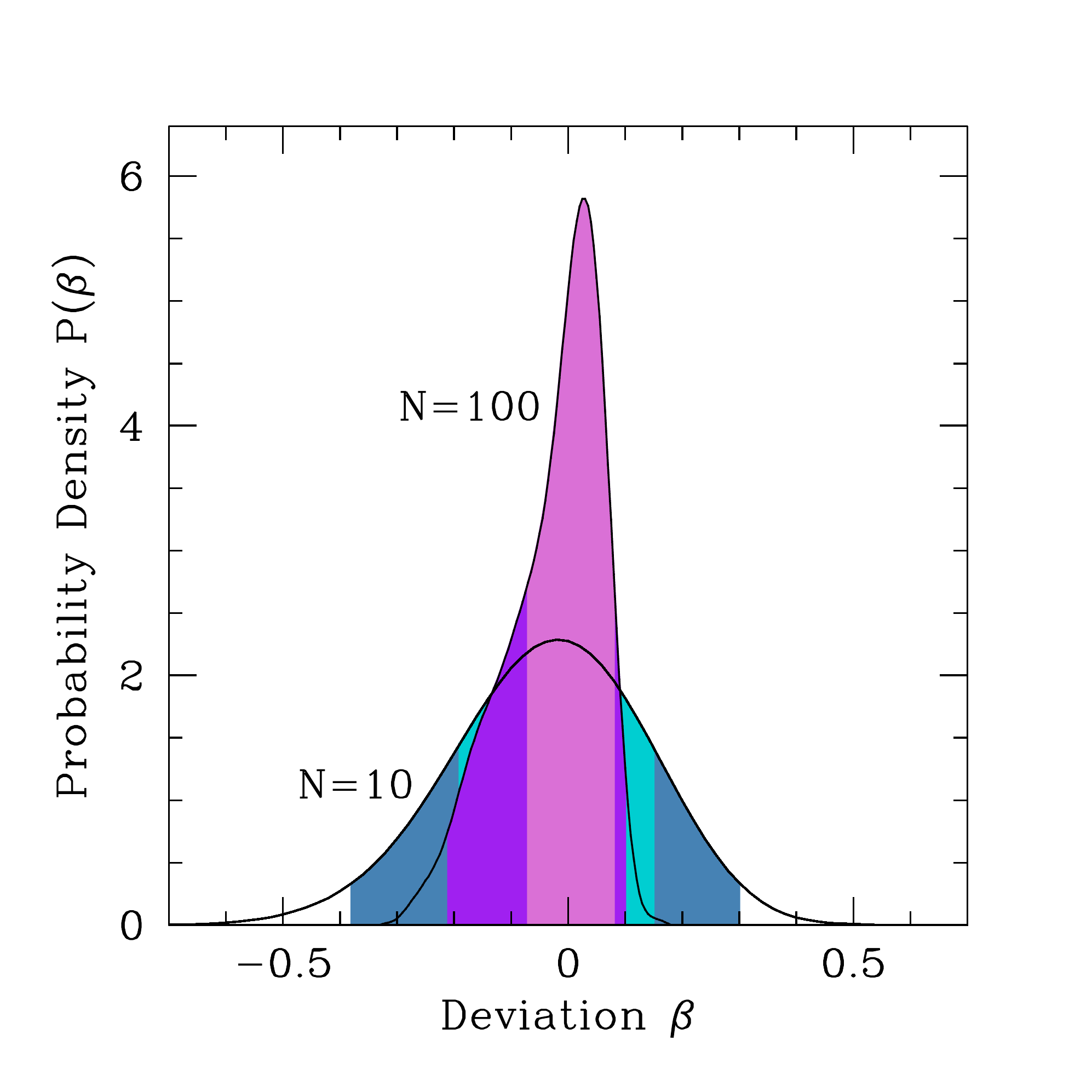,width=2.03in}
\end{center}
\caption{Left panel: $1\sigma$ and $2\sigma$ confidence contours of the probability density of the mass and distance of Sgr~A$^\ast$ for existing measurements (S-stars, ``G09''~\cite{Gillessen09}; masers, ``R14''~\cite{Reid14}; star cluster, ``C15''~\cite{Chatzopoulos15}), a simulated measurement of the shadow size of Sgr~A$^\ast$ for $N=10$ observations with a seven-station EHT array (``EHT''), and several combinations thereof. The simulated EHT measurement improves the other constraints on the mass and distance significantly. Center and right panels: Simulated $1\sigma$ and $2\sigma$ confidence contours of the probability density of the deviation parameters $\alpha_{13}$ and $\beta$, respectively, corresponding to $N=10$ and $N=100$ EHT observations, each marginalized over the mass and distance using the combination of all data sets (``all'') in the $N=10$ case and of simulated stellar-orbit observations from a 30m-class telescope~\cite{Weinberg05} in the $N=100$ case. Taken from Ref.~\cite{JohannsenPRL}.}
\label{fig:PRLconstraints}
\end{figure*}

Reference~\cite{JohannsenPRL} combined the above simulated EHT measurement of the angular shadow radius of Sgr~A$^\ast$ with existing measurements of its mass and distance assuming a nearly circular shape of the shadow and a Gaussian distribution $P_{\rm EHT}({\rm data}|M,D,a,\theta,\alpha_{13},\beta)$ of the angular radius with an uncertainty $\sigma=1.5~{\rm \mu as}$ and a mean corresponding to the maximum of the distribution $P_{\rm prior}(M,D)$ of the combined measurements of Refs.~\cite{Gillessen09,Chatzopoulos15,Reid14} assuming a Kerr black hole with spin $a=0.5r_g$ and inclination $\theta=60^\circ$. Then, they used Bayes' theorem to express the likelihood of the mass, distance, and deviation parameters given the data as $P(M,D,\alpha_{13},\beta|{\rm data}) = C\,P_{\rm EHT}({\rm data}|M,D,\alpha_{13},\beta)P_{\rm prior}(M,D)$, where $C$ is a normalization constant and where the likelihood was marginalized over the spin and inclination which only affect the size of the shadow marginally (see Figs.~\ref{fig:shadowshapes} and \ref{fig:rings09}).

Figure~\ref{fig:PRLconstraints} shows the $1\sigma$ and $2\sigma$ confidence contours of the probability density of the mass and distance and of the deviation parameters, respectively, for 10 EHT observations. Figure~\ref{fig:PRLconstraints} also shows the constraints on the deviation parameters for future measurements of the mass and distance of Sgr~A$^\ast$ obtainable with a 30m-class telescope with estimated uncertainties $\Delta M,~\Delta D\sim0.1\%$~\cite{Weinberg05} combined with 100 EHT observations. Here, all EHT measurements are assumed to be independent and identical so that their uncertainty can be reduced by a factor of $\sqrt{N}$. 

\begin{table}[ht]
\begin{center}
\footnotesize
\begin{tabular}{lcc}
\multicolumn{3}{c}{}\\
Data   & ~~Mass ($10^6\,M_\odot$) & ~~Distance (kpc) \\
\hline
EHT+G09 & ~~$4.16^{+0.18+0.38}_{-0.16-0.31}$ & ~~$8.18^{+0.19+0.39}_{-0.19-0.37}$ \\
EHT+R14 & ~~$4.22^{+0.13+0.28}_{-0.13-0.24}$ & ~~$8.34^{+0.16+0.32}_{-0.15-0.31}$ \\
EHT+C15 & ~~$4.17^{+0.11+0.22}_{-0.11-0.21}$ & ~~$8.38^{+0.11+0.21}_{-0.11-0.21}$ \\
All     & ~~$4.22^{+0.09+0.20}_{-0.09-0.17}$ & ~~$8.33^{+0.08+0.17}_{-0.08-0.15}$ \\
\hline
\end{tabular}
\caption{Simulated mass and distance measurements using existing data (G09~\cite{Gillessen09}; R14~\cite{Reid14}; C15~\cite{Chatzopoulos15}) as priors. Taken from Ref.~\cite{JohannsenPRL}.}
\label{tab:massdistance}
\end{center}
\end{table}

In this setup, the EHT alone can measure the mass-distance ratio (in units of $10^6\,M_\odot/{\rm kpc}$) $M/R=0.505^{+0.013+0.029}_{-0.011-0.020}$ for $N=10$ observations and $M/R=0.502^{+0.010+0.026}_{-0.005-0.007}$ for $N=100$ observations, respectively. Table~\ref{tab:massdistance} lists constraints on the mass and distance corresponding to various combinations of the EHT measurements for 10 observations with existing data showing significant improvements. In particular, combining the EHT result with the parallax measurement by Ref.~\cite{Reid14} is comparable to the mass and distance measurements from stellar orbits including the combined result of Refs.~\cite{Gillessen09,Chatzopoulos15}. If all data sets are combined as shown in the left panel of Fig.~\ref{fig:PRLconstraints}, Ref.~\cite{JohannsenPRL} obtained the constraints on the deviation parameters $\alpha_{13}=0.1^{+0.7+1.5}_{-0.8-1.4}$, $\beta=-0.02^{+0.17+0.32}_{-0.17-0.36}$ in the $N=10$ case, while, in the $N=100$ case, they found $\alpha_{13}=-0.13^{+0.43+0.90}_{-0.21-0.34}$, $\beta=0.03^{+0.05+0.07}_{-0.10-0.24}$; the uncertainties of the mass and distance remained at the $\sim0.1\%$ level. Here, all results are quoted with $1\sigma$ and $2\sigma$ error bars, respectively.

The simulated constraints on the deviation parameters $\alpha_{13}$ and $\beta$ also translate into specific constraints on the parameters of known black-hole metrics in other theories of gravity (RS2, MOG, EdGB, Bardeen; see Sec.~\ref{subsec:metrics}). These constraints are listed in Table~\ref{tab:devconstraints}. Note, however, that the coupling in quadratic gravity theories (i.e., theories that are quadratic in the Riemann tensor) such as EdGB has units proportional to an inverse length squared (or inverse mass squared in gravitational units). Therefore, much stronger constraints on such couplings can be obtained from observations of stellar-mass compact objects which have much lower masses and much stronger spacetime curvatures than supermassive black holes~\cite{Braneworld3,Maselli15}. While the shadow size also depends on the parameter $\alpha_{22}$, its effect is too weak to yield meaningful constraints in this scenario.

\begin{table}[h]
\begin{center}
\footnotesize
\begin{tabular}{lll}
\multicolumn{3}{c}{}\\
Theory  & ~Constraints ($N=10$) & ~Constraints ($N=100$)  \\
\hline
RS2	&  ~$\beta_{\rm tidal}=-0.02^{+0.17+0.32}_{-0.17-0.36}$  & ~$\beta_{\rm tidal}=0.03^{+0.05+0.07}_{-0.10-0.24}$  \\
MOG     &  ~$\alpha=-0.02^{+0.13+0.22}_{-0.13-0.24}$             & ~$\alpha=0.03^{+0.05+0.06}_{-0.08-0.17}$  \\
EdGB    &  ~$\zeta_{\rm EdGB}\approx0^{+0.1+0.2}_{-0.1-0.3}$     & ~$\zeta_{\rm EdGB}\approx0.022^{+0.035+0.057}_{-0.072-0.150}$ \\
Bardeen &  ~$g^2/r_g^2\approx-0.1^{+0.6+1.0}_{-0.4-0.9}$           & ~$g^2/r_g^2\approx0.09^{+0.14+0.22}_{-0.29-0.60}$ \\
\hline
\end{tabular}
\caption{$1\sigma$ and $2\sigma$ constraints on the parameters of black holes in specific theories of modified gravity (RS2~\cite{RS2BH}; MOG~\cite{MOG}; EdGB~\cite{Mignemi93,Kanti96,YS11,Pani11,AY14,M15}; Bardeen~\cite{BardeenBH,BambiModesto13}) implied by the simulation of Ref.~\cite{JohannsenPRL}.}
\label{tab:devconstraints}
\end{center}
\end{table}

At least in the case when the metric of Ref.~\cite{Jmetric} is interpreted as a vacuum solution in $f(R)$ gravity, the constraint on the parameter $\beta$ would imply a constraint on the quadrupole moment of Sgr~A$^\ast$ given by the expression $M_2=-M\sqrt{1-\beta}a^2$ [in gravitational units; see Eq.~(\ref{eq:betamult}) and Ref.~\cite{Suvorov15}]. Consequently, the above measurement of the shadow size would infer the quadrupole moment of Sgr~A$^\ast$ with a precision of $\sim9\%$ and $\sim5\%$ at the $1\sigma$ level in the $N=10$ and $N=100$ cases, respectively.

The analysis of Ref.~\cite{JohannsenPRL} estimated the shadow radius from an image of Sgr~A$^\ast$ that is constant, thus neglecting small-scale variability in the image. This variability will originate first from the accretion flow itself with a characteristic timescale that is comparable to the period of the ISCO, which ranges from about half an hour for a Schwarzschild black hole to approximately four minutes for a maximally rotating Kerr black hole. Second, electron scatter broadening of the image will blur the image, although this effect is largely invertible as shown in Ref.~\cite{deblurring}. Electron scattering will also introduce refractive substructure into the apparent image with a characteristic timescale of approximately one day, which can also cause image distortions that will vary stochastically from epoch to epoch~\cite{deblurring,Johnson15}. However, since Ref.~\cite{JohannsenPRL} fit the brightness along the chords with Gaussians, their estimate of the shadow radius is insensitive to remaining uncertainties in the interstellar scattering law. Therefore, in practice, one image of a quiescent accretion flow as the one shown in Fig.~\ref{fig:ringimages} likely corresponds to an average of several EHT observations, over which time the source variability will average out~\cite{Lu15} (but see Ref.~\cite{Medeiros16}). Likewise, the effects of different realizations of refractive substructure on different observing days will average out.

The results of Ref.~\cite{JohannsenPRL} will also be affected moderately by uncertainties in the calibration of the EHT array and in the accretion flow model used for the image reconstruction. The former imposed a $5\%$ uncertainty in early EHT observations with a three-station array, estimated from calibration for their visibility amplitudes~\cite{Fish11}. For larger telescope arrays such as the seven-station array used in this simulation, however, many more internal cross-checks will be available to improve the relative calibration of stations (the absolute calibration is not important). In particular, the use of three individual phased interferometers (Hawaii, CARMA, ALMA) that simultaneously record conventional interferometric data will permit scan-by-scan cross calibration of the amplitude scale of the array. In addition, measurements of closure phases and closure amplitudes along different telescope triangles and quadrangles are immune to calibration errors.

In contrast to other accretion flow studies (see, e.g., Fig.~\ref{fig:RIAF3}), the uncertainties regarding the employed accretion flow model likely only play a subdominant role in this simulation as long as a (nearly circular) shadow is clearly visible in the image, because the size and shape of the shadow are almost entirely determined by the underlying spacetime alone (see Sec.~\ref{subsec:shadows}). Although the method of Ref.~\cite{JohannsenPRL} relies on the presence of an accretion flow which emits the radiation that comprises the bright ring surrounding the shadow, the brightness profiles along the different chords in the image will have local peaks near the location of the shadow corresponding to the longest optical photon path in the accretion flow irrespective of the details of the accretion flow itself~\cite{JohannsenPRL}.

Interpreting combined data sets as in the analysis of Ref.~\cite{JohannsenPRL} must be done with great care, because it can be difficult to properly include their independent systematic uncertainties, which are likely to dominate their error budgets. For example, the current orbit-based measurements of Refs.~\cite{Gillessen09,Meyer12} nearly disagree at a statistically-significant level and Ref.~\cite{Reid14} neglects systematic errors arising from their choice of outlier removal and possible deviations from an axisymmetric velocity field. Before a 30m-class optical telescope will be available, the uncertainties of mass and distance measurements based on stellar orbits will be further reduced by continued monitoring and the expected improvement in astrometry with the second-generation instrument GRAVITY for the Very Large Telescope Interferometer~\cite{GRAVITY} (see Sec.~\ref{sec:stars}).

\begin{figure*}[ht]
\begin{center}
\psfig{figure=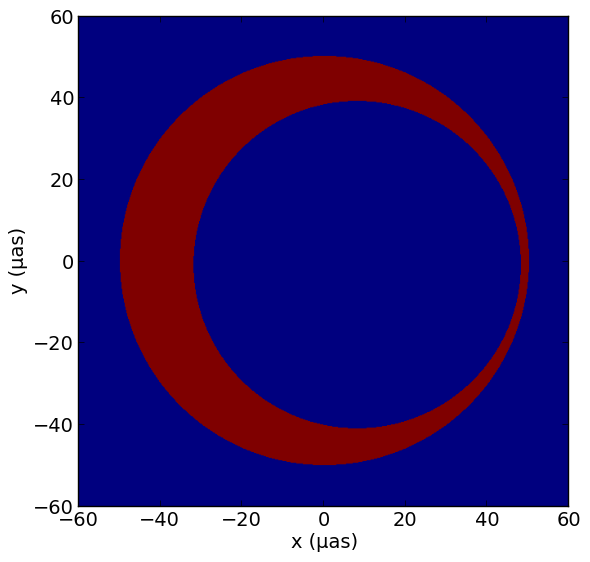,width=1.88in}
\psfig{figure=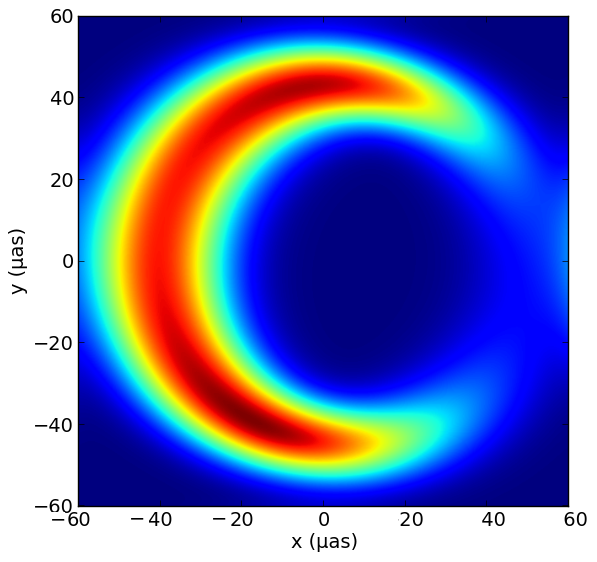,width=1.9in}
\psfig{figure=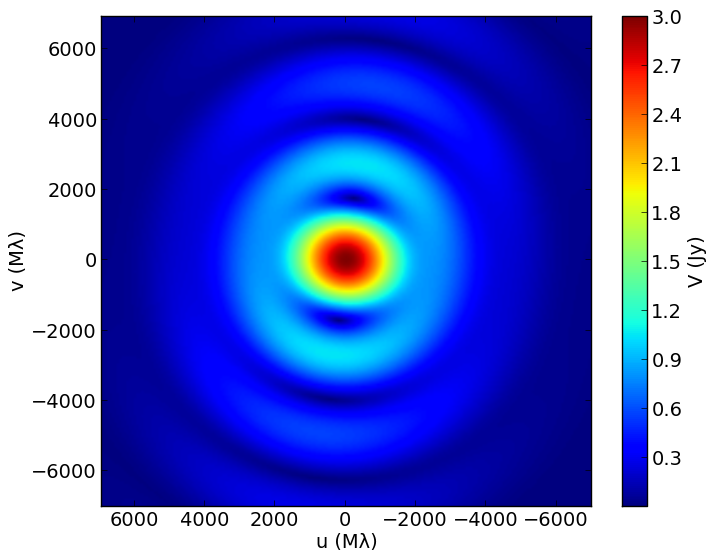,width=2.3in}
\end{center}
\caption{Sample crescent image (left panel), corresponding blurred crescent image (center panel), and visibility amplitudes of the blurred image (right panel) for a crescent model with radii $50~{\rm \mu as}$ and $40~{\rm \mu as}$ of the outer and inner circles, resectively, where the inner circle is centered at the coordinates $(8~{\rm \mu as},1~{\rm \mu as})$ in the image shown in the left panel. Taken from Ref.~\cite{crescent13}.}
\label{fig:crescent}
\end{figure*}

\begin{figure*}[ht]
\begin{center}
\psfig{figure=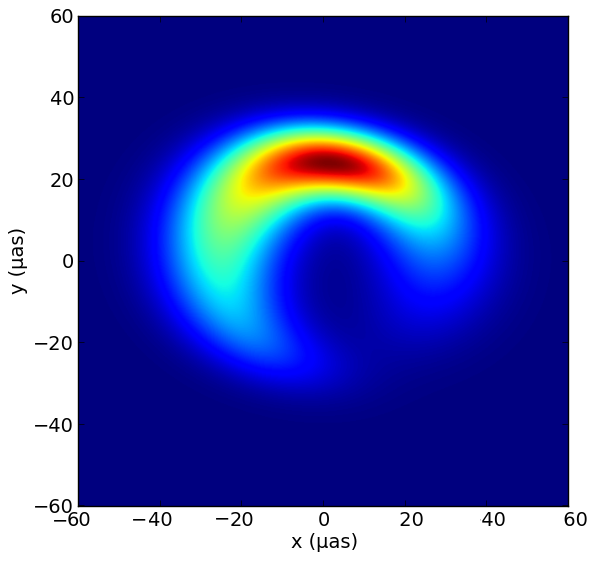,width=2.4in}
\psfig{figure=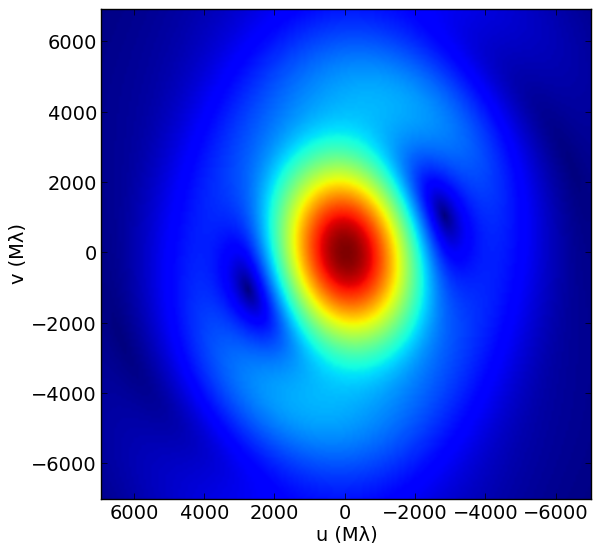,width=2.4in}
\end{center}
\caption{Most likely image of Sgr~A$^\ast$ in the (left panel) image and (right panel) $uv$-planes inferred from a fit of the crescent model of Ref.~\cite{crescent13} to early EHT data~\cite{Doele08,Fish11}. Taken from Ref.~\cite{crescent13}.}
\label{fig:crescentfit}
\end{figure*}

References~\cite{Psaltis14,JohannsenPRL} obtained estimates of the shadow radius from images in the image plane instead of inferring the radius directly from the observed EHT data in the $uv$-plane, where $u$ and $v$ are the visibility amplitudes of the image. In order to avoid the systematic uncertainties of the Fourier transform of the $uv$ data during the image reconstruction, Ref.~\cite{crescent13} constructed a analytical geometric four-parameter crescent model of the shadow in the $uv$-plane, where the surface brightness across the two overlapping disks in the image is constant. Figure~\ref{fig:crescent} shows an example of a crescent image together with the corresponding scatter-broadened images in the image and $uv$-planes. Figure~\ref{fig:crescentfit} shows the most likely image of Sgr~A$^\ast$ in the image and $uv$-planes, respectively, inferred from a fit of the crescent model to the early EHT data~\cite{Doele08,Fish11}. Identifying the inner circle as the shadow of the black hole, this model could provide another estimate of the shadow radius. More sophisticated models with a varying surface brightness across the disks have been explored in Ref.~\cite{Benkevitch}.

\section{Variability}
\label{sec:variability}

Variability in the emission of Sgr~A$^\ast$ has been observed at NIR/mm/sub-mm (e.g., \cite{Zylka95,FalckeProc99,ZhaoYoung03,Genzeletal03,Herrnstein04,Eckart04,Miyazaki04,Mauerhan05,Eckart06,Yusef-Zadeh06,Marrone08,Dodds-Eden09,Trap11,Fish11,Eckart12,Witzel12,Rauch16}) and x-ray~\cite{Baganoff01,Porquet03,Eckart04,Eckart06,Yusef-Zadeh06,Marrone08,Dodds-Eden09,Trap11,Eckart12,Nowak12,Degenaar13,Neilsen13,Barrierre13} wavelengths. While the exact mechanism which causes the observed variability remains unclear, several models for such flares have been proposed. These include the sudden heating of hot electrons in a jet~\cite{Markoff01}, compact flaring structures (``hot spots'') on nearly circular orbits in the accretion flow around Sgr~A$^\ast$~\cite{Bro06a,Bro06b,Hamaus09} (c.f., \cite{StellaVietri98,Abramowicz01,Abramowicz03}), the ejection of a plasma blob out of the accretion flow~\cite{vanderLan66,YusefZadehQPO06}, magnetohydrodynamic turbulence along with density fluctuations~\cite{Goldston05,Chan06,Ball16} and particle accelerations due to Rossby wave instabilities~\cite{TaggerMelia06,Falanga07} (c.f., Ref.~\cite{TaggerVarniere06}), and red noise~\cite{Do09}. Infalling material such as the gas cloud G2~\cite{Gillessen12,Pfuhl15}, perhaps the product of a binary star merger~\cite{Witzel14}, could also lead to a substantial flux increase over several months~\cite{Schartmann12}.

For a Kerr black hole, a measurement of the orbital period of a hot spot can be used to infer the spin of Sgr~A$^\ast$ and several authors have argued that Sgr~A$^\ast$ must be rotating based on observed rapid periodicities. Reference~\cite{AschenbachQPO04} found quasi-periodic variability at different periods ranging from $\sim100~{\rm s}$ to $\sim2250~{\rm s}$ and obtained values of the mass $M=2.72^{+0.12}_{-0.19}\times 10^6\,M_\odot$ and the spin $a=0.9939^{+0.0026}_{-0.0074}r_g$ for Sgr~A$^\ast$ at $1\sigma$ confidence. Reference~\cite{Genzeletal03} and Ref.~\cite{BelangerQPO06} identified quasi-periodic variability with periods of $\sim17~{\rm min}$ and $\sim22~{\rm min}$, respectively, and inferred corresponding values of the spin of $a\approx0.52r_g$ and $a\approx0.22r_g$ assuming that the emission originates from the ISCO (see, also, Ref.~\cite{Howard15}). Since the Keplerian frequency of a hot spot is highest at the ISCO, Ref.~\cite{Trippe07} argued that the spin of Sgr~A$^\ast$ has a value $a\gtrsim0.7r_g$ based on a flare with a $\sim13~{\rm min}$ period. Reference~\cite{MeyerQPO06} analyzed the variability detections of Ref.~\cite{Eckart06} in a two component hot spot/ring model, within which the hot spot travels on top of a ring-like truncated disk, and found values of the spin $0.4r_g\leq a\leq r_g$ and inclination $\theta\gtrsim35^\circ$ at $3\sigma$ confidence. On the other hand, Rossby wave instabilities may naturally produce periodicities on the order of tens of minutes even if Sgr~A$^\ast$ is not spinning~\cite{TaggerMelia06,Falanga07}. Therefore, these estimates of the parameters of Sgr~A$^\ast$ and, in particular, of the spin, remain uncertain and the underlying emission mechanism of flares must be better understood.

\begin{figure*}[ht]
\begin{center}
\psfig{figure=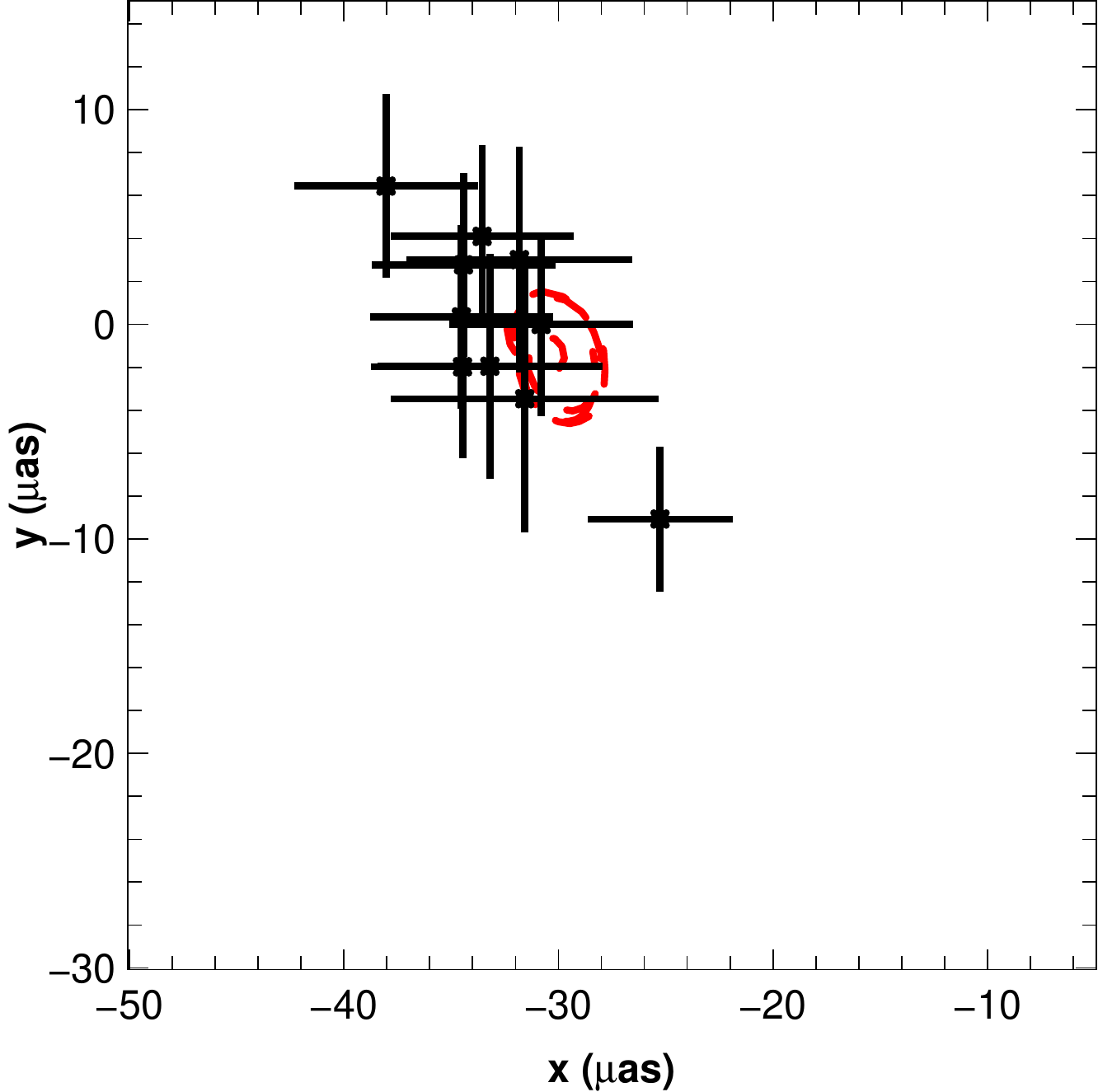,width=2in}
\psfig{figure=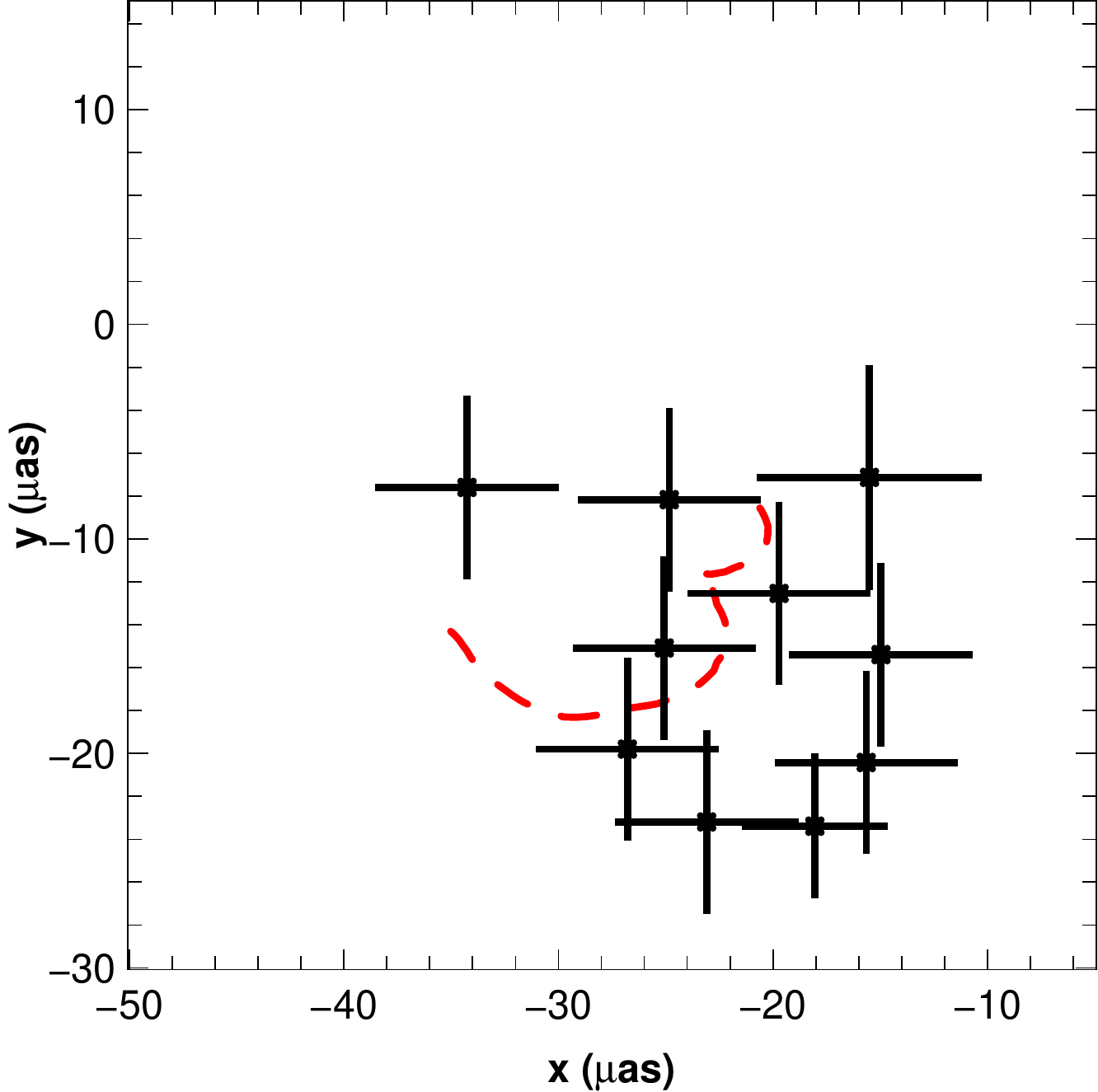,width=2in}
\psfig{figure=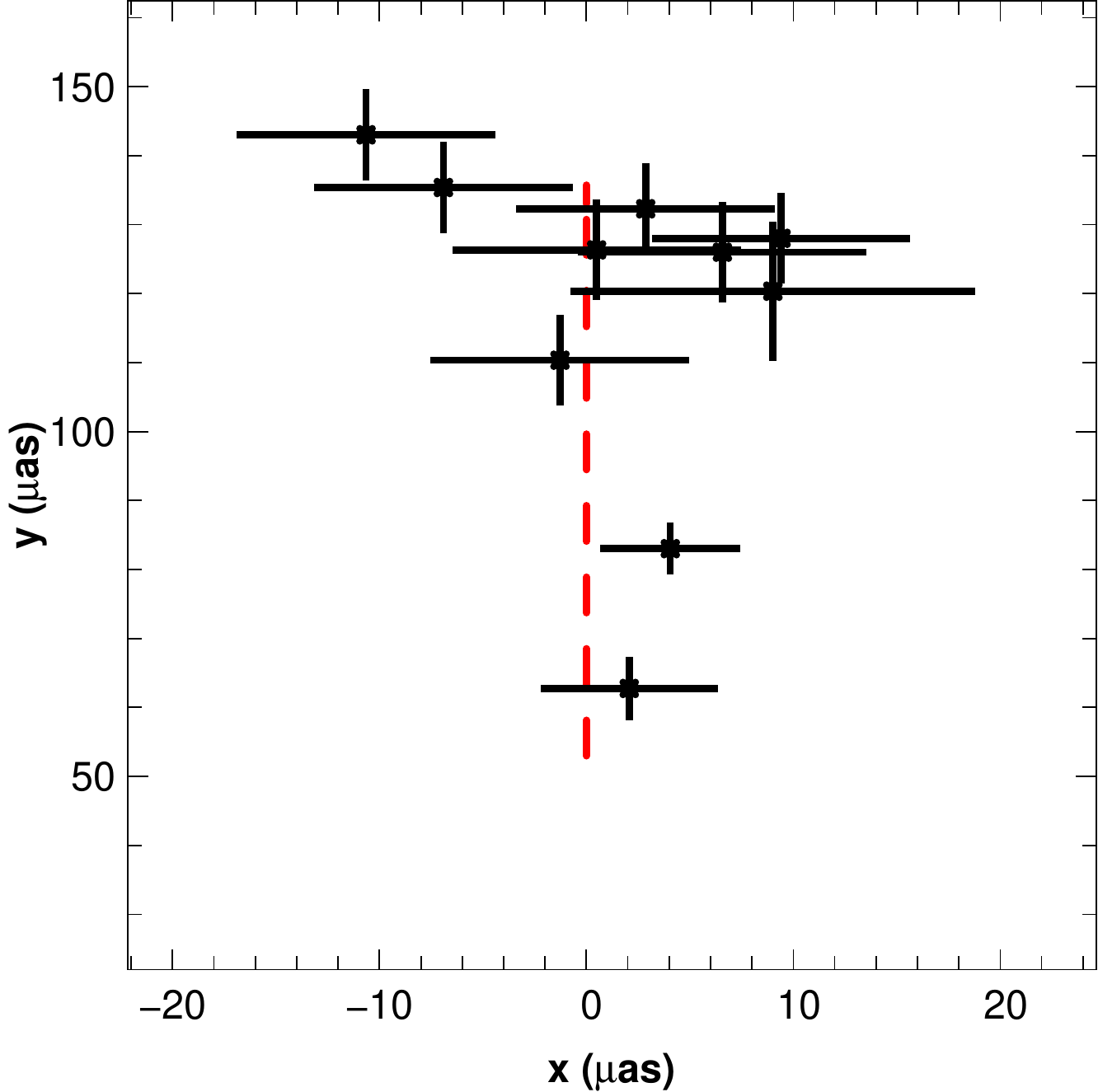,width=2in}
\psfig{figure=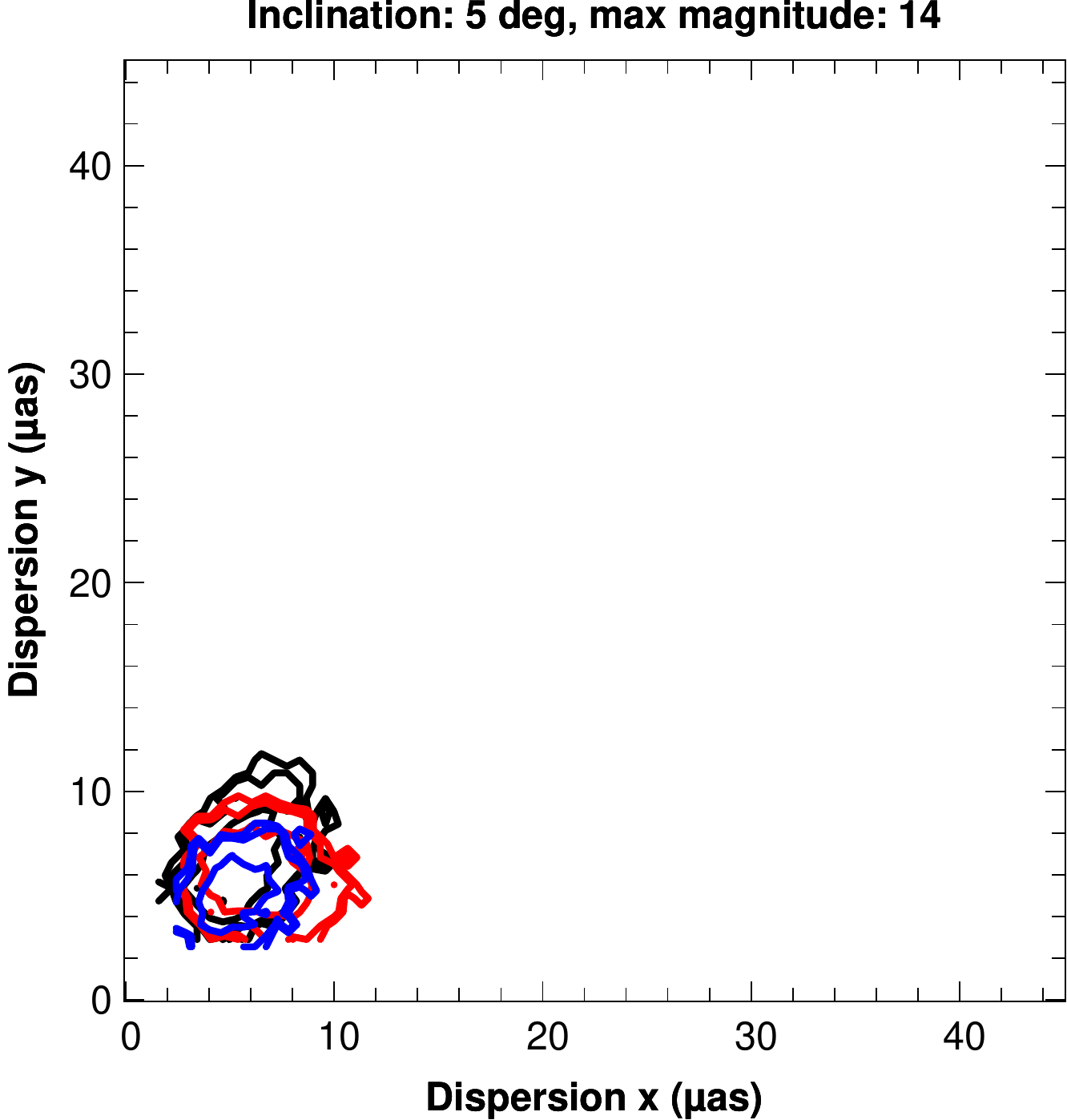,width=2in}
\psfig{figure=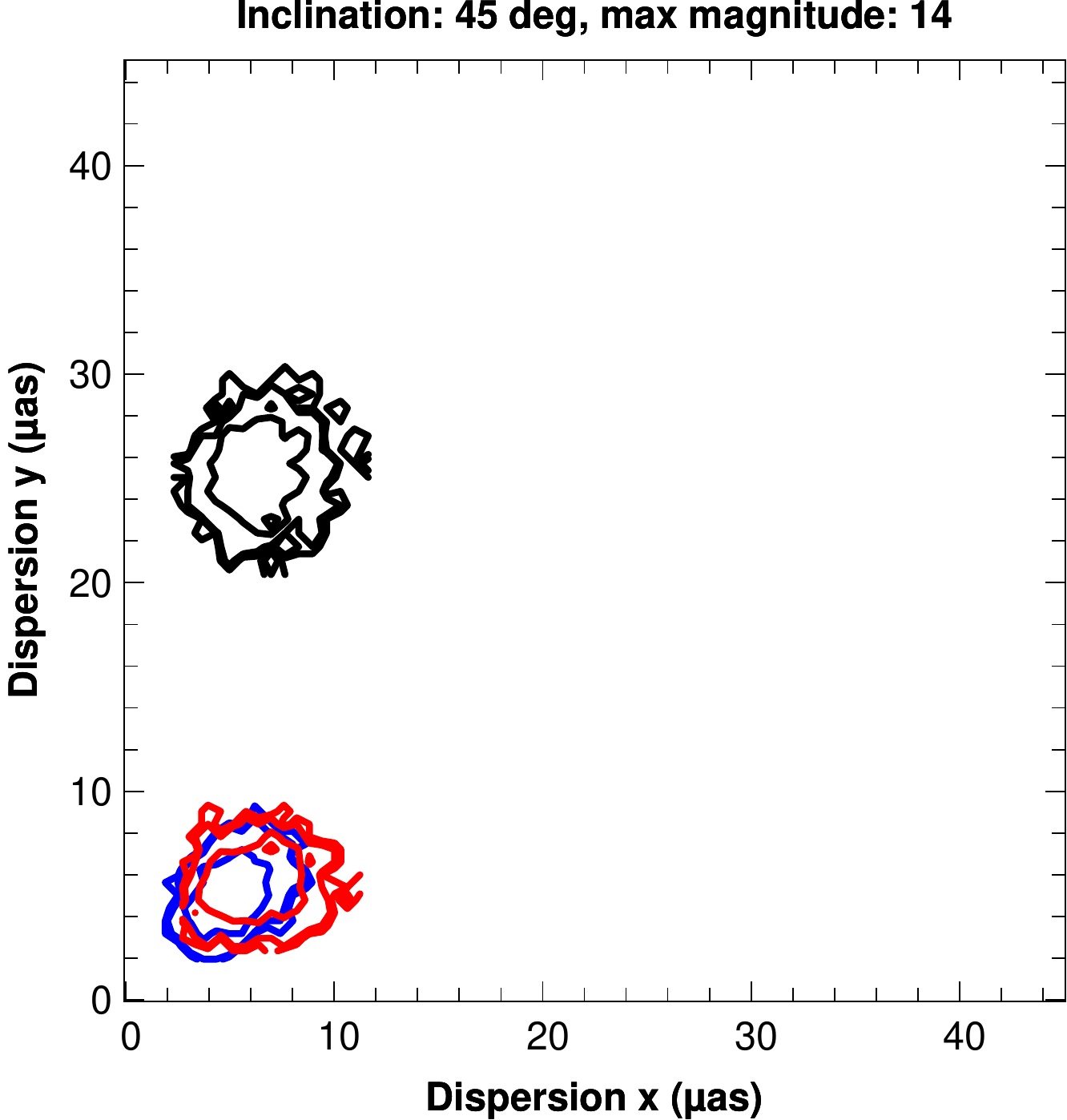,width=2in}
\psfig{figure=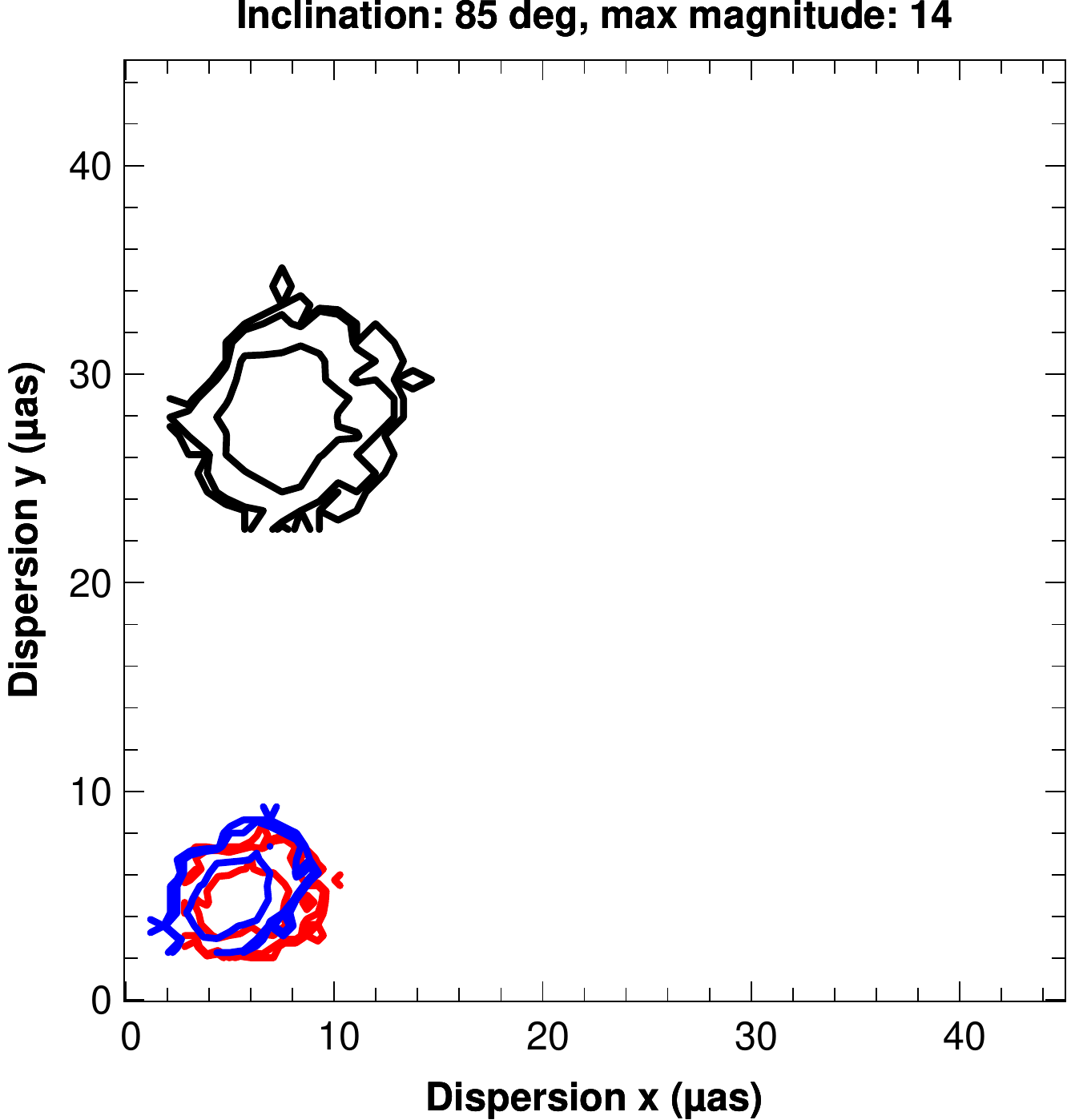,width=2in}
\end{center}
\caption{The panels in the top row show simulated one-night GRAVITY observations of a flare caused by Rossby wave instability (left panel), red noise (center panel), or a plasma blob ejected from the accretion flow at an angle of $45^\circ$ (right panel) assuming realistic astrometric performances of the instrument with an integration time of 100~s. The dashed red line in each panel shows the theoretical centroid track of the flare. The panels in the bottom row show 68\%, 95\%, and 99\% confidence contours of the dispersions of the measured $x$ and $y$ locations for the flares observed with GRAVITY over 1000 nights at inclinations of $5^\circ$, $45^\circ$, and $85^\circ$ (left to right panels), corresponding to flares in the Rossby wave instability (blue contours), red noise (red contours), and ejected blob (black contours) models. The ejected blob model can easily be distinguished from the two other models at medium and high inclination, while the other two models cannot be distinguished regardless of the inclination. Taken from Ref.~\cite{VincentGRAVITY14}.}
\label{fig:Vincent}
\end{figure*}

Deeper insight into the structure of flares is expected to be gained by observations with instruments such as GRAVITY and with the EHT. References~\cite{VincentGRAVITY11,VincentGRAVITY14} simulated GRAVITY observations of such flares in different models and showed that moving and non-moving flares located at the ISCO radius can be distinghed even for faint flares with a K-band magintude of 15 and that flares originating from a blob ejected from the accretion flow can be distinguished from other flare models if the blob is ejected at an inclination larger than $\sim45^\circ$ and the flare has a duration of $\gtrsim1.5~{\rm h}$ and a K-band magnitude roughly between 14 and 15. Figure~\ref{fig:Vincent} shows simulated GRAVITY observations of flares over one night employing three different flare models (Rossby wave instability, red noise, and ejected blob models) and assuming realistic astrometric performances of the instrument with an integration time of 100~s and simulated measurement errors of $8~{\rm \mu as}$. Figure~\ref{fig:Vincent} also shows confidence contours of the directional dispersion of the simulated flare locations for the three models at different inclinations.

References~\cite{Bro06a,Bro06b} designed a 3D hot spot model with a Gaussian density profile of an overdensity of non-thermal electrons in the accretion flow with an extent of a few gravitational radii. The EHT is expected to be able to detect such flares and their orbital periods via closure phase/closure amplitude analysis~\cite{Doelehotspot09} and via polarization measurements~\cite{FishHotspot09}. Reference~\cite{Johnsonhotspot14} estimated that the EHT can make such detections with a precision of $\sim5~{\rm \mu as}$ on timescales of minutes, which is comparable to the anticipated precision of GRAVITY for similar observations~\cite{GRAVITY,VincentGRAVITY11}. Reference~\cite{JohnsonAxis15} analyzed the lagged covariance between interferometric baselines of similar lengths but slightly different orientations and demonstrated that the peak in the lagged covariance indicates the direction and angular velocity of the accretion flow, thus enabling the EHT to measure these quantities.

Reference~\cite{PaperIII} pointed out that measurements of the orbital period of hot spots should be able to measure the spin of Sgr~A$^\ast$ in that model even if the no-hair theorem is violated, because the Keplerian frequency of a given hot spot at a fixed radius depends only weakly on deviations from the Kerr metric (see Table~\ref{tab:devparams}). In addition to a measurement of the distance of the hot spot from the black hole with GRAVITY~\cite{VincentGRAVITY14} or polarimetric VLBI observations~\cite{Johnsonhotspot14}, EHT observations could also determine that distance either in combination with an analysis of the data similar to the ones of Refs.~\cite{Bro09a,Bro11a,Bro14} or, perhaps, via observations of one hot spot and its tidal deformation~\cite{ChristianLoeb15}. Combined observations of several hot spots at different radii with GRAVITY or the EHT could be used as tracers of the spacetime. The spin of Sgr~A$^\ast$ may also be measured by observations of infalling gas inside of the ISCO~\cite{Moriyama15}. These techniques could potentially also constrain deviations from the Kerr metric should they exist.

\begin{figure*}[ht]
\begin{center}
\psfig{figure=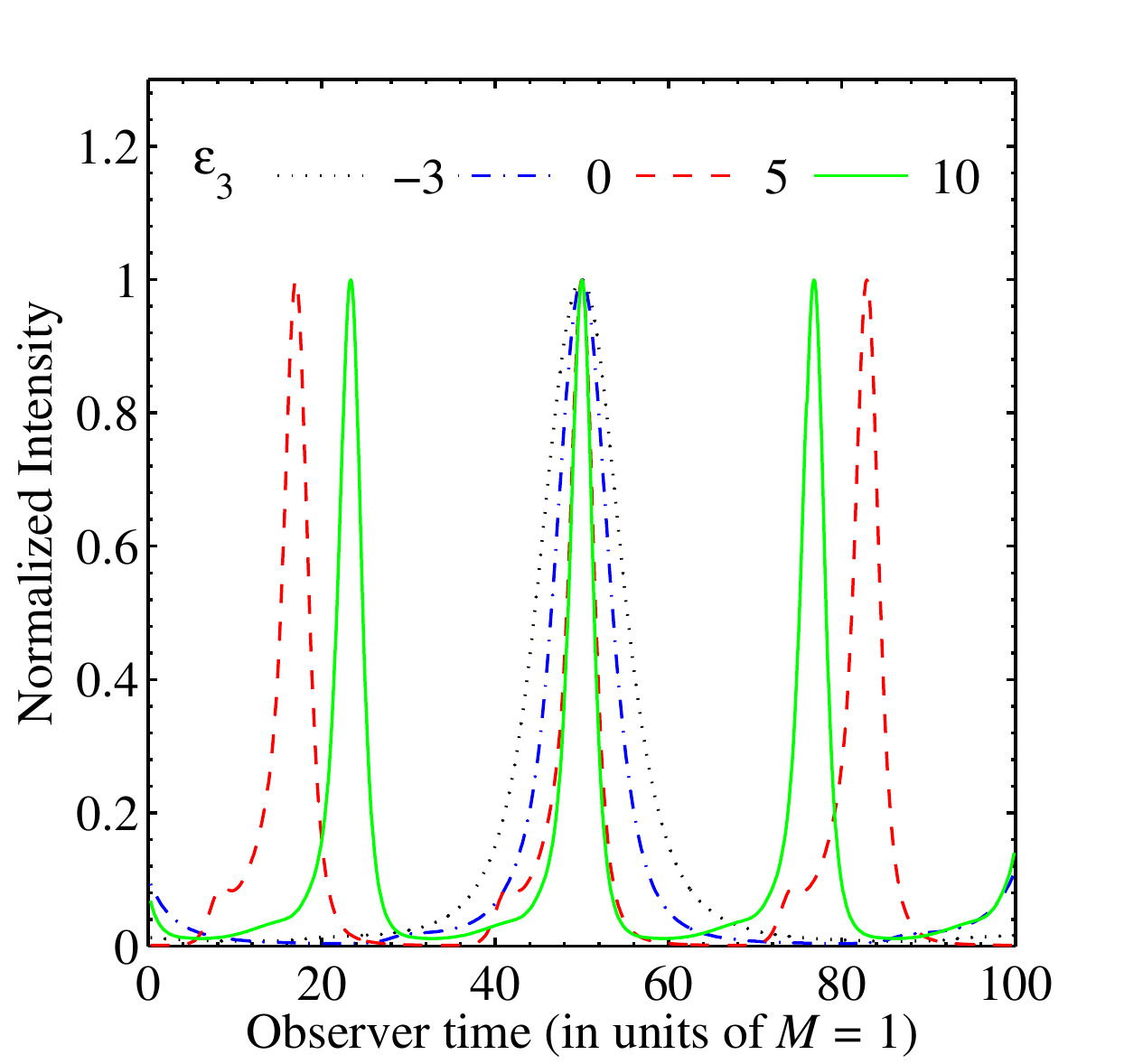,width=1.9in}
\psfig{figure=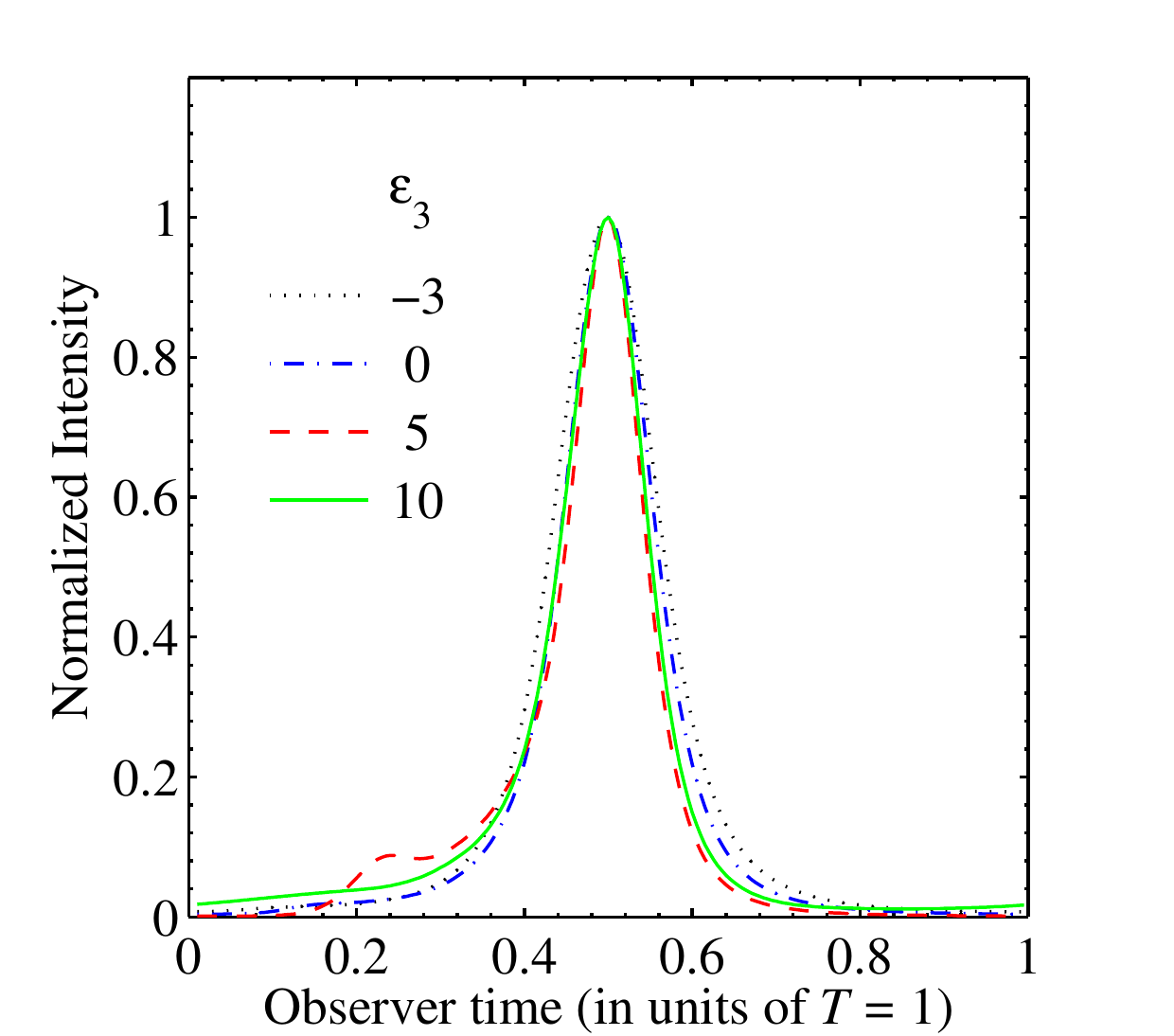,width=2.04in}
\psfig{figure=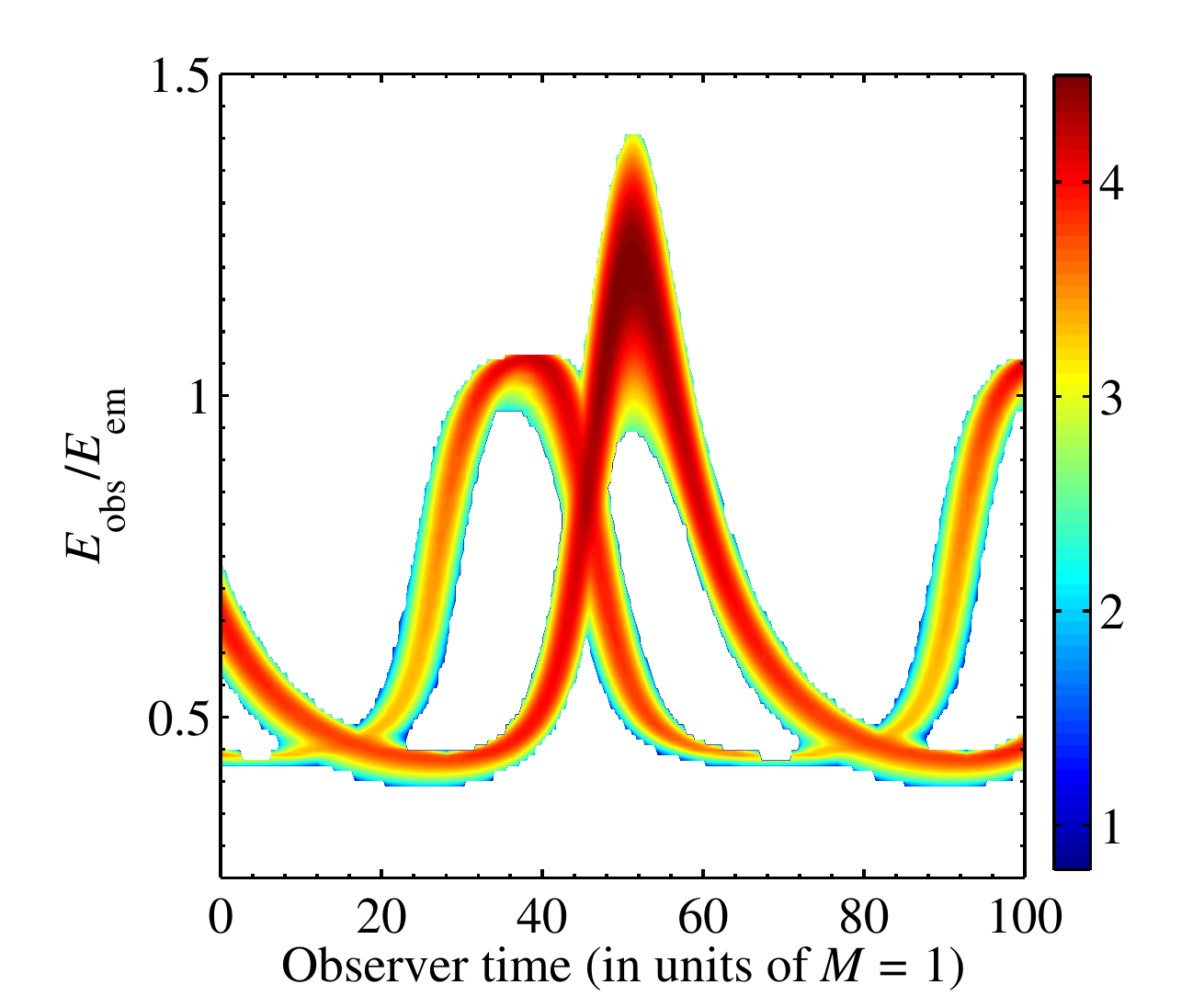,width=2.15in}
\end{center}
\caption{The left and center panels show light curves of a compact hotspot centered at the ISCO radius orbiting around the compact object with values of the spin $a=0.5r_g$ and inclination $\theta=60^\circ$ in the metric of Ref.~\cite{JPmetric} for different values of the deviation parameter $\epsilon_3$, respectively measured in gravitational units and in units of the orbital period. The width of the light curve decreases and the frequency of the hot spot increases for increasing values of the parameter $\epsilon_3$, primarily because the ISCO radius decreases. The right panel shows a spectrogram (observed to emitted energy ratio as a function of time) of a similar hot spot orbiting around a compact object with values of the spin $a=0.72r_g$ and deviation parameter $\epsilon_3=-3$ viewed at the same inclination. The curve with the larger amplitude corresponds to the primary image of the hotspot, while the curve with the smaller amplitude corresponds to the secondary image of the hot spot. Taken from Ref.~\cite{BambiHotspot1}.}
\label{fig:hotspot1}
\end{figure*}

Reference~\cite{BambiHotspot1} considered a 2D hotspot with a Gaussian density profile located in the equatorial plane of the compact object in the metric of Ref.~\cite{JPmetric} assuming monochromatic emission. Figure~\ref{fig:hotspot1} shows light curves and spectrograms of hotspots in this model for different values of the spin and deviation parameter $\epsilon_3$ of the compact object. For increasing values of the parameter $\epsilon_3$, the width of the light curve decreases and the frequency of the hot spot increases, which is caused primarily by the corresponding decrease of the ISCO radius. For hot spots orbiting at the same ISCO radius around compact objects with different sets of values of the spin and deviation parameter that correspond to that radius, there is a slight phase shift between the primary and secondary curves in the spectrogram potentially allowing these signals to be distinguished if the ISCO can be determined independently~\cite{BambiHotspot1}. Reference~\cite{BambiHotspot2} considered a similar model, where the hotspot is located at a fixed (small) height above or below the equatorial plane and found slight changes of the brightness of the hot spot depending on its position. Reference~\cite{BambiHotspot3} applied this model to wormholes.

\section{Discussion}
\label{sec:conclusions}

At present, general relativity remains the standard theory of gravity. Its validity has been confirmed by a number of experiments in the weak-field regime~\cite{Will14} and none of the few strong-field tests have, so far, detected any deviation from it, neither with observations of neutron stars~(see Ref.~\cite{DeDeo03}; \cite{Psaltis08,Antoniadisetal13,Yagi14a,Yagi14b,Zhu15}) nor of black holes~\cite{Psaltis07,Braneworld1,Braneworld2,BambiBarausse11,BambiEfficiency,Braneworld3,Braneworld4,Bro14,Kong14}; c.f., Ref.~\cite{Psaltis08}. Likewise, there is no indication of a violation of general relativity on cosmological scales~\cite{Koyama15}.

However, general relativity is expected to break down at some level for several theoretical reasons such as its nonrenormalizability in a grand-unification scheme (see, e.g., Ref.~\cite{Wald84}), the cosmological constant problem (see, e.g., Ref.~\cite{Weinberg89}), and the hierarchy problem (see, e.g., Ref.~\cite{Maartens04}. Moreover, in $\Lambda$CDM, the general-relativistic standard model of cosmology, dark matter and dark energy make up about 26\% and 69\% of the total mass-energy content of the universe, respectively~\cite{Reiss11a,Reiss11b,Beutler11,Anderson14,Betoule14,Planck,Ross15}, but the nature of dark matter and, especially, of dark energy still remains largely mysterious.

Thus far, we have barely begun to probe the strong-field regime of general relativity found around compact objects (as well as the cosmological regime) and great progress is expected to be made in the coming years and decades (c.f., Fig.~\ref{fig:parameterspace}). Tests of general relativity in both regimes require an appropriate underlying framework. For weak-field tests, a description of observables in terms of a parameterized post-Newtonian approach is sufficient~\cite{Will93}, while for (model-independent) strong-field tests the spacetime itself has to be modelled carefully based on a Kerr-like metric (e.g., \cite{MN92,CH04,GB06,VH10,JPmetric,VYS11,Jmetric,CPR14,Lin16}).

The nature of black holes as encapsulated by the (general-relativistic) no-hair theorem provides the basis for unprecedented tests of general relativity with strong-field and weak-field probes. Sgr~A$^\ast$ is a prime target for such tests and three different experiments have high promise for a test of the no-hair theorem in the next few years and decades. NIR monitoring of stars orbiting around Sgr~A$^\ast$ has already led to precise measurements of the mass and distance of Sgr~A$^\ast$~\cite{Ghez08,Gillessen09,Gillessen09b,Schoedel09,Meyer12,Do13,Chatzopoulos15}. Continued monitoring as well as the expected instrumental improvement with GRAVITY~\cite{GRAVITY} and future 30m-class optical telescopes (see Ref.~\cite{Weinberg05}) will further improve upon these measurements. Such observations will most likely detect orbital precessions and radial velocity corrections of stars induced by post-Newtonian effects including frame-dragging or even those caused by the quadrupole moment of Sgr~A$^\ast$, in particular during pericenter passages of the star S2 (which will take place in 2018) or of other S-stars (e.g.,~\cite{Will08,Merritt10,Angelil10,Zhang15}).

Timing observations of radio pulsars on orbits around Sgr~A$^\ast$ could provide another precise measurement of the mass, spin, and quadrupole moment of Sgr~A$^\ast$~\cite{WK99,Pfahl04,Liu12} or of stellar-mass black holes in binaries with pulsars~\cite{Liu14}. For Sgr~A$^\ast$, such observations, carried out over about five years with an SKA-like telescope, could detect frame-dragging at the $10^{-3}$ level and test the no-hair theorem at the $10^{-2}$ level~\cite{Liu12}. The recent discovery of a magnetar at a distance of only $\sim0.1~{\rm pc}$ from Sgr~A$^\ast$~\cite{Kennea13,Mori13,Rea13,Eatough13,Zadeh15} has spurred the hopes of finding a suitable pulsar that is close enough to the Galactic center. Although many pulsar searches have been conducted at high observing frequencies over several years~\cite{Kramer00,Johnston06,Deneva09,Macquart10,Bates11,Eatough13,Siemion13}, the discovery of such a pulsar may require targeted surveys with the SKA~\cite{Kramer15,Keane15,Shao15}. Both of these methods (NIR and timing observations) track the orbits of stars or pulsars around Sgr~A$^\ast$, which may be perturbed by surrounding stars~\cite{Merritt10}, drag forces~\cite{Psaltisdrag12}, stellar winds and tidal disruptions~\cite{Antonini10,Antonini13,Psaltiswinds13}, or other effects~\cite{Sadeghian13}.

The EHT is expected to probe Sgr~A$^\ast$ on event-horizon scales and to take the first direct image of a black hole. The size and shape of the shadow of Sgr~A$^\ast$ (or any other black hole) depends directly on the properties of the underlying spacetime, i.e., on the mass and spin for Kerr black hole (e.g., \cite{Falcke00,Chan13}) as well as on potential deviations from the Kerr metric if the no-hair theorem is violated (e.g., \cite{PaperII,AE12,J13rings}). In addition, deviations from the Kerr metric can modify the properties of the accretion flow surrounding the black hole (e.g., \cite{PaperI,BambiBarausse11,PaperIII,BambiTorus15}). Early EHT observations of Sgr~A$^\ast$ in 2007--2013 with a three-station array have resolved structure~\cite{Doele08} and detected variability~\cite{Fish11} as well as polarized emission~\cite{JohnsonScience15} on event horizon scales. Furthermore, Refs.~\cite{Fish11,Fish15} detected a number of closure phases along the initial CARMA--SMTO--Hawaii telescope array. As of 2015, the EHT is comprised by eight different sites, and VLBI observations with telescope arrays that include more than three stations are scheduled to begin in spring 2016.

Within the context of RIAF model images, the early EHT data favor small values of the spin~\cite{Bro09a,Bro11a,Bro14,Brodericketal15}, while constraints on deviations from the Kerr metric remain weak~\cite{Bro14}. However, such constraints within a specific RIAF model will improve dramatically with EHT observations using larger telescope arrays~\cite{Johannsenetal15}.
Since the shadow itself is largely independent from the properties of the accretion flow, measurements of its size and shape in combination with existing stellar-orbits data can be used to improve upon current measurements of the mass and distance of Sgr~A$^\ast$ and to infer potential deviations from the Kerr metric with high precision~\cite{Psaltis14,JohannsenPRL}.

Both NIR observations with instruments such as GRAVITY and VLBI observations with the EHT should also probe regions of quasi-periodic emission in the accretion flow of Sgr~A$^\ast$ with unprecedented precision~\cite{Doelehotspot09,FishHotspot09,GRAVITY,VincentGRAVITY11,VincentGRAVITY14}. Such observations may also distinguish between different models~\cite{VincentGRAVITY11,VincentGRAVITY14} and infer other properties of Sgr~A$^\ast$ and its accretion flow (e.g., \cite{VincentGRAVITY14,Johnsonhotspot14,ChristianLoeb15}). In particular, measurements of the orbital period of flares can be used to constrain the spin (and perhaps even the quadrupole moment) of Sgr~A$^\ast$ (e.g., \cite{Genzeletal03,AschenbachQPO04,MeyerQPO06,Eckart06,Trippe07}). Current spin estimates based on detections of variability, however, are uncertain and cover practically the entire range of spin values from $\approx0$~\cite{TaggerMelia06,Falanga07} to $\approx1$~\cite{AschenbachQPO04}. Further theoretical and observational studies of variability are required in order to better understand the underlying emission mechanism of flares.

The observables of the experiments discussed here are very different and will, therefore, be affected by different systematics. Given these uncertainties, a measurement of the spin and a potential violation of the no-hair theorem by more than one experiment would be very convincing if these measurements were to agree.

A remaining theoretical challenge, however, is the proper combination of tests of the no-hair theorem with weak-field and strong-field probes, because they typically use different coordinate systems which complicates the direct comparison of their results. Although vacuum spacetimes in general relativity can be characterized by a set of scalar multipole moments if they are also stationary and axisymmetric~\cite{BeigSimon80,BeigSimon81}, it remains unclear if such a property also holds in other gravity theories beyond the perturbative regime [c.f., Eq.~(\ref{eq:mult})], especially if they do not satisfy the Laplace equation in the far-field at orders of the radial distance from the black hole that involve deviations from general relativity.

So far, (non-perturbative) multipole moments outside of general relativity have only been defined for stationary, asymptotically flat spacetimes in certain scalar-tensor theories of graviy~\cite{Pappas15} and $f(R)$ gravity~\cite{Suvorov15}. One solution to this issue is perhaps the approach of Ref.~\cite{Zhang15} who analyzed stellar orbits in the Kerr spacetime using a ray-tracing algorithm. Performing such an analysis in a Kerr-like spacetime would directly link weak-field probes of stars orbiting around the Galactic center with strong-field observables such as the shadow or the accretion flow of Sgr~A$^\ast$.

In addition to Sgr~A$^\ast$, the supermassive black hole at the center of M87 is another prime target of the EHT and early EHT observations at 230~GHz with three-station telescope arrays have already detected structure on the order of $\approx5.5$ Schwarzschild radii~\cite{DoeleM87,Akiyama15} and measured a number of closure phases~\cite{Akiyama15}. Reference~\cite{Hada16} reported VLBI observations of M87 at 86~GHz using the Very Long Baseline Array and the Green Bank Telescope. In contrast to Sgr~A$^\ast$, EHT observations of this supermassive black hole do not face the same challenges with scattering or refractive time scales that require extra analysis effort. Given its much greater mass ($\sim4-6\times10^9M_\odot$~\cite{Murphy11,Walsh13}), the time scales for M87 are also longer and the rotation of the Earth is less of a challenge. In addition, the spatial scales of strong-gravity signatures are approximately comparable to those in Sgr~A$^\ast$, but the time scales for strong-gravity effects such as the orbital period of matter particles near the ISCO are much longer and, therefore, tractable via time sequenced EHT observations that allow full imaging fidelity in each epoch~\cite{DoeleM87,Akiyama15}.

Reference~\cite{Luetal14} simulated images of the supermassive black hole at the center of M87 at 230~GHz and 345~GHz based on a 7--8 station EHT array assuming realistic measurement conditions. Reference~\cite{Luetal14} showed that such an array would have a resolution of $20-30~{\rm \mu as}$ ($2-4$ Schwarzschild radii) and is capable of resolving the shadow of the black hole and of tracing real-time structural changes on scales of a few Schwarzschild radii.

Other sources exist, but these have smaller angular sizes of their respective shadows~\cite{SMBHmasses} (see Fig.~\ref{fig:ehtsources}). Recently, Refs.~\cite{vandenBosch12,Walsh15} measured a mass of about $5\times10^9M_\odot$ for the supermassive black hole in NGC~1277, which, therefore, has an angular shadow size of roughly $7~{\rm \mu as}$ in the sky (assuming a distance of $71~{\rm Mpc}$ as in Ref.~\cite{Walsh15}). These black holes may be resolvable on horizon scales with future VLBI stations in space~\cite{Hirabayashi98,Zhakarov05,Kardashev09,Wild09}. In any case, the prospects for a test of the no-hair theorem and, thereby, of general relativity within the coming years are great.


I thank P. Cunha, S. Doeleman, C. Herdeiro, and P. Pani for useful comments. This work was supported in part by Perimeter Institute for Theoretical Physics. Research at Perimeter Institute is supported through Industry Canada and by the Province of Ontario through the Ministry of Research \& Innovation.

\begin{appendix}
\section{Mappings to Black Holes in Alternative Gravity Theories}
\label{appendix}

Here, I summarize the mappings of the metric of Ref.~\cite{Jmetric} [see Eq.~(\ref{eq:Jmetric})] to the analytically known black hole solutions of EdGB and dCS gravity, as well as to other Kerr-like metrics. Some of these mappings were derived in Ref.~\cite{Jmetric}.

\subsection{Einstein-dilaton-Gauss-Bonnet Gravity}

Static and slowly-rotating black holes in gravity theories described by Lagrangians modified from the standard Einstein-Hilbert form by scalar fields coupled to quadratic curvature invariants were investigated in Refs.~\cite{Mignemi93,Kanti96,YS11,Pani11,AY14,M15}. In the limit of small deviations, the Kerr metric is modified by a perturbation $h_{\alpha\beta}$, which, to linear order in the spin $a$ and the parameter $\zeta_{\rm EdGB}$, has the nonzero components
\ba
h_{tt}^{\rm EDBG} &=& -\frac{\zeta_{\rm EdGB}}{3}\frac{M^3}{r^3} \left( 1 + \frac{26M}{r}+\frac{66}{5}\frac{M^2}{r^2}+\frac{96}{5}\frac{M^3}{r^3}-\frac{80M^4}{r^4} \right), \\
h_{rr}^{\rm EdGB} &=&  - \zeta_{\rm EdGB} \left(1-\frac{2M}{r}\right)^{-2} \frac{M^2}{r^2} \nn \\
&& \times \left( 1 + \frac{M}{r} + \frac{52}{3}\frac{M^2}{r^2} + \frac{2M^3}{r^3} + \frac{16}{5}\frac{M^4}{r^4} - \frac{368}{3}\frac{M^5}{r^5} \right), \\
h_{t\phi}^{\rm EdGB} &=& \frac{3}{5}\zeta_{\rm EdBG} \frac{aM^3\sin^2\theta}{r^3} \nn \\
&& \times \left( 1 + \frac{140}{9}\frac{M}{r}+\frac{10M^2}{r^2}+\frac{16M^3}{r^3}-\frac{400}{9}\frac{M^4}{r^4} \right).
\ea
Here, the parameter $\zeta_{\rm EdGB}$ is defined by the equation
\be
\zeta_{\rm EdGB} \equiv \frac{16\pi\alpha_{\rm EdGB}^2}{\beta_{\rm EdGB} M^4},
\ee
where $\alpha_{\rm EdGB}$ and $\beta_{\rm EdGB}$ are the coupling constants of the theory; see, e.g., Ref.~\cite{AY14}.

The mapping is, then, given by the equations
\ba
\sum_{n=3}^\infty \alpha_{1n} \left( \frac{M}{r} \right)^n &=& -\frac{M^3 \zeta_{\rm EdGB}}{30(r-2M)r^6} \nn \\ 
&& \times (5r^4 + 130Mr^3 + 66M^2r^2 + 96M^3r - 400M^4), \\
\sum_{n=2}^\infty \alpha_{2n} \left( \frac{M}{r} \right)^n &=& - \frac{M^3 \zeta_{\rm EdGB}}{30(r-2M)r^6} (13r^4+134Mr^3+74M^2r^2 \nn \\
&& +96M^3r-592M^4), \\
\sum_{n=2}^\infty \alpha_{5n} \left( \frac{M}{r} \right)^n &=& \frac{M^2 \zeta_{\rm EdGB}}{15(r-2M)r^6} (15r^5+15Mr^4+260M^2r^3 \nn \\
&& +30M^3r^2+48M^4r-1840M^5), \\
\ea
and the lowest-order coefficients are:
\ba
\alpha_{13} &=& -\frac{1}{6}\zeta_{\rm EdGB},~~~\alpha_{14} = -\frac{14}{3}\zeta_{\rm EdGB},~~~\alpha_{15} = -\frac{173}{15}\zeta_{\rm EdGB}, \nn \\
\alpha_{23} &=& -\frac{13}{30}\zeta_{\rm EdGB},~~~\alpha_{24} = -\frac{16}{3}\zeta_{\rm EdGB},~~~\alpha_{25} = -\frac{197}{15}\zeta_{\rm EdGB}, \nn \\
\alpha_{52} &=& \zeta_{\rm EdGB},~~~\alpha_{53} = 3\zeta_{\rm EdGB},~~~\alpha_{54} = \frac{70}{3}\zeta_{\rm EdGB}.
\ea

Reference~\cite{Jmetric} only contains the mapping to the static black hole solution. In this case, $A_2(r)=1$ and all parameters $\alpha_{2i}$, $i\geq2$, vanish, leaving $A_1(r)$ and $A_5(r)$ as the remaining deviation functions. Note, however, that in Ref.~\cite{Jmetric} the mapping of the deviation function $A_1(r)$ is missing and that Eqs.~(133) and (134) contain a typo. As shown in Ref.~\cite{AY14}, the black hole metric in EdGB gravity is not integrable at ${\cal O}(a^2)$. Therefore, it can only be mapped to the metric of Ref.~\cite{Jmetric} up to ${\cal O}(a)$.

\subsection{Dynamical Chern-Simons Gravity}

Slowly rotating black holes in dCS gravity were first analyzed in Ref.~\cite{YP09}. In these solutions, only the $(t,\phi)$ component of the metric is modified, which is given by the expression
\be
h_{t\phi}^{\rm dCS} = \frac{5}{8}\zeta_{\rm dCS}\frac{aM^4}{r^4}\sin^2\theta \left(1+\frac{12M}{7r}+\frac{27M^2}{10r^2}\right).
\ee
In this case, the mapping is
\ba
\alpha_{24} &=& \frac{5}{8}\zeta_{\rm dCS}, \\
\alpha_{25} &=& \frac{15}{14}\zeta_{\rm dCS}, \\
\alpha_{26} &=& \frac{27}{16}\zeta_{\rm dCS}.
\ea
All other deviation parameters vanish. Note that the metric to ${\cal O}(a^2)$ found in Ref.~\cite{YYT12} is likewise not integrable and, thus, cannot be mapped to the metric of Ref.~\cite{Jmetric}.

\subsection{The Modified Gravity Bumpy Kerr Metric}

Reference~\cite{GY11} constructed an explicit form of the modified gravity bumpy Kerr metric~\cite{VYS11}, given by the equation
\be
g_{\rm \mu\nu}^{\rm MGBK} = g_{\rm \mu\nu}^{\rm K} + h_{\rm \mu\nu}^{\rm MGBK},
\label{eq:MGBKmetric}
\ee
where the $g_{\rm \mu\nu}^{\rm K}$ is the Kerr metric in Eq.~(\ref{eq:kerr}). The correction $h_{\rm \mu\nu}^{\rm MGBK}$ depends on three (nonzero) deviation functions given by the expansions
\ba
\gamma_A &\equiv& \sum_{n=0}^\infty \gamma_{A,n} \left(\frac{M}{r}\right)^n,~~~A=1,4, \\
\gamma_3 &\equiv& \frac{1}{r} \sum_{n=0}^\infty \gamma_{3,n} \left(\frac{M}{r}\right)^n,
\ea
where $\gamma_{1,0}=\gamma_{1,1}=\gamma_{3,0}=\gamma_{3,2}=\gamma_{4,0}=\gamma_{4,1}=0$~\cite{GY11} as well as (preferentially) $\gamma_{1,2}=\gamma_{3,1}=\gamma_{4,2}=0$~\cite{Jmetric}. The fourth deviation function, $\Theta_3(\theta)$, is set to zero~\cite{GY11}.

This metric can be mapped to the metric of Ref.~\cite{Jmetric} (expanded to linear order in the deviation parameters) via the relations
\ba
\epsilon_n &=& 0,~~~n\geq3,
\label{eq:mapepMGBK} \\
\alpha_{5n} &=& \gamma_{1,n},~~~n\geq2
\label{eq:mapalpha5MGBK}
\ea
as well as 
\ba
\sum_{n=3}^\infty \alpha_{1n} \left(\frac{M}{r}\right)^n &=& \frac{1}{4(r^2+a^2)\Delta} \{ 8aMrh_{t\phi}^{\rm MGBK} \nn \\
&& +[2r^4+a^4 +a^2r(3r+4M)  \\
&& +a^2(\Delta-2Mr)\cos2\theta]h_{tt}^{\rm MGBK}  \}, 
\label{eq:mapalpha1MGBK} \\
\sum_{n=2}^\infty \alpha_{2n} \left(\frac{M}{r}\right)^n &=& -\frac{1}{2a\Delta} [a(\Sigma-4Mr)h_{tt}^{\rm MGBK} \nn \\
&& + 2(\Sigma-2Mr)\csc^2\theta h_{t\phi}^{\rm MGBK} ].
\label{eq:mapalpha2MGBK} 
\ea
At least up to order $n=5$, the latter two equations can be written in the form
\ba
\alpha_{1n} - \frac{\alpha_{5n}}{2} &=& -2\gamma_{4,n-1} + \gamma_{4,n}, \nn \\
\alpha_{2n} - \frac{\alpha_{5n}}{2} &=& \frac{a}{M}\gamma_{3,n-1} - \frac{M}{a}(2\gamma_{3,n}-\gamma_{3,n+1}),
\ea
where the first equation holds for $n\geq3$, while the second equation holds for $n\geq2$~\cite{Jmetric}.

\subsection{The Metric of Ref.~\cite{Lin16}}
\label{appLin16}

The metric of Ref.~\cite{Lin16} has the following nonvanishing elements:
\ba
g_{tt}&=&- \left(1 - \frac{2 m_1(r) r}{\Sigma}\right) , \nn \\
g_{rr}&=&\frac{\Sigma}{\Delta_2} , \nn \\
g_{\theta\theta}&=&\Sigma, \nn \\
g_{\phi\phi}&=&\left(r^2 + a^2 + \frac{2 a^2 m_1(r) r \sin^2\theta}{\Sigma}\right) \sin^2\theta , \nn \\
g_{t\phi}&=&- \frac{2 a m_1(r) r \sin^2\theta}{\Sigma}  \label{eq:LTGBmetric},
\ea
where
\be
\Delta_2 \equiv r^2 - 2 m_2(r) r + a^2
\label{eq:Delta2}
\ee
and where $\Sigma$ is given by Eq.~(\ref{eq:sigma}).

There is no direct mapping between this metric and the Kerr-like metric of Ref.~\cite{Jmetric} except for the trivial Kerr (or Kerr-Newman) case. Whether or not there exists a coordinate transformation that can relate these metrics is unclear.

The above metric depends on two deviation functions $m_1(r)$ and $m_2(r)$, where the former occurs in the $(t,t)$, $(t,\phi)$, and $(\phi,\phi)$ elements and the latter in the $(r,r)$ element; the $(\theta,\theta)$ element is unmodified. It is straightforward to generalize this metric by introducing deviation functions $m_i(r,\theta)$, $i=1,\ldots,4$, and $\tilde{\Sigma}\equiv\Sigma+f(r)$ as defined in Eq.~(\ref{eq:Sigmatilde}) and writing the metric elements in the form
\ba
g_{tt}&=&- \left(1 - \frac{2 m_1(r) r}{\tilde{\Sigma}}\right) , \nn \\
g_{rr}&=&\frac{\tilde{\Sigma}}{\Delta_2} , \nn \\
g_{\theta\theta}&=&\tilde{\Sigma}, \nn \\
g_{\phi\phi}&=&\left(r^2 + a^2 + \frac{2 a^2 m_3(r) r \sin^2\theta}{\tilde{\Sigma}}\right) \sin^2\theta , \nn \\
g_{t\phi}&=&- \frac{2 a m_4(r) r \sin^2\theta}{\tilde{\Sigma}}  \label{eq:LTGBmetric2}.
\ea

The resulting metric can be mapped to the metric of Ref.~\cite{Jmetric} by the choices
\ba
m_1 &=& \frac{\tilde{\Sigma}}{2r} \left[ 1+\frac{\tilde{\Sigma} \left( a^2A_2^2\sin^2\theta-\bar{\Delta} \right)}{\left[ A_1\left(r^2+a^2\right)-a^2A_2\sin^2\theta \right]^2} \right], \nn \\
m_2 &=& \frac{r^2+a^2-\bar{\Delta}A_5}{2r}, \nn \\
m_3 &=& \frac{\tilde{\Sigma}}{2a^2r\sin^2\theta} \left[ \frac{\tilde{\Sigma} \left( A_1^2\left(r^2+a^2\right)^2-a^2\bar{\Delta}\sin^2\theta \right)}{\left[ A_1\left(r^2+a^2\right)-a^2A_2\sin^2\theta \right]^2} -r^2-a^2 \right], \nn \\
m_4 &=& \frac{\tilde{\Sigma}^2 \left[ \left(r^2+a^2\right)A_1A_2-\bar{\Delta} \right]}{2r\left[ A_1\left(r^2+a^2\right)-a^2A_2\sin^2\theta \right]^2}
\label{eq:generalmap}
\ea
for the deviation functions $m_i(r,\theta)$, $i=1,\ldots,4$, as can be shown by equating the corresponding elements of both metrics and solving for the deviation functions in the metric in Eq.~(\ref{eq:LTGBmetric2}). Here, $\bar{\Delta}$ is defined in Eq.~(\ref{eq:beta}). Expanding the RHS  in powers of $M/r$ [including the functions $A_1$, $A_2$, $A_5$, and $f$ as in Eqs.~(\ref{eq:A1})--(\ref{eq:A5}) and (\ref{eq:f})], the functions $m_i(r,\theta)$, $i=1,\ldots,4$, at the lowest two nonvanishing orders in the deviation parameters are giving by the expressions
\ba
m_1 &=& M-\frac{\beta M^2}{2r} +\frac{(2\alpha_{13}-\epsilon_3)M^3}{2r^2}, \nn \\
m_2 &=& M-\frac{(\alpha_{52}+\beta)M^2}{2r} +\frac{(2\alpha_{52}-\alpha_{53})M^3}{2r^2}, \nn \\
m_3 &=& M+\frac{\epsilon_3M^3}{2a^2\sin^2\theta} +\frac{a^2M^2(2\alpha_{22}-\beta)+\frac{\epsilon_4M^4}{\sin^2\theta}}{2a^2r} + \frac{ \left( 2a^2\alpha_{23}+\frac{a^2\epsilon_3+M^2\epsilon_5}{\sin^2\theta} \right)M^3 }{2a^2r^2}, \nn \\
m_4 &=& M+\frac{(\alpha_{22}-\beta)M^2}{2r} +\frac{(\alpha_{13}+\alpha_{23})M^3}{2r^2},
\label{eq:msorder2}
\ea
assuming $a\neq0$ (as well as $\sin\theta\neq0$) in the latter two equations. If $a=0$ (or $\sin\theta=0)$, then the metric no longer depends on the functions $m_3$ and $m_4$. Note that the function $m_2$ can have a term of order $M/r$ (which is not ruled out by weak-field constraints as claimed in Ref.~\cite{Lin16}) and that $m_3$ contains a zeroth-order term.

\subsubsection{The Metric of Ref.~\cite{BambiModesto13}~~~~~~~~~~~~~~~~~~~~~~~~~~~~~~~~~~~~~~~~~~~~~~~~~~~~~~~~~~~~~~~~~~~~~~~~~~~~}
\label{app:BambiModesto13}

In the rotating generalization of the static Bardeen metric~\cite{BardeenBH} constructed in Ref.~\cite{BambiModesto13}, the deviation functions $m_i(r)$, $i=1,\ldots,4$, take the form
\be
m_1=\ldots=m_4=\frac{r^3}{\left(r^2+g^2\right)^{3/2}}M.
\ee
Therefore, using the mapping in Eq.~(\ref{eq:msorder2}) up to ${\mathcal O}(M^4/r^4)$, the nonvanishing deviation parameters are given by the equations
\ba
\alpha_{13} &=& -\frac{3}{2}\frac{g^2}{M^2},~\alpha_{14}=2\alpha_{13},~\alpha_{15}=\frac{3g^2\left(4a^2+5g^2-16M^2\right)}{8M^4}, \nn \\
\alpha_{23} &=& \alpha_{13},~\alpha_{24}=\alpha_{14},~\alpha_{25}=\alpha_{15}, \nn \\
\alpha_{53} &=& -2\alpha_{13},~\alpha_{54}=-2\alpha_{14},~\alpha_{55}=-2\alpha_{15}.
\label{eq:Bardeenmap}
\ea
This seems to suggest a general mapping of the form $A_1(r)=A_2(r)=-\frac{1}{2}A_5(r)$, $f(r)=0$ as well as $\beta=0$. However, imposing this general mapping in Eq.~(\ref{eq:generalmap}) leads to the equation $A_1(r)=-2^{1/3}$ which contradicts Eq.~(\ref{eq:A1}). Therefore, this general mapping cannot be correct.

\subsection{The Metric of Ref.~\cite{Antoci01}}
\label{appAntoci01}

Reference~\cite{Antoci01} propsed a modified Schwarzschild metric of the form
\ba
g_{tt}&=& -\left(1-\frac{2M}{F}\right), \nn \\
g_{rr}&=& \frac{F'^2}{1-\frac{2M}{F}}, \nn \\
g_{\theta\theta}&=& \frac{F^2}{r^2}, \nn \\
g_{\phi\phi}&=& \frac{F^2}{r^2}\sin^2\theta,  \label{eq:Antocimetric}
\ea
where $F$ is a function of radius fulfilling suitable boundary conditions in the limit $r\rightarrow\infty$ and $F'$ is its derivative.

Setting $a=0$ and $\beta=0$ in the metric of Ref.~\cite{Jmetric} and comparing the respective metric elements, I arrive at the equations
\ba
F(r) &=& r^2\left[r^2+f(r)\right], \nn \\
F(r) &=& -\left[ 9Mr^5(r-2M)^2 A_1(r)^2 + \sqrt{3r^{10}(r-2M)^3 \left[ 27M^2r-54M^3-r^5A_1(r)^2 \right]} \right]^{-\frac{1}{3}} \nn \\
&& \times \frac{1}{ 3(r-2M)  } \bigg[ 3^{\frac{2}{3}}r^5(r-2M)A_1(r)^2 + 3^{\frac{1}{3}}\bigg[ 9Mr^5(r-2M)^2A_1(r)^2 \nn \\
&& + \sqrt{3r^{10}(r-2M)^3 \left[ 27M^2r-54M^3-r^5A_1(r)^2 \right]} \bigg]^{\frac{2}{3}} \bigg], \nn \\
F'(r) &=& \pm \sqrt{ \frac{ F(r)[F(r)-2M] }{ r^3(r-2M)A_5(r) } },
\ea
which can be rewritten as complicated relations between the deviation functions $f$, $A_1$, and $A_5$. The deviation function $A_2$ does not occur in the metric elements due to the condition $a=0$.

\section{Approximate Expressions of the Displacement, Diameter, and Asymmetry of the Shadow}
\label{approx}

The diameter, displacement, and asymmetry of a shadow around a black hole (or other compact object) can be determined approximately from fits of a large set of simulated shadows. If the metric possesses a third constant of motion, the locations and shapes of these shadows can be calculated analytically. Otherwise, they have to be calculated numerically.

The diameter $L_{\rm K}$, displacement $D_{\rm K}$, and asymmetry $A_{\rm K}$ of a Kerr black hole are given by the approximate expressions~\cite{J13rings}
\ba
L_{\rm K}(\theta,a) &\approx& L_1^{\rm K} + L_2^{\rm K}\cos\left(n_1^{\rm K} \theta+\varphi^{\rm K}\right),
\label{eq:L_Kerrfit} \\
D_{\rm K}(\theta,a) &\approx& D_1^{\rm K} \sin\left(n_2^{\rm K} \theta\right),
\label{eq:D_Kerrfit} \\
A_{\rm K}(\theta,a) &\approx& \frac{1}{A_1^{\rm K} \theta^{n_3^{\rm K}} + A_2^{\rm K} \theta^{n_4^{\rm K}}},
\label{eq:A_Kerrfit}
\ea
where
\ba
L_1^{\rm K} &=& 10.3907 - 0.520134a^{2.19644}-0.109423a^{11.5509}, \nn \\
L_2^{\rm K} &=& 0.141644a^{2.06874}+0.077576a^{2.80628}-0.305764a^{4.44562} \nn \\
&& +0.603316a^{6.58314}-0.801615a^{9.97291} +0.892182a^{15.1866} \nn \\
&& -0.508699a^{18.8103}, \nn \\
n_1^{\rm K} &=& -2.00026-0.0424525a^{6.43112}+0.0503142a^{36.5515}+0.118102a^{153.450}, \nn \\
\varphi^{\rm K} &=& 3.14246+0.0970934a^{5.40872}+0.0964881a^{13.3813}-0.0990121a^{196.868}, \nn \\
D_1^{\rm K} &=& 2a + 0.188324a^{3.29029}+0.153158a^{10.4479}+0.0982701a^{48.1825} \nn \\
&& +0.0500847a^{328.089}, \nn \\
n_2^{\rm K} &=& 1-0.0149883a^{0.656102}+0.0188647a^{0.729558}+0.0370128a^{2.49086} \nn \\
&& +0.147758a^{30.2625} -0.163190a^{31.1614} -0.0148721a^{259.142}, \nn \\
A_1^{\rm K} &=& -\frac{3.29826}{(1-a)^{0.0708700}}+\frac{4.06676}{a^{1.96447}}+\frac{1.69824}{a^{2.06984}}+0.535821a^{58.9636} \nn \\
&& +0.801151a^{495.375}, \nn \\
n_3^{\rm K} &=& -2.55018+0.679338(1-a)^{0.306884}+0.263229a^{1.12755} \nn \\
&& +0.113085a^{2.87034}-0.0657765a^{19.0048} -0.0497249a^{60.8532}, \nn \\
A_2^{\rm K} &=& \frac{0.0482563}{(1-a)^{0.291648}}+\frac{1.17042}{a^{1.99594}}-0.259672a^{7.55857}-0.306290a^{55.7992} \nn \\
&& -0.323404a^{475.326}, \nn \\
n_4^{\rm K} &=& 0.235657+1.39539(1-a)^{0.755126}+1.33468a^{1.12230}-0.842482a^{1.85033} \nn \\
&& +0.0962717a^{30.9888} +0.140634a^{119.303} +0.127251a^{506.431}.
\ea

These fits are valid for values of the spin $0r_g\leq a\leq0.999r_g$. The fit formula of the shadow diameter is accurate to $<0.09\%$ for spin values $0r_g \leq a\leq0.997r_g$ and to $<0.16\%$ for spins $0r_g\leq a\leq0.999r_g$. The fit formula of the displacement is accurate to $<1.1\%$ for spin values $0r_g\leq a\leq0.97r_g$, to $<2.4\%$ for spins $0r_g\leq a\leq0.98r+g$, and to $<3.7\%$ for spins $0r_g\leq a\leq0.999r_g$. For values of the asymmetry $A\geq0.01r_g$, the asymmetry fit is accurate to $<1\%$ for spin values $0r_g\leq a\leq0.6r_g$, to $<4.8\%$ for spins $0r_g\leq a\leq0.98r_g$, and to $<19.4\%$ for spins $0r_g\leq a\leq0.999r_g$. The largest uncertainties in the fit occur at low inclinations, and the fit is accurate at all spin values to $<2.5\%$ for inclinations $\theta\geq22.4^\circ$. Values of the ring asymmetry $A<0.01r_g$, which occur only at spin values $a\sim0$ and inclinations $\theta\sim0$, were partly affected by numerical uncertainty and, therefore, neglected in the fit. Since these values of the asymmetry are very small, this affects the fit only marginally. The asymmetry can also be fitted with an ansatz of the form $A_0\sin^n \theta$ as in Refs.~\cite{PaperII,Chan13}. Such a fit, however, introduces comparatively large errors at high spins, where the asymmetry deviates significantly from a sinosoidal form~\cite{J13rings}.

For the shadows around the compact objects described by the quasi-Kerr metric~\cite{GB06}, the displacement $D_\epsilon$ and asymmetry $A_\epsilon$ are given by the expressions~\cite{PaperII}
\ba
D_\epsilon &\approx& 2a \sin \theta(1-0.41\epsilon\sin^2 \theta), \label{eq:QKdisp} \\
A_\epsilon &\approx& \left[0.84\epsilon +0.36\left(\frac{a}{r_g}\right)^3\right]\sin^{3/2} \theta, \label{eq:QKasym}
\ea
which are valid for $0r_g \leq a \leq0.4r_g$ and $0\leq \epsilon \leq 0.5$.

The diameter $L_\alpha$, displacement $D_\alpha$, and asymmetry $A_\alpha$ of a Kerr-like black hole described by the metric of Ref.~\cite{Jmetric} as a function of the parameters $\alpha_{13}$ and $\alpha_{22}$ are given by the approximate expressions~\cite{J13rings}
\ba
L_\alpha(\theta,a,\alpha_{13},\alpha_{22}) &\approx& L_1^{\rm K}+L_1^\alpha \alpha_{13}+L_2^\alpha \alpha_{13}^2 +L_3^\alpha \alpha_{22} +L_4^\alpha \alpha_{22}^2 \nn \\
&& +L_2^{\rm K}(1+L_5^\alpha \alpha_{13}) (1+L_6^\alpha \alpha_{22}) \cos\left(n_1^{\rm K}\theta+\varphi^{\rm K}\right),
\label{eq:L_fit1} \\
D_\alpha(\theta,a,\alpha_{13},\alpha_{22}) &\approx& D_1^{\rm K} \left(1+D_1^\alpha \alpha_{13}+D_2^\alpha \alpha_{13}^2\right) \left(1+D_3^\alpha \alpha_{22}+D_4^\alpha \alpha_{22}^2\right) \nn \\
&& \times \sin\left[ n_2^{\rm K}(1+n_1^\alpha \alpha_{13}+n_2^\alpha \alpha_{13}^2) (1+n_3^\alpha \alpha_{22}+n_4^\alpha \alpha_{22}^2)\theta \right],
\label{eq:D_fit1} \\
A_\alpha(\theta,a,\alpha_{13},\alpha_{22}) &\approx& (1+A_1^\alpha \alpha_{13}+A_2^\alpha \alpha_{13}^2+A_3^\alpha \alpha_{13}^3) (1+A_4^\alpha \alpha_{22}+A_5^\alpha \alpha_{22}^2+A_6^\alpha \alpha_{22}^3) \nn \\
&& \times \left[ A_1^{\rm K}(1+A_7^\alpha \alpha_{13})(1+A_8^\alpha \alpha_{22}) \theta^{n_3^\alpha} \right. \nn \\
&& \left. + A_2^{\rm K}(1+A_9^\alpha \alpha_{13})(1+A_{10}^\alpha \alpha_{22}) \theta^{n_4^\alpha}\right]^{-1}, \nn \\
\label{eq:A_fit1}
\ea
where
\ba
L_1^\alpha  &\equiv& 0.390575+0.289090a^{3.03136}+0.351485a^{26.0507}, \nn \\
L_2^\alpha  &\equiv& -0.0210469-0.0704973a^{3.96055}-0.156894a^{30.0636}, \nn \\
L_3^\alpha  &\equiv& -0.0771662a^{3.30096}-0.112522a^{19.3585}, \nn \\
L_4^\alpha  &\equiv& -0.0232177a^{2.34745}-0.0474855a^{498.043}, \nn \\
L_5^\alpha  &\equiv& -0.0910695+0.597861a^{34.2332}, \nn \\
L_6^\alpha  &\equiv& -0.665419a^{1.47464}, \nn \\
D_1^\alpha  &\equiv& -0.0471772-0.0714876 a^{4.67811}-2.80254 a^{507.201}, \nn \\
D_2^\alpha  &\equiv& 0.0100617+0.197316 a^{77.0341}+0.173694a^{194.820}, \nn \\
D_3^\alpha  &\equiv& 0.175643+0.107683 a^{6.64090}-0.309073 a^{50.4534}, \nn \\
D_4^\alpha  &\equiv& 0.0144201+0.323182a^{30.8217}-0.377931a^{232.160}, \nn \\
n_1^\alpha  &\equiv& -0.0617871-0.151930a^{37.0587}+2.21857 a^{131.299}, \nn \\
n_2^\alpha  &\equiv& 0.00753226-0.00323579 a^{17.8087}+1.40633 a^{97.0069}, \nn \\
n_3^\alpha  &\equiv& 0.0180080-0.0970060a^{25.9451}+0.439863 a^{104.732}, \nn \\
n_4^\alpha  &\equiv& -0.0183142-0.228196a^{23.7779}+0.0583285 a^{96.0575}, \nn \\
A_1^\alpha  &\equiv& 0.0403176-0.458487a^{3.74974}, \nn \\
A_2^\alpha  &\equiv& -0.0649604+0.0847185a^{1.56286}, \nn \\
A_3^\alpha  &\equiv& 0.00650092-0.00620249a^{116.004}, \nn \\
A_4^\alpha  &\equiv& 1.24363+0.211407a^{20.7016}, \nn \\
A_5^\alpha  &\equiv& 0.281333-0.00508372a^{2.06092}, \nn \\
A_6^\alpha  &\equiv& 0.0287109-0.266727a^{3.89586}, \nn \\
A_7^\alpha  &\equiv& 0.286452+0.640543a^{22.2178}, \nn \\
A_8^\alpha  &\equiv& 0.014822-0.407110a^{6.69441}, \nn \\
A_9^\alpha  &\equiv& 0.274555-0.722911a^{3.25899}, \nn \\
A_{10}^\alpha  &\equiv& 0.208396+0.799466a^{22.7461}.
\ea

These fits are valid for values of the spin $0r_g \leq a\leq0.998r_g$ and of the deviation parameters $-1 \leq \alpha_{13},~\alpha_{22} \leq 2$. The fit of the diameter is accurate to $<7.5\%$ for spins $0r_g\leq a\leq0.95r_g$ and to $<28.5\%$ for spins $0r_g \leq a\leq0.998r_g$. The fit of the displacement is accurate to $< 13.3\%$ in spin range $0.1r_g \leq a \leq 0.85r_g$ with an average accuracy of $2\%$. The accuracy is significantly smaller at high and very low spins and the error can exceed $100\%$ in some cases. Finally, the fit of the asymmetry has an average accuracy of $<4.6\%$, but it can deviate from the simulated data set by factors of order unity, especially at very high spins~\cite{J13rings}.

The diameter $L_\beta$, displacement $D_\beta$, and asymmetry $A_\beta$ of a Kerr-like black hole described by the metric of Ref.~\cite{Jmetric} as a function of the parameters $\beta$ are given by the approximate expressions
\ba
L_\beta(i,a,\beta) &\approx& L_1^{\rm K}+L_1^\beta \beta+L_2^\beta \beta^2 +L_2^{\rm K}(1+L_3^\beta \beta) \cos\left(n_1^{\rm K}i+\varphi^{\rm K}\right),
\label{eq:L_fit2} \\
D_\beta(i,a,\beta) &\approx& D_1^{\rm K} \left(1+D_1^\beta \beta+D_2^\beta \beta^2\right) \sin\left( n_2^{\rm K} i \right),
\label{eq:D_fit2} \\
A_\beta(i,a,\beta) &\approx& (1+A_1^\beta \beta +A_2^\beta \beta^2+A_3^\beta \beta^3) \nn \\
&& \times \left[ A_1^{\rm K}(1+A_4^\beta \beta) i^{n_3^{\rm K}} + A_2^{\rm K}(1+A_5^\beta \beta) i^{n_4^{\rm K}}\right]^{-1}, \nn \\
\label{eq:A_fit2}
\ea
where
\ba
L_1^\beta  &\equiv& -1.74543-0.119076 a^{2.51927}-0.0721428 a^{12.8512}, \nn \\
L_2^\beta  &\equiv& -0.189261+0.00796316 a^{0.611402}, \nn \\
L_3^\beta  &\equiv& 0.278320-0.298400 a^{24.0068}, \nn \\
D_1^\beta  &\equiv& 0.176305-0.0462988 a^{0.258073}, \nn \\
D_2^\beta  &\equiv& 0.411830-0.390541 a^{0.0106134}, \nn \\
A_1^\beta  &\equiv& 0.0756928-1.14075 a^{1.72148}, \nn \\
A_2^\beta  &\equiv& 10.0124-11.2256 a^{0.121151}, \nn \\
A_3^\beta  &\equiv& 0.294512-1.18541 a^{1.54349}, \nn \\
A_4^\beta  &\equiv& 0.831976-2.61358 a^{0.379280}, \nn \\
A_5^\beta  &\equiv& -0.902760-10.2187 a^{5.01912}.
\ea

These fits are valid for values of the spin $0r_g \leq a\leq0.998r_g$ and of the deviation parameter $-2.8 \leq \beta \leq 1$ as long as the condition in Eq.~(\ref{eq:lowerbounds1}) is fulfilled. The fit of the diameter is accurate to $<5.3\%$. The fit of the displacement has an average accuracy of $<3\%$, while the fit of the asymmetry has an average accuracy of $<16\%$. The accuracy is smallest at high and very low spins and the error can exceed $100\%$ in some cases, especially for values of the parameter $\beta$ close to the upper bound defined in Eq.~(\ref{eq:KNhor}).

\end{appendix}

\end{document}